\newif\ifDissLarge
\newif\ifDissDim
\newif\ifDissHead

\DissLargetrue  
\DissDimtrue     
\DissHeadtrue    

\ifDissLarge
  \documentclass[12pt,twoside,a4paper]{report}
  \IfFileExists{a4.cls}{\usepackage{a4}}{\relax}
\else
  \documentclass[11pt,twoside,a4paper]{report}
  \IfFileExists{a4.cls}{\usepackage{a4}}{\relax}
\fi

\newcommand{\MaquaHeadings}
{
  \ifDissHead  
    \pagestyle{fancyplain}
    \lhead[\fancyplain{}{\bfseries\thepage}]
          {\fancyplain{}{\bfseries\sffamily\let\uppercase\relax\leftmark}}
    \rhead[\fancyplain{}{\bfseries\sffamily\let\uppercase\relax\rightmark}]
          {\fancyplain{}{\bfseries\thepage}}
    \cfoot{\fancyplain{\bfseries\thepage}{}}
  \else
    \pagestyle{headings}
  \fi
}

\ifDissDim
   \IfFileExists{typearea.sty}{\relax}{\DissDimfalse}
\fi
\ifDissDim
  \usepackage[a4paper,headinclude,DIV14,BCOR5mm]{typearea}
\else
\fi

\ifDissHead
   \IfFileExists{fancyheadings.sty}{\relax}{\DissHeadfalse}
\fi
\ifDissHead
  \usepackage{fancyheadings}
\fi
\MaquaHeadings

\usepackage{latexsym}
\usepackage[dvips]{graphicx}
\usepackage{bbm}     
\usepackage{ifthen}                   
\usepackage{exscale}                  
\usepackage[intlimits]{amsmath}       
\usepackage{amsfonts}
\usepackage{amssymb,amscd}
\usepackage[dvips]{epsfig}                   
\usepackage{array}
\usepackage{wrapfig}
\usepackage{graphics}

\newcommand{\tr}{\mbox{tr}}

\newcommand{\dslash}{\partial \hskip -0.5em /}

\newcommand{\Dslash}{D \hskip -0.7em /}

\newcommand{\pslash}{p \hskip -0.5em /}
\newcommand{\qslash}{q \hskip -0.5em /}
\newcommand{\beqa}{\begin{eqnarray}}
\newcommand{\eeqa}{\end{eqnarray}}
\newcommand{\beq}{\begin{equation}}
\newcommand{\eeq}{\end{equation}}

\newcommand{\cZ}{\mathcal{Z}}

\newcommand{\cL}{\mathcal{L}}
\newcommand{\cS}{\mathcal{S}}
\newcommand{\cV}{\mathcal{V}}

\begin{document} 
\thispagestyle{empty}
{\small
\rightline{http://xxx.lanl.gov/abs/hep-ph/0304220}}

\vspace{0cm}

\begin{center}
\begin{large}
{\bf Non-perturbative Propagators, Running Coupling and\\
Dynamical Mass Generation in Ghost-Antighost \\ Symmetric Gauges in QCD}\\
\end{large}
\vspace{0.5cm}
{ 
Christian\ S.\ Fischer
\\}
\vspace{0.2cm}
{\it Institute for Theoretical Physics, T\"ubingen University \\
Auf der Morgenstelle 14, D-72076 T\"ubingen, Germany
}
\end{center}
\vspace{0.3cm}
\centerline{\large Abstract}
\normalsize
\noindent
We present approximate non-perturbative solutions for the propagators
as well as the running coupling of QCD. We solve a 
coupled system of renormalised, truncated Dyson--Schwinger equations 
for the ghost, gluon and quark propagators in flat Euclidean space-time. 
We employ {\it ans\"atze} for the dressed vertices such that the running 
coupling and the quark mass function are 
independent of the renormalisation point. At large momenta we obtain the 
correct one-loop anomalous dimensions for all propagators. Our solutions 
are in good agreement with the results of recent lattice calculations.

In the Yang-Mills sector of Landau gauge QCD we find a weakly vanishing 
gluon propagator in the infrared and a ghost propagator more singular than 
a simple pole. This is in accordance with Zwanziger's horizon condition 
and the Kugo-Ojima confinement criterion. The running coupling possesses an 
infrared fixed point at $\alpha(0)=8.92/N_c$. To investigate the influence 
of boundary conditions on the propagators we solved the ghost and gluon 
DSEs also on a four-torus. Our results show typical finite volume effects 
but are still close to the continuum solutions for sufficiently large volumes. 

In general ghost-antighost symmetric gauges we study the infrared behaviour
of the ghost and gluon propagators. No power-like solutions exist when 
replacing all dressed verticed with bare ones. The results of the Landau 
gauge limit are recovered from a different direction in gauge parameter space.

For the quark propagator we find dynamically generated quark masses that 
agree well with phenomenological values and corresponding results from
lattice calculations. The effects of unquenching the system are found to be 
small. In particular the infrared behaviour of the ghost and gluon dressing 
functions found in pure Yang-Mills theory is almost unchanged as long as 
the number of light flavors is smaller than four. 

%
\vspace{0.3cm}

\thispagestyle{empty}

\noindent
{\bf \Large Note added to the eprint-version\\}

This PhD-thesis has been submitted to the University of T\"ubingen
in November 2002 and has been accepted in March 2003. Chapters 3 and 5
of this thesis are based on refs.[83,84] (hep-ph/0202195, hep-ph/0202202), 
however some material has been reordered and more detailed explanations 
have been added. The contents of chapter 6 are published in 
ref.[140] (hep-th/0301094). Chapter 4 represents the research status 
of November 2002. The contents of this chapter together with recent 
results on ghost-antighost symmetric gauges can be found in 
ref.[109] (hep-th/0304134).

\thispagestyle{empty}

\begin{titlepage}
\begin{center}
\vspace*{1cm}
{\bf \LARGE Non-perturbative Propagators, Running} 

\vspace*{3mm}
{\bf \LARGE Coupling and Dynamical Mass Generation }

\vspace*{3mm}
{\bf \LARGE in Ghost-Antighost Symmetric Gauges in }

\vspace*{3mm}
{\bf \LARGE QCD}

\vspace*{3cm}
{\bf \large Dissertation}\\
\vspace*{2cm}
zur Erlangung des Grades eines\\
Doktors der Naturwissenschaften\\
\vspace*{2cm}
der Fakult\"at f\"ur Mathematik und Physik\\
der Eberhard-Karls-Universit\"at zu T\"ubingen\\
\vspace*{2cm}
vorgelegt von\\
{\bf Christian S. Fischer}\\
aus Illertissen\\
\vspace*{0.5cm}
2003\\

\end{center}
\end{titlepage}

\pagenumbering{roman}
\tableofcontents
\chapter{Introduction}
\pagenumbering{arabic}

Quantum Chromo Dynamics (QCD) is the quantum field theory which describes the 
{\it strong interaction} of the fundamental building blocks of matter, 
the {\it quarks} and {\it gluons} 
\cite{Itzykson:1980rh,Rivers:1987hi,Cheng:1985bj,Weinberg:1995mt,Peskin:1995ev,Muta:1998vi}.
In contrast to Abelian gauge theories like Quantum Electro Dynamics (QED), the  
non-Abelian nature of the gauge symmetry of QCD not only induces interactions
between quarks and gluons but also among gluons themselves. 
This last effect is expected to generate the phenomenon of {\it confinement}, {\it i.e.}
the permanent inclusion of all colour charges in colour neutral objects, the {\it hadrons}.

{\it 'Quantum chromo dynamics is a Lagrangian field theory in 
search of a solution.'} This statement, quoted from the classical review
of Marciano and Pagels on QCD \cite{Marciano:1978su}, has not lost  
its relevance since it has been written down in 1977. Although in the meantime a lot of 
progress has been made it is still not clear how the plethora of observed bound state objects, 
the hadrons, can arise from the fundamental quark and gluon fields of QCD. 
In the last thirty years a lot of different strategies have been employed to explore 
both the large and small momentum properties of hadrons. The physical phenomena encountered
at large momentum transfers are very well described by perturbation theory.
{\it Asymptotic freedom} means that the interaction strength of QCD 
tends to zero at small distances. High energy probes therefore picture hadrons as
quark and gluon lumps with definite quantum numbers described by so called structure functions.
This picture, however, starts to break down at energies around 1-2 GeV and is surely inadequate 
at length scales corresponding to the size of the nucleon. At such scales the strong interaction 
is strong enough to invalidate perturbation theory and one has to employ completely different 
methods to deal with what is called {\it Strong QCD}. 

There are two phenomena of QCD which are important for this work:
the mechanism of {\it confinement} and that of {\it dynamical chiral symmetry breaking}, 
{\it i.e.} the generation 
of quark masses via interactions. Neither of these phenomena can be accounted for in perturbation theory,
thus they are genuine effects of Strong QCD. 
Interestingly, both phenomena appear to be
connected. From finite temperature studies of QCD we infer that both effects seem to disappear at
roughly the same temperature but the reasons for this are yet unclear \cite{Karsch:2001vs}.

The framework chosen in this work to investigate the small momentum regime of QCD 
are the Dyson-Schwinger equations of motion for correlation functions of the fields. 
Certainly a great step forward in understanding QCD would be the detailed 
knowledge of the basic correlation functions, the propagators.  
Information on certain confinement mechanisms is encoded in these two-point functions. 
Furthermore the mechanism of dynamical chiral symmetry breaking can be studied directly in 
the Dyson-Schwinger equation for the quark propagator, which is the gap equation of QCD.
Besides being related to the fundamental issues of QCD, the quark and gluon propagators 
are vital ingredients for phenomenological models describing low and medium energy hadron physics.
Bound state calculations based on the Bethe-Salpeter equations for mesons or on the Faddeev equations for
baryons (for reviews see \cite{Roberts:2000aa,Alkofer:2000wg})
might one day be capable to bridge the gap between the fundamental theory, QCD, and phenomenology.

Throughout this work we will compare the results obtained from Dyson-Schwinger equations with those of lattice 
Monte Carlo simulations (see {\it e.g.} \cite{Creutz:2000ji}). 
Combining the strengths and weaknesses of both approaches allows
one to make quite definite statements for the propagators in a large momentum range.
Lattice Monte Carlo simulations include all non-perturbative physics of Yang-Mills theories and 
are therefore the only ab initio calculation method available so far. However, the simulations
suffer from limitations at small momenta due to finite volume effects. One has to rely on 
extrapolation methods to obtain the infinite volume limit. Furthermore calculations
including quarks are subtle on the lattice, as it is very difficult to implement fermions with
small bare masses. On the other hand Dyson-Schwinger equations can be solved analytically in the infrared and are
the proper tool to assess the effects of dynamical quarks. However,
in order to obtain a closed system of equations one has to employ {\it ans\"atze} for 
higher correlations functions. The quality of these truncations can be ascertained by
comparison with lattice results. 

We will recall in the next chapter some basic aspects of Strong QCD.
Based on the symmetries of the (generalised) QCD Lagrangian certain aspects of
dynamical chiral symmetry breaking as well as confinement are reviewed. In particular
we will recall how information on the so called {\it Kugo-Ojima confinement mechanism},
the notion of {\it positivity} and Zwanziger's {\it horizon condition} are encoded 
in the propagators of QCD. From the generalised QCD Lagrangian of Baulieu and Thierry-Mieg
\cite{Baulieu:1982sb}
we will derive a modified Dyson-Schwinger equation for the ghost propagator, which together
with the corresponding equations for the gluon and quark propagators are the basic tools of our investigation.

The third chapter is devoted to Landau gauge, where a novel truncation scheme is introduced
that allows to solve the coupled ghost and gluon Dyson-Schwinger equations of pure Yang-Mills theory. Contrary to 
earlier attempts this is done without any angular approximations in the loop integrals of the
equations. Besides the numerical solutions for general momenta we obtain analytical 
results for the ultraviolet and infrared region of momentum. We are thus able to show
that the anomalous dimensions of one-loop perturbation theory are reproduced by our
solutions for the full ghost and gluon propagators. For small momenta the ghost and gluon dressing 
functions follow power laws, which are in accordance with the Kugo-Ojima confinement
criterion as well as Zwanziger's horizon condition. We are able to show that only one out of
two infrared solutions already found in earlier investigations is connected to the numerical solutions for
general momenta. The resulting running coupling possesses an infrared fixed point. Our results
for the propagators are in nice agreement to recently obtained lattice calculations.

Landau gauge is a special gauge in the sense that it allows for surprisingly simple vertex 
{\it ans\"atze} in the truncation of the Dyson-Schwinger equations. In chapter four we will explore whether a
truncation employing bare vertices can be extended to general gauges. 
We solve the corresponding Dyson-Schwinger equations analytically in the infrared employing power laws for the 
dressing functions. The main results of Landau gauge, an infrared vanishing gluon 
propagator and an infrared diverging ghost turn out to be persistent for the class of 
linear covariant gauges. However, no power law solutions for general ghost-antighost
symmetric gauges can be found. Numerical solutions for the limit of Landau gauge from
different directions in gauge parameter space turn out to be stable.

In chapter five we change the base manifold and investigate the Dyson-Schwinger equations of Landau gauge
Yang-Mills theory on a four-torus. There are three ideas motivating such a change.
The {\it first} idea stems from the observation that torus calculations share the finite volume problem with  
lattice Monte Carlo simulations. However, as the continuum limit is known in the framework of 
Dyson-Schwinger equations,
one is able to judge extrapolation methods on a torus. Our solution
for the gluon propagator on a torus resembles closely the one found in the continuum 
whereas the ghost dressing function deviates in the very infrared. The {\it second} idea
is a technical one. Although the treatment is inverted in this work, 
we first have been able to obtain results without angular approximations on the torus
and only subsequently in the continuum. This is due to the torus acting as an effective regulator 
in the infrared. The {\it third} idea, which remains for future work, is to include
topological obstructions like twisted boundary conditions in Dyson-Schwinger equations on a torus.  

The last chapter of this work focuses on the quark propagator. We present solutions for the
quenched system of quark Dyson-Schwinger equations using the results of chapter three as input, and compare to
lattice calculations. We then go one step further and solve the unquenched coupled 
system of equations for the ghost, gluon and quark propagators of QCD. This is done without
any angular approximations. Our results again reproduce the anomalous dimensions of one-loop
perturbation theory in the ultraviolet. The effects of unquenching the system, {\it i.e.} including
the quark loop in the gluon equation, is found to be small if the number of
light flavours is small, $N_f^{light}<4$. In particular the 
infrared behaviour of the ghost and gluon dressing functions remain unchanged. The quark masses
generated by dynamical chiral symmetry breaking are found to be close to phenomenological values.

\chapter{Aspects of Strong QCD}
 
 
\section{The generating functional of QCD \label{gen-func}}

Working in Euclidean space-time\footnote{We will adopt Euclidean metric throughout this work.
A justification of this choice will be given in subsection \ref{positivity}.}
the generating functional of the quantum field
theory of quarks and gluons is given by
\beqa
Z[J,\eta,\bar{\eta}] &=& 
\int {\cal D} [A \bar{\Psi} \Psi] 
\nonumber\\
 &&\hspace*{-0.3cm}\exp\left\{-\int d^4x \left(\bar{\Psi} \left( -\Dslash + m\right) \Psi + 
\frac{1}{4} F_{\mu \nu}^2\right) + \int d^4x 
\left(A^a_\mu J^a_\mu+\bar{\eta}\Psi+\bar{\Psi}\eta \right)\right\}, \nonumber\\
\label{genfunc}
\eeqa 
where we have introduced the Grassmann valued sources $\bar{\eta}$ and $\eta$ for the  
quark fields $\Psi$ and $\bar{\Psi}$ and the source $J^a_\mu$ for the gauge field
$A^a_\mu$. Furthermore we used the abbreviation $\Dslash = \gamma_\mu D_\mu$ with
Euclidean $\gamma$-matrices\footnote{We
use hermitian $\gamma$-matrices defined in appendix \ref{gamma}.} 
and the covariant derivative $D_\mu$ given in eq.~(\ref{covder}).
The quark fields are spin-1/2 fermions which transform according to a 
fundamental representation of the gauge group $SU(N_c)$. The central 
focus of this work is QCD, {\it i.e.} the gauge group $SU(3)$. However in the course
of our investigations we will run across some results, that are valid for general
gauge group $SU(N_c)$. 
Some comparisons with lattice calculations will be done for $SU(2)$.

The non-Abelian gluon fields, $A^a_\mu$, transform according to the adjoint representation of the
gauge group. The corresponding field strength tensor and the covariant derivative
in the adjoint representation are given by    
\beqa
F_{\mu \nu}^a &=& \partial_\mu A_\nu^a - \partial_\nu A_\mu^a -g f^{abc} A_\mu^b
A_\nu^c \;, \\
D_\mu^{ab} &=& \partial_\mu \delta^{ab} + g f^{abc} 
A_\mu^c .
\eeqa
Here $g$ is the (unrenormalised) coupling constant of the theory and $f^{abc}$ are the
structure constants of the gauge group. With the help of 
the generators $t^a$ of $SU(N_c)$ we can rewrite the covariant derivative in the
fundamental representation
\beq
D_\mu = \partial_\mu + igA_\mu \;, \label{covder}
\eeq 
with $A_\mu = A_\mu^a t^a$ and $[t^a,t^b] = if^{abc}t^c$.
The Lagrangian ${\cal{L}}=\bar{\Psi} \left( -\Dslash + m\right) \Psi + 
\frac{1}{4} F_{\mu \nu}^2$ of our theory is invariant under local gauge transformations.

One of the most intricate tasks in the quantisation of a field theory is the separation
into physical and non-physical degrees of freedom, which is a prerequisite for the definition
of the physical state space of the theory. The integration over all possible gauge field 
configurations $A$ in the generating functional (\ref{genfunc}) includes the ones that 
are gauge equivalent. Therefore the integration generates an infinite
constant, the volume of the gauge group ${\cal G}$, which has to be absorbed in the 
normalisation. More important, the gauge freedom implies that the quadratic part of the
gauge field Lagrangian has zero eigenvalues and therefore cannot be inverted\footnote{
If one avoids the generating functional and employs canonical quantisation
this problem manifests itself on the level of commutation relations of the fields. 
These are usually fixed at time zero and should then determine the commutators for all
times. However, this cannot be the whole story, since one can always gauge transform
to a field that vanishes at time zero. So one has to remove the freedom of gauge 
transformations here as well \cite{Marciano:1978su}.}. 
This prevents the definition of a perturbative gauge field propagator \cite{Pokorski:1987ed}.

In order to single out one representative configuration from each gauge 
orbit 
\beq
[A^U] := \left\{A^U = UAU^\dagger + UdU^\dagger : U(x) \in SU(N_c) \right\}
\eeq
one has to impose the gauge fixing condition $f^a(A)=0$ on the generating functional. 
This is conveniently done by inserting the identity
\beq
1=\Delta[A] \int {\cal D}g \: \delta(f^a(A)) \label{FP}
\eeq
into the generating functional (\ref{genfunc}) and absorbing the group integration $\int {\cal D}g$ 
in a suitable 
normalisation \cite{Faddeev:1967fc}. We will discuss problems with this gauge fixing prescription
in more detail in subsection \ref{gribov}.

In linear covariant gauges the Faddeev-Popov determinant $\Delta[A]$ reads explicitly
\beq
\Delta[A] = \mbox{Det}\left(-\partial_\mu D_\mu^{ab}\right) 
\eeq
and can be written as a functional integral over two new Grassmann valued fields $c$ and $\bar{c}$.
Furthermore the gauge fixing condition $f^a(A)=\partial_\mu A^\mu - i\lambda B^a=0$
employing the auxiliary field $B$ 
can be represented by a Gaussian integral centred around $i\lambda B^a=0$ (see {\it e.g.}
\cite{Peskin:1995ev} for details). We then have
\beqa
\Delta[A]\delta(f^a(A)) &=& \int {\cal D}[\bar{c},c,B] \exp\left\{-\int d^4x \left( 
- i \partial_\mu \bar{c} D_\mu c + i B^a \partial_\mu A^a_\mu + \frac{\lambda}{2} B^a B^a
\right) \right\} \hspace*{4mm}\nonumber\\
&=& \int {\cal D}[\bar{c},c] \exp\left\{-\int d^4x  \left(
- i \partial_\mu \bar{c} D_\mu c -\frac{(\partial_\mu A_\mu^a)^2}{2 \lambda} \right) \right\}.
\label{gauge-fix}
\eeqa

Introducing the sources $\sigma$ and $\bar{\sigma}$ for the antighost and ghost field
respectively we arrive at the {\it gauge fixed generating functional}
\beqa
Z[J,\sigma,\bar{\sigma},\eta,\bar{\eta}] &=& 
\int {\cal D} [A \bar{\Psi} \Psi c \bar{c}]  \nonumber\\
 &&\exp\left\{-\int d^4x {\cal L}_{eff} + \int d^4x 
\left(A^a_\mu J^a_\mu+\bar{\eta}\Psi+\bar{\Psi}\eta + \bar{\sigma}c + \bar{c}\sigma 
\right) \right\}, \hspace*{4mm} 
\label{genfunc-gf}
\eeqa 
with the effective (unrenormalised) Lagrangian 
\beq
{\cal{L}}_{eff} = {\cal{L}}_M + {\cal{L}}_A + {\cal{L}}_{GF}  
=  \bar{\Psi} \left( -\Dslash + m\right) \Psi + 
\frac{1}{4} F_{\mu \nu}^2 + \frac{\left(\partial_\mu A_\mu \right)^2}
{2 \lambda} 
- i \partial_\mu \bar{c} D_\mu c .
\label{Lagrangian}
\eeq
Note that a factor of $i$ appears in front of the ghost terms as we have used {\it real} ghost and
antighost fields. We will see the importance of this choice in section \ref{conf}.

In effect we have modified the Lagrangian of our theory with a term containing the 
(unphysical) ghost and antighost fields $c$ and $\bar{c}$ which transform according to the
adjoint representation of the gauge group and the gauge fixing part which may 
or may not be written with the help of the Nakanishi-Lautrup auxiliary field $B^a$. 
 
Apart from the convention of real ghost fields
the Lagrangian (\ref{Lagrangian}) is the usual one employed in perturbation theory. It has some
important properties:
\begin{itemize}
\item[(i)] it is of dimension 4,
\item[(ii)] it is Lorenz invariant and globally gauge invariant,
\item[(iii)] it is BRS and anti-BRS invariant\footnote{The explicit definitions of the BRS and 
anti-BRS transformations used in this work are given in subsection \ref{gg-symm}.},
\item[(iv)]  and hermitian.
\end{itemize}
The number of dimensions and Lorenz invariance are certainly dictated by experiment. The further
symmetries, global gauge invariance and BRS symmetry, are discussed in section \ref{symm-qcd}.
Hermiticity is necessary to define the physical S-matrix of the theory. We will say a little more 
about this issue in subsection \ref{Kugo-Ojima}. 

The Lagrangian (\ref{Lagrangian}) arises from a specific gauge fixing procedure, the Faddeev-Popov
method. This, however, is not the only gauge fixing procedure that has been employed so far.
Indeed, as will be discussed in more detail in subsection \ref{gribov}, the Faddeev-Popov method is
not capable to fix the gauge completely. Although it is currently not known to what extend
this poses a problem for strong QCD, it is desirable to develop alternatives which do not suffer
from such a deficiency. Examples for different gauge fixing procedures are 
topological gauge fixing \cite{Baulieu:1998rp,Schaden:1998xw}, where the aim is to represent the
partition function of QCD by a topological invariant, or 
stochastic gauge fixing \cite{Parisi:1981ys,Zwanziger:2002ia}, which employs
a Fokker-Planck equation for the probability distribution in gauge field space\footnote{A pedagogical 
treatment of this topic can be found in \cite{Rivers:1987hi}.}. 

On the other hand, one could reverse arguments and claim the properties (i)-(iv) to be crucial for
the quantum field theory of strong interaction. Without bothering about the details of the gauge fixing procedure 
one could then search for the most general Lagrangian satisfying (i)-(iv). This view has been adopted in
\cite{Baulieu:1982sb,Thierry-Mieg:1985yv}. It has been shown that, omitting topological terms, 
the most general polynomial 
in the fields $A_\mu$, $c$, $\bar{c}$, $\Psi$ and $\bar{\Psi}$ satisfying (i)-(iv) can be written
\beqa
\cal{L} &=& \bar{\Psi} \left( -\Dslash + m\right) \Psi 
+ \frac{1}{4} F_{\mu \nu}^2 + \frac{\left(\partial_\mu A_\mu \right)^2}{2 \lambda} \nonumber\\ 
&&+ \frac{\alpha}{2} \left(1-\frac{\alpha}{2}\right) \frac{\lambda}{2}
(\bar{c}\times c)^2 
- i\frac{\alpha}{2} D_\mu \bar{c} \partial_\mu c
- i\left(1-\frac{\alpha}{2} \right) \partial_\mu \bar{c} D_\mu c .
\label{Lagrangian-gen}
\eeqa
Here the abbreviation $(\bar{c} \times c)^a= gf^{abc} \: \bar{c}^b c^c$ is used.
Again both ghost fields, $\bar{c}$ and $c$, are chosen to be real, which is
necessary here to maintain the hermiticity of the Lagrangian for all values of the gauge 
parameters $\lambda$ and $\alpha$, see {\it e.g.} \cite{Nakanishi:1990qm} and
references therein. 

We easily see, that the new Lagrangian (\ref{Lagrangian-gen}) is a generalisation of the 
Faddeev-Popov Lagrangian (\ref{Lagrangian}) with a new, second gauge parameter $\alpha$. 
This gauge parameter controls the symmetry properties of the ghost content of the Lagrangian. 
For the cases $\alpha=0$ and $\alpha=2$ one recovers the usual Faddeev--Popov Lagrangian 
(\ref{Lagrangian}) and its mirror image, respectively, where the role of ghost and antighost 
have been interchanged. For the value $\alpha=1$ the Lagrangian (\ref{Lagrangian-gen}) 
is completely symmetric in the ghost and antighost fields. Compared to the Faddeev-Popov case 
the four ghost interaction is an additional term in the theory. Note that such a term
is {\it e.g.} found in topological gauge fixing scenarios \cite{Schaden:1998xw} or as a 
result of partial gauge fixing in maximal Abelian gauges 
\cite{Quandt:1998rw,Quandt:1998rg,Schaden:2000fv}. 

In ref.~\cite{Baulieu:1982sb} it has been shown that the S-matrix of the theory
(\ref{Lagrangian-gen}) is invariant under
variation of the gauge parameters $\lambda$ and $\alpha$. Therefore gauge invariance
of physical observables is ensured. One-loop calculations confirm in particular the
independence of the first nontrivial coefficient of the $\beta$-function from the gauge parameters.

Furthermore, the existence of a renormalised BRS-algebra
has been proven \cite{Baulieu:1982sb}, thus the theory given by (\ref{Lagrangian-gen}) is
multiplicatively renormalisable.
From one-loop calculations one finds that the Faddeev-Popov values of the gauge parameters, 
$\alpha=0$ and $\alpha=2$, are
fixed points under the renormalisation procedure. The same is true for the ghost-antighost
symmetric case $\alpha=1$. The case of Landau gauge, $\lambda=0$, 
corresponds to a fixed point as well, because the constraint $\partial_\mu A_\mu=0$ is not affected 
by a rescaling of the gluon field.  

The correspondence between the bare Lagrangian (\ref{Lagrangian-gen}) and its renormalised
version including counterterms is given by the following rescaling transformations
\beqa
A_\mu^a &\rightarrow& \sqrt{Z_3}A_\mu^a, \hspace*{1cm} \bar{c}^a c^b \rightarrow \tilde{Z}_3 \bar{c}^a c^b,
\hspace*{1cm} \bar{\Psi}\Psi \rightarrow Z_2 \bar{\Psi}\Psi, \\
g &\rightarrow& Z_g g, \hspace*{2.1cm} 
\alpha \rightarrow Z_\alpha \alpha, \hspace*{1.8cm} \lambda \rightarrow Z_\lambda \lambda, 
\label{rescaling}
\eeqa
where six independent renormalisation constants $Z_3,\tilde{Z}_3,Z_2,Z_g,Z_\alpha$ and 
$Z_\lambda$ have been introduced. 
Furthermore five additional (vertex-) renormalisation constants are related to these via 
Slavnov--Taylor identities,
\beq
Z_1 = Z_g Z_3^{3/2}, \hspace{0.4cm} \tilde{Z}_1=Z_g \tilde{Z}_3 Z_3^{1/2}, \hspace{0.4cm} 
Z_{1F} = Z_g Z_3^{1/2} Z_2, \hspace{0.4cm}
 Z_4=Z_g^2 Z_3^2, \hspace{0.4cm} \tilde{Z}_4=Z_g^2 \tilde{Z}_3^2 . \hspace{0.4cm}
\eeq

Most of the calculations performed in this work are done in Landau gauge. By partial integration
it is easy to see that in Landau gauge the additional gauge parameter $\alpha$ drops out of the 
Lagrangian (\ref{Lagrangian-gen}) due to the condition $\partial_\mu A_\mu=0$. Our results are therefore
completely insensitive to a decision between the usual Faddeev-Popov version of QCD or the generalised 
version. The only exception occurs in chapter \ref{symm-chap}, where we investigate
the infrared behaviour of the ghost and gluon propagators for general values of the gauge parameters
$\lambda$ and $\alpha$.    

\section{Symmetries of QCD \label{symm-qcd}}

In the following we will recall some of the symmetries of the generalised Lagrangian
(\ref{Lagrangian-gen}). While ghost number symmetry and BRS symmetry are
vital ingredients for the Kugo-Ojima confinement scenario discussed in the next
section, chiral symmetry and its dynamical breaking will be important in chapter \ref{quark},
when we solve the Dyson-Schwinger equation of the quark propagator.
All considerations in this section are independent of
the specific value of the gauge parameters $\lambda$ and $\alpha$ of our
general Lagrangian (\ref{Lagrangian-gen}).

\subsection{Chiral symmetry}

Let us first discuss the quark sector of the Lagrangian (\ref{Lagrangian-gen}). The
approximate chiral symmetry of the quark terms in the Lagrangian proves to be very fruitful
to generate low energy expansions of QCD (for reviews see {\it e.g.}
\cite{Leutwyler:1994fi,Ecker:1995gg,Manohar:1996cq,Ecker:1998ai}).
A lot of qualitative and quantitative properties of
hadrons have been inferred from the concept of approximate chiral symmetry. Effective models
like the global colour model \cite{Tandy:1997qf} or the NJL model 
(\cite{Nambu:1961tp,Nambu:1961fr},
see also \cite{Alkofer:1995mv} for a review) try to capture the basic properties of QCD
by approximating the gluon sector with a simple effective interaction
but maintaining the chiral symmetry aspects of the quarks. 

From eq.~(\ref{Lagrangian-gen}) we have the quark part of the QCD Lagrangian
\beq
{\cal{L}_M} = \bar{\Psi} \left( -\Dslash + m\right) \Psi \;, 
\eeq
where $m$ is a diagonal matrix containing the masses of six different flavours of quarks,
generated in the electroweak sector of the standard model (see {\it e.g.}
\cite{Hollik:1999uy}). The chiral limit, $m=0$, is appropriate for the three light quarks
up, down and strange, which are considered only in the following. In the chiral case
the quark fields can be split in left- and right-handed Weyl-spinors
\beq
\Psi_R = \frac{1+\gamma_5}{2} \Psi \;, \hspace{1cm} \Psi_L = \frac{1-\gamma_5}{2} \Psi. 
\eeq
The resulting Lagrangian is symmetric under the global unitary transformation 
SU(3) $\times$ SU(3) $\times$ U(1) $\times$ U(1), which generates the currents
\beqa
j_\mu   &=& \bar{\Psi} \gamma_\mu \Psi \;, \\ 
j_\mu^5 &=& \bar{\Psi} \gamma_\mu \gamma_5 \Psi \;, \label{axial}\\ 
j_\mu^f &=& \bar{\Psi} \gamma_\mu t^f \Psi \;, \\ 
j_\mu^{5f} &=& \bar{\Psi} \gamma_\mu \gamma_5 t^f \Psi \;,  
\eeqa
where $t^f=\frac{\lambda^f}{2}$ denote the generators of SU(3) flavour transformations given
by the Gell-Mann matrices $\lambda^f$.
These currents are conserved on the classical level of the theory. Quantum corrections,
however, spoil the conservation law for the axial current, eq.~(\ref{axial}).
This effect is known as Adler-Bell-Jackiw anomaly and has the observable consequence of allowing
the otherwise forbidden decay of the uncharged pion $\pi^0$ into two photons. 

Under the presence of a non-vanishing mass matrix $m$ and including the effects of the anomaly we
have the divergences of these currents given by
\beqa
\partial^\mu j_\mu   &=& 0 \label{baryon-current} \;,\\ 
\partial^\mu j_\mu^5 &=& 2 i \bar{\Psi} m \gamma_5 \Psi - \frac{g^2}{16 \pi^2}  
\epsilon^{\mu \nu \sigma \rho} F_{\mu \nu}^a F_{\sigma \rho}^a  \;,\\ 
\partial^\mu j_\mu^f &=& \bar{\Psi} \left[ t^f, m \right] \Psi \label{vector} \;,\\ 
\partial^\mu j_\mu^{5f} &=& \bar{\Psi} \left\{ t^f, m \right\}\Psi \;. \label{axial-vector} 
\eeqa
Thus only one current, eq.~(\ref{baryon-current}), is conserved
and describes baryon number conservation in strong interaction processes.
The vector current, eq.~(\ref{vector}), is conserved in the case of identical
quark masses and thus describes the approximate flavour symmetry in the light quark
sector of QCD. 

The axial vector current, eq.~(\ref{axial-vector}), is broken if we have a  
non-vanishing quark mass matrix in the Lagrangian of QCD. This situation is 
called {\it explicit chiral symmetry breaking}. Since the current quark masses of at least the 
up and down quark are very small, we still expect approximately degenerate 
parity partners of the lowest lying hadron spectra, if the current masses were the only reason
for broken chiral symmetry. However, such parity partners are not observed in nature. 

The solution of this puzzle is the effect of {\it dynamical chiral
symmetry breaking} described by the Dyson-Schwinger equation for the quark
propagator. We will see in detail in chapter \ref{quark}, how the
strong interaction in the quark equation generates physical quark masses of the order of
several hundred MeV even in the {\it chiral limit} of zero bare masses in the
Lagrangian of our theory.

\subsection{Ghost number symmetry}

The conserved Faddeev-Popov {\it ghost number} $N_{FP}$ is generated by the scale
transformation
\beqa
c^a &\rightarrow& e^{\Theta} c^a \;, \nonumber\\
\bar{c}^a &\rightarrow& e^{-\Theta} \bar{c}^a \;, \label{ghost-trafo}
\eeqa
with a real parameter $\Theta$. Via the Noether-theorem this symmetry leads to a conserved
current and a conserved charge $Q_c$, the {\it FP ghost charge}. The charge leaves all
other fields invariant and acts on the ghost fields as
\beqa
{[iQ_c,c^a(x)]} &=& c^a(x) \;, \nonumber\\
{[iQ_c,\bar{c}^a(x)]} &=& -\bar{c}^a(x) \;. 
\eeqa
The ghost number $N_{FP}$ is then identified with the eigenvalue of the operator $Q_c$
multiplied by $i$. Note that the appearance of the hermitian operator $Q_c$ with purely 
imaginary eigenvalues is perfectly consistent in the presence of indefinite metric, which 
seems to be the case in quantum field theories \cite{Nakanishi:1990qm}. Note further,
that only the scale transformation (\ref{ghost-trafo}) and not the 
corresponding phase transformation is compatible with the choice of real ghost and 
antighost fields. Ghost number conservation and the BRS-symmetry discussed in the next
subsection play an important role in the Kugo-Ojima confinement scenario summarised
in subsection \ref{Kugo-Ojima}.

\subsection{Global gauge symmetry and BRS-symmetry \label{gg-symm}}
 
Although the Lagrangian (\ref{Lagrangian-gen}) is gauge fixed such that local gauge
invariance is not present any more, there are two gauge symmetries
left: The global gauge symmetry and the so called BRS-symmetry which has been
found by Becchi, Rouet and Stora\footnote{At the same time the symmetry has been discovered 
independently by Tyutin, see \cite{Iofa:1976je}.} \cite{Becchi:1976nq}. 

The {\it global gauge transformations} of the gauge field and the quark field are given by 
\beqa
A_\mu &\rightarrow& A_\mu^\prime = e^{i t^a \Lambda^a} A_\mu e^{-i t^a \Lambda^a} \;, \\
\Psi &\rightarrow& \Psi^\prime = e^{i t^a \Lambda^a}\Psi \;,
\eeqa
with space-time independent parameters $\Lambda^a$ and the generators $t^a$ of the gauge group. Although
the global gauge transformation is a symmetry of the Lagrangian it is not clear whether
a corresponding well defined charge exists, {\it i.e.} whether global gauge symmetry
is spontaneously broken or not. 
This will play an important role in the discussion of
the Kugo-Ojima confinement criterion in subsection \ref{Kugo-Ojima}. 
Note that there are
no fundamental reasons why the global gauge symmetry should be unbroken, as Elitzurs
theorem of unbroken gauge symmetry only applies to local transformations \cite{Elitzur:1975im}.

The most efficient way to introduce {\it BRS-symmetry} is by means of the nilpotent 
BRS-operator $s$. The BRS-transformation of the gluon, ghost and quark fields
as well as the auxiliary field $B$ are given by
\beqa
s \Psi &=& -ig t^a c^a \Psi \;, \label{quark-brs}\\
s A_\mu^a &=&  D_\mu^{ab} c^b \;, \\
s c^a &=&  -\frac{g}{2} f^{abc} c^b c^c \;, \\
s \bar{c}^a &=&  i B^a \;, \\
s B^a &=&  0 \;. \label{BRS-trafo} 
\eeqa
Note that the Nakanishi-Lautrup auxiliary field $B$ can be eliminated from the 
BRS-transformations by using its equation of motion, $\partial_\mu A^\mu - i\lambda B^a=0$. 
Note furthermore that the application of the 
BRS-operator $s$ on a field increases the ghost number by $+1$, thus we can assign the value
$N_{FP}=+1$ to the BRS-operator itself.  

The BRS-transformations (\ref{BRS-trafo}) can be seen as (local)
gauge transformations with the ghost field $c(x)$ as parameter. Thus the transformations
describe a global symmetry, since one is not free to treat different space-time points
independently. Similar to global gauge symmetry, it is not clear whether the BRS-symmetry
generates a well defined BRS-charge $Q_B$. Indeed, it has been argued 
\cite{Fujikawa:1983ss,Baulieu:1998rp} that BRS-symmetry is
broken as a consequence of the presence of Gribov copies\footnote{The problem of Gribov copies 
is discussed in some more detail in subsection \ref{gribov}.}.
The Kugo-Ojima confinement scenario, discussed in the next section, assumes a
well defined, {\it i.e.} unbroken, BRS-charge $Q_B$.

With the help of the BRS-operator $s$ the Lagrangian (\ref{Lagrangian-gen}) can be 
written as
\beqa
\cal{L} &=& {\cal{L}}_M + {\cal{L}}_A + {\cal{L}}_{GF} \\
        &=& {\cal{L}}_M + {\cal{L}}_A - i s \left( \partial_\mu \bar{c}^a A_\mu^a
  - \frac{\lambda}{2} \bar{c}^a B^a + i \frac{\lambda \alpha}{4} g f^{abc}  c^a \bar{c}^b \bar{c}^c \right).  
\label{Lagrangian-gen-brs}
\eeqa
Thus the BRS-invariance of the Lagrangian can be inferred from the local gauge invariance
of the part ${\cal{L}}_M + {\cal{L}}_A$ and the nilpotency of the BRS-operator, $s^2=0$. 

Under the assumption of its existence the BRS-charge $Q_B$
together with the ghost charge $Q_c$ constitute a simple algebraic
structure characterised by the following relations
\beqa
\{ Q_B, Q_B \} &=& 2 (Q_B)^2 = 0 \;, \\
{[iQ_c, Q_B ]} &=& Q_B \;, \\
{[iQ_c, Q_c ]} &=& 0 \;. \label{BRS-algebra} 
\eeqa
This algebra is called BRS-algebra in what follows.
 
It is interesting to note that the Lagrangian (\ref{Lagrangian-gen}) is also invariant under
the anti-BRS transformations defined by
\beqa
\bar{s} \Psi &=&  -ig t^a \bar{c}^a \Psi \;, \\
\bar{s} A_\mu^a &=&  D_\mu^{ab} \bar{c}^b \;, \\
\bar{s} \bar{c}^a &=&  -\frac{g}{2} f^{abc} \bar{c}^b \bar{c}^c \;, \\
\bar{s} {c}^a &=&  -i B^a-\frac{g}{2} f^{abc} \bar{c}^b {c}^c \;, \\
\bar{s} B^a &=&  -g f^{abc} \bar{c}^b B^c \;. 
\eeqa
Similar to the BRS-operator $s$ the anti-BRS operator $\bar{s}$ is nilpotent and both operators are 
related by
\beq
s\bar{s} + \bar{s}s = 0
\eeq
It has been argued, however, that although adding structure to the mathematical framework of the theory 
the presence of anti-BRS symmetry has no influence on the physical content
\cite{Nakanishi:1990qm}. Consequently this symmetry will play no role in the following discussions.  


\section{Aspects of confinement \label{conf}}

In this section we will summarise some aspects of confinement. All three topics discussed, 
the Kugo-Ojima confinement scenario, the notion of positivity and Zwanziger's
horizon condition generate testable predictions for the behaviour of the propagators of QCD.

\subsection{The Kugo-Ojima confinement scenario \label{Kugo-Ojima}}

It has been stated already above that one of the most intricate problems in quantum field 
theories is the separation of physical and unphysical degrees of freedom. In QCD this
problem is directly connected with the issue of {\it confinement}, since we are searching
for the mechanism which eliminates the coloured degrees of freedom from the physical
state space $\cV_{phys}$ of the theory, which is supposed to contain the colourless hadronic
states observed in experiment. 

From a theoretical point of view to be able to define a physical S-matrix
between the physical states of the theory three conditions should be 
satisfied \cite{Nakanishi:1990qm}:
\begin{itemize}
\item The Hamiltonian $H$ corresponding to the Lagrangian of the theory should be hermitian.
\item The physical subspace $\cV_{phys}$ of the state space of the theory should be invariant under
time evolution, {\it i.e.} $H \cV_{phys} \subseteq \cV_{phys}$.
\item The physical subspace should be positive semidefinite, {\it i.e.} 
$\langle \Psi|\Psi\rangle \ge 0$, if $|\Psi\rangle \in \cV_{phys}$. 
\end{itemize}
Certainly, the first of these criteria holds, because the general 
Lagrangian (\ref{Lagrangian-gen}) is hermitian due to our choice of real ghost 
fields\footnote{Note that for general values of the
gauge parameters $\lambda$ and $\alpha$ this is not the case in the original version
of the Lagrangian in ref.~\cite{Baulieu:1982sb}, where complex ghost fields have been chosen.} 
$c$ and $\bar{c}$. The second 
criterion suggests the definition of the physical subspace via a conserved charge, which
commutes with the Hamiltonian of the theory and thus guarantees the invariance of the
subspace under time evolution. The third criterion is necessary to allow for the
usual probabilistic interpretation of the quantum theory. In general the complete state space
$\cV$ has indefinite metric and one therefore has to prove explicitly, that the third 
criterion holds.  

To proceed we recall briefly on an intuitive level what is meant by the notion of 
asymptotic states\footnote{See 
\cite{Haag:1992hx} for a mathematical rigorous introduction into the concept of
asymptotic states and the problems related with asymptotic bound states as well as
asymptotic massless particles.}. What is really observed in particle
physics are not fields but particles. Such particles are present long before and after scattering
processes and are described by so called asymptotic states. These states $|\Phi_{as}\rangle$
are created by asymptotic fields $\Phi_{as}$ which are defined by the weak operator limit 
of the corresponding field operators $\Phi$ for large absolute 
times \cite{Lehmann:1955rq}:
\beq
\langle f | \Phi(x) - \Phi_{as}(x) | g \rangle \:\: \longrightarrow \:\: 0 \hspace{1cm} \mbox{if} 
\hspace{0.3cm} x^0 \rightarrow \pm \infty,
\eeq
for any two states $|f \rangle , |g \rangle \in \cV$. 

The asymptotic states constitute two asymptotic
state spaces, $\cV_{in}$ for $x^0 \rightarrow -\infty$ and $\cV_{out}$
for $x^0 \rightarrow +\infty$, which can be shown to be isomorphic to the Fock space of
free fields. One of the crucial postulates of axiomatic field theory, called {\it asymptotic 
completeness}, states the equivalence between the asymptotic and the complete state spaces of the 
theory:
\beq
\cV_{in} = \cV = \cV_{out}
\eeq
This is most important to define the S-matrix as unitary transformation between the in- and 
out-states of the theory. Whereas the complete S-matrix acts on the whole
state space $\cV$, the {\it physical S-matrix} acts on the space $\cV_{phys} \subset \cV$ 
of physical states only.  
   
Once one has succeeded to define $\cV_{phys}$, it is necessary to show that it only 
contains colourless states according to the confinement hypothesis.
Based on symmetries described in the last section the 
{\it Kugo-Ojima confinement scenario} describes a mechanism, 
by which such a positive (semi-)definite state space containing only colourless 
states is generated \cite{Kugo:1979gm}.

One of the basic assumptions of the scenario is the existence of the conserved
BRS-charge $Q_B$, which is used to postulate the physical subspace ${\cV}_{phys}$ of the 
state space ${\cV}$ 
by\footnote{The corresponding construction $\partial_\mu
A_\mu |phys \rangle =0$ in QED is known as Gupta-Bleuler condition.}
\beq
{\cV_{phys}} =
{ \left\{|phys\rangle : Q_B|phys\rangle=0\right\}}.
\eeq
This space can be shown to have a positive semidefinite metric \cite{Kugo:1979gm}.
Recalling the BRS-transformations given in the last section the space ${\cV}_{phys}$
contains two different sorts of states. The {\it first} ones are the so called 
{\it BRS-daughter states}, $|\phi\rangle$. Each of these states  
can be generated by applying the BRS-operator to a corresponding {\it parent state} 
$|\pi\rangle$, which is not element of ${\cV}_{phys}$. We thus have  
$|\phi\rangle = s|\pi\rangle$. The {\it second} ones   
are {\it BRS-singlet states} for which no such parent states 
exist\footnote{In geometrical language the space of
BRS-singlets, $H(Q_B,\cV)$, is called a cohomology, the physical state space 
${\cV}_{phys}=Z(Q_B,\cV)$ is denoted as cocycle space and the space of 
BRS-daughter states, $B(Q_B,\cV)$, is called a coboundary space.
The cocycle space ${\cV}_{phys}$ contains the closed forms with 
respect to the BRS-charge, whereas the coboundary space contains the exact forms
(see {\it e.g.} \cite{Baez:1995sj}).}.

Let us first discuss the BRS-daughter states. The annihilation of these states
by the BRS-charge $Q_B$ is a trivial consequence of the nilpotency of the BRS-operator,
which implies that $Q_B Q_B=0$. 
Due to the BRS-algebra (\ref{BRS-algebra}) of the ghost charge $Q_c$ and the
BRS-charge $Q_B$ 
daughter states $|\phi\rangle$ and their parents $|\pi\rangle$ always occur in 
pairs, {\it i.e.} two daughters and two parents form a so called {\it BRS-quartet}.
Denoting the ghost number by $N_{FP}$ we have the quartet 
$\left(|\pi,{N_{FP}}\rangle,|\phi,{N_{FP}+1}\rangle,|\pi,{-N_{FP}-1}\rangle,|\phi,{-N_{FP}}\rangle\right)$ 
related by the BRS- and ghost number-transformations:
\beqa
Q_B |\pi,{N_{FP}}\rangle &=& |\phi,{N_{FP}+1}\rangle \\ 
Q_B |\pi,{-N_{FP}-1}\rangle &=& |\phi,{-N_{FP}}\rangle \\ 
Q_c |\pi,{N_{FP}}\rangle &=& |\phi,{-N_{FP}}\rangle \\ 
Q_c |\pi,{-N_{FP}-1}\rangle &=& |\phi,{N_{FP}+1}\rangle  
\eeqa
It is easy to show, that the BRS-daughter states $|\phi \rangle$ are orthogonal to all states 
$| \Psi \rangle$ 
of $\cV_{phys}$ as
\beq
\langle \Psi | \phi \rangle = \langle \Psi | Q_B | \pi \rangle = 0.
\eeq
As a consequence the daughter states do not contribute to the physical S-matrix, {\it i.e.} the 
corresponding asymptotic states are not part of the physical spectrum of the theory.
This is also true for the asymptotic states of parent states \cite{Kugo:1979gm}. We therefore
have the result, that the asymptotic states of all members of BRS-quartets do not
correspond to physical particles. This {\it confinement of quartet states} is known 
as {\it quartet mechanism}.

As an example the so called {\it elementary quartet} can be constructed, which consists of the
parent states $|A_\mu^a \rangle,|\bar{c}^a \rangle$ and the daughters 
$|D_\mu^{ab} c^b \rangle,|B^a \rangle$. From these states corresponding asymptotic states of
massless particles can be inferred \cite{Kugo:1979gm}\footnote{In \cite{Kugo:1979gm} 
this is done by a thorough analysis of the correlation functions 
$\langle D_\mu^{ab} c^b \bar{c}^a \rangle$ and $\langle A_\mu^a B^a \rangle$. The intuitive
argument given in \cite{Peskin:1995ev} is based on the free field limit $g=0$, which is
certainly not appropriate in a strong coupling gauge theory.}:
\beqa
{\bar{c}^{as,a}(x)} &=& 
{\bar{\gamma}^a(x)+ \ldots} \label{cbar-as}\\
{A_\mu^{as,a}(x)} &=& 
{\partial_\mu \chi^a(x)+ \ldots} \\
{(D_\mu c)^{as,a}(x)} &=& 
{\partial_\mu \gamma^a(x)+ \ldots} \\
{B^{as,a}(x)} &=&  
{\beta^a(x)+ \ldots} 
\eeqa
It turns out, that these asymptotic states describe ghosts, antighosts and 
longitudinally polarised gluons, which are therefore confined by the quartet mechanism.
 
We now return to the remaining states in $\cV_{phys}$, the BRS-singlets $|\Phi \rangle$.
The asymptotic states of the BRS-singlets are candidates to describe the physical
particles of the theory, namely baryons and mesons.
It can be argued \cite{Kugo:1979gm}, that all BRS-singlet states have vanishing ghost 
number $N_{FP}$, as it is expected for physical states. In order to have {\it confinement}
one has to show that 
\begin{itemize}
\item there is a well defined, {\it i.e. unbroken}, global colour charge $Q^a$ with
\beq
\langle \Phi | Q^a | \Phi^\prime \rangle = 0 \label{confinement}
\eeq
for all BRS-singlet states $| \Phi \rangle, | \Phi^\prime \rangle \in \cV_{phys}$, 
\item the {\it cluster decomposition property} is violated for these states.
\end{itemize}

As the cluster decomposition property is not investigated in this work we briefly explain
what it means and put it aside afterwards. Speaking intuitively, cluster decomposition means the
possibility to divide each given lump of particles into subsets which can be torn apart. For QCD 
this is not what is observed in experiment, as the cluster decomposition property
would imply the possibility to split
an observable colourless object into observable coloured objects.
A rigorous mathematical formulation of the cluster decomposition property is given in \cite{Strocchi:1978ci}.
Here we just mention that the cluster decomposition property may fail only for a quantum field theory
with an indefinite metric and without a mass gap in the whole state space $\cV$ 
\cite{Nakanishi:1990qm}. This certainly not excludes a mass gap in $\cV_{phys}$ as is expected 
in the case of confinement in QCD\cite{Feynman:1981ss}.    

Let us now come back to the global colour charge $Q^a$. This charge can be defined by
\beq
Q^a = \int d^3x J_0^a(x),
\eeq
where the divergence free current $J_\mu^a$ is generated by the unfixed global gauge symmetry
{\it c.f.} subsection \ref{gg-symm}.
With the help of the equations of motion the current $J_\mu^a$ can be written as\footnote{Our
treatment follows closely ref.~\cite{Kugo:1995km}.} 
\beq
gJ_\mu^a = \partial_\nu F^a_{\mu \nu} + \left\{ Q_B, D_\mu^{ab} \bar{c}^b \right\} \label{maxwell}.  
\eeq
For obvious reasons this equation is called {\it quantum Maxwell equation}. One can thus separate 
the global colour charge $Q^a$ into two different charges $G^a$ and $N^a$ corresponding to the two
terms of eq.~(\ref{maxwell}):
\beq
Q^a = \int d^3x \frac{1}{g} \left(
\partial_i F^a_{0 i} + \left\{ Q_B, D_0^{ab} \bar{c}^b \right\}\right) = G^a + N^a.
\label{charge1}
\eeq   
One crucial point with a Noether current is that it is only defined up to an arbitrary term
of the form of a total derivative. Thus if eq.~(\ref{charge1}) is well defined, one can
redefine the global colour charge,
\beqa
 Q^a &=& \int d^3x \left( J_0^a - \frac{1}{g} \partial_\nu F_{0 \nu}^a \right) \nonumber \\
 &=& \int d^3x \frac{1}{g} \left\{ Q_B, D_0^{ab} \bar{c}^b \right\} =  N^a.
\label{charge2}
\eeqa   
If the second equation is well defined, we immediately satisfy eq.~(\ref{confinement}) due
to the nilpotency of the BRS-charge, $(Q_B)^2=0$.

The problem is, however, that there is the possibility that the 
three dimensional integral in eq.~(\ref{charge2}) does not converge 
{\it i.e.} global gauge symmetry is spontaneously broken. One version of the Goldstone theorem
states the equivalence of the following conditions on a conserved current
$J_\mu$ and its global charge $Q$ \cite{Nakanishi:1990qm}:
\begin{itemize}
\item[(a)] $Q = \int d^3x J_0$ does {\it not} suffer from spontaneous symmetry breakdown;
\item[(b)] $J_\mu$ contains {\it no discrete massless spectrum}: 
$\langle 0 | J_\mu | \Psi(p^2=0) \rangle = 0$.
\end{itemize}

However, we already know from eq.~(\ref{cbar-as}) that the antighost $\bar{c}$ is a member
of the elementary quartet and  
contains a one-particle contribution from the massless asymptotic field $\bar{\gamma}$.
We therefore have
\beq
D_\mu \bar{c}^a = \partial_\mu \bar{c}^a + g f^{abc} A_\mu^c \bar{c}^b \hspace{0.5cm} 
\longrightarrow \hspace{0.5cm} \left(1+u\right) \partial_\mu \bar{\gamma}^a
\hspace*{1cm} \mbox{if} \hspace*{2mm} x_0 \rightarrow \pm \infty.
\eeq
Here the proportionality factor $1$ stems from the first term of the covariant 
derivative and the proportionality factor $u$ from
the term containing the gauge field. As the global colour charge is proportional to $D_\mu \bar{c}^a$,
see eq.~(\ref{charge2}), we arrive at the condition 
\beq
1+u = 0 \label{1+u}
\eeq
for the global colour charge to be well defined \cite{Kugo:1979gm}.

This condition has been derived for general linear covariant gauges\footnote{As the quantum
Maxwell equation (\ref{maxwell}) can be written down for the general case $\alpha \not = 0$
as well, we believe it to hold for general values of the gauge parameters $\lambda$ and $\alpha$.
}, $\alpha=0$.
A special case is Landau gauge, which is our preferred choice for most of the investigations
in the next chapters of this thesis. In Landau gauge
it has been shown, that the condition (\ref{1+u})
is connected to the ghost dressing function $G(p^2)$. With the definition
\beq
D^{ab}_{ghost} = - \frac{\delta^{ab} G(p^2)}{p^2} \label{ghost-prop}
\eeq    
for the ghost propagator one obtains the relation \cite{Kugo:1995km}
\beq
\setlength{\fboxsep}{3mm}
\fbox{$ \displaystyle \frac{1}{G(0)} = 1+u = 0 \label{1+u-landau}.$}
\eeq
We are thus in the position to check the {\it Kugo-Ojima confinement criterion},
eq.~(\ref{1+u}), by calculating the ghost dressing function $G(p^2)$ in the infrared.
This is most conveniently done in the framework of Dyson-Schwinger equations, as analytic expressions 
for the infrared behaviour of the propagators can be obtained. 

The condition eq.~(\ref{1+u-landau}) has also been subject to various investigations on the lattice.
There one aims either at the direct determination\footnote{This is done using
the function $U(p^2)$ which is defined as
\beq
\int d^4x e^{ipx} \langle 0 | T D_\mu c^a(x) D_\nu \bar{c}^b(0) |0\rangle = 
\left[ (\delta_{\mu \nu}-\frac{p_\mu p_\nu}{p^2}) U(p^2) - \frac{p_\mu p_\nu}{p^2} \right] \delta_{ab}
\eeq
Here the symbol $T$ denotes time ordering. It can be shown that the parameter $u$ is 
given by $u=U(p^2=0)$ \cite{Kugo:1979gm}.} 
of the parameter $u$
\cite{Nakajima:2000kh,Nakajima:2000mp,Nakajima:2002kh}, or determines the ghost dressing function
in the lattice analogue to Landau gauge \cite{Suman:1996zg,Langfeld:2002dd,Bloch:2002we}. Whereas 
exploratory calculations of the first class give a result of 
$u \approx -0.8$,
the ghost dressing functions obtained in the second class of investigations indicate a divergence
when extrapolated to zero momentum and thus are in agreement with the criterion eq.~(\ref{1+u-landau}).
However, one has to keep in mind that infinite volume extrapolations from lattice results might 
suffer from fundamental problems.

\subsection{Positivity in an Euclidean quantum field theory \label{positivity}}

In the Kugo-Ojima confinement scenario the conserved BRS-charge $Q_B$ is employed to define
a subspace $\cV_{phys}$ from the state space of QCD, which can be shown to be positive semidefinite.
However, this is only one particular mechanism to ensure the probabilistic interpretation of the quantum
theory. Even if the Kugo and Ojima scenario eventually will turn out not to apply for QCD, 
there has to be {\it some} mechanism which singles out a physical, positive semidefinite 
subspace in QCD due to the arguments given at the beginning of subsection \ref{Kugo-Ojima}.  

This suggests another, quite general criterion for confinement, namely {\it violation of positivity}.
If a certain degree of freedom has negative norm contributions in its propagator it cannot be
part of the physical content of the theory\footnote{If the Kugo-Ojima construction is appropriate,
this implies that the corresponding particle is a member of an asymptotic BRS-quartet.}.
In the following we will briefly explain what {\it positivity} means in the context of a 
Euclidian quantum field theory. In chapter \ref{quark} we will search for negative norm contributions
in our solutions for the quark and gluon propagators.

Quantum field theories are completely described in terms of {\it correlation functions}, which can be
ordered by an infinite hierarchy with respect to the number of contributing space-time points.
Basic examples of such correlation functions are propagators (two point correlations)
and vertices (three point correlations). Both are the central objects investigated in this work.
These correlation functions are subject to mathematical properties described by
the axioms of axiomatic quantum field theory\footnote{An introductory overview on axiomatic
quantum field theory is given in the book of Haag \cite{Haag:1992hx}.}. 
A set of such axioms has been given first in 1957 by Wightman \cite{Wightman} 
for field theories 
formulated in Minkowski space. The Euclidean counterpart of the Wightman axioms have been 
found by Osterwalder and Schrader \cite{Osterwalder:1973dx,Osterwalder:1975tc} in 1973. 

This second set of axioms will be relevant for us, as we will work in Euclidean space-time
throughout this thesis. There are two reasons for this choice: one technical and one physical.
The technical aspect is the absence of poles on the real positive $p^2$-axis for the propagators
of Euclidean field theory. This is a huge practical simplification when dealing with
Dyson-Schwinger equations\footnote{Only some exploratory calculations of Dyson-Schwinger equations
in Minkowski space can be found in the literature, see {\it e.g.} 
\cite{Sauli:2001cj,Sauli:2002tk}.}. Maybe even more
important is the physical aspect. In the course of this work we will run across several results
which are most advantageously compared to lattice Monte Carlo simulations, which are performed in
Euclidean space-time. Whereas both methods alone suffer from specific problems detailed in later 
chapters, the interplay between lattice simulations and Dyson-Schwinger calculations leads to 
well justified statements on the infrared behaviour of QCD correlation functions. 
 
The Euclidean counterpart to the notion
of positivity in Minkowski space is the Oster\-walder-Schrader axiom of {\it reflection
positivity}. For a thorough mathematical formulation of the axiom the reader is 
referred to refs.~\cite{Haag:1992hx,Osterwalder:1973dx,Osterwalder:1975tc}. 
We are interested in the 
special case of a two point correlation function, {\it i.e.} a propagator $S(x-y)$, for which the
condition of reflection positivity can be written as 
\beq
\int d^4x \; d^4y \; \bar{f}(\vec{x},-x_0) \; S(x-y) \; {f}(\vec{y},y_0) \; \ge 0 \;.
\eeq
where $f$ is a complex valued test function with support in $\{(\vec{x},x_0) \: : \: x_0 > 0 \}$
and time ordered arguments.
After three-dimensional Fourier transformation this condition implies
\beq
\int_0^\infty dt \; dt^\prime \; \bar{f}(t^\prime,\vec{p}) \; S(-(t+t^\prime),\vec{p}) \;
 f(t,\vec{p}) \; \ge 0
\eeq
where $S(x_0,\vec{p}) := \int d^3x \; S(\vec{x},x_0) \; e^{i\vec{p}\vec{x}}$. The momentum dependence
of the corresponding Fourier transform of the test function $f$ has been chosen to provide a 
suitable smearing around the three-momentum $\vec{p}$. This condition will be tested for
the quark and gluon propagator in chapter \ref{quark}.   

\subsection{The horizon condition \label{gribov}}

The last aspect of confinement summarised in this section is the horizon condition formulated by 
Zwanziger \cite{Zwanziger:1992ac,Zwanziger:1993qr}. 
This brings us back to the gauge fixing procedure, described in section 
\ref{gen-func}. Recall that the problem of fixing a gauge is equivalent to singling one
representative configuration from each gauge orbit 
\beq
[A^U] := \left\{A^U = UAU^\dagger + UdU^\dagger : U(x) \in SU(N_c) \right\}.
\eeq 
It has been shown by Gribov \cite{Gribov:1978wm}, that the simple Faddeev-Popov procedure
generates a hyperplane $\Gamma$ in gauge field configuration space which still contains gauge field
configurations connected by a gauge transformation. 
These multiple intersection points of a gauge orbit with $\Gamma$ are called 
{\it Gribov copies}.
An almost unique representative of each gauge orbit is obtained, if one restricts the hyperplane 
$\Gamma$ to the so called {\it Gribov region} $\Omega$. This is conveniently done by minimising
the following $L^2$-norm of the vector potential along the gauge orbit \cite{vanBaal:1997gu}:
\beq
F_A(U) := ||A^U||^2 = ||A||^2 - 2 i \int d^4 x \; \tr(\omega \partial A ) 
+ \int d^4x \; \tr( \omega FP(A) \omega) + O(\omega^3) \;,
\eeq  
with the gauge transformation $U=\exp(i\omega(x))$ and the Faddeev-Popov operator
\beq
FP(A) = -\partial D(A)^{ab} = -\partial^2 \delta^{ab}- g f^{abc}\partial_\mu A_\mu^c \;.
\eeq
Any local minimum thus implements strictly the Landau gauge condition $\partial A=0$, and the
Faddeev-Popov operator has to be a positive operator.  
The Gribov region $\Omega$ defined by this prescription can be shown to be convex, 
to contain at least one intersection point with each gauge orbit and to be bounded in every 
direction of the hyperplane $\Gamma$ \cite{Zwanziger:1982na}. Furthermore the lowest eigenvalue
of the Faddeev-Popov operator approaches zero at the boundary $\partial \Omega$, the 
{\it first Gribov horizon}.   

\begin{figure}[t]
\begin{center}
\epsfig{file=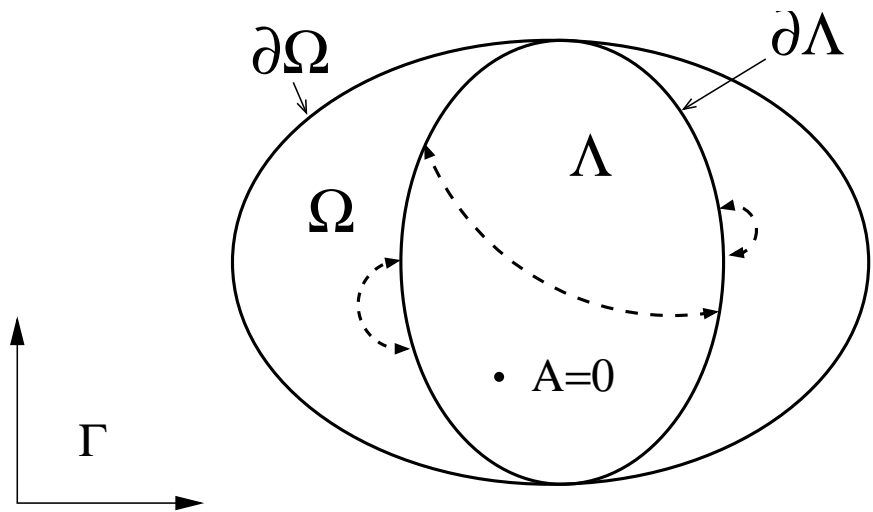,width=11.5cm}
\end{center}
\caption{\sf Sketch of the hyperplane $\Gamma$ in gauge field configuration space 
obtained by the Faddeev-Popov gauge fixing method. Shown are the first Gribov region $\Omega$,
and the fundamental modular region $\Lambda$ containing the trivial configuration $A=0$. The arrows
pointing to the boundary $\partial \Lambda$ of the fundamental modular region indicate 
that different boundary points have to be identified thus generating topological obstructions.}
\label{FOM}
\end{figure}

The absolute minimum of the $F_A$ defines the {\it fundamental modular region} $\Lambda$.
It can be shown that $\Lambda$ is convex as well, bounded in $\Omega$ and contains
the origin $A=0$ at a finite distance of the boundary $\partial \Lambda$.
Furthermore points on the boundary $\partial \Lambda$ have to be identified to 
completely fix the gauge \cite{vanBaal:1992zw,vanBaal:1997gu}. The set $\partial \Omega \cap
\partial \Lambda$ contains the so
called {\it singular boundary points} which are related by infinitesimal gauge transformations.
This whole situation is sketched in figure \ref{FOM}.

The horizon condition has been formulated in an attempt to 
restrict the generating functional of the gauge fixed theory to the fundamental modular
region $\Lambda$. 
It was shown for lattice gauge theory in the thermodynamical limit 
\cite{Zwanziger:1992ac,Zwanziger:1994dh} 
as well as for continuum theory \cite{Zwanziger:1993qr}, that such a 
restriction might be possible, if the probability distribution inside the fundamental modular region is
concentrated at the Gribov horizon, {\it i.e.} at the region $\partial \Omega \cap \partial \Lambda$.
Entropy arguments have been employed to
argue for this condition. Furthermore it has been argued, that this implies 
the quantum field theory to be in the nontrivial, confining 
phase \cite{Zwanziger:1992ac}.

Interestingly enough, the horizon condition originally formulated in different terms can be 
connected to the ghost dressing function $G(p^2)$ 
\cite{Zwanziger:1993qr,Zwanziger:1994dh,Zwanziger:2001kw}, which has been defined in the last 
subsection, eq.~(\ref{ghost-prop}). Due to the proximity of infrared ({\it i.e.} almost constant) 
gauge field configurations to the Gribov horizon \cite{Gribov:1978wm} the horizon condition
is equivalent to 
\beq
\lim_{p\rightarrow 0}[G(p^2)]^{-1} = 0 \;. \label{horizon}
\eeq
This is the same condition as the Kugo-Ojima confinement criterion, eq.~(\ref{1+u-landau}).

Furthermore, the same entropy arguments have been been employed to argue for a vanishing
gluon propagator $D(p^2)$ in the infrared \cite{Zwanziger:1991gz,Zwanziger:1992ac}:      
\beq
\lim_{p\rightarrow 0}[D(p^2)] = 0 \;. \label{horizon2}
\eeq
Both of these conditions can be checked by solving the corresponding 
Dyson-Schwinger equations (DSEs) for the ghost and gluon propagators in Landau gauge. 
Turning the argument round, eqs.~(\ref{horizon}),(\ref{horizon2})
are appropriate boundary conditions for the DSEs to generate solutions corresponding to
a restriction of the generating functional (\ref{genfunc}) 
to the Gribov region $\Omega$ \cite{Zwanziger:2001kw}.
In the remaining  chapters of this thesis we will demonstrate the agreement of our
solutions of the DSEs with the conditions (\ref{horizon}), (\ref{horizon2}). 

\section{The Dyson-Schwinger equations for the QCD propagators \label{DSE-sec}}

Having summarised some general aspects of strong QCD which are relevant for the discussion of
the propagators of the theory we focus our attention on the Dyson-Schwinger equations of motion
for the ghost, gluon and quark propagator. These are a coupled set of integral equations which
are derived from the generating functional (\ref{genfunc-gf}) together with the generalised Lagrangian
(\ref{Lagrangian-gen}). We concentrate on the derivation of the ghost DSE, as the four-ghost interaction 
term of our generalised Lagrangian
generates new loops different from those of standard Faddeev-Popov gauges. As the derivation
is rather lengthy we just give a summary in this chapter, deferring details to appendix 
\ref{qcd-appendix}.
The formal structures of the gluon and quark
DSEs remain unchanged compared to the Faddeev-Popov case and we therefore just give the
results at the end of the section. 

With the conventions defined in appendix \ref{Def-app} the
Dyson-Schwinger equation of motion for the ghost propagator reads 
\beq
\left\langle \frac{\delta S}{\delta \bar{c}^c(z)} \bar{c}^b(y) 
\right\rangle = \delta(z-y) \delta_{cb}.
\label{ghostdse_main}
\eeq
Here the brackets $\langle.\rangle$ indicate the expectation value of the enclosed field operators. 
The derivative of the action $S[J,c,\bar{c}]=\int d^4x \; {\cal{L}}$ is given by
\beqa
\frac{\delta S}{\delta \bar{c}^c(z)} &=&  \partial^2 c^c(z)
+ \frac{\alpha}{2} \left(1-\frac{\alpha}{2}\right) \frac{\lambda}{2}
g^2 f^{cde} f^{fge} \bar{c}^d(z) c^f(z) c^g(z) \nonumber\\
&&+ i \left(1-\frac{\alpha}{2} \right) g f^{cde} \partial_\mu \left(
A^e_\mu(z) c^d(z)\right)
+ i\frac{\alpha}{2} g f^{cde} A^e_\mu(z) \partial_\mu 
c^d(z).
\label{ghostder_main}
\eeqa
We now decompose the full four-ghost correlation function into connected parts and use
the relation
\beq
\delta(y-x)\delta^{ab} 
= \int d^4z \left[D_G^{db}(z-y) \right]^{-1} D_G^{ad}(x-z) \nonumber\\
\eeq
for the ghost propagator $D_G$ to arrive at
\beqa
[D_G^{ab}(x-y)]^{-1}
&=&\partial^2 \delta(x-y) \delta^{ab} \nonumber\\ 
&-& \frac{\alpha}{2} \left(1-\frac{\alpha}{2}\right) \frac{\lambda}{2}
g^2 f^{cde} f^{fge} \int d^4z \; [D_G^{ac}(x-z)]^{-1} \times \nonumber\\ 
&&\hspace*{0.5cm}\left\{ \langle \bar{c}^b(y) \bar{c}^d(z) c^f(z) c^g(z)\rangle 
+ \langle \bar{c}^b(y)c^g(z)\rangle\langle \bar{c}^d(z) c^f(z)\rangle \right. \nonumber\\ 
&& \hspace*{4.7cm} - \left. \langle \bar{c}^b(y)c^f(z)\rangle\langle 
\bar{c}^d(z)c^g(z)\rangle \right\} \nonumber\\ 
&-& i \left(1-\frac{\alpha}{2} \right) g f^{cde} 
\int d^4z \; [D_G^{ac}(x-z)]^{-1}\langle \bar{c}^b(y)\partial_\mu \left(
A^e_\mu(z) c^d(z)\right)\rangle \nonumber\\
&-& i \frac{\alpha}{2} g f^{cde} \int d^4z \; [D_G^{ac}(x-z)]^{-1}\langle 
\bar{c}^b(y) A^e_\mu(z) \partial_\mu c^d(z) \rangle. 
 \label{DSEb_main}
\eeqa

The remaining task is to decompose the connected Green's functions into 
one-particle irreducible ones. Plugging in the definitions of the bare ghost-gluon
and the bare four-ghost vertex defined in appendix \ref{Def-app} we arrive at the ghost
Dyson--Schwinger equation in coordinate space:
\beqa
[D_G^{ab}(x-y)]^{-1}
&=&[D_G^{(0) ab}(x-y)]^{-1} \nonumber\\
&-& \int d^4[zuvz_1z_2z_3] \; 
\Gamma^{(0) bde}_\mu (y,u,v) \;  
D_{\mu \nu}^{ef}(v-z_1) \;\Gamma_\nu^{fha}(z_1,z_3,x) \; D_G^{hd}(u-z_3)
\nonumber\\  
&-&  \int d^4[uv] \; \Gamma_{4gh}^{(0) bdfa}(x,u,v,y) \; D_G^{fd}(v-u) \nonumber\\
&-& \frac{1}{2} \int d^4[zuvu_1u_2u_3u_4] \; 
\Gamma^{(0) bdgf}_{4gh}(y,z,v,u)\; D_G^{fe}(u-u_4) \times \nonumber\\
&& \hspace*{4cm} 
D_G^{gi}(v-u_2)\; \Gamma_{4gh}^{jaei}(u_3,x,u_4,u_2)\; D_G^{jd}(u_3-z) \nonumber\\ 
&-& \frac{1}{2} \int d^4[zuvu_1u_2u_3u_4u_5] \; \Gamma^{(0) bdgf}_{4gh}(y,z,v,u)
\; D_{\mu \nu}^{ek}(u_1-u_4) \times
\nonumber\\
&& \hspace*{0.4cm} D_G^{fl}(u-u_5) \; \Gamma^{kal}_\nu(u_4,x,u_5)\; D_G^{gi}(v-u_2) \;
\Gamma^{eij}_\mu(u_1,u_3,u_2) \;D_G^{jd}(u_3-z). \nonumber\\ 
\label{DSEe_main}
\eeqa
Here we used the abbreviation $d^4[xyz] := d^4x \; d^4y \; d^4z$. The ghost-gluon vertex is denoted
by $\Gamma_\nu$ and the four-ghost vertex by $\Gamma_{4gh}$. The superscript $(0)$ indicates a bare 
vertex or propagator.  

\begin{figure}[t]
\epsfig{file=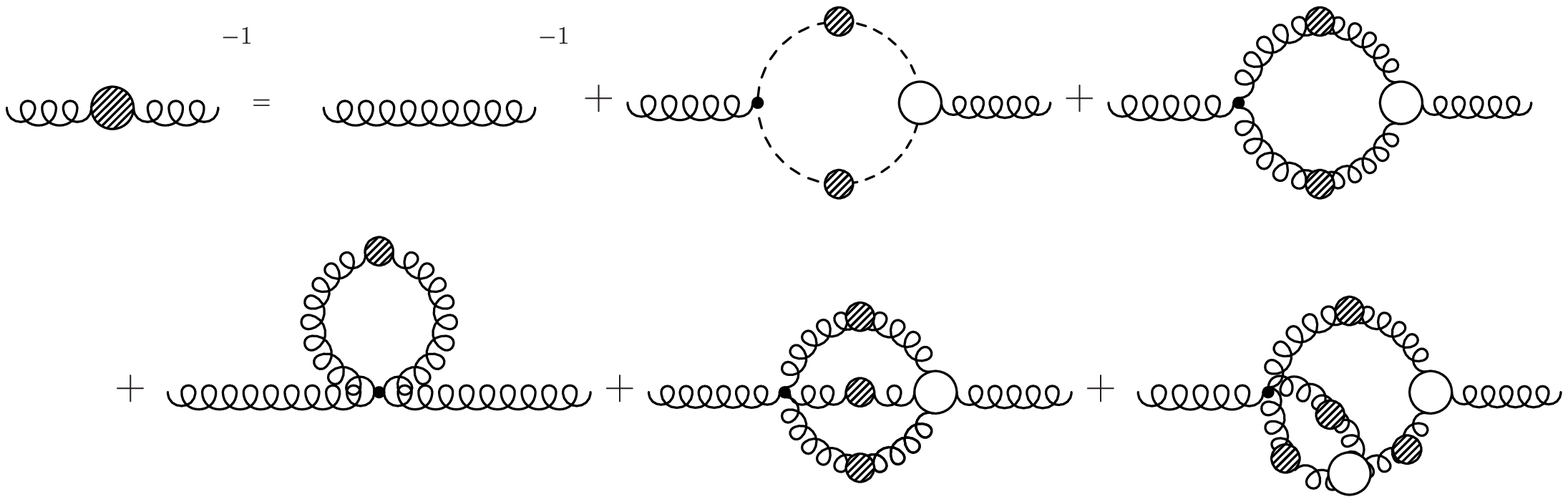,width=16cm} 

\vspace*{1cm}
\epsfig{file=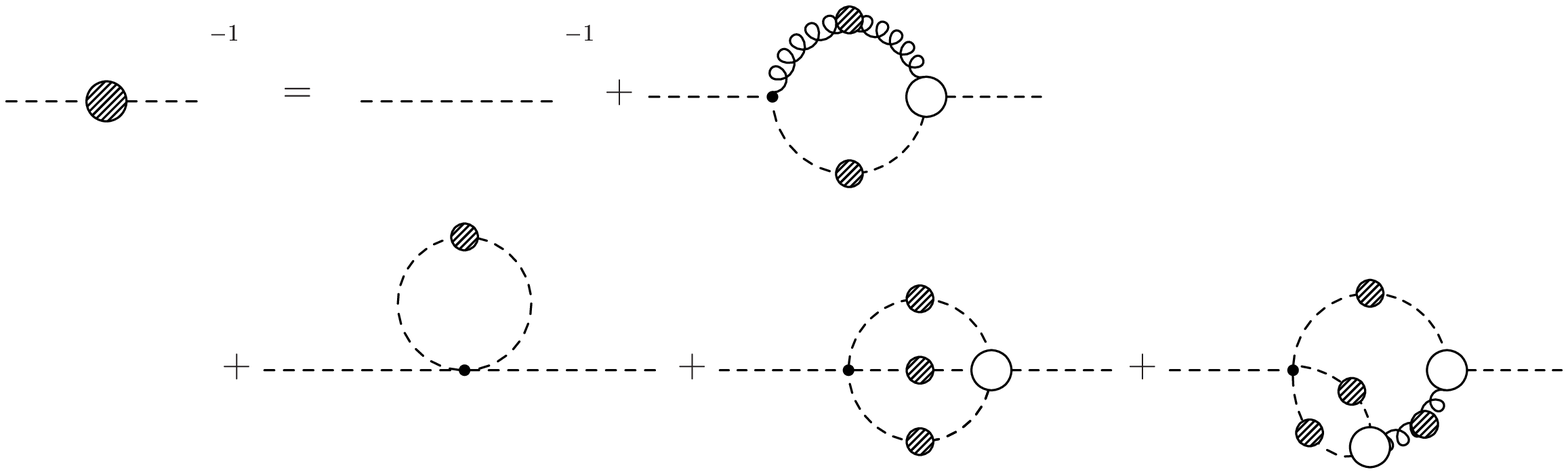,width=16cm} 

\vspace*{1cm}
\epsfig{file=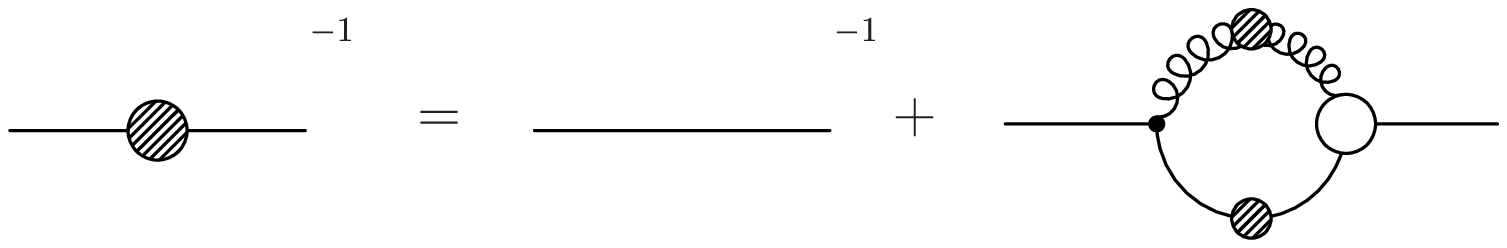,width=11cm}
\caption{\sf The coupled Dyson--Schwinger equations for
the gluon, ghost and quark propagators from a BRS and Anti-BRS symmetric
Lagrangian. Each of the 
equations for the gluon and ghost propagators 
contains genuine one-loop diagrams, a tadpole contribution and a sunset and a squint diagram.
All signs and weight factors have been absorbed in the diagrams.}
\label{DSEpic}
\end{figure}
We now perform a Fourier transformation of eq.~(\ref{DSEe_main}) and finally introduce renormalisation
factors at the appropriate places:
\setlength{\jot}{4mm}
\beqa
[D_G(p)]^{-1}
&=& \tilde{Z}_3 [D_G^{(0)}(p)]^{-1} \nonumber\\ 
&-& \tilde{Z}_1 \frac{g^2 N_c}{(2\pi)^4}\int d^4q \:\Gamma^{(0)}_\mu (p,q) \: 
D_{\mu \nu}(p-q) \:\Gamma_\nu(q,p)D_G(q) \nonumber\\ 
&-& \tilde{Z}_4 \frac{g^2 N_c}{(2\pi)^4} \int d^4q \: 
\Gamma_{4gh}^{(0)} \:  D_G(q) \nonumber\\
&+& \tilde{Z}_4 \frac{1}{2} \frac{g^4 N_c^2}{(2\pi)^8}\int d^4[q_1q_2] \Gamma^{(0)}_{4gh} \:
D_G(q_1) \:D_G(p-q_1-q_2) \:\Gamma_{4gh}(p,q_1,q_2)\: D_G(q_2) 
\nonumber\\ 
&-& \tilde{Z}_4 \frac{1}{4} \frac{g^4 N_c^2}{(2\pi)^8} \int d^4[q_1q_2] \: \Gamma^{(0)}_{4gh}\:  
D_{\mu \nu}(p-q_1) \: D_G(q1) \times \nonumber\\
&&  \hspace*{3cm}  \Gamma_\nu(p,q_1)\:  D_G(q_2)\: 
\Gamma_\mu(-p+q_1+q_2,q_2)\: D_G(p-q_1-q_2). \nonumber\\ 
\label{DSE-ghost-ren}
\eeqa
The colour traces have already been carried out and the reduced vertices defined in appendix 
\ref{Def-app} have been used.
The four-ghost interaction generates three new diagrams in the ghost equation, a tadpole contribution and two
two-loop diagrams. Furthermore the bare ghost-gluon vertex depends on the gauge parameter $\alpha$,
\setlength{\jot}{5mm}
\beqa
\Gamma_\mu^{(0) abc}(k,p,q) &=& g f^{abc} (2 \pi)^4 \delta^4(k+q-p) \Gamma_\mu^{(0)}(p,q) \nonumber\\
\Gamma_\mu^{(0)}(p,q) &=& \left[ \left(1-\frac{\alpha}{2}\right)q_\mu + \frac{\alpha}{2} p_\mu \right].
\eeqa
Note the symmetry between the ghost momentum $p_\mu$ and the
antighost momentum $q_\mu$, when the gauge parameter $\alpha$ is set to one. 
\setlength{\jot}{0mm}

The respective equation for the gluon propagator is formally the same as in the Faddeev-Popov 
case, where the equations have been derived in ref.~\cite{Eichten:1974et}.
Differences occur in the explicit form of the bare ghost-gluon vertex and the dressed vertices  
in general depend on the gauge parameters. As the derivation of the DSE from the
generating functional brings nothing new we refrain from displaying it explicitly and
merely give the result: 
\beqa
[D(p)]^{-1}_{\mu \nu}
&=& Z_3 [D^{(0)}(p)]^{-1}_{\mu \nu} \nonumber\\ 
&+& \tilde{Z}_1 \frac{g^2 N_c}{(2\pi)^4}\int d^4q \:\Gamma^{(0)}_\mu (p,q) \: 
D_G(p-q) \:\Gamma_\nu(q,p)D_G(q)
\nonumber\\ 
&-& Z_1 \frac{1}{2}\frac{g^2 N_c}{(2\pi)^4}\int d^4q \:\Gamma^{(0)}_{\mu \rho \sigma} (p,q) \: 
D_{\rho \rho^\prime}(p-q) \:\Gamma_{\rho^\prime \nu \sigma^\prime}(q,p)D_{\sigma \sigma^\prime}(q) \nonumber\\
&-& Z_4  \frac{1}{2} \frac{g^2 N_c}{(2\pi)^4} \int d^4q \: \Gamma^{(0)}_{\mu \nu \rho \sigma}
 \:  D_{\rho \sigma}(q) \nonumber\\
&-&  Z_4 \frac{1}{6} \frac{g^4 N_c^2}{(2\pi)^8}\int d^4[q_1q_2] \: 
\Gamma^{(0)}_{\mu \rho \sigma \lambda} \:
D_{\rho \rho^\prime}(q_2) \:D_{\sigma \sigma^\prime}(p-q_2-q_1) 
\times \nonumber\\
&& \hspace*{8cm} \Gamma_{\rho^\prime \nu \lambda^\prime \sigma^\prime}(p,q_1,q_2)\: 
D_{\lambda \lambda^\prime}(q_1) \nonumber\\ 
&-& Z_4  \frac{1}{2} \frac{g^4 N_c^2}{(2\pi)^8} \int d^4[q_1q_2] \: 
\Gamma^{(0)}_{\mu \rho \sigma \lambda}\:  
D_{\rho \rho^\prime}(p-q_1-q_2) \: D_{\sigma \sigma^\prime}(q_2) \times \nonumber\\
&&  \hspace*{2cm}  \Gamma_{\rho^\prime \zeta \sigma^\prime}(p-q_1-q_2,q_2)\:  
D_{\zeta \zeta^\prime}(p-q_1)\: \Gamma_{\zeta^\prime \nu \lambda^\prime}(p-q_1,q_1)
\: D_{\lambda \lambda^\prime}(q_1). \nonumber\\ 
\label{DSE-gluon-ren}
\eeqa
The definitions and conventions for the gluon propagator $D_{\mu \nu}$, the three-gluon vertex 
$\Gamma_{\mu \nu \rho}$ and the four-gluon vertex $\Gamma_{\mu \nu \rho \sigma}$ are 
given in appendix \ref{Def-app}.

The Dyson-Schwinger equation for the quark propagator is derived in a similar way from
the generating functional and reads explicitly
\beq
[S(p)]^{-1} = [S^{(0)}(p)]^{-1} - Z_{1F} \frac{g^2 C_f}{(2\pi)^4} \int d^4q \;
\Gamma^{(0)q}_\mu(p,q) \; D_{\mu \nu}(p-q) \; S(q) \; \Gamma^q_\mu(p,q) \;, \hspace*{0.2cm} 
\eeq
where $S$ denotes the quark propagator and $\Gamma^q_\mu$ the quark-gluon vertex.
The factor $C_f=(N_c^2-1)/(2N_c)$ stems from the colour trace which has already been carried out.

All three Dyson-Schwinger equations are shown diagrammatically in figure \ref{DSEpic}.
One clearly sees the striking similarity between the ghost and the gluon equation once a 
four-ghost interaction has been introduced. Both equations have bare and one
loop parts, a tadpole contribution, a sunset and a squint diagram.

\chapter{Propagators of Landau gauge Yang-Mills theory \label{YM}}

In this chapter we investigate the Dyson--Schwinger equations for the propagators of
Yang--Mills theory. The knowledge of the two point
functions of Yang--Mills theory, the ghost and gluon propagator,
might shed light on those fundamental properties of QCD, which are generated in the gauge sector
(for a recent review see \cite{Alkofer:2000wg}). This is clearly the case
for the phenomenon of confinement, as can be inferred from lattice 
calculations\footnote{In these simulations the string tension between a 
pair of static quarks has been calculated and found to be linearly rising as is expected
for confined quarks. No dynamical quarks are involved in the calculations, therefore one concludes
that the string tension is generated by the gauge field only. There is even evidence
that only very particular gauge field configurations, namely center vortices, are
responsible for the linearly rising potential \cite{Langfeld:2001cz}. This might shed new light on the
origin of the violation of the cluster decomposition principle discussed in 
subsection \ref{Kugo-Ojima}.}.
Furthermore the knowledge of the interaction strength in the gauge sector of QCD
provides the basis for a successful description of hadronic physics
\cite{Alkofer:2000wg,Roberts:2000aa}.
Based on the idea of infrared slavery older works on this subject assumed a
gluon propagator that is strongly singular in the
infrared. Recent studies based either
on Dyson--Schwinger equations
\cite{vonSmekal:1997is,Atkinson:1998tu,Watson:2001yv,Alkofer:2001iw,Zwanziger:2001kw,Lerche:2002ep,Zwanziger:2002ia}
or Monte-Carlo lattice calculations
\cite{Mandula:1999nj,Bonnet:2000kw,Bonnet:2001uh,Langfeld:2001cz,Cucchieri:1999ky,Cucchieri:2000kw,Cucchieri:1998fy}
in Landau gauge indicate quite the opposite: an infrared finite or even 
infrared vanishing gluon propagator.

Lattice simulations and the Dyson-Schwinger approach are complementary in the following sense:
On the one hand, lattice calculations include all non-perturbative 
physics of Yang--Mills theories but cannot make definite statements about 
the far infrared due to the finite lattice volume. On the other hand, 
Dyson--Schwinger equations (DSEs) allow one to extract the leading infrared behaviour
analytically and the general non-perturbative behaviour with moderate numerical
effort. However, the infinite tower of coupled 
non-linear integral DSEs has to be truncated in order to be manageable.
As we will see in the course of this chapter, the propagators  
of SU(2) and SU(3) Landau gauge Yang--Mills theory 
coincide for these two different approaches reasonably well.
Thus we are confident that our results for the qualitative features of
these propagators are trustworthy. 

Throughout this chapter 
we stay in the framework of ordinary Faddeev--Popov quantisation. 
Interesting enough, some of
our results can be directly compared with recent calculations 
obtained in a framework employing stochastic quantisation \cite{Zwanziger:2002ia}.
We will
thus be able to check for systematic errors connected to the appearance
of Gribov copies\footnote{A corresponding investigation of the influence of Gribov copies on the 
gluon propagator in lattice simulations can be found in 
ref.~\cite{Cucchieri:1997dx,Alexandrou:2000ja}} \cite{Gribov:1978wm}, c.f. the discussion in section \ref{gribov}.   

Landau gauge, which has been chosen for all DSE studies of Yang-Mills theory so far,
is special for a number of
reasons. {\it First}, it is a fixed point under the
renormalisation procedure. This means that the gauge parameter $\lambda$ 
is not renormalised when $\lambda=0$, a fact which simplifies the 
renormalisation of the DSEs considerably. 
{\it Second}, as we saw in the last chapter, Landau gauge is a ghost-antighost
symmetric gauge. This is of principal interest as we
are then allowed to interpret ghost and antighost as (unphysical) particle
and antiparticle. On the other hand it simplifies matters if one attempts to
construct a non-perturbative dressed ghost-gluon vertex, as one is guided
by a symmetry. This has been exploited in  
references \cite{vonSmekal:1997is,vonSmekal:1998is,hauck}. {\it Third}, the
ghost-gluon vertex does not suffer from ultraviolet divergences in Landau gauge,
as has been shown by Taylor
\cite{Taylor:1971ff}. Again, this simplifies the search for a suitable {\it ansatz} for this 
vertex considerably. Indeed, one is even allowed to use the bare ghost-gluon 
vertex, as we will see in the course of this chapter. 

This chapter is organised as follows:
We first give a brief summary of previously employed truncation and 
approximation schemes for the coupled gluon and ghost DSEs in Landau gauge.
We discuss the key role of the ghost-gluon vertex in these truncations and
show how a non-perturbative definition of the running coupling can be inferred. 
One of the obstacles encountered in providing numerical solutions 
are the angular integrals inherent to these equations. Therefore 
approximated treatments of the angular integrals have been applied so far
\cite{vonSmekal:1997is,Atkinson:1998tu}. In general these angular
approximations proved to be good for high momenta but less trustworthy in the
infrared. Recent studies \cite{Atkinson:1998zc,Zwanziger:2001kw,Lerche:2002ep} 
therefore concentrated on the infrared analysis, where exact results have been
gained for the limit of vanishing momentum.
However, as we will see in the course of this chapter, not every extracted
infrared solution is connected to a numerical solution for finite values
of momenta. The main part of this chapter is therefore devoted to the
construction of a novel truncation scheme, which 
allows to overcome the angular approximation for the whole range of momenta
\cite{Fischer:2002eq,Fischer:2002hn}.
Thus we are able to single out the physical infrared solutions of the 
schemes used in \cite{Atkinson:1998zc,Zwanziger:2001kw,Lerche:2002ep}.
In the ultraviolet region of momentum
we obtain the correct one-loop anomalous dimensions of the propagators 
known from resummed perturbation theory. 

\section{Gluon and ghost 
Dyson--Schwinger equations in flat Euclidean space-time \label{gluonghost}}

The coupled set of gluon and ghost Dyson--Schwinger
equations, which has been given diagrammatically  
in Fig.~\ref{DSEpic} in the last chapter,
loses a considerable amount of complexity in Landau gauge.
There, the four-ghost vertex vanishes and we are left with one dressing loop in the
ghost equation. As we will only be concerned with pure
Yang--Mills theory in this chapter the quark loop in the gluon equation
disappears as well. Furthermore in Landau gauge the tadpole term provides 
an (ultraviolet divergent) constant only and will drop out during renormalisation. 
Thus we will neglect this contribution from the very beginning. 
All these simplifications are due to our choice of Landau gauge.
The resulting system of equations is still very complicated as it contains
full two-loop diagrams in the gluon equation. The first assumption of all 
truncation schemes up to now is, that contributions from these two-loop diagrams 
may safely be neglected\footnote{The only exception is the scheme discussed
in \cite{Bloch:2001wz}, which has not been solved yet.}. 
We will join in this assumption and
provide some arguments for its validity in subsection \ref{YM-results}.

\begin{figure}[t]
  \centerline{ \epsfig{file=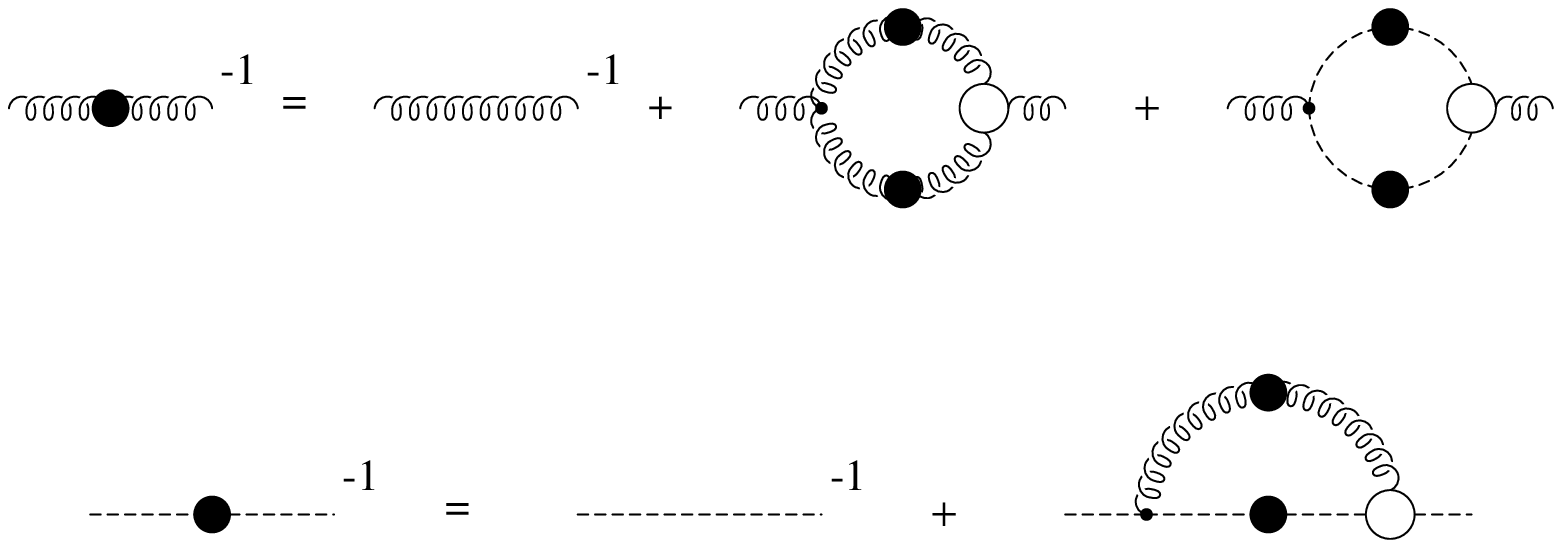,width=0.98\linewidth} }
  \vskip 3mm
  \caption{\sf Diagrammatic representation of the truncated Landau gauge
  gluon and ghost DSEs studied in this chapter. In the gluon 
  Dyson--Schwinger equation terms with four--gluon vertices and quarks
  have been dismissed.}
  \label{GluonGhost}
\end{figure}

Thus, we will effectively study the
coupled system of equations as depicted in Fig.\ \ref{GluonGhost}.
The corresponding equations are given by
\setlength{\jot}{3mm}
\beqa
[D_G(p)]^{-1}
&=& \tilde{Z_3} [D_G^{(0)}(p)]^{-1}  
- \tilde{Z_1} \frac{g^2 N_c}{(2\pi)^4}\int d^4q \:\Gamma^{(0)}_\mu (p,q) \: 
D_{\mu \nu}(p-q) \:\Gamma_\nu(q,p)D_G(q) \;, \hspace*{1.5cm} 
\label{YM-DSE-ghost} \\
{[D(p)]^{-1}_{\mu \nu}}
&=& Z_3 [D^{(0)}(p)]^{-1}_{\mu \nu}  
+ \tilde{Z_1} \frac{g^2 N_c}{(2\pi)^4}\int d^4q \:\Gamma^{(0)}_\mu (p,q) \: 
D_G(p-q) \:\Gamma_\nu(q,p)D_G(q)
\nonumber\\ 
&&\hspace{1cm}- Z_1 \frac{1}{2}\frac{g^2 N_c}{(2\pi)^4}\int d^4q 
\:\Gamma^{(0)}_{\mu \rho \sigma} (p,q) \: 
D_{\rho \rho^\prime}(p-q) \:\Gamma_{\rho^\prime \nu \sigma^\prime}(q,p)
D_{\sigma \sigma^\prime}(q). 
\label{YM-DSE-gluon}
\eeqa
Here the ghost-gluon vertex is denoted by the symbol $\Gamma_\nu(q,p)$,
whereas the three-gluon vertex is given by $\Gamma_{\rho \nu \sigma}(q,p)$.
Furthermore we have the coupling $g$ and the number of colours $N_c$ stemming from the
colour trace of the respective loops.
Suppressing colour indices the explicit expressions for the ghost and gluon propagators as well as the
inverse of the gluon propagator are given by
\beqa
D_{G}(p) &=& - 
\frac {G(p^2)}{p^2} \;, \label{Gh-prop}\\
D_{\mu \nu}(p) &=& 
\left(\delta_{\mu \nu} - \frac{p_\mu p_\nu}{p^2}
\right) \frac{Z(p^2)}{p^2} + \lambda \frac{p_\mu p_\nu}{p^4} 
\;, \label{Gl-prop} \\
\left[ D_{\mu \nu}(p)\right]^{-1} &=& 
\left(\delta_{\mu \nu} - \frac{p_\mu p_\nu}{p^2}
\right) \frac{p^2}{Z(p^2)} + \frac{1}{\lambda} p_\mu p_\nu 
\;. \label{Gl-prop-inv}
\eeqa 
For all linear covariant gauges the longitudinal parts of the full and bare inverse gluon
propagators cancel each other in the gluon equation (\ref{YM-DSE-gluon}). Furthermore
in Landau gauge we have $\lambda=0$.  
\setlength{\jot}{0mm}

At this stage of treating the equations we spot a problem: The left hand
side of the gluon equation (\ref{YM-DSE-gluon}) is transverse to the gluon momentum
therefore the right hand side of this equation should be transverse as well. This is certainly 
the case in the exact theory. However in an approximate treatment the gluon
polarisation on the right hand side may acquire spurious
longitudinal terms due to breaking gauge
invariance. In general there are two possible sources for this violation: The
{\it first} one is the 
use of vertices which violate the corresponding Slavnov-Taylor identity. The
{\it second} one is the use of a regularisation scheme which breaks gauge invariance, such as
a cutoff in the radial momentum integral. 

We postpone the problem of the gauge invariance of the vertex {\it ansatz} to the next subsection 
and discuss the regularisation problem first.
In the following we outline some rather abstract arguments which will become 
more transparent in section \ref{truncation}, where we investigate the infrared and 
ultraviolet properties of our new truncation scheme in detail.

In general a cutoff in the loop integrals can lead to quadratically ultraviolet
divergent terms in the gluon equation. Such terms are scheme dependent and therefore
unphysical. Furthermore they are highly ambiguous because they depend on the 
momentum routing in the loop integral. 
Unfortunately, a gauge invariant regularisation scheme avoiding these terms is
hard to implement in Dyson--Schwinger studies\footnote{For the corresponding
use of dimensional regularisation see {\it e.g.\/} 
refs.~\cite{vonSmekal:1991fp,Gusynin:1998se,Schreiber:1998ht}).}. 

It has been argued though \cite{Brown:1988bm,Brown:1989bn}, that quadratic divergences 
can occur only in
that part of the right hand side of the equation which is proportional to
the metric $\delta_{\mu\nu}$. 
Therefore an alternative procedure to avoid quadratic divergences
is to contract the equation with the tensor \cite{Brown:1988bm}
\begin{equation}
{\mathcal R}_{\mu\nu} (p) =
\delta_{\mu\nu} - 4 \,\frac{p_\mu p_\nu}{p^2} \, , 
\label{Rproj}
\end{equation} 
which is constructed such that ${\mathcal R}_{\mu\nu} (p) \,\delta^{\mu\nu} =0$.
However, as has become obvious recently \cite{Lerche:2002ep}, 
the use of the tensor (\ref{Rproj}) interferes 
with the infrared analysis of the coupled gluon-ghost system
(see also ref. \cite{Hauck:1998sm} for a corresponding discussion in a much
simpler truncation scheme). 

In order to study this problem more carefully we will contract the Lorentz
indices of eq.~(\ref{YM-DSE-gluon}) with the one-parameter family of tensors
\begin{equation}
{\mathcal P}^{(\zeta )}_{\mu\nu} (p) = \delta_{\mu\nu} - 
\zeta \,\frac{p_\mu p_\nu}{p^2} 
\, . 
\label{Paproj}
\end{equation} 
This allows us to interpolate continuously from the tensor (\ref{Rproj}) to
the transversal one (with $\zeta =1$). We will then encounter quadratic divergences
proportional to the factor $(4-\zeta)$, which can be identified unambiguously
and removed by hand. We will be able to show that this procedure restores
the correct perturbative behaviour of the equations even with a finite cutoff $\Lambda$.

Having removed all quadratic divergences we are then in a position to evaluate 
the remaining degree of breaking gauge invariance. As a completely transversal
right hand side would be independent of $\zeta$ after contraction with the projector
(\ref{Paproj}), the variation of our solutions with $\zeta$ is a measure for
the influence of the artificial longitudinal terms on the right hand side of the equation. 

\section{The ghost-gluon vertex in DSE studies and the running coupling \label{gg-vertex-sec}}

We now focus
on two issues that are connected with the ghost-gluon vertex of Landau gauge. 
We will argue for the surprising fact, that one can safely
use a bare ghost-gluon vertex even in the non-perturbative momentum region
of the DSEs.
Furthermore we describe, how one is able to relate
the running coupling of the strong interaction to the ghost and gluon dressing
functions by the renormalisation properties of the ghost-gluon vertex.

\subsection{Dressing the ghost-gluon vertex \label{vertex-dressing}}
\begin{figure}[t]
\vspace{0.5cm}
\centerline{
\epsfig{file=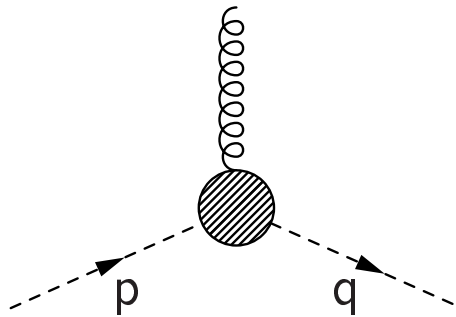}
}
\caption{\sf \label{gg-vertex} A diagrammatical representation of the ghost-gluon vertex.}
\end{figure}
For our further treatment of the ghost and gluon system of equations, 
(\ref{YM-DSE-ghost}) and (\ref{YM-DSE-gluon}), we have to specify explicit forms
for the dressed ghost-gluon vertex $\Gamma_\nu(q,p)$ and the dressed three-gluon
vertex $\Gamma_{\rho \nu \sigma}(q,p)$. As has already been mentioned in the introduction
to this chapter, the ghost-gluon vertex does not attribute an independent
ultraviolet divergence in Landau gauge, {\it i.e.\/} one has $\widetilde Z_1 = 1$ 
\cite{Taylor:1971ff}. Therefore a truncation based on the tree-level form
for the ghost-gluon vertex function, 
\beq
\Gamma_\mu (q,p) = iq_\mu 
\label{bare-gg-vertex}
\eeq
is compatible
with the desired short distance behaviour of the solutions. Here the momentum
$q_\mu$ is the momentum of the outgoing ghost, see Fig. \ref{gg-vertex}.
Thus we obtain the correct
ultraviolet behaviour of the ghost loop in the gluon equation (\ref{YM-DSE-gluon})
and the dressing loop in the ghost equation (\ref{YM-DSE-ghost}),
as will be shown explicitly below. 

However, as the effects of non-perturbative vertex dressing are supposed to
be most pronounced in the infrared, one might wonder whether the tree-level
form of the ghost-gluon vertex leads to a sensible infrared behaviour of these equations.
Furthermore the bare vertex violates the {\it Slavnov-Taylor identity} (STI),
which restricts
that part of the ghost-gluon vertex which is longitudinal in the gluon momentum.
Such an identity is the manifestation of gauge invariance and 
can be derived using the BRS-invariance of the gauge-fixed Lagrangian of Yang--Mills theory. 
 
Considering this, obviously the best way to obtain a
properly dressed {\it ansatz} for the ghost-gluon vertex is to solve the corresponding STI.
This strategy has been followed by 
von Smekal, Hauck and Alkofer in \cite{vonSmekal:1997is,vonSmekal:1998is}.
On the level of connected Green's functions the STI for the 
ghost-gluon vertex in general linear covariant gauges reads \cite{vonSmekal:1998is}
\beq
\frac{1}{\lambda} \langle c^c(z) \bar{c}^b(y) \partial A^a(x) \rangle
-\frac{1}{\lambda} \langle c^c(z) \bar{c}^a(x) \partial A^b(y) \rangle = 
-\frac{g}{2} f^{cde} \langle c^d(z) c^e(z) \bar{c}^a(x) \bar{c}^b(y) \rangle.
\label{ghost-gluon-STI}
\eeq
Here we have the spatial coordinates $x$, $y$ and $z$, the bare coupling $g$, the gauge
parameter $\lambda$ and the
real structure constant $f^{cde}$ of the gauge group SU(3). 
The left hand side of this equation can be decomposed into the full ghost-gluon vertex
and respective propagators. However, the right hand side contains the connected ghost-ghost
scattering kernel, which is completely unknown. 

By neglecting this irreducible correlation
von Smekal, Hauck and Alkofer were able to construct a vertex
{\it ansatz} which solves the resulting approximate STI. Together with a similar construction for
the three-gluon vertex they obtained a closed system of equations which has been solved
numerically using an angular approximation (c.f. subsection \ref{lorenz-trunc}). 
The results, an infrared vanishing 
gluon propagator and an highly singular ghost in the infrared
have been confirmed since in other DSE-calculations
as well as lattice Monte-Carlo simulations 
\cite{Mandula:1999nj,Bonnet:2000kw,Bonnet:2001uh,Langfeld:2001cz}. 
Thus the old idea of {\it infrared slavery} based on the notion of an 
infrared divergent gluon propagator has been abandoned since.    

However, as became clear later, the vertex construction of  
refs.~\cite{vonSmekal:1997is,vonSmekal:1998is} is somewhat problematic. It has been
shown in ref.~\cite{Atkinson:1998tu} that this vertex causes inconsistencies in the infrared
behaviour of the ghost equation once the angular integrals of the dressing loop are treated
exactly. Furthermore it has been shown in refs.~\cite{Watson:1999ha,watson} that the neglection of
the irreducible ghost-ghost scattering kernel in the identity (\ref{ghost-gluon-STI})
is at odds with perturbation theory. On the other hand it is hard to see, 
how one could include this irreducible correlation and thus improve the construction
of refs.~\cite{vonSmekal:1997is,vonSmekal:1998is}. 

Therefore Atkinson and Bloch chose a different strategy  
and employed a bare ghost-gluon vertex in their truncation scheme 
\cite{Atkinson:1998tu,Atkinson:1998zc}, which
keeps only the ghost loop in the gluon equation\footnote{Although there is an
attempt in ref.~\cite{Atkinson:1998tu} to include the gluon loop as well, the authors themselves
note an inconsistency between their construction 
and perturbation theory. Thus the focus
was mainly on the 'ghost-loop only' situation.}. The numerical calculations in this scheme
are also obtained using angular approximations in the integrals. 
Amazingly, though, the bare vertex scheme and the one from 
\cite{vonSmekal:1997is,vonSmekal:1998is} provide results with qualitatively similar
infrared behaviour: the gluon propagator vanishes in the infrared and the ghost propagator
is highly singular there. 

The surprising conclusion from the comparison of these two truncation schemes is, that
a bare ghost-gluon vertex (\ref{bare-gg-vertex}) 
is not only capable of providing the correct
ultraviolet behaviour of the ghost loop in the gluon equation (\ref{YM-DSE-gluon})
and the dressing loop in the ghost equation (\ref{YM-DSE-ghost}), but in addition leads
to a satisfactory infrared behaviour of the equations in accordance with
lattice Monte--Carlo simulations. Consequently recent analytical
infrared investigations concentrated on the bare vertex 
\cite{Atkinson:1998zc,Zwanziger:2001kw}. The influence of multiplicative
corrections to the bare vertex has been assessed in refs.~\cite{Lerche:2002ep,Lerche}
and are found to be irrelevant for the qualitative behaviour in the infrared.

For our novel truncation scheme, detailed in section \ref{truncation}, we will therefore
use the bare ghost-gluon vertex (\ref{bare-gg-vertex}), keeping in mind that we have
to check for the influence of artificial longitudinal terms due to the violation
of the Slavnov-Taylor identity (\ref{ghost-gluon-STI}). As a major improvement to previous
calculations we will be able to overcome the angular approximation and give solutions
for the ghost and gluon dressing functions which include the full angular dependence
of the loops in the equations.

\subsection{The running coupling of the strong interaction \label{coupling-sec}}

Before we give the details of our truncation scheme there is still more to be learned from the 
ghost-gluon vertex in Landau gauge. In the following we consider the renormalisation of the vertex
term in the unrenormalised Lagrangian, eq.~(\ref{Lagrangian-gen}). In Landau gauge we have
\beq
{\cal{L}}_{ghost-gluon} = g f^{abc} \partial_\mu \bar{c}^a A_\mu^c c^b \;,
\label{L-bare}   
\eeq
which is identical to the respective term in the Faddeev-Popov Lagrangian, eq.~(\ref{Lagrangian}).
Multiplicative
renormalisability means that it is possible to render all Green's functions finite
by renormalising the fields and parameters of the Lagrangian without changing
its form. Recall that the coupling $g$, the gluon field and the ghost field 
are renormalised according to ({\it c.f.} eq.~(\ref{rescaling}))
\beqa
A_\mu^a &\rightarrow& \sqrt{Z_3}A_\mu^a \,, \label{a-field-ren}\\
\bar{c}^a&\rightarrow& \sqrt{\tilde{Z}_3} \bar{c}^a \,, \\
c^a &\rightarrow& \sqrt{\tilde{Z}_3} c^a \,, \label{c-field-ren}\\
g &\rightarrow& Z_g g \,, \label{g-ren}
\eeqa
where we have unrenormalised objects on the left hand side and renormalised ones
of the right hand side of the relations. Furthermore, by definition, the
vertex renormalisation constant $\tilde{Z}_1$ is related to the renormalisation
constants of the constituent fields 
of the vertex by
\beq
\tilde{Z}_1=Z_g \tilde{Z}_3 Z_3^{1/2}.
\eeq
We then have the renormalised ghost-gluon vertex part of the Lagrangian
\beq
{\cal{L}}_{ghost-gluon}^R = \tilde{Z}_1 g f^{abc} \partial_\mu \bar{c}^a A_\mu^c c^b \,,   
\eeq
which has the same form as the respective term (\ref{L-bare}) in the bare Lagrangian. 

We are now able to exploit the fact again that $\tilde{Z}_1=1$ in Landau gauge 
\cite{Taylor:1971ff}. As the strong running coupling is defined by 
$\alpha = g^2/(4\pi)$, we obtain the relation
\beq
\alpha(\Lambda^2) = \frac{\alpha(\mu^2)}{\tilde{Z}_3^2(\mu^2,\Lambda^2) Z_3(\mu^2,\Lambda^2)}
\label{coup}
\eeq
from the renormalisation of the coupling, eq.~(\ref{g-ren}). Here we gave the explicit
arguments of the renormalised coupling $\alpha(\mu^2)$, evaluated at the 
renormalisation point $\mu$, and the bare 
coupling $\alpha(\Lambda^2)$, which depends on the cutoff $\Lambda$ of our theory. 
From the relations (\ref{a-field-ren}) and (\ref{c-field-ren}) we furthermore
infer that the ghost and gluon dressing functions, $G(p^2)$ and $Z(p^2)$,
are renormalised according to
\beqa
G(p^2,\Lambda^2) &=& G(p^2,\mu^2) \,\tilde{Z}_3(\mu^2,\Lambda^2) \,, \nonumber\\  
Z(p^2,\Lambda^2) &=& Z(p^2,\mu^2)\, {Z}_3(\mu^2,\Lambda^2) \,, 
\label{zg-ren}  
\eeqa
where the unrenormalised quantities are on the left hand side of the equations.
Certainly the renormalisation point $\mu$ is completely free. We are thus
allowed to substitute these relations into eq.~(\ref{coup}) 
and renormalise once at an arbitrary point $\mu$ and once at the specific value 
$p$. We obtain
\beq
\alpha(\Lambda^2)\, G^2(p^2,\Lambda^2)\, Z(p^2,\Lambda^2) 
= \alpha(\mu^2)\, G^2(p^2,\mu^2) \, Z(p^2,\mu^2)
= \alpha(p^2) \, G^2(p^2,p^2) \, Z(p^2,p^2)
\label{alp}
\eeq
It suffices now to impose the non-perturbative renormalisation condition
\beq
G^2(p^2,p^2)\,Z(p^2,p^2)=G^2(\mu^2,\mu^2)\,Z(\mu^2,\mu^2)=1
\eeq 
on equation (\ref{alp}) to arrive at
\beq
\setlength{\fboxsep}{3mm}
\fbox{$ \displaystyle \alpha(p^2) = \alpha(\mu^2) \, G^2(p^2,\mu^2) \, Z(p^2,\mu^2)$}
\label{coupling}
\eeq
This is a defining equation for the {\it non-perturbative running coupling}
$\alpha(p^2)$ of Landau gauge QCD \cite{vonSmekal:1997is,vonSmekal:1998is}. Note that the
running coupling $\alpha(p^2)$ defined this way does not depend on the arbitrary 
renormalisation point $\mu$. This is equivalent to saying 
that the right hand side of this equation 
is a renormalisation group invariant \cite{Mandelstam:1979xd}.

\section{The truncation scheme \label{truncation}}

Having spent some time with the discussion of Landau gauge we now
return to the coupled set of eqs.~(\ref{YM-DSE-ghost}) and (\ref{YM-DSE-gluon})
and specify explicit expressions for the ghost-gluon vertex $\Gamma_\nu(q,p)$
and the three-gluon vertex $\Gamma_{\rho \nu \sigma}(q,p)$. In the following we
will show, that the bare ghost-gluon vertex
\beq
\Gamma_\mu (q,p) = iq_\mu 
\label{bare-gg-vertex2}
\eeq
and the construction
\beq
\Gamma_{\rho \nu \sigma}(q,p) = \frac{1}{Z_1(\mu^2,\Lambda^2)} \, 
\frac{G(q^2)^{(1-a/\delta-2a)}}{Z(q^2)^{(1+a)}}
\frac{G((q-p)^2)^{(1-b/\delta-2b)}}{Z((q-p)^2)^{(1+b)}} \, \Gamma^{(0)}_{\rho \nu \sigma}(q,p)
\label{3g-vertex}
\eeq
with the bare three-gluon vertex $\Gamma^{(0)}_{\rho \nu \sigma}$ given in 
eq.~(\ref{three-gluon-app})
and the new parameters $a$ and $b$ lead to the correct one-loop anomalous dimensions of the
dressing functions in the ultraviolet, provided the quadratic divergences have been removed.
This is true for arbitrary values of the
parameters $a$ and $b$, although we will later
argue for the specific values $a=b=3\delta$, where $\delta$ is the anomalous dimension
of the ghost. 

For simplicity we introduce the abbreviations $x:=p^2$, $y:=q^2$ and $z:=k^2=(q-p)^2$ for
the squared momenta appearing as arguments of the dressing functions. 
Furthermore $s:=\mu^2$ and $L:=\Lambda^2$ denote the squared 
renormalisation point and the squared momentum cutoff of the theory.  
Substituting the two vertices in the ghost and gluon system (\ref{YM-DSE-ghost}) and
(\ref{YM-DSE-gluon}) we then arrive at
\setlength{\jot}{1mm}
\begin{eqnarray} 
\frac{1}{G(x)} &=& \tilde{Z}_3 - g^2N_c \int \frac{d^4q}{(2 \pi)^4}
\frac{K(x,y,z)}{xy}
G(y) Z(z) \; , \label{ghostbare} \\ 
\frac{1}{Z(x)} &=& {Z}_3 + g^2\frac{N_c}{3} 
\int \frac{d^4q}{(2 \pi)^4} \frac{M(x,y,z)}{xy} G(y) G(z) \nonumber\\
&&\hspace{0.6cm}+ 
 g^2 \frac{N_c}{3} \int \frac{d^4q}{(2 \pi)^4} 
\frac{Q(x,y,z)}{xy} \frac{G(y)^{(1-a/\delta-2a)}}{Z(y)^{a}}
\frac{G(z)^{(1-b/\delta-2b)}}{Z(z)^{b}} 
 \; .
\label{gluonbare} 
\end{eqnarray} 
The kernels ordered with respect to powers of $z$ have the form:
\begin{eqnarray}
K(x,y,z) &=& \frac{1}{z^2}\left(-\frac{(x-y)^2}{4}\right) + 
\frac{1}{z}\left(\frac{x+y}{2}\right)-\frac{1}{4} \,,\nonumber\\
M(x,y,z) &=& \frac{1}{z} \left( \frac{\zeta-2}{4}x + 
\frac{y}{2} - \frac{\zeta}{4}\frac{y^2}{x}\right)
+\frac{1}{2} + \frac{\zeta}{2}\frac{y}{x} - \frac{\zeta}{4}\frac{z}{x} \,, \nonumber\\
Q(x,y,z) &=& \frac{1}{z^2} 
\left( \frac{1}{8}\frac{x^3}{y} + x^2 -\frac{19-\zeta}{8}xy + 
\frac{5-\zeta}{4}y^2
+\frac{\zeta}{8}\frac{y^3}{x} \right)\nonumber\\
&& +\frac{1}{z} \left( \frac{x^2}{y} - \frac{15+\zeta}{4}x-
\frac{17-\zeta}{4}y+\zeta\frac{y^2}{x}\right)\nonumber\\
&& - \left( \frac{19-\zeta}{8}\frac{x}{y}+\frac{17-\zeta}{4}+
\frac{9\zeta}{4}\frac{y}{x} \right) 
+ z\left(\frac{\zeta}{x}+\frac{5-\zeta}{4y}\right) + z^2\frac{\zeta}{8xy} \,.
\hspace*{0.5cm} \label{new_kernels}
\end{eqnarray}
 
\subsection{Ultraviolet analysis \label{ultraviolet}}

In order to identify the quadratically divergent terms in the kernels $K$, $M$ and $Q$ we
now analyse eqs.~(\ref{ghostbare}) and (\ref{gluonbare}) in the limit of large momenta $x$.
It is known from resummed perturbation theory (see e.g. \cite{Muta:1998vi})
that the behaviour of the dressing functions
for large Euclidean momenta can be described as 
\setlength{\jot}{0mm}
\begin{eqnarray}
Z(x) &=& Z(s) \left[ \omega \log\left(\frac{x}{s}\right)+1 \right]^\gamma  \; ,
\label{gluon_uv}\\
G(x) &=& G(s) \left[ \omega \log\left(\frac{x}{s}\right)+1 \right]^\delta  \; .
\label{ghost_uv}
\end{eqnarray}
Here $Z(s)$ and $G(s)$ denote the value of the dressing functions at some
renormalisation point $s:=\mu^2$ and $\gamma$ and $\delta$ are the respective 
anomalous dimensions. To one loop one has $\delta = - 9/44$ and $\gamma = - 1
-2\delta=-13/22$ for arbitrary number of colours $N_c$ and no quarks, $N_f=0$. 
Furthermore, $\omega = {11N_c}\alpha(s)/{12\pi}$.

The slowly varying logarithmic behaviour of the dressing functions in the ultraviolet 
justifies the angular approximation
\beq
Z(z),G(z) \longrightarrow Z(y),G(y) \hspace*{1cm} \mbox{for} \:\:\:  y>x
\eeq
in the ultraviolet. The angular integrals in eqs.~(\ref{ghostbare}) and (\ref{gluonbare})
can then be trivially calculated using the angular integration formulae of appendix \ref{angular}.
Furthermore, as the cutoff $L=\Lambda^2$ can be chosen arbitrary large, the 
integrals will be dominated by the part from $x$ to $L$. We then obtain
\begin{eqnarray} 
\frac{1}{G(x)} &=& \tilde{Z}_3 - g^2\frac{N_c}{16 \pi^2} \int \frac{dy}{x}
\frac{3}{4y} G(y) Z(y) \; , \label{ghost-UV} \\ 
\frac{1}{Z(x)} &=& {Z}_3 + g^2\frac{N_c}{48 \pi^2} 
\int \frac{dy}{x} \left[\left(\frac{4-\zeta}{4} + \frac{\zeta-2}{4}\frac{x}{y}
\right) G^2(y) \right. \nonumber \\
&&\hspace*{2cm}+\left.\left(\frac{-3(4-\zeta)}{2} - 
\frac{\zeta+24}{4}\frac{x}{y}
+\frac{7}{8}\frac{x^2}{y^2}\right) 
\frac{G(y)^{(2-(a+b)/\delta-2(a+b))}}{Z(y)^{(a+b)}} \right]
 \, . \hspace{0.8cm}
\label{gluon-UV} 
\end{eqnarray} 
We are now able to identify the quadratic divergences in the gluon equation, which are the
two terms independent of the integration momentum. Both, the one from the ghost loop and the
one from the gluon loop, are proportional to $(4-\zeta)$ and therefore vanish when we use the
Brown-Pennington projector, eq.~(\ref{Rproj}). For general values of $\zeta$ we have to 
subtract
these terms by hand. However, this cannot be done straightforwardly at the
level of integrands: Such a procedure would disturb the infrared properties of
the Dyson--Schwinger equations. Since we anticipate from previous studies and
analytic work \cite{Zwanziger:2001kw,Lerche:2002ep} that the ghost loop is the
leading contribution in the infrared the natural place to subtract the
quadratically ultraviolet divergent constant is the gluon loop. We do this 
by employing the substitution 
\begin{equation}
Q(x,y,z) \,\, \to \,\,\tilde{Q} (x,y,z) = Q(x,y,z) + \frac 5 4 (4-\zeta )
\label{Q-tilde}
\end{equation}
in eq.\ (\ref{gluonbare}). Only the problematic terms in eq.~(\ref{gluon-UV}) then disappear.

Having subtracted the quadratic divergences from the ghost and gluon system we now
check for the logarithmic divergences, which should match to perturbation theory. 
We choose the perturbative renormalisation condition $G(s)=Z(s)=1$ at a large
Euclidean renormalisation point $s=\mu^2$ and plug 
the perturbative expressions (\ref{gluon_uv}) and (\ref{ghost_uv}) in eqs.~(\ref{ghost-UV}),
(\ref{gluon-UV}). Thus we arrive at
\beqa
\left[ \omega \log\left(\frac{x}{s}\right)+1 \right]^{-\delta} &=&
\tilde{Z}_3 - \frac{3 N_c g^2}{64 \pi^2 \omega(\gamma+\delta+1)} 
\times \nonumber\\
&&\hspace*{0.8cm}\left\{
\left[ \omega \log\left(\frac{L}{s}\right)+1 \right]^{\gamma+\delta+1} - 
\left[ \omega \log\left(\frac{x}{s}\right)+1 \right]^{\gamma+\delta+1} \right\}\,, 
\hspace*{0.9cm} \nonumber\\
\left[ \omega \log\left(\frac{x}{s}\right)+1 \right]^{-\gamma} &=&
Z_3 + \left(\frac{N_c g^2}{96 \pi^2 \omega (2\delta+1)} 
- \frac{7 N_c g^2}{48 \pi^2 \omega (2\delta+1)}\right)\times \nonumber \\
&&\hspace*{0.8cm}\left\{
\left[ \omega \log\left(\frac{L}{s}\right)+1 \right]^{2\delta+1} -
\left[ \omega \log\left(\frac{x}{s}\right)+1 \right]^{2\delta+1} \right\} \,. 
\label{gluoneq_uv}
\end{eqnarray}
Note that these equations are completely independent of the parameters $a$ and $b$ 
in the three-gluon vertex, eq.~(\ref{3g-vertex}). Note also, that the ultraviolet 
behaviour of the equations is independent of the parameter $\zeta$ in the projector
(\ref{Paproj}). We are therefore left with a transversal structure in the gluon
equation for ultraviolet momenta. The renormalisation constants $Z_3(s,L)$ and 
$\tilde{Z}_3(s,L)$ cancel the cutoff dependence, {\it i.e.} the respective first terms in the
brackets. Thus, the power and the prefactor of the second term have to match with
the left hand side of the equations. This leads to three conditions:
\begin{eqnarray}
\gamma+2\delta+1 &=& 0 \; , \label{gd1}\\
\frac{3}{4 (\gamma+\delta+1)} \frac{N_c g^2}{ 16\pi^2 \omega} &=& 1 \;, 
\label{go1}\\
\frac{13}{6 \: (2\delta+1)} \frac{N_c g^2}{16\pi^2 \omega} &=& 1 \; .
\end{eqnarray}
Eq.\ (\ref{gd1}) is of course nothing else but consistency of the ghost
equation with one-loop scaling. All three equations together result in the correct
anomalous dimensions $\gamma=-13/22$ and $\delta=-9/44$ for an arbitrary number of
colours and zero flavours. 

Having established this result let us pause for a moment and reflect our construction
(\ref{3g-vertex}) for the three-gluon vertex. There are two possible ways to
obtain this construction and we leave it to the taste of the reader which philosophy
to prefer. The {\it first} way to see eq.~(\ref{3g-vertex}) is to take it at face value as
a minimally dressed vertex {\it ansatz}, constructed in such a way as to obtain the correct 
one-loop scaling of the gluon loop. Then the parameters $a$ and $b$ are completely
free. Note that the choice
$a=b=0$ corresponds to the truncation scheme of \cite{Bloch:2001wz} 
whereas $a=3\delta, b=0$ together with the appropriate vertex dressings 
reproduces case c) of ref.\ \cite{vonSmekal:1997is}. 

The {\it second} point of view, which is to be found in reference 
\cite{vonSmekal:1997is}, is to employ a bare three-gluon vertex,
$\Gamma^0_{\rho \nu \sigma}(q,p)$, and then ask the question how the renormalisation
constant $Z_1$ has to behave, given that the theory should have the correct perturbative limit.
Certainly, as both vertices violate their respective Slavnov-Taylor identities, there is no
reason why the vertex renormalisation constant $Z_1$ should obey the corresponding
identity $Z_1=Z_3/\tilde{Z}_3$. For the present truncation scheme the answer is, that the 
constant $Z_1$ acquires a momentum dependence according to   
\begin{equation} 
Z_1 \;\; \longrightarrow \;\; {\cZ}_1(x,y,z;s) = 
\frac{G(y,s)^{(1-a/\delta-2a)}}{Z(y,s)^{(1+a)}}
\frac{G(z,s)^{(1-b/\delta-2b)}}{Z(z,s)^{(1+b)}} \, .
\label{Z1_ansatz}
\end{equation}
This is precisely the form required to transform our vertex (\ref{3g-vertex}) to the bare one and
nevertheless obtain eqs.~(\ref{ghostbare}) and (\ref{gluonbare}).   

A reasonable choice of parameters is then one which keeps ${\cZ}_1$ as
weakly dependent as possible on the momenta $y$ and $z$,
{\it c.f.}~Fig.~\ref{z1.dat} in appendix \ref{one-loop}. 
The infrared behaviour of the gluon loop in the gluon equation depends
strongly on $a$ and $b$. Setting $b=0$ one can distinguish three
cases: For $a<0$ the gluon loop is subleading in the infrared, for
$a=0$ as in ref.\ \cite{Bloch:2001wz} the gluon loop produces the same power as
the ghost loop, for $a>0$ the gluon loop becomes the leading term in the
infrared. In the latter case we did not find a solution to the coupled
gluon-ghost system.
In Appendix \ref{one-loop} we demonstrate that $a=b=3\delta$ minimises the momentum 
dependence of ${\cZ}_1$. Thus we use these values except stated
otherwise explicitly.

\subsection{Infrared analysis \label{infrared}}

The leading infrared behaviour of the propagator functions in this truncation scheme 
for the special case of the transverse projector ($\zeta=1$) has been determined recently 
\cite{Zwanziger:2001kw,Lerche:2002ep}. Our analysis in this subsection is valid for
general values of the parameter $\zeta$ and furthermore includes subleading
contributions \cite{Fischer:2002hn}. The general assumption at the beginning 
of all analytic infrared investigations is, that the ghost and gluon dressing functions,
G and Z, behave like power laws in the infrared:
\beqa
Z(x) &=& A x^{\kappa_1}, \nonumber\\
G(x) &=& B x^{\kappa_2}. \label{zg-power}
\eeqa
This assumption is justified, if we are able to show that the {\it ansatz} (\ref{zg-power}) solves
the ghost and gluon system, eqs.~(\ref{ghostbare}), (\ref{gluonbare}), 
self-consistently in the infrared. As all loop integrals in the equations are dominated by
contributions around the external momentum $x=p^2$, we are allowed to substitute the power laws for
the whole momentum range up to the cutoff $L=\Lambda^2$. Errors due to this approximation
are subleading in the infrared. Furthermore, as has been shown 
in detail in reference \cite{Lerche:2002ep}, the renormalisation constants $Z_3$ and $\tilde{Z}_3$
can be dropped for very small momenta $x$: They are either subleading in the infrared
(gluon equation) or have to be zero when the renormalisation takes place at
$\mu=0$ (ghost equation). We will explain this point more precisely later on.  
Plugging the power laws (\ref{zg-power}) 
into eqs.~(\ref{ghostbare}), (\ref{gluonbare}) we then arrive at
\setlength{\jot}{2mm}
\begin{eqnarray} 
\frac{1}{B x^{\kappa_2}} &=& 
- g^2N_c AB \int \frac{d^4q}{(2 \pi)^4}
\frac{K(x,y,z)}{xy}
 z^{\kappa_1} y^{\kappa_2}  \,, \label{ghostbare-ir}\\  
\frac{1}{A x^{\kappa_1}} &=& 
g^2\frac{N_c}{3} B^2 
\int \frac{d^4q}{(2 \pi)^4} \frac{M(x,y,z)}{xy} y^{\kappa_2} z^{\kappa_2}  \nonumber\\
&&+ 
 g^2 \frac{N_c}{3} B^{-4-12\delta} A^{-6\delta}\int \frac{d^4q}{(2 \pi)^4} 
\frac{\tilde{Q}(x,y,z)}{xy} y^{(-2-6\delta){\kappa_2} - 3\delta {\kappa_1}}
z^{(-2-6\delta){\kappa_2} - 3\delta {\kappa_1}} \,, \hspace*{0.8cm}
\label{gluonbare-ir} 
\end{eqnarray} 
with the kernels $K$, $M$ and $\tilde{Q}$ given by eqs.~(\ref{new_kernels}), (\ref{Q-tilde}) and
the parameters $a=b=3\delta$ as motivated in the last subsection. 
\setlength{\jot}{0mm}

We first investigate the ghost equation (\ref{ghostbare-ir}). Shifting the cutoff $L=\Lambda^2$ 
to infinity
we are able to use the 
formula \cite{Lerche:2002ep}
\begin{equation}
\int d^4q \: y^a z^b = \pi^2 x^{2+a+b} \frac{\Gamma(2+a)\Gamma(2+b)\Gamma(-a-b-2)}
{\Gamma(-a)\Gamma(-b)\Gamma(4+a+b)}
\label{irintegral}
\end{equation}
for the integration of the dressing loop on the right hand side of the equation. The straightforward
but tedious algebra is done with the help of
the algebraic manipulation program FORM \cite{Vermaseren:2000nd}. We obtain
\beqa
\frac{1}{B x^{\kappa_2}} &=&  
-  x^{\kappa_1+\kappa_2} \:
\frac{g^2N_c AB}{16 \pi^2} \frac {3} 
{2\,(\kappa_1  + \kappa_2 )\,( - 1 + \kappa_1  + \kappa_2 )} \times \nonumber\\
&& \hspace*{5cm} \times \frac{
\Gamma (2 - \kappa_1  - \kappa_2 
)\,\Gamma (1 + \kappa_1 )\,\Gamma (2 + \kappa_2 )}{\Gamma (3 + \kappa_1 
 + \kappa_2 )\,\Gamma (2 - \kappa_1 )\,\Gamma (1 - \kappa_2 )}. \hspace{1cm}{ }
\label{ghostbare-ir2} 
\eeqa
Matching powers of $x$ on both sides of the equation we arrive at the condition
\beq
\kappa_1 = -2\kappa_2 \; .
\label{kappa-relation}
\eeq 
With the definition $\kappa:=-\kappa_2$ we thus have from eqs.~(\ref{zg-power}) the power laws
\beq
\setlength{\fboxsep}{2mm}
\fbox{$ \displaystyle \begin{array}{c} Z(x) = A x^{2\kappa} , \\ G(x) = B x^{-\kappa}. \end{array} $}
\label{zg-power-kappa}
\eeq
Our derivation of these power laws used the special form of a bare ghost-gluon vertex. It has been
shown, however, that relation (\ref{kappa-relation}) holds under the general 
assumption that the ghost-gluon vertex can be expanded in a power series \cite{Watson:2001yv}.
We thus have the interesting situation that the infrared divergence of one of our dressing functions,
$Z$ or $G$, is always connected to the vanishing of the other. This has interesting
consequences for the running coupling $\alpha(x)$. Recall the definition
\beq
\alpha(x) = \alpha(s) \, G^2(x,s) \, Z(x,s)
\label{coupling2}
\eeq
from eq.~(\ref{coupling}) with $x=p^2$ and $s=\mu^2$ and $\alpha(s)=g^2/{4 \pi}$. 
Substituting the power laws 
(\ref{zg-power-kappa}) we obtain
\beq
\alpha(0) =  \frac{g^2}{4 \pi}AB^2.
\eeq
The coupling approaches a finite value in the infrared.
Within the framework of Dyson-Schwinger studies such a behaviour of the running coupling 
has first been found in ref.~\cite{vonSmekal:1997is}.
An infrared fixed point is also found in the flow equation study in ref.~\cite{Gies:2002af} 
and analytic perturbation theory \cite{Milton:1997fc,Milton:1998us,Milton:2000fi,Shirkov:2001sm}. 

Consider now the gluon equation (\ref{gluonbare-ir}). After some algebra using 
formula (\ref{irintegral}) we find, that the ghost loop is proportional to $x^{2\kappa_2}$, whereas
the gluon loop is subleading in the infrared. We thus again obtain relation (\ref{kappa-relation}),
$\kappa_1 = -2\kappa_2$,
as matching condition for the ghost loop and the leading power on the 
left hand side of the equation in the infrared. 
Together with eq.~(\ref{ghostbare-ir2}) we arrive at the
two conditions  
\begin{eqnarray} 
\frac{1}{18}
\frac{(2+\kappa)(1+\kappa)}{(3-2\kappa)}
 &=& \frac{\Gamma^2(2-\kappa)\Gamma(2\kappa)}{\Gamma(4-2\kappa) \Gamma^2(1+\kappa)}
\: \frac{g^2N_c}{48 \pi^2}{AB^2} \; ,
\label{kappa1}\\
\frac{4\kappa-2}{4\zeta \kappa - 4\kappa + 6 - 3\zeta}
&=& \frac{\Gamma^2(2-\kappa)\Gamma(2\kappa)}{\Gamma(4-2\kappa) \Gamma^2(1+\kappa)}
\:\frac{g^2N_c}{48 \pi^2}{AB^2} \; .
\label{kappa2}
\end{eqnarray} 
\begin{figure}[t]
\vspace*{1cm}
\centerline{
\epsfig{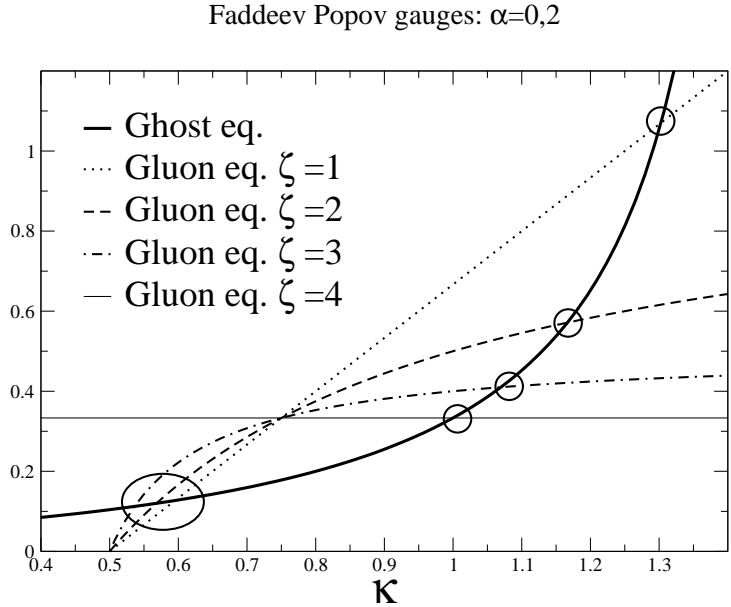}
}
\caption{\sf \label{kappa.dat}
Here the graphical solution to equations (\ref{kappa1}) and (\ref{kappa2}) is shown. 
The thick line represents the
left hand side of equation (\ref{kappa1}), whereas the other four curves depict 
the left hand side of equation 
(\ref{kappa2}) for different values of the parameter $\zeta$. The ellipse marks the 
bulk of solutions between
$\kappa=0.5$ and $\kappa=0.6$ for different $\zeta$, whereas the circles show the 
movement of the solution $\kappa=1.3$ 
for a transverse projector to $\kappa=1$ for the Brown-Pennington case, $\zeta=4$.
}
\end{figure}
from the ghost and gluon equations in the infrared. Equating both left and right hand sides
we are able to determine $\kappa$, see Fig.~\ref{kappa.dat}. For the Brown--Pennington projector, 
{\it i.e.} $\zeta=4$, one then
finds the known solution $\kappa =1 $ \cite{Atkinson:1998tu}. However,
as can be seen immediately, the left hand side of the second equation 
possesses a zero for $\kappa=1/2$ which is cancelled by a pole only for $\zeta=4$. 
Lowering $\zeta$ only slightly a further solution with $\kappa$
slightly larger than 0.5 exists. For the transverse projector, {\it i.e.}
$\zeta=1$, this latter solution becomes 
\beq
\kappa = \frac{93-\sqrt{1201}}{98} \approx 0.595353
\eeq
in accordance with
refs.~\cite{Zwanziger:2001kw,Lerche:2002ep}. Also the solution $\kappa =1 $
changes continuously when lowering $\zeta$. The corresponding $\kappa$  are
then all larger than 1 and contradict the masslessness condition, see section 5.2
of ref.~\cite{Alkofer:2000wg} for a discussion of this condition. One of the main
results of the numerical section of this chapter is that the infrared 
behaviour $\kappa \approx 0.5$ matches
to a numerical solution whereas no numerical solutions could be found with the
infrared behaviour $\kappa \ge 1 $.
This is in perfect accordance with the Kugo-Ojima criterion and Zwanziger's
horizon condition discussed in section \ref{conf}.  

Now let us come back again to the running coupling, eq.~(\ref{coupling2}). 
As can be seen directly from eqs.~(\ref{kappa1}), (\ref{kappa2}),  
the product $N_c g^2 AB^2$ is constant for given
$\kappa$. With $\alpha(0) = {g^2}AB^2/{4 \pi}$ one concludes immediately that
$\alpha(x)$ is proportional to $N_c^{-1}$.  Furthermore, the ghost and gluon
dressing functions $Z(x)$ and $G(x)$ are independent of the number of colours:
$N_c$ enters the Dyson-Schwinger equations only in the combination 
$g^2 N_c$ at our
level of truncation. From the solution $\kappa = 0.595$ of the infrared
analysis with the transverse projector $\zeta=1$ one determines the infrared
fixed point of the running coupling to be 
\beq
\alpha(0) = \frac{2 \pi}{3 \, N_c} \frac{\Gamma(3-2\kappa) \; \Gamma(3+\kappa) \; \Gamma(1+\kappa)}
{\Gamma^2(2-\kappa) \; \Gamma(2\kappa)} \approx  8.915/N_c
\eeq
for general numbers of colours $N_c$.

Having determined the leading infrared behaviour of the ghost and gluon system we
now check for subleading contributions, and judge their importance for the numerical
treatment of the coupled system of equations.
An obvious extension of the power law (\ref{zg-power-kappa}) is the {\it ansatz}
\begin{eqnarray}
Z(x) &=& A x^{2 \kappa} (1+f x^{\rho}) \nonumber\\
G(x) &=& B x^{- \kappa} (1+g x^{\rho}) 
\label{IRansatz}
\end{eqnarray}
which is substituted in the eqs.~(\ref{ghostbare}) and (\ref{gluonbare}).
After integration the conditions on the leading term 
remain unchanged.   
Matching subleading powers leads to the coupled set of homogeneous equations for $f,g$ and $\rho$:
\setlength{\jot}{3mm}
\begin{eqnarray}
\left(3 \nu 
\frac{6 \kappa \,(\kappa-1) \,(-3+2\kappa)}{(\kappa+\rho-2)\,(\kappa+\rho-1)\,(\kappa+\rho)}
\frac{\Gamma(2\kappa)\,\Gamma(2-\kappa+\rho)\,\Gamma(3-\kappa-\rho)}
{\Gamma(4-2\kappa)\,\Gamma(1+\kappa-\rho)\,\Gamma(3+\kappa+\rho)}
-1\right) g && \hspace{1.0cm}\nonumber\\
+ \left( 3 \nu
\frac{3 \,(-2+2\kappa+\rho)}{2\,(\kappa+\rho-2)\,(\kappa+\rho-1)\,(\kappa+\rho)}
\frac{\Gamma(2-\kappa)\,\Gamma(2\kappa+\rho+1)\,\Gamma(3-\kappa-\rho)}
{\Gamma(3-2\kappa-\rho)\,\Gamma(1+\kappa)\,\Gamma(3+\kappa+\rho)}
\right) f &=& 0 \; , \nonumber\\
\left(\nu \frac{4 \zeta \kappa - 4\kappa + 2\rho -2 \zeta \rho +6 - 3\alpha}{2\kappa-\rho-1}
\frac{\Gamma(2-\kappa)\,\Gamma(2\kappa-\rho)\, \Gamma(2-\kappa+\rho)}
{\Gamma(1+\kappa)\,\Gamma(1+\kappa-\rho)\,\Gamma(4-2\kappa+\rho)}
\right) g + f &=& 0 \; . \nonumber \\
\end{eqnarray}
Here $\nu={N_c g^2 A B^2}/{48 \pi^2}={\alpha(0)}/{4\pi}$. There is either the trivial solution $f=g=0$ 
or one has to set the determinant of these linear equations
to zero. We then obtain the results $\rho_{(1)}=0, \rho_{(2)}=0.58377, 
\rho_{(3)}=1.20300$ and several other solutions with higher values of the power.
The solution $\rho_{(1)}=0$ corresponds to the pure power solution. The lowest non-vanishing 
solution, $\rho_{(2)}=0.58377$, is sufficiently high that we safely may neglect it in the numerical 
treatment of the infrared part of the equations. This will be detailed in the next section.
\setlength{\jot}{0mm}

Finally, let us come back to the role of the renormalisation constants $\tilde{Z}_3$ and
${Z}_3$ in eq.~(\ref{ghostbare-ir}), (\ref{gluonbare-ir}). 
There are two possible situations, which have been clarified in 
\cite{Lerche:2002ep}. {\it First}, 
consider an infrared vanishing dressing function, as is the case in the gluon equation.
Then the left hand side of eq.~(\ref{gluonbare-ir}) as well as the loop integral on
the right hand side are diverging, and the $x$-independent constant ${Z}_3$
is subleading and therefore negligible in the infrared. 
{\it Second}, in the ghost equation we have the situation that 
the ghost dressing function $G(x)$ diverges in the infrared. Therefore  
the equation is not renormalisable at the point $\mu^2=0$.
Correspondingly the behaviour of the renormalisation constant $\tilde{Z}_3(\mu^2,\Lambda^2)$ 
is such that for a given cutoff $\Lambda^2$ the renormalisation constant tends to zero as 
$\mu^2 \rightarrow 0$. At the point $\mu^2=0$ we have $\tilde{Z}_3(0,\Lambda^2)=0$, 
the renormalisation process breaks down and
no scale can be generated. The power solution (\ref{zg-power-kappa}) is then not only
an infrared approximation but a solution for the whole momentum range up to infinity.
This in turn implies that the power solution can be determined with $\tilde{Z}_3$ set to zero. 
We conclude that both renormalisation constants, $Z_3$ and $\tilde{Z}_3$, play no role
in the infrared analysis of the ghost and gluon system in accordance with our assumption
above eqs.~(\ref{gluonbare-ir}), (\ref{ghostbare-ir}).  
  
\section{Renormalisation and numerical results \label{YM-ren-results}}

Corresponding to the analytical solution of the equations (\ref{ghostbare}) and (\ref{gluonbare})
in the infrared, our numerical treatment is done without the help of any angular approximations. 
Thus the present investigation is the first calculation of the ghost and gluon dressing functions
which takes into account the full angular dependence in the loops for all momenta 
\cite{Fischer:2002hn}.
The technical details of the necessary numerical procedures to solve 
eqs.~(\ref{ghostbare}), (\ref{gluonbare}) are given in appendix \ref{numerics}.
In the following subsections we describe the renormalisation scheme employed in our calculations
and give the numerical solutions for the ghost and gluon propagators as well as the running coupling.

\subsection{The renormalisation scheme \label{subtraction}}

We apply a MOM regularisation scheme similar to the ones used previously in 
refs.~\cite{vonSmekal:1997is,Atkinson:1998tu}. In such a scheme the equations for the
ghost and gluon dressing functions for the external momentum $x$ and a fixed subtraction scale
$t_{ghost},t_{glue}$ are subtracted from each other. If we write the 
equations (\ref{ghostbare}) and (\ref{gluonbare}) symbolically as
\beqa
\frac{1}{G(x)} &=& \tilde{Z}_3 + \Pi_{ghost}(x) \,, \\
\frac{1}{Z(x)} &=& {Z}_3 + \Pi_{glue}(x) \,,
\eeqa
this procedure yields
\beqa
\frac{1}{G(x)} &=& \frac{1}{G(t_{ghost})} + \Pi_{ghost}(x) - \Pi_{ghost}(t_{ghost}) \,, \\
\frac{1}{Z(x)} &=& \frac{1}{Z(t_{glue})} + \Pi_{glue}(x) - \Pi_{ghost}(t_{glue}).
\eeqa
We now see that the unknown renormalisation constants
$Z_3$ and $\tilde{Z}_3$ drop out and instead of them we have to specify the two
input variables $G(t_{ghost})$ and $Z(t_{glue})$. For numerical reasons it is favourable
to subtract the ghost equation at a very small momentum, $t_{ghost} \rightarrow 0$ 
and the gluon equation at a perturbative scale $l$. 
We thus have to specify values for $Z(l)$ and the 
parameter $B$ in the power law (\ref{zg-power-kappa}) which
describes the infrared behaviour of the ghost dressing function.
In practice we encounter a smooth transition at the infrared matching point $\epsilon^2$ 
(c.f. appendix \ref{numerics}) of our numerical integrals only if $B$ and $Z(l)$ are uniquely
related. This corresponds to the fact that we are not able to implement the
perturbative renormalisation condition $G(\mu^2)=Z(\mu^2)=1$ for general renormalisation
points $\mu$ but only the weaker condition $Z(\mu^2)\,G^2(\mu^2)=1$. A similar observation has
been made in \cite{vonSmekal:1997is,vonSmekal:1998is}. In the actual
calculation we use the arbitrary value $Z(l=(174 \,\mbox{GeV})^2)=0.83$. The value of the 
renormalisation point $\mu$ is given implicitly by specifying the coupling 
$\alpha(\mu^2)=g^2/16\pi^2=0.97$ entering eqs.~(\ref{ghostbare}), (\ref{gluonbare}).

\subsection{Numerical results \label{YM-results}}

For our numerical results we have fixed the momentum scale by calculating
the running coupling for the colour group SU(3), and requiring 
the experimental value
$\alpha(x)=0.118$ at $x=M_Z^2=(91.187 \, \mbox{GeV})^2$ \cite{Hagiwara:2002pw}.   
\begin{figure}[th!]
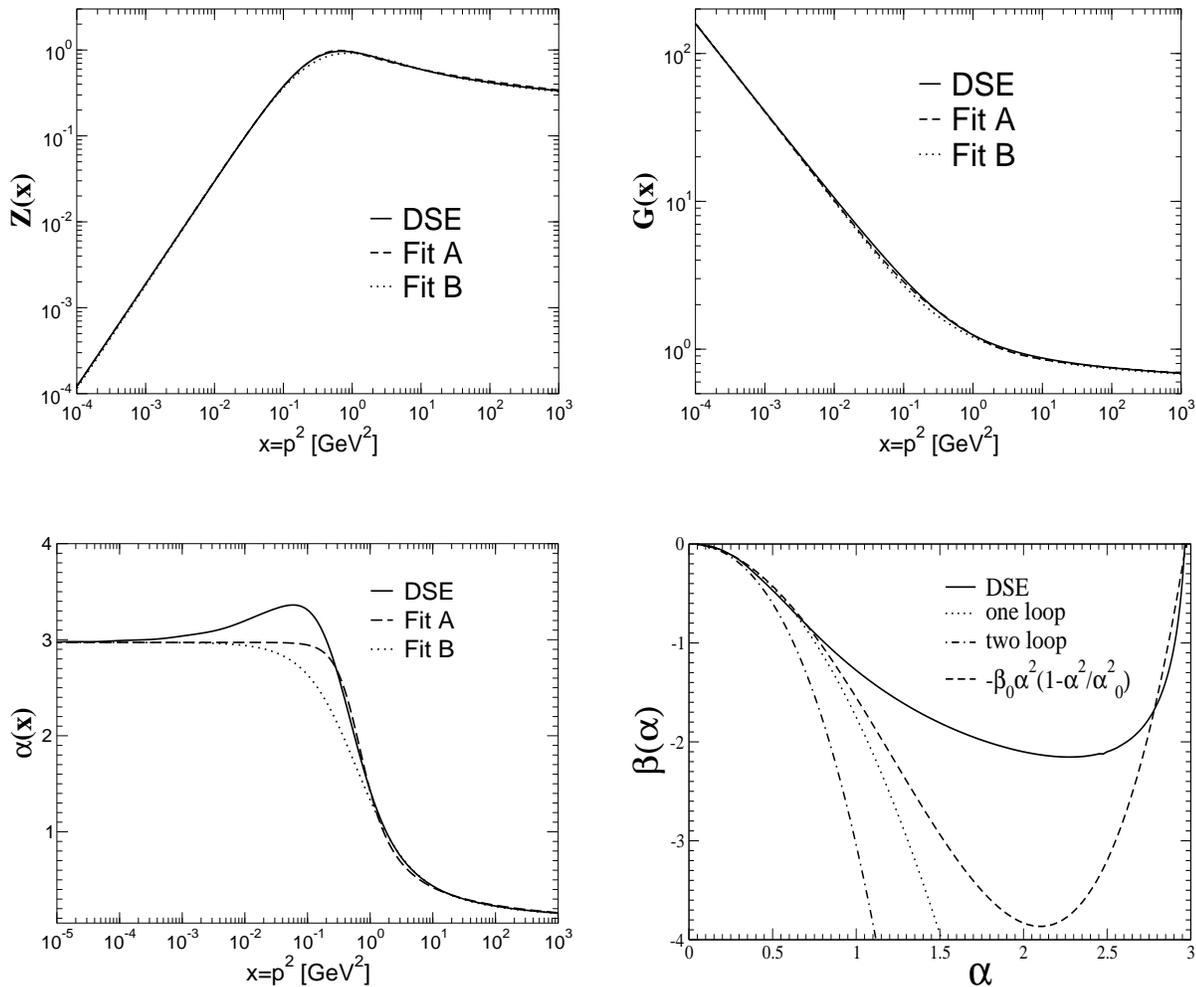

\vspace{0.5cm}
\centerline{
\epsfig{file=cont.main.z.eps,width=7.5cm,height=6cm}
\hspace{0.5cm}
\epsfig{file=cont.main.g.eps,width=7.5cm,height=6cm}
}
\vspace{1cm}
\centerline{
\epsfig{file=cont.main.a.eps,width=7.5cm,height=6cm}
\hspace{0.5cm}
\epsfig{file=beta.eps,width=7.5cm,height=6cm}
}
\caption{\sf \label{new.dat}
Shown are the results for the gluon dressing function $Z$, the ghost dressing 
function $G$ and the running coupling $\alpha$ using a transverse projector, $\zeta=1$. 
The two sets of fit functions are given in eqs.~(\ref{fitA}) and (\ref{fitB}).
The $\beta$-function corresponding to our DSE-solution is compared to the one- and two-loop
expressions as well as to a polynomial in $\alpha$.}
\end{figure}

In Fig.~\ref{new.dat} we show the results for the transverse projector,
$\zeta=1$. The gluon and ghost dressing functions behave power-like for low
momenta with $\kappa=0.595$ and obey one-loop scaling in the ultraviolet as expected.
We found no numerical solution for the second analytical infrared power from
subsection \ref{infrared}, $\kappa \approx 1.3$.
We thus conclude, that only the infrared power $\kappa=0.595$ connects to the numerical solution
for all momenta. This result will be corroborated in chapter \ref{torus}. As has been mentioned in the 
introduction to this chapter first results have been obtained in stochastic quantisation, which
is supposed to solve the Gribov problem \cite{Zwanziger:2002ia}. The infrared analysis of a bare
vertex truncation scheme in stochastic Landau gauge leads to the power $\kappa=0.521$ for the
transversal part of the gluon propagator. This is surprisingly close to the value in Faddeev-Popov
quantisation and suggests the influence of Gribov copies within the Gribov region to be small.   
 
According to the power solution of the dressing functions 
the running coupling has a fixed point in the infrared. Furthermore it decreases
logarithmically in the perturbative regime above several GeV in accordance with perturbation theory.
For intermediate momenta the behaviour of our running coupling induces a node in
the $\beta$-function around  $(100 \mbox{MeV})^2$. This, however,
corresponds to a double valued $\beta$-function, which is a somewhat strange result not expected from
renormalisation group analysis. 
Therefore we regard it as an artefact of our truncation scheme. Recall from section \ref{gluonghost}
that we have omitted the two-loop diagrams in the gluon equation in all our calculations.
Since the two-loop diagrams contain gluon dressing functions only they are subleading in the
infrared in the present framework. Furthermore they are subleading in the ultraviolet 
as can be seen from perturbation theory. Thus the only region where the inclusion of the 
two-loop diagrams could lead to qualitative corrections of our solutions is the intermediate 
momentum regime. This is the region where the bump in the coupling appears.

The $\beta$-functions in the bottom right diagram are defined in appendix
\ref{alpha}. Compared to the one- and two-loop $\beta$-functions of perturbation theory
the $\beta$-function
from our DSE-solution \cite{Alkofer:2002ne}
resembles the scaling behaviour of the one-loop result in the ultraviolet,
that is for small values of $\alpha$. 

The asymptotic behaviour of the solutions can also be seen from the
functional form of our fits. We employ two different fit functions \cite{Alkofer:2002aa} for the running
coupling $\alpha(x)$:
\setlength{\jot}{2mm}
\begin{eqnarray}
\mbox{Fit A:} \quad
\alpha(x) &=& \frac{\alpha(0)}{\ln[e+a_1 (x/\Lambda^2_{QCD})^{a_2}+
b_1(x/\Lambda^2_{QCD})^{b_2}]}\,,
\label{fitA}\\ 
\mbox{Fit B:} \quad
\alpha(x) &=& \frac{1}{a+(x/\Lambda^2_{QCD})^b} 
\bigg[ a \: \alpha(0) + \nonumber\\
&&\hspace*{2cm} \left.\frac{4 \pi}{\beta_0} \left(\frac{1}{\ln(x/\Lambda^2_{QCD})}
- \frac{1}{x/\Lambda_{QCD}^2 -1}\right)(x/\Lambda^2_{QCD})^b \right]\,. \hspace*{0.5cm}  
\label{fitB}
\end{eqnarray}
The value $\alpha(0)=8.915/N_c$ is known from 
the infrared analysis. In both fits the ultraviolet behaviour 
of the solution fixes the scale,  $\Lambda=0.714 \,\mbox{GeV}$. 
Note that we have employed
a MOM scheme, and thus $\Lambda_{QCD}$ has to be interpreted as
$\Lambda_{MOM}^{N_f=0}$, {\it i.e.} this scale has the expected magnitude.
Fit A employs the four additional parameters: 
$a_1=1.106$, $a_2=2.324$,
$b_1=0.004$, $b_2=3.169$.
Fit B has only two free parameters:
$a=1.020$, $b=1.052$. 
The dressing functions $Z(x)$ and $G(x)$ are then described by
\setlength{\jot}{3mm}
\begin{eqnarray}
R(x) &=& \frac{c \,(x/\Lambda^2_{QCD})^{\kappa}+d \,(x/\Lambda^2_{QCD})^{2\kappa}}
{1+ c \,(x/\Lambda^2_{QCD})^{\kappa}+d \,(x/\Lambda^2_{QCD})^{2\kappa}} \,, \nonumber\\
Z(x) &=& \left( \frac{\alpha(x)}{\alpha(\mu)} \right)^{1+2\delta} R^2(x) \,, \nonumber\\ 
G(x) &=& \left( \frac{\alpha(x)}{\alpha(\mu)} \right)^{-\delta} R^{-1}(x)  \,,
\label{zg-fit}
\end{eqnarray}
where $c,d$ are fitting parameters for the auxiliary function $R(x)$. 
They are given by $c=1.269$ and  $d=2.105$. Recall 
from subsection \ref{ultraviolet} that
the anomalous dimension $\gamma$ of the gluon is related to the 
anomalous dimension $\delta$ of the ghost by $\gamma=-1-2\delta$ and $\delta=-9/44$ for
the number of flavours $N_f=0$.
\setlength{\jot}{0mm}
\begin{figure}[th!]
\centerline{
\epsfig{file=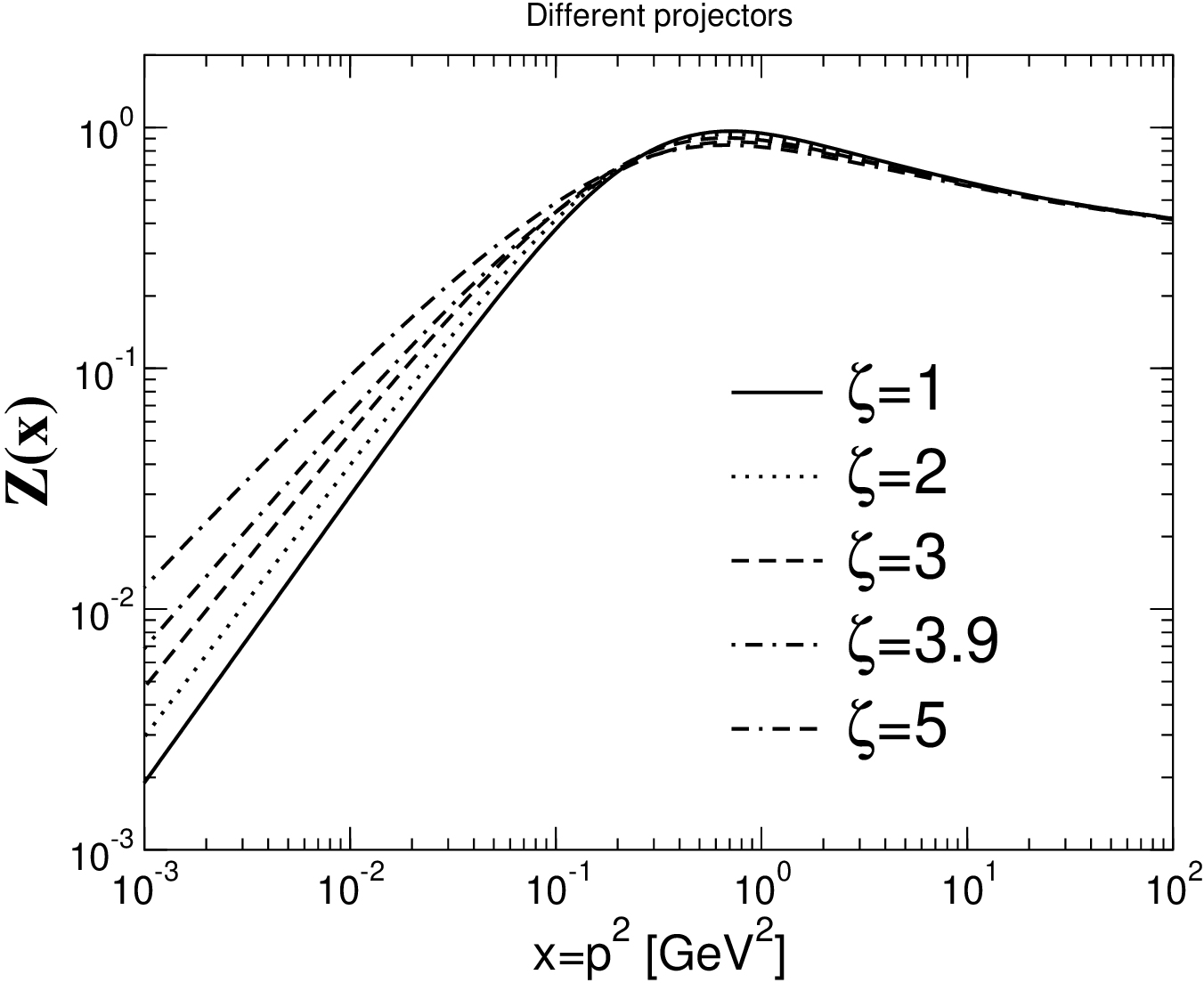,width=7.5cm,height=6cm}
\hspace{0.5cm}
\epsfig{file=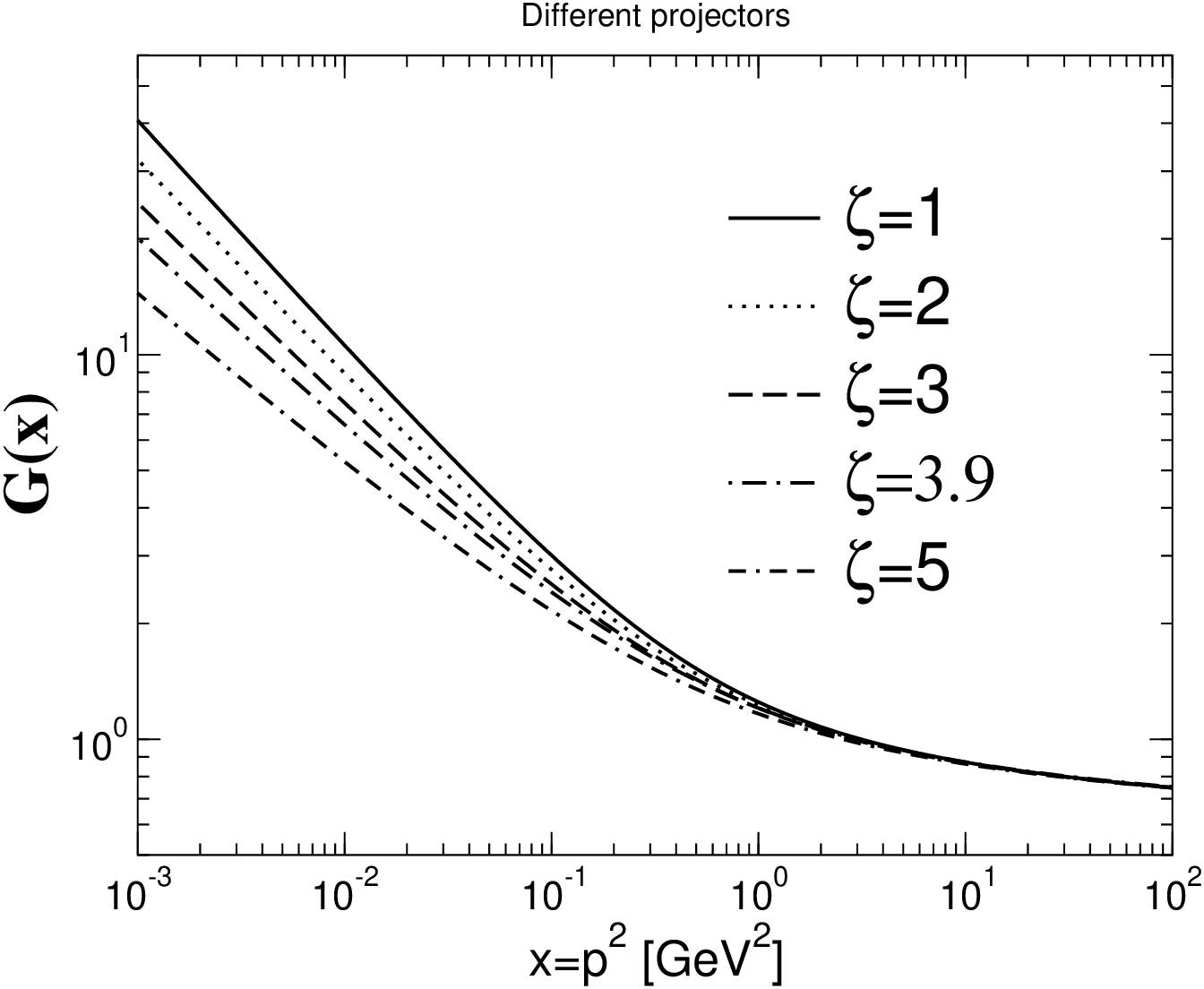,width=7.5cm,height=6cm}
}
\vspace{1cm}
\centerline{
\epsfig{file=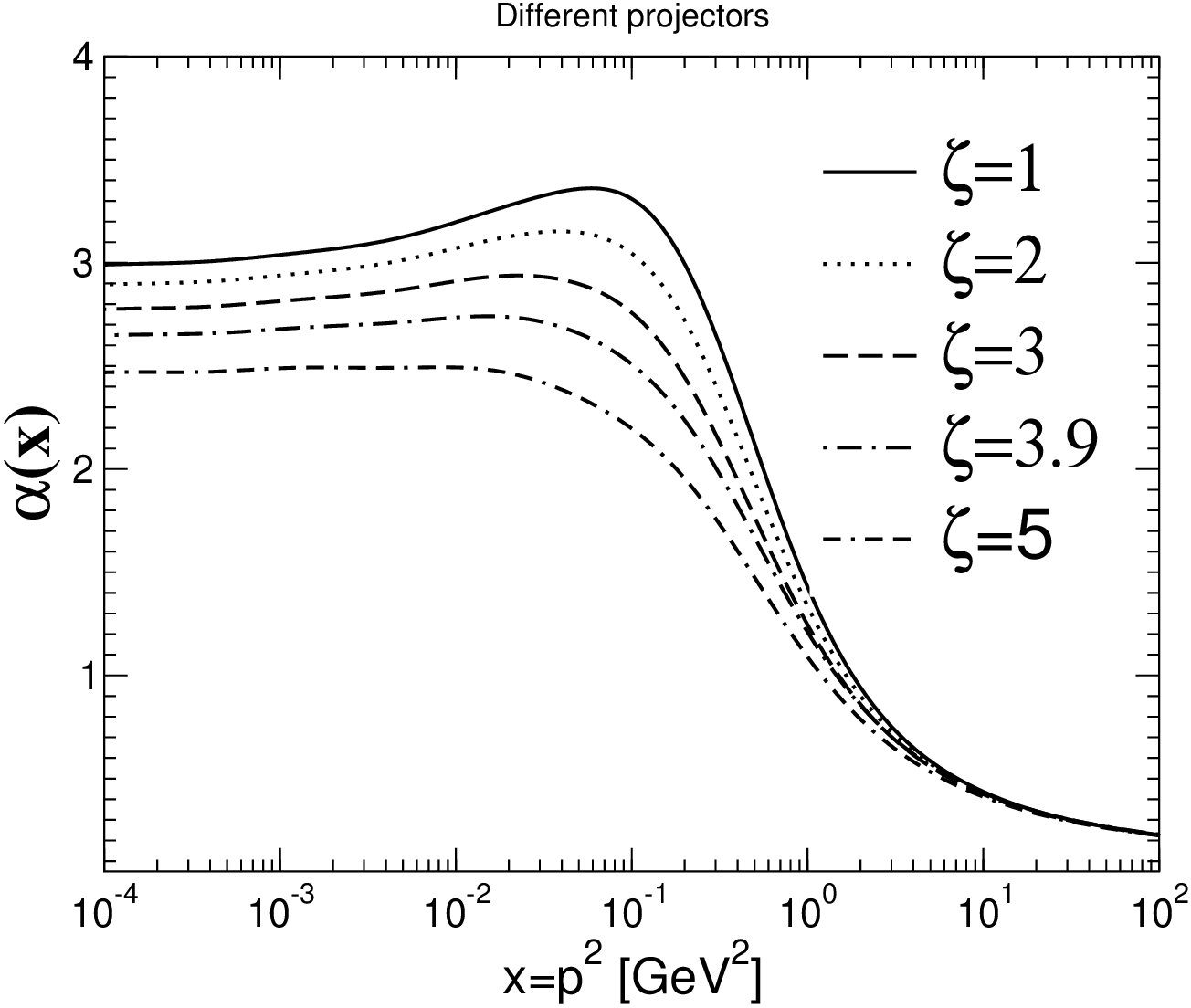,width=7.5cm,height=6cm}
}
\caption{\sf \label{proj.dat}
Shown are the results for the gluon dressing function, the ghost dressing 
function and the running coupling, {\it c.f.} Fig.\ \protect{\ref{new.dat}},
for different projectors.}
\end{figure}
 
Whereas Fit A is better in the region $0.3 \,\mbox{GeV}^2 < x<1 \, \mbox{GeV}^2$
where $\alpha$ is strongly rising, Fit B is slightly better in the region 
$1 \, \mbox{GeV}^2 < x < 10 \, \mbox{GeV}^2$, where hadronic $\tau$-decay takes place 
\cite{Ackerstaff:1998yj,Geshkenbein:2002ri}. As can
be seen in Fig.~\ref{new.dat} both fits works very well and can be used as input
for phenomenological calculations in future work. 

Our results for different
values of the parameter $\zeta$ ({\it c.f.}~eq.~(\ref{Paproj})) are shown in Fig.~\ref{proj.dat}. In accordance
with the infrared analysis the power $\kappa$ changes from $\kappa=0.5953$ for $\zeta=1$ to $\kappa=0.4610$
for $\zeta=5$. The perturbative properties of the solutions remain unchanged. The bump in the
running coupling gets smaller but does not disappear even for $\zeta=5$. It has already been stated above
that the dressing functions would be independent of $\zeta$ in a complete treatment of the gluon equation.
As all our solutions are very similar even on a quantitative level we conclude that transversality is 
lost only to a moderate extent. This is a somewhat surprising result in such a simple truncation scheme 
as the one at hand.
\begin{figure}[th!]
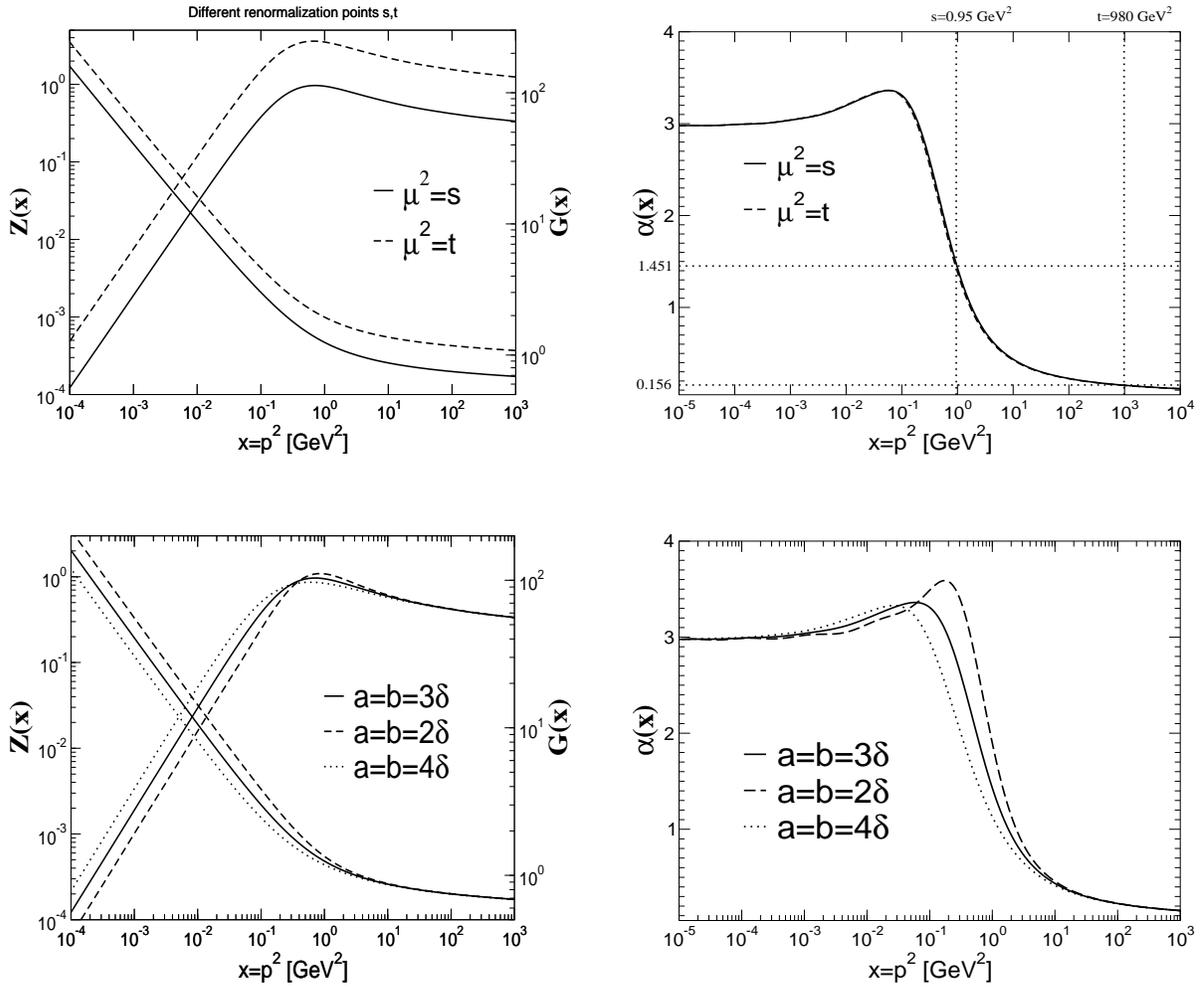

\centerline{
\epsfig{file=cont.s.zg.eps,width=7.5cm,height=6cm}
\hspace{0.5cm}
\epsfig{file=cont.s.a.eps,width=7.5cm,height=6cm}
}

\vspace{1cm}
\centerline{
\epsfig{file=cont.ab.zg.eps,width=7.5cm,height=6cm}
\hspace{0.5cm}
\epsfig{file=cont.ab.a.eps,width=7.5cm,height=6cm}
}
\caption{\sf \label{tech.dat}
Here we display two technical issues. The upper panel shows the ghost and gluon dressing
function as well as the running coupling for two different renormalisation points 
$\mu^2=s=0.9 \, \mbox{GeV}^2$ and $\nu^2=t=900 \, \mbox{GeV}^2$.
The independence of the running coupling on the renormalisation point is clearly demonstrated.
The lower panel shows the variation of the dressing functions with the parameters $a$ and $b$
from the construction in eq.~(\ref{3g-vertex}). Recall $\delta=-9/44$.
}
\end{figure}

The Brown--Pennington projector, $\zeta=4$, is an exceptional case as can be
seen from eqs.~(\ref{kappa1}), (\ref{kappa2}).  Here the $\kappa$-dependence of the second term
cancels and only one solution, $\kappa=1$, can be found ({\it
c.f.}~ref.~\cite{Atkinson:1998zc}). We found no numerical solutions for this
case. However, within the limit of numerical accuracy, solutions for $\zeta$
slightly different from 4 can be found leading to a value for $\kappa$ slightly
different from 1/2. {\it E.g.} in Fig.~\ref{proj.dat} the case $\zeta=3.9$
leading to $\kappa=0.5038$  is depicted.

In Fig.~\ref{tech.dat} we discuss two technical issues. First, we demonstrate what happens
if we choose two different renormalisation points by specifying two different values for
$\alpha(s) = g^2/4\pi$. Starting from eqs.~(\ref{zg-ren}) we can easily see that a change from
the renormalisation point $s=\mu^2$ to the new value $t$ is performed by
\beqa
G(x,t) &=& G(x,s) \frac{\tilde{Z}_3(s,L)}{\tilde{Z}_3(t,L)} \,, \nonumber\\
Z(x,t) &=& Z(x,s) \frac{{Z}_3(s,L)}{{Z}_3(t,L)} \,.
\eeqa
As the renormalisation constants $\tilde{Z}_3$ and $Z_3$ are independent of momentum, this
results in the mere multiplication of the dressing functions by a constant number. Our numerical
results obey exactly this behaviour, as can be seen in the upper right diagram of 
Fig. \ref{tech.dat}. The running coupling, however, is independent of the renormalisation
point, c.f. eq.~(\ref{coupling}), as is clearly demonstrated in the upper right diagram
of Fig. \ref{tech.dat}. 

In the lower panel of Fig. \ref{tech.dat} we show what happens when we vary the 
parameters $a$ and $b$ in our construction of the three-gluon vertex (\ref{3g-vertex}). Clearly,
the qualitative behaviour of the curves does not change. In particular the ultraviolet behaviour of the
solutions is independent of the values of $a$ and $b$ in accordance with our analysis in subsection 
\ref{ultraviolet}. However, we recognise the rising bump
in the running coupling when we lower the absolute values of $a$ and $b$ from $a=b=3\delta$
to $a=b=2\delta$. This corresponds to the equations becoming more and more unstable and no
solutions are found for even lower values of $a$ and $b$. The extreme case $a=b=0$, where
the gluon loop becomes leading in the infrared, has been 
investigated in reference \cite{Bloch:2001wz} with the same negative result: no 
solutions have been found.
\begin{figure}[t]
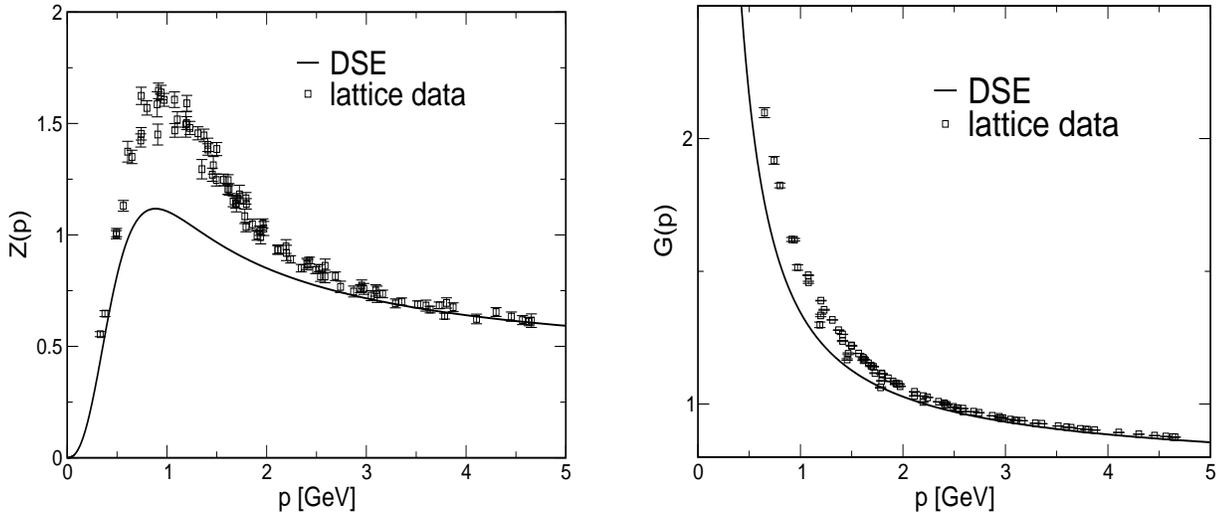

\vspace{1cm}
\centerline{
\epsfig{file=latt_glue_cont.eps,width=7.5cm,height=6.8cm}
\hspace{0.8cm}
\epsfig{file=latt_ghost_cont.eps,width=7.5cm,height=6.8cm}
}
\caption{\sf \label{lattice.dat}
Solutions of the Dyson-Schwinger equations compared to recent lattice 
results for two colours \cite{Langfeld:2002dd,Bloch:2002we}.}
\end{figure}

Finally, we compare our results to recent SU(2)
lattice calculations\footnote{Note that the same set of lattice data
have been analysed differently in refs.~\cite{Langfeld:2001cz,Langfeld:2002bg}. A comparison of our results
with the differently analysed data can be found in ref.~\cite{Fischer:2002hn}.} 
\cite{Bloch:2002we,Langfeld:2002dd}. As has already been stated above 
the ghost and gluon dressing functions from Dyson-Schwinger equations are independent
of the numbers of colours at least to our level of truncation. The only caveat in
comparing our results with the lattice ones is the adjustment of the momentum
scale. We used the lattice result $\alpha_{SU(2)}(x)=0.68$ at $x=10 \,\mbox{GeV}^2$ as input,
which leads to a slightly different scale than the one used in the case of SU(3).
The two graphs in figure (\ref{lattice.dat}) show that the main
qualitative features, the infrared suppression of the gluon dressing function and the divergence of
the ghost dressing function are common properties of both, the
lattice solutions and the one from Dyson-Schwinger equations. Even the power 
$\kappa \approx 0.595$ of the gluon dressing function from the DSEs is very close 
to the one that can be extracted from the lattice fit to be $\kappa \approx 0.5$.
The main difference between the two approaches is in the medium energy region around one
GeV, where the Dyson-Schwinger solutions suffer from the missing two-loop contributions
that are certainly present in lattice Monte-Carlo simulations\footnote{Despite this shortcome
the DSE-solutions are surprisingly good also in this momentum regime. Of particular importance is
the reproduction of the bump in the gluon propagator at $p\approx 0.8$ GeV. In lattice calculations
center vortices have been identified as promising candidates for field configurations which are
responsible for confinement \cite{Langfeld:2001cz}. Recent investigations on the lattice
suggest \cite{Langfeld:2001fd} that the bump of the gluon propagator is induced by such 
field configurations.}.
The combined evidence of the two methods points
strongly towards an infrared vanishing or finite gluon propagator and an infrared
singular ghost propagator in Landau gauge. 

\section{Summary}

In this chapter we have presented approximate non-perturbative solutions for the
gluon and the ghost propagators as well as the running coupling in Landau
gauge. We obtained these solutions for the Dyson--Schwinger equations in a 
truncation scheme working with a bare ghost-gluon vertex and an {\it ansatz} for the 
three-gluon vertex such that we reproduce the correct one-loop scaling of
the ghost and gluon dressing functions. We attempt to obtain two-loop scaling 
by the inclusion of the diagrams that involve four-gluon vertices in future work.

An important improvement to previous
treatments has been the explicit numerical calculation of all angular
integrals thus overcoming the angular approximations that have been made so
far. We could show that for a given projector only one out of two 
analytical solutions in the infrared can be connected to a numerical solution 
for finite momenta. For a transversal projector, $\zeta=1$, we found that the gluon
propagator is only weakly infrared vanishing, $D_{gluon}(p^2) \propto
(p^2)^{2\kappa -1}$, $\kappa =0.595\ldots$, and the ghost propagator is 
highly infrared singular, $D_{ghost}(p^2) \propto (p^2)^{-\kappa -1}$. This is in accordance
with the Kugo-Ojima confinement criterion and Zwanziger's horizon condition. 
The running coupling possesses an infrared fixed point with the value 
$\alpha(0) \approx 2.97$ (or, for a general number $N_c$  of colours, 
$\alpha(0) \approx 8.92/N_c$). 

Despite the simplicity of the truncation our solutions agree
remarkably well with recent lattice calculations performed
for two colours. Due to the finite lattice volume the lattice results cannot, of
course, be extended into the far infrared. In this respect our results are
complementary to the lattice ones: We do obtain the infrared behaviour
analytically. On the other hand, lattice calculations include, at least in
principle, all non-perturbative effects whereas we had to rely on truncations.
{\it E.g.\/} the deviations for the gluon renormalisation functions at
intermediate momenta depicted in Fig.\ \ref{lattice.dat}
might be due to the neglect of the four-gluon vertex
function in our calculations.

\chapter{Towards general gauges in the DSEs of Yang--Mills theory \label{symm-chap}}

In the last chapter we have exploited some properties of Landau gauge
that simplify the Dyson-Schwinger equations of the propagators
of Yang--Mills theory considerably. The fact that the 
ghost-gluon vertex is not renormalised in Landau gauge has turned out
to be the key for the formulation of our truncation scheme and 
leads to a useful definition of the non-perturbative running coupling. 
In other gauges matters are much more complicated and no one
has been able to solve the DSEs for general gauges yet. As a first step
towards such a solution we present an exploratory study in this chapter.
We investigate what happens, when we extend the simple truncation scheme
employing bare vertices to the whole family of general gauges which are
given by the gauge parameters $\alpha$ and $\lambda$ of the general
Lagrangian, eq.~(\ref{Lagrangian-gen}).

Away from the Landau gauge limit no direct connection between
the Kugo--Ojima confinement criterion and the infrared behaviour of the ghost
dressing function can be seen. This opens the possibility that other degrees
of freedom like the longitudinal gluon may take over the infrared dominant role
of the ghost dressing function in Landau gauge. As a matter of fact, infrared
dominance of longitudinal gluons is seen if stochastic quantisation is used
instead of Faddeev--Popov quantisation \cite{Zwanziger:2002ia}.

This chapter is organised as follows: In the next section we project the
ghost and gluon Dyson-Schwinger equations
for general gauges onto the respective dressing functions. Then we
perform an infrared analysis for the bare vertex truncation of these DSEs.
We show that in general ghost-antighost symmetric gauges the
genuine two--loop terms (generalised squint and sunset diagram) in the gluon
and the ghost DSEs become important in the infrared. This is different from
Landau gauge at least in the framework of the bare vertex truncation. 
In general  ghost-antighost symmetric gauges the infrared behaviour of the two-loop terms
exclude power-behaved solutions for the gluon and ghost propagators when bare
vertices are employed. In section \ref{symm-sol} 
we will provide numerical solutions for the
DSEs in the Landau gauge limit of the ghost--antighost symmetric case of the
Lagrangian and recover the solutions found in the last chapter from a
different direction in the two dimensional gauge parameter space \cite{Alkofer}. 
 
\section{Projection of the gluon equation}

\begin{figure}[t]
\epsfig{file=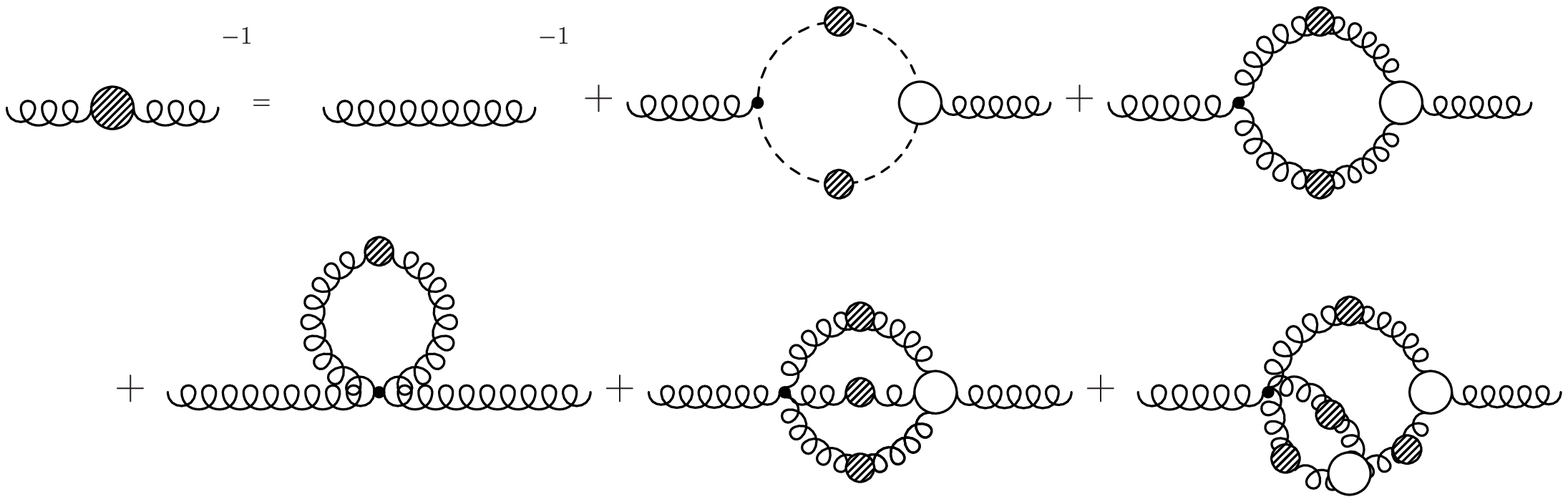,width=15cm}
\epsfig{file=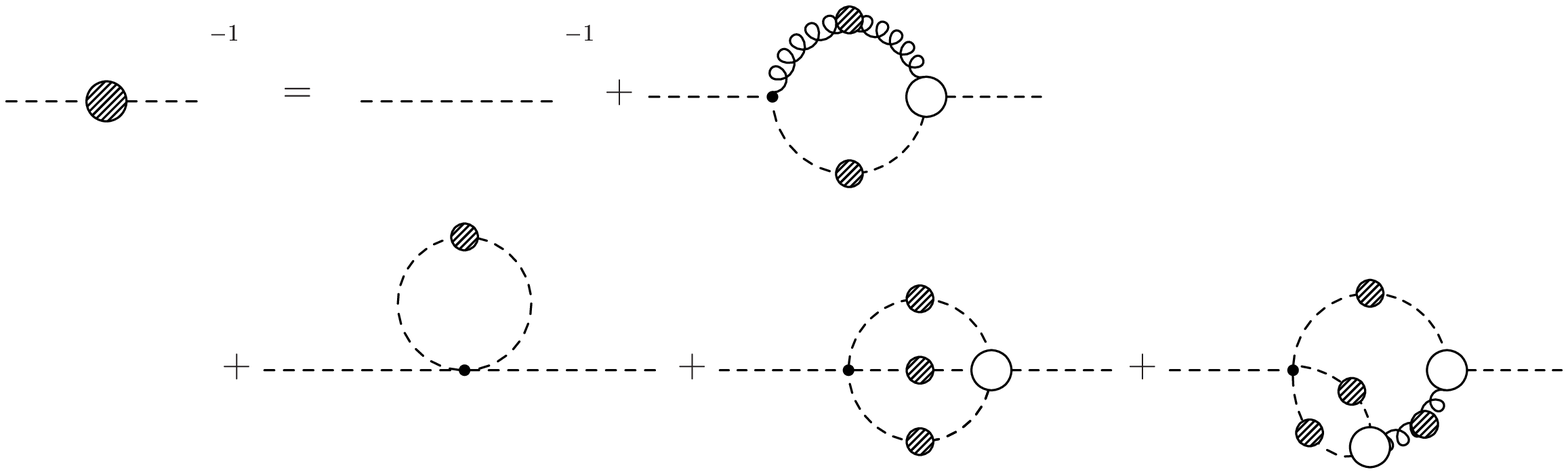,width=15cm}
\caption{\sf The coupled gluon and ghost Dyson--Schwinger equations from a BRS and 
Anti-BRS symmetric Lagrangian. Each equation contains one-loop diagrams, a 
tadpole contribution and a sunset and a squint diagram.}
\label{DSEpic2}
\end{figure}
In Fig.~\ref{DSEpic2} we show again the gluon and ghost Dyson-Schwinger equations
for general gauges in diagrammatical notation, {\it c.f.} section \ref{DSE-sec}.
The ghost and gluon equations are remarkably similar, both having tadpole
and non-perturbative one-loop and two-loop contributions. 

\begin{figure}
\epsfig{file=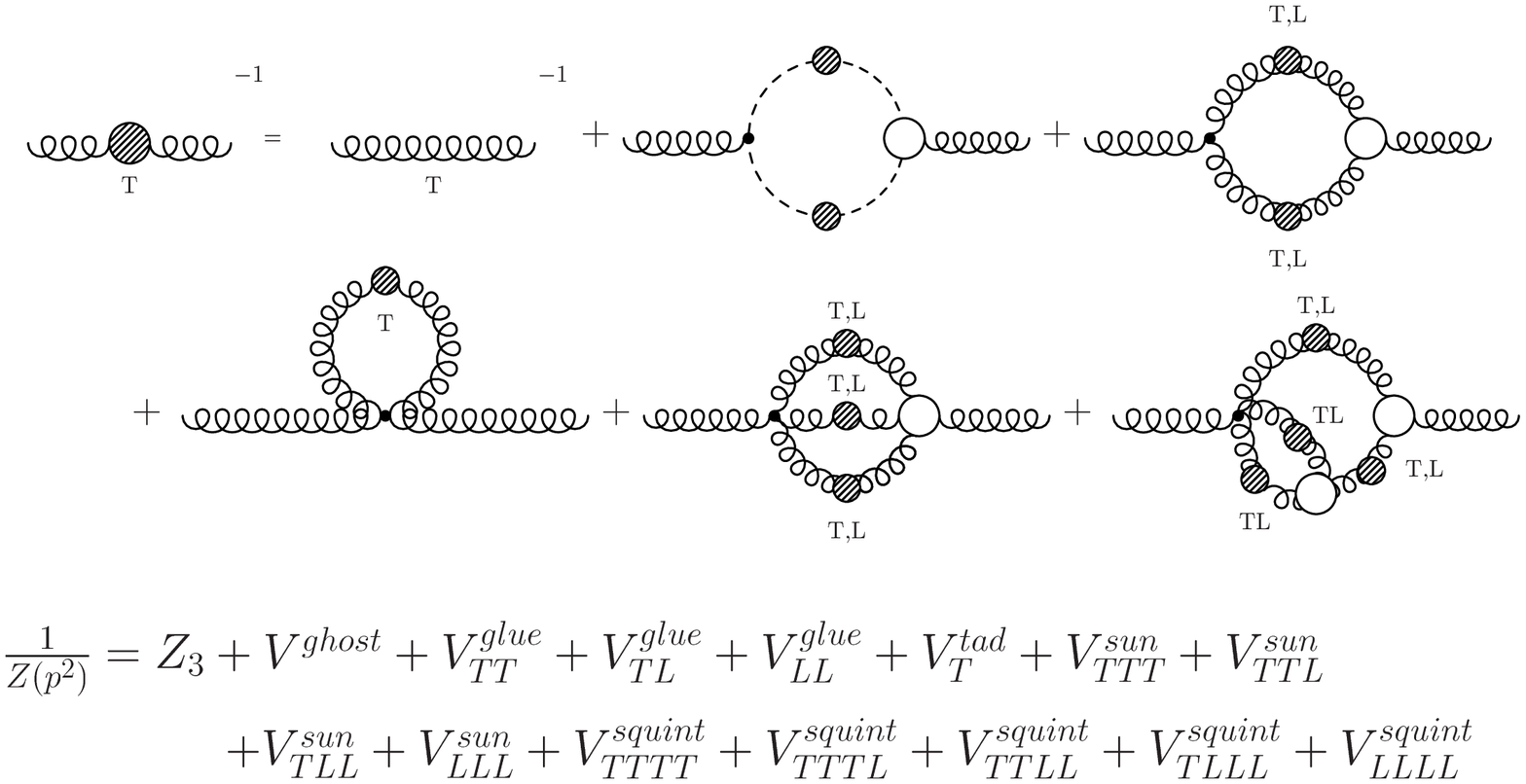,width=15cm}
\epsfig{file=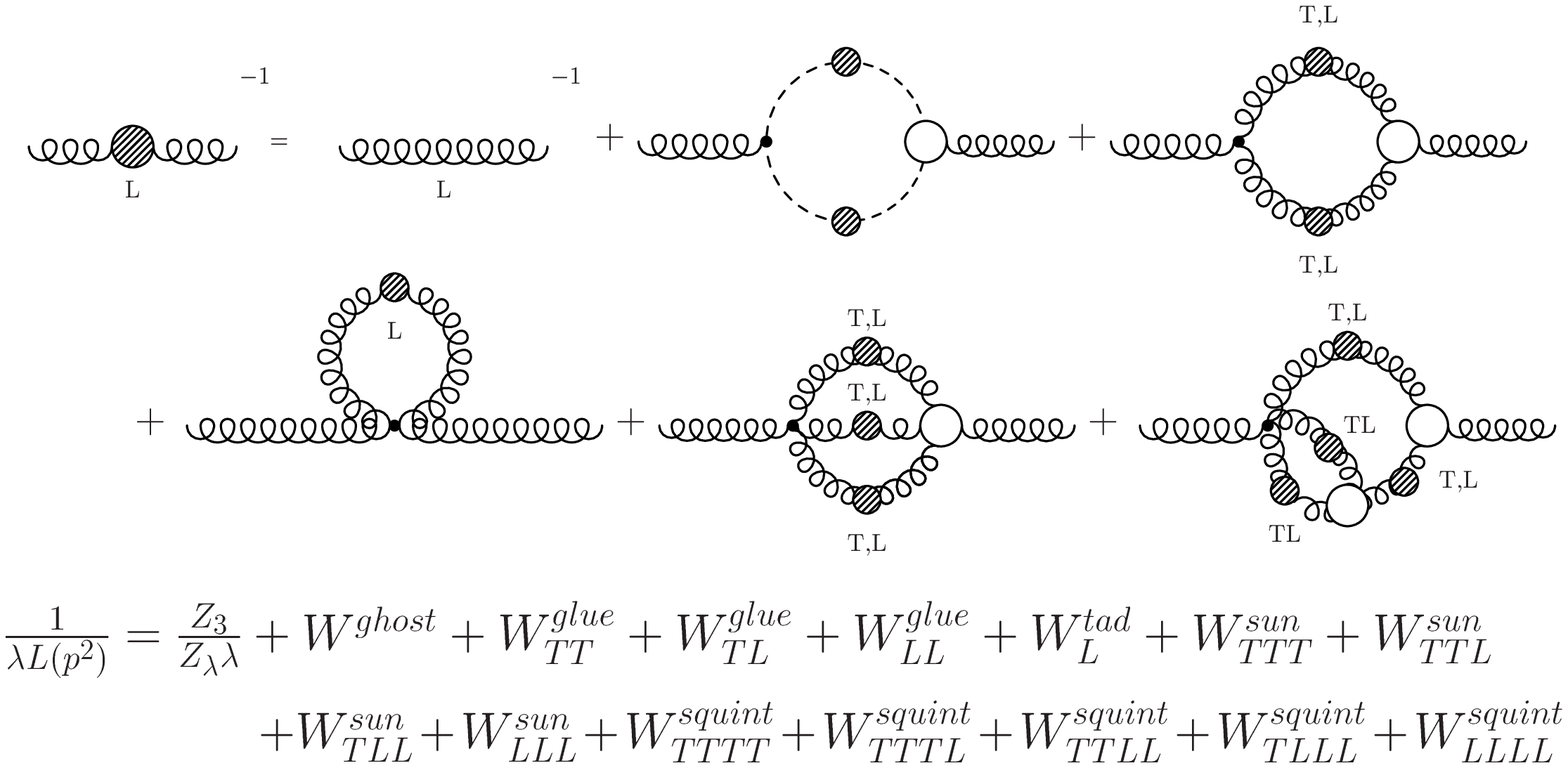,width=15cm}
\epsfig{file=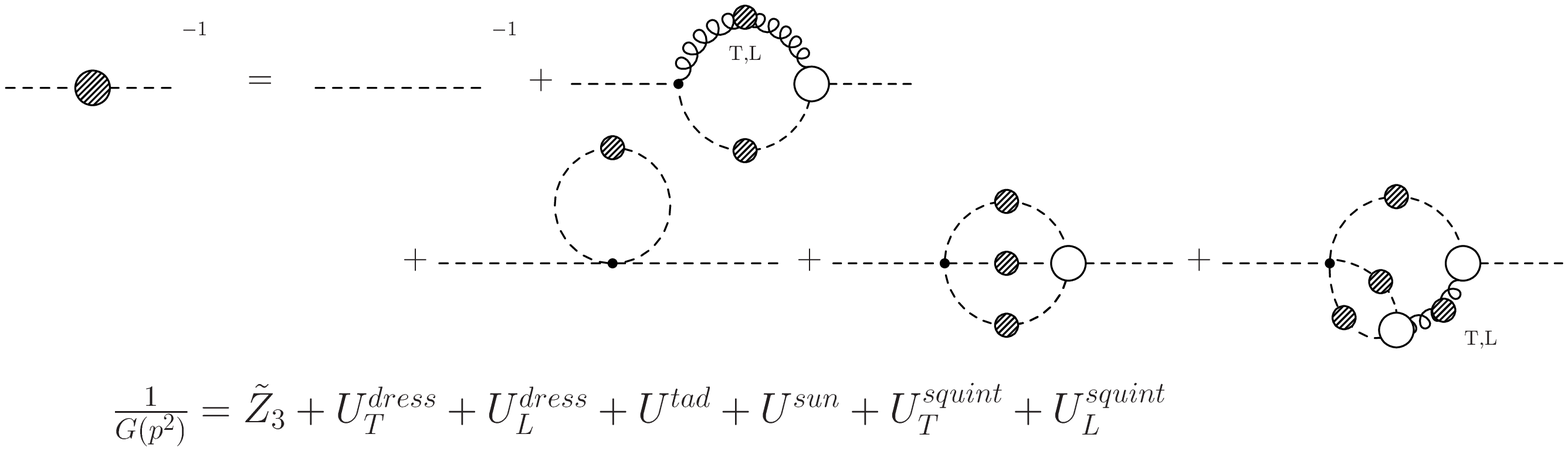,width=15cm}
\caption{\sf Various contributions from the respective diagrams in the transversal (T) 
and longitudinal (L) gluon equation and the equation for the ghost dressing function.}
\label{DSE_proj}
\end{figure}
In order to sort the various contributions of the gluon equation to the inverse of the
gluon propagator on the left hand side we project the equation on its longitudinal 
and transverse parts. It is well known that for linear covariant gauges, {\it i.e.} $\alpha=0$, 
the longitudinal part of the gluon propagator remains undressed \cite{Alkofer:2000wg}. 
However, away from linear covariant gauges
this is not the case as can be seen from the Slavnov-Taylor identity derived in 
\cite{Baulieu:1982sb}. We then have three dressing functions, $Z,L,G$, in the general case and 
the propagators are given by
\setlength{\jot}{3mm}
\beqa
D_{\mu \nu}(p) &=& \left[D_{\mu \nu}(p)\right]_T + \left[D_{\mu \nu}(p)\right]_L \nonumber \\
&=& \left( \delta_{\mu \nu} - \frac{p_\mu p_\nu}{p^2}\right)
\frac{Z(p^2)}{p^2} + \lambda L(p^2) \frac{p_\mu p_\nu}{p^4} \,, \\
D_G(p) &=& -\frac{G(p)}{p^2} \,. 
\eeqa
The transversal and longitudinal gluon dressing functions $Z(p^2)$ and $L(p^2)$ can be extracted by contracting
the gluon equation with the transversal and longitudinal projector respectively. The results are given
graphically in Fig.~\ref{DSE_proj}, where we also specify our notation for the different contributions
being analysed in the next section. Contributions in the transversal part of the gluon equation are denoted by
the symbol $V$, contributions in the longitudinal part by $W$ and the ones in the ghost equation by $U$.
The subscripts $T$ and $L$ indicate the respective parts of the gluon propagator running around in the loops
of the diagrams and abbreviations for the names of the diagrams are used. For example the symbol $W^{sun}_{LLT}$ denotes
a contribution from the sunset diagram to the longitudinal gluon equation with two longitudinal and one 
transverse part of the gluon propagator running in the loop. To isolate the dressing functions the left
hand sides of the equations have already been divided by factors of $3 p^2$ and $p^2$, respectively. 
\setlength{\jot}{0mm}

\section{Infrared analysis with bare vertices for arbitrary gauge parameters \label{infrared-a}}

We are now ready to determine the behaviour of the two-point functions at small momenta $p^2$, in a way
very similar to our analysis in the last chapter for Landau gauge. In section \ref{vertex-dressing} we
discussed the central observation in Landau gauge, that there is no
qualitative difference of the solutions found with bare vertices or with vertices dressed
by the use of Slavnov-Taylor identities. This has been shown recently for a 
range of possible vertex dressings in a truncation scheme without any angular approximations 
\cite{Lerche:2002ep}. The reason for
this somewhat surprising result has been attributed to the non-renormalisation of the ghost-gluon vertex in
Landau gauge, {\it i.e.} $\tilde{Z}_1=1$. It seems as if the violation of gauge invariance using a bare vertex is
not that severe in Landau gauge such that the resulting equations still provide meaningful results.
In the following we will explore to what extent such a simple truncation idea is applicable in
other gauges.

Our power law ansatz for the dressing functions is
\beq
G(x) = Bx^\beta, \:\:\:\:
Z(x) = Ax^\sigma, \:\:\:\:
L(x) = Cx^\rho,
\eeq 
where $x=p^2$ has been used. Together with the expressions for the bare vertices given in appendix 
\ref{Def-app}
we plug the power laws into the ghost and the gluon equation. The formulae for the various integrals 
are given in appendix \ref{tensor}. The straightforward but tedious algebra is done
with the help of the algebraic manipulation program FORM \cite{Vermaseren:2000nd}. 
In section \ref{infrared} we have argued 
that the renormalisation constants $Z_3$ and $\tilde{Z}_3$ can be dropped in the infrared analysis.
Furthermore the tadpoles just give constant 
contributions to the respective propagators which vanish in the process of renormalisation. 
Thus we can safely omit them in the present investigation.

For general gauges, $\alpha \ne 0$ and $\lambda \ne 0$, we obtain the following structure:
\setlength{\jot}{3mm}
\beqa
B^{-1}x^{-\beta} &=& x^{\sigma+\beta}(U^\prime)^{dress}_{T} + x^{\rho+\beta}(U^\prime)^{dress}_{L} 
  + x^{3\beta}(U^\prime)^{sun} \nonumber\\
&&+x^{\sigma+3\beta} (U^\prime)^{squint}_{T} +x^{\rho+3\beta}  (U^\prime)^{squint}_{L} \,, \label{gheq1}\\
A^{-1}x^{-\sigma} &=& x^{2\beta}(V^\prime)^{ghost} +x^{2\sigma} (V^\prime)^{glue}_{TT} 
  + x^{\sigma+\rho}(V^\prime)^{glue}_{TL} + x^{2\rho}(V^\prime)^{glue}_{LL} \nonumber\\
&&+ x^{3\sigma}(V^\prime)^{sun}_{TTT} + x^{2\sigma+\rho}(V^\prime)^{sun}_{TTL}
  + x^{\sigma+2\rho}(V^\prime)^{sun}_{TLL}+ x^{3\rho}(V^\prime)^{sun}_{LLL}\nonumber\\
&&+ x^{4\sigma}(V^\prime)^{squint}_{TTTT} + x^{3\sigma+\rho}(V^\prime)^{squint}_{TTTL}
  + x^{2\sigma+2\rho}(V^\prime)^{squint}_{TTLL} \nonumber\\
&&+ x^{\sigma+3\rho}(V^\prime)^{squint}_{TLLL}
  + x^{4\rho}(V^\prime)^{squint}_{LLLL} \,, \label{gl_trans1}\\
\left(C \lambda\right)^{-1} x^{-\rho}
&=& + x^{2\beta}(W^\prime)^{ghost} + x^{2\sigma}(W^\prime)^{glue}_{TT} 
  + x^{\sigma+\rho}(W^\prime)^{glue}_{TL}  \nonumber\\
&&+ x^{3\sigma}(W^\prime)^{sun}_{TTT} + x^{2\sigma+\rho}(W^\prime)^{sun}_{TTL}
  + x^{\sigma+2\rho}(W^\prime)^{sun}_{TLL}+ x^{3\rho}(W^\prime)^{sun}_{LLL}\nonumber\\
&&+ x^{4\sigma}(W^\prime)^{squint}_{TTTT} + x^{3\sigma+\rho}(W^\prime)^{squint}_{TTTL}
  + x^{2\sigma+2\rho}(W^\prime)^{squint}_{TTLL} \nonumber\\
&&+ x^{\sigma+3\rho}(W^\prime)^{squint}_{TLLL}  \,. \label{gl_long1}
\eeqa
Here the primed quantities $U^\prime$, $V^\prime$ and $W^\prime$ are momentum independent 
functions of the exponents $\beta$, $\sigma$ and $\rho$. The corresponding
momentum dependent quantities $U$, $V$ and $W$ have been introduced in Fig.~\ref{DSE_proj}.
The pattern of the equation is such that each primed factor on the right hand side is
accompanied by the squared momentum $x$ to the power of the dressing function content of the respective diagram.
In appendix \ref{app-diag} we demonstrate how such a pattern emerges for the example of 
the sunset diagram in the ghost equation, $(U)^{sun}$. Note that the contributions $(W^\prime)^{glue}_{LL}$
and $(W^\prime)^{squint}_{LLLL}$ are zero and therefore missing
in the longitudinal gluon equation (\ref{gl_long1}) as momentum conservation cannot hold with three
longitudinal gluons in the three gluon vertex. 

For the following argument we focus on one particular contribution on each right hand side 
of the equations, keeping in mind that all other contributions have no explicit minus
sign in the exponents:
\beqa
B^{-1}x^{-\beta} &=&  x^{3\beta}(U^\prime)^{sun} + \ldots \,, \label{gheq2}\\
A^{-1}x^{-\sigma} &=& x^{4\sigma}(V^\prime)^{squint}_{TTTT} + \ldots  \,,\label{gl_trans2} \\
\left(C \lambda\right)^{-1} x^{-\rho}
&=& x^{3\rho}(W^\prime)^{sun}_{LLL} + \ldots \,. \label{gl_long2}
\eeqa
In general the coefficients $(U^\prime)^{sun}$, $(V^\prime)^{squint}_{TTTT}$ and $(W^\prime)^{sun}_{LLL}$ are nonzero
and explicitly given in appendix \ref{app-diag}. 
{\it First}, it is easy to see from equations (\ref{gheq2}), (\ref{gl_trans2}) and (\ref{gl_long2}) that neither 
$\beta$ nor $\sigma$ nor $\rho$ can be negative. If one of these powers would be negative the
limit $x \rightarrow 0$ would lead
to a vanishing left hand side of the respective equation whereas the right hand side is singular in this limit.
This is a contradiction as the power on the left hand side of the equation should match the leading power on
the right hand side.  
{\it Second}, if one of the exponents 
$\beta$, $\sigma$ or $\rho$ would be positive, then the diverging left hand side
of the respective equation would require a diverging term on the right hand side as well.
However, we already saw that $\beta$, $\sigma$ or $\rho$ cannot be negative and therefore 
we have no diverging term on the right hand side of the equations. Thus none of the exponents
$\beta$, $\sigma$ and $\rho$ can be positive. 
The remaining possibility is $\beta=\sigma=\rho=0$, but then we have logarithms
on the right hand side of the equation which do not match the constant on the left hand side.
Thus we arrive at the conclusion that there is no 
power solution for general gauges when bare
vertices are used.   

There are two limits for the gauge parameters $\alpha$ and $\lambda$ in which the situation changes.
The first one is $\alpha=0$, that are ordinary linear covariant gauges. Due to the corresponding
Slavnov-Taylor identity the
longitudinal part of the gluon propagator remains undressed, $L(p^2)=1$ \cite{Alkofer:2000wg}.
The longitudinal gluon equation
becomes trivial as the Slavnov-Taylor identity forces all nontrivial contributions
on the right hand side to cancel each other. In the ghost equation the squint as well as
the sunset diagram disappear, {\it i.e.} $(U^\prime)^{sun}=(U^\prime)^{squint}=0$. 
The term of the dressing loop which contains the longitudinal 
part of the gluon propagator, $(U^\prime)^{dress}_L$, vanishes as well, 
{\it c.f.} appendix \ref{app-diag}.
We are left with the contribution $U^{dress}_T \sim x^{(\beta  + \sigma )}$, similar to Landau gauge. 
This is the reason why we escape the argument given below eq.~(\ref{gl_long2}).
The explicit expression for the ghost equation is given by 
\beqa
B^{-1}x^{-\beta} &=&  x^{(\beta  + \sigma )} \frac {g^2 N_c \tilde{Z}_1 AB}{16 \pi^2} \, \frac {-3} 
{2\,(\beta  + \sigma )\,( - 1 + \beta  + \sigma )} \times \nonumber\\
&& \hspace*{5cm} \times \frac{
\Gamma (2 - \beta  - \sigma 
)\,\Gamma (1 + \sigma )\,\Gamma (2 + \beta )}{\Gamma (3 + \beta 
 + \sigma )\,\Gamma (2 - \sigma )\,\Gamma (1 - \beta )}.
\eeqa
which can be compared with the Landau gauge expression, eq.~(\ref{ghostbare-ir2}).
Matching left and right hand sides we conclude that 
the Landau gauge result
\beq
\sigma=-2\beta
\label{kappa}
\eeq
is valid for all linear covariant gauges when bare vertices are used. 
\setlength{\jot}{0mm}

Compared to general gauges the gluon equation does not change in structure when $\alpha=0$. 
Therefore we still have $\sigma > 0$, according to the argument given below
eq.~(\ref{gl_long2}). The identity (\ref{kappa}) thus requires $\beta < 0$ and we have 
a divergent ghost and a vanishing gluon dressing function in the infrared similar to Landau gauge.
With bare vertices this implies that the ghost loop is the dominant contribution 
in the gluon equation for small momenta. Since the ghost loop is independent of the gauge parameter
$\lambda$, our Landau gauge results of section \ref{infrared} for the exponent $\kappa=-\beta=\sigma/2$
are valid for all linear covariant gauges.
However, a word of caution is in order. In Landau gauge there are indications
\cite{vonSmekal:1997is,Watson:2001yv} that the general result (\ref{kappa})
does not change when the vertices are dressed. This has been confirmed recently for a 
range of possible vertex dressings \cite{Lerche:2002ep}. It is a completely open question whether this is the 
same for $\lambda \ne 0$.

Having addressed the case of linear covariant gauges with $\alpha=0$ we now turn to 
the other interesting limit, {\it i.e.} $\lambda=0$ while $\alpha \ne 0$. It is easy to see that the
$\alpha$-dependence of
the Lagrangian (\ref{Lagrangian-gen}) can be eliminated in this case by partial integration using
the constraint $\partial_\mu A_\mu=0$. However, on the level of the DSEs with
bare vertices there remain spurious $\alpha$-dependent terms on the right hand side of the gluon equation. 
In the next section we will investigate the dependence of the Landau
gauge solution on the gauge parameter $\alpha$ if these spurious terms are present.

\section{Solutions in Landau gauge \label{symm-sol}}

To assess the influence of the spurious $\alpha$-terms in Landau gauge
we use the truncation scheme of section \ref{truncation}. Again we employ the general tensor
\begin{equation}
{\mathcal P}^{(\zeta )}_{\mu\nu} (p) = \delta_{\mu\nu} - 
\zeta \,\frac{p_\mu p_\nu}{p^2} 
 \end{equation} 
to contract the Lorenz indices in the gluon equation. By varying the parameter $\zeta$ we are able to 
test additionally for spurious longitudinal terms.
For $\lambda=0$ and general values of the gauge parameter $\alpha$
the coupled set of equations for the ghost and gluon dressing functions are 
(c.f. eqs.~(\ref{ghostbare}), (\ref{gluonbare}), (\ref{new_kernels}), (\ref{Q-tilde}), 
where $\alpha=0$)
\setlength{\jot}{3mm}
\begin{eqnarray} 
\frac{1}{G(x)} &=& Z_3 - g^2N_c \int \frac{d^4q}{(2 \pi)^4}
\frac{K(x,y,z)}{xy}
G(y) Z(z) \; , \label{ghostbare-a} \\ 
\frac{1}{Z(x)} &=& \tilde{Z}_3 + g^2\frac{N_c}{3} 
\int \frac{d^4q}{(2 \pi)^4} \frac{M(x,y,z)}{xy} G(y) G(z) \nonumber\\
&&\hspace{0.6cm}+ 
 g^2 \frac{N_c}{3} \int \frac{d^4q}{(2 \pi)^4} 
\frac{Q(x,y,z)}{xy} \frac{G(y)^{(1-a/\delta-2a)}}{Z(y)^{a}}
\frac{G(z)^{(1-b/\delta-2b)}}{Z(z)^{b}} \; .
\label{gluonbare-a} 
\end{eqnarray} 
The kernels ordered with respect to powers of $z:=p^2=(k-q)^2$ have the form:
\begin{eqnarray}
K(x,y,z) &=& \frac{1}{z^2}\left(-\frac{(x-y)^2}{4}\right) + 
\frac{1}{z}\left(\frac{x+y}{2}\right)-\frac{1}{4} \,, \nonumber\\
M(x,y,z) &=& \frac{1}{z} \left( \frac{(\zeta-1)\alpha^2-(\zeta-1)2\alpha+\zeta-2}{4}x + 
\frac{y}{2} - \frac{\zeta}{4}\frac{y^2}{x}\right)
+\frac{1}{2} + \frac{\zeta}{2}\frac{y}{x} - \frac{\zeta}{4}\frac{z}{x} \,, \hspace*{1cm} \nonumber\\ 
Q(x,y,z) &=& \frac{1}{z^2} 
\left( \frac{1}{8}\frac{x^3}{y} + x^2 -\frac{19-\zeta}{8}xy + 
\frac{5-\zeta}{4}y^2
+\frac{\zeta}{8}\frac{y^3}{x} \right)\nonumber\\
&& +\frac{1}{z} \left( \frac{x^2}{y} - \frac{15+\zeta}{4}x-
\frac{17-\zeta}{4}y+\zeta\frac{y^2}{x}\right)\nonumber\\
&& - \left( \frac{19-\zeta}{8}\frac{x}{y}+\frac{17-\zeta}{4}+
\frac{9\zeta}{4}\frac{y}{x} \right) \nonumber\\
&& + z\left(\frac{\zeta}{x}+\frac{5-\zeta}{4y}\right) + z^2\frac{\zeta}{8xy} +\frac{5}{4}(4-\zeta) \,.
\label{new_kernels-a}
\end{eqnarray}
\setlength{\jot}{0mm}
\begin{figure}
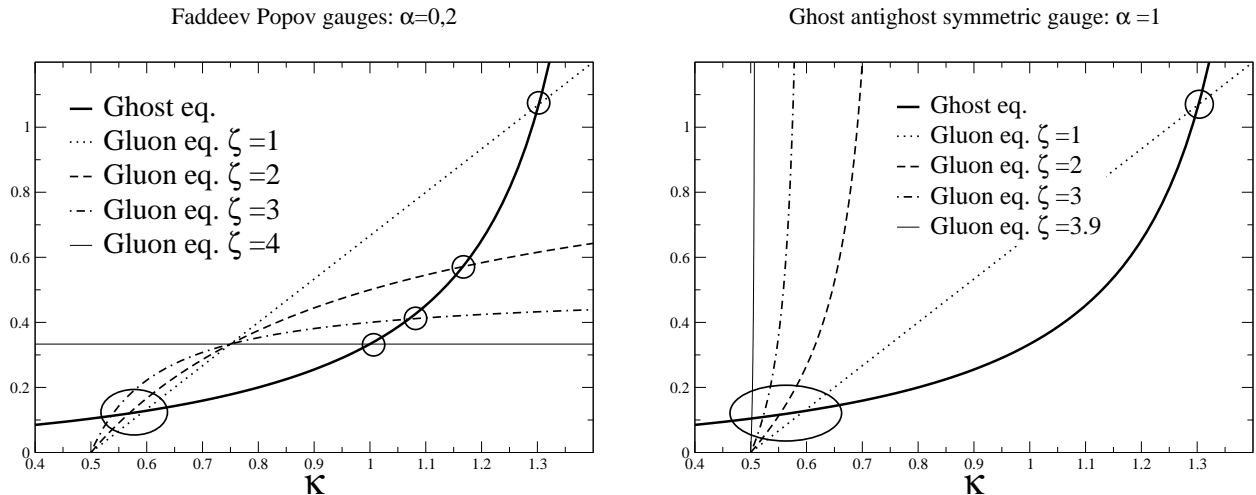

\centerline{
\epsfig{file=kappa_s.eps,height=6.5cm} \hspace*{0.7cm}
\epsfig{file=kappa2.eps,height=6.5cm}
}
\caption{\sf \label{kappa.a.dat}
Here the graphical solution to equation (\ref{kappa3}) is shown. 
The thick line represents the
left hand side of equation (\ref{kappa3}), whereas the other curves depict 
the right hand side for different values of the parameters $\zeta$. The left figure shows results for
$\alpha=0$ and $\alpha=2$, whereas in the figure on the right we have $\alpha=1$. 
The ellipses mark the bulk of solutions between
$\kappa=0.5$ and $\kappa=0.6$, whereas the circles show spurious solutions with $\kappa \ge 1$.
}
\end{figure}

First we accomplish the infrared analysis. With $\kappa=-\beta=\sigma/2$ we employ the ansatz
\beq
Z(x) = Ax^{2 \kappa} \,, \:\:\:\:\:\:\: G(x) = Bx^{-\kappa} 
\eeq
in the equations (\ref{ghostbare-a}) and (\ref{gluonbare-a}). After integration we match coefficients of equal 
powers on both sides of the equations and obtain
\begin{eqnarray} 
&&\frac{1}{18}
\frac{(2+\kappa)(1+\kappa)}{(3-2\kappa)}
    \nonumber\\
&&=\frac{(4\kappa-2)\,( - 1 + \kappa)}{(\zeta  - 1)\left[4\,\kappa
^{2}\,(\alpha ^{2} - 2\,\alpha  + 1) + 8\,\kappa\,\alpha
 \,(2 - \alpha )+ 3\,\alpha \,(\alpha  - 2)\right] + \kappa\,(10 - 7\,
\zeta )  - 6 + 3\,\zeta }
\; . \nonumber\\ \nonumber\\
\label{kappa3}
\end{eqnarray} 
The values of $\kappa$ for different tensors $\mathcal{P}^{(\zeta)}$ can be read off
Fig.~\ref{kappa.a.dat}.
The curve given by the fully drawn line represents the term on the left hand side of eq.~(\ref{kappa3}),
whereas the other lines depict the right hand side for several values of the parameter $\zeta$.
The solutions between $\kappa=0.5$ and $\kappa=0.6$ remain nearly unchanged when
$\alpha$ is varied, whereas the solutions for $\kappa >1$ are only present when $\alpha=0,2$.
In the last chapter we found that only the solutions $\kappa \approx 0.5$ 
connect to numerical results for finite momenta. It is satisfactory to observe that by 
varying the gauge parameter $\alpha$ such solutions can be identified as spurious already
on the level of the infrared analysis.
\begin{figure}[th!]
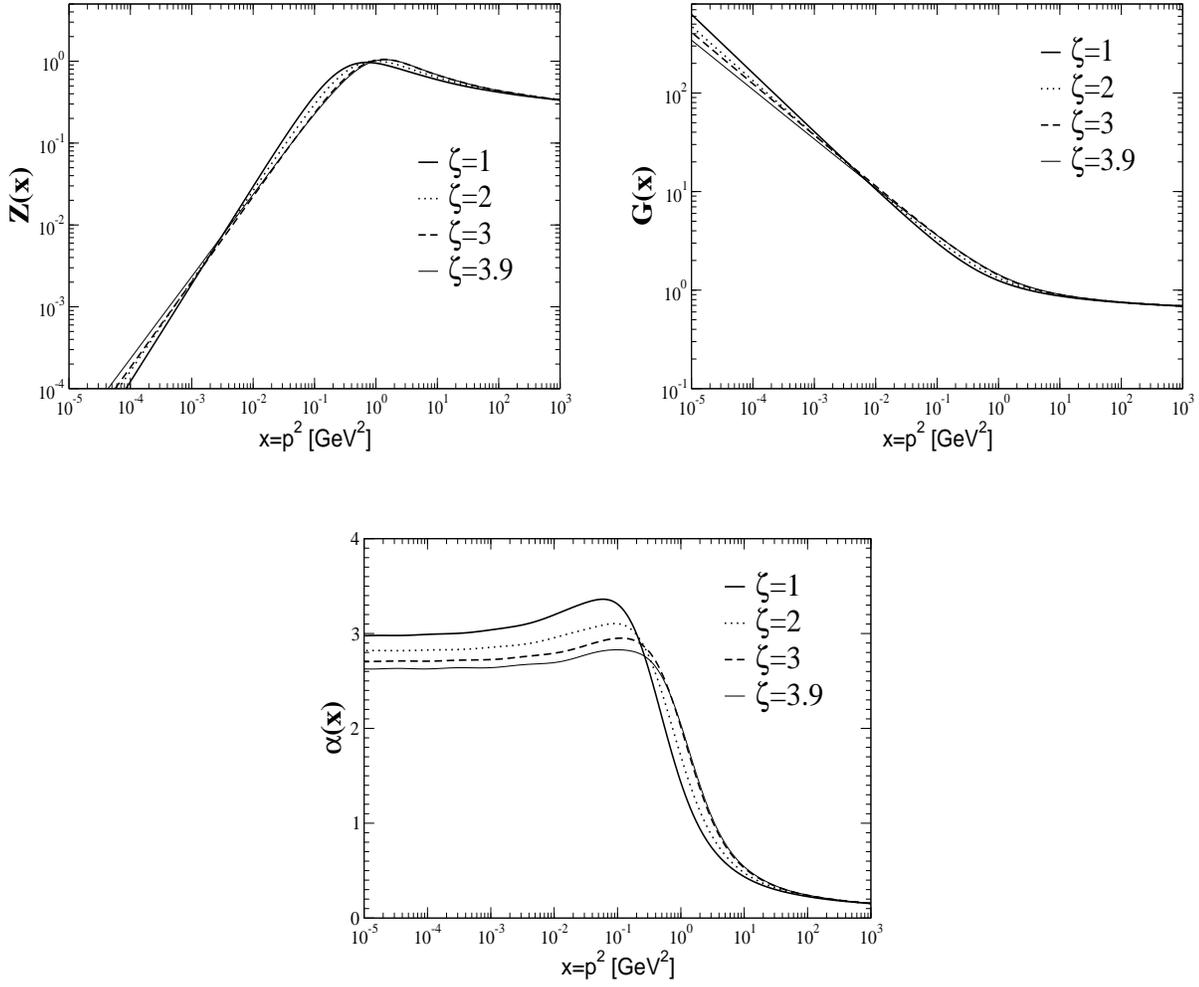

\vspace{0.5cm}
\centerline{
\epsfig{file=z_symm.eps,width=7.5cm,height=6cm}
\hspace{0.5cm}
\epsfig{file=g_symm.eps,width=7.5cm,height=6cm}
}
\vspace{1cm}
\centerline{
\epsfig{file=a_symm.eps,width=7.5cm,height=6cm}
}
\caption{\sf \label{new.a.dat}
Here we display the gluon dressing function, $Z(x)$, the ghost dressing function, $G(x)$ and
the running coupling $\alpha(x)$ in the truncation scheme of chapter \ref{YM} for the gauge parameters 
$\alpha=1$ and $\lambda=0$ and different tensors $\mathcal{P}^{(\zeta)}$ employed in the gluon equation.}
\end{figure}

We now explore the impact of the spurious $\alpha$ term on the behaviour of the solutions
for all momenta $x$. We solved the equations 
(\ref{ghostbare-a}) and (\ref{gluonbare-a}) numerically 
using the technique described in appendix \ref{numerics}. Compared to the usual Landau gauge, $\alpha=0$,
we obtain the greatest deviations for the ghost-antighost symmetric case, $\alpha=1$. The results can be seen in Fig.
(\ref{new.a.dat}). As the dependence of the kernel of the ghost loop on $\alpha$ vanishes in the case of the
transverse projector, $\zeta=1$, this solution is identical with the one displayed in Fig.~\ref{new.dat}
in subsection \ref{YM-results}.
For the other cases the power $\kappa$ changes from $0.5953$ for $\zeta=1$ to $0.5020$ for $\zeta=3.9$ in accordance
with the infrared analysis. The ultraviolet properties of the solutions are slightly disturbed
compared to the case $\alpha=0,2$. An analytical ultraviolet analysis similar to the one in 
subsection \ref{ultraviolet} 
reveals that the $\alpha$-term in the ghost loop induces a spurious dependence
of the anomalous dimensions on the parameter $\zeta$:
\setlength{\jot}{3mm}
\beqa
\gamma &=& \frac{-26-(\zeta-1)\,\alpha\,(2-\alpha)}{44+(\zeta-1)\,\alpha\,(2-\alpha)} \nonumber\\
\delta &=& \frac{-9}{44+(\zeta-1)\,\alpha\,(2 - \alpha)}
\eeqa
For general $\alpha$ only the transverse projector removes the alpha-term in the 
ghost equation and leads to the correct one loop scaling of the equations, that is $\delta=-9/44$ 
for the ghost and $\gamma=-13/22$ for the gluon dressing function for an arbitrary number of colours and
zero flavours.
\setlength{\jot}{0mm}

\section{Summary}

We have studied the infrared behaviour of the ghost and gluon propagators in
covariant ghost-antighost symmetric gauges. We derived the corresponding 
Dyson--Schwinger equations for these propagators including the ones of linear
covariant gauges as the limit of a vanishing gauge parameter. Note that
ghost-antighost symmetric gauges are particularly interesting as they allow an
interpretation of the antighost field being the antiparticle of the ghost which
includes also the possibility of ghost-antighost condensates. Due to the
emergence of a four-ghost interaction term in the Lagrangian for general values
of the gauge parameters the Dyson--Schwinger equation of the ghost propagator
displays a rich  structure very similar to the one of the gluon equation. On
the other hand, in the gluon equation we obtain the same structure as in linear
covariant gauges apart from the fact that the gluon propagator acquires a
nontrivial longitudinal part which appears in turn in all diagrams. The gluon
and ghost equations depend therefore on three independent dressing functions,
one for the ghost, one for the transversal part of the gluon propagator and one
for the longitudinal one.

We then employed a truncation scheme for the Dyson--Schwinger equations that
uses bare vertices in place of the dressed ones. The success of this particular
truncation scheme in Landau gauge has been attributed to the
non-renormalisation of the ghost-gluon vertex, {\it i.e.} $\widetilde Z_1=1$.  We
addressed the infrared behaviour of the ghost and gluon propagators for general
gauges by employing power law {\it ans\"atze} for the respective dressing
functions. We then have been able to evaluate the infrared behaviour of the
gluon and ghost equations analytically. 

For all linear covariant gauges we find a similar result as compared to the
one  in Landau gauge: An infrared suppressed gluon propagator and an infrared
enhanced ghost. Whereas in Landau gauge there are indications that this generic
result is not changed when the vertices are dressed \cite{Lerche:2002ep}, it
remains an open question whether this is the case in linear covariant gauges in
general. Away from linear covariant gauges, {\it i.e.} in the general case $\alpha
\not=0$ and $\lambda \not=0$, we do not find power solutions for the dressing
functions. Again, this might be altered significantly by  appropriate vertex
dressings. Nevertheless, it remains to be emphasised that therefore also
the occurrence of a ghost-antighost vacuum condensate is excluded in this 
specific truncation scheme within this class of gauges.

A special case among all gauges considered here is Landau gauge. In the limit
$\lambda=0$ the general Lagrangian (\ref{Lagrangian-gen}) becomes independent of
the second gauge parameter $\alpha$, thus Landau gauge is also a special case
of ghost-antighost symmetric gauges. Although the Lagrangian of the theory is
independent of the gauge parameter $\alpha$, our simple truncation scheme
breaks this invariance and spurious  $\alpha$-dependent terms arise in the
ghost loop of the gluon Dyson--Schwinger equation. Examining the case
$\alpha=1$ we showed that the influence of these spurious terms is very small.
We determined solutions for the ghost and gluon dressing functions both
analytically in the infrared and numerically for finite momenta and  found
solutions close to the ones of chapter \ref{YM}. We thus recovered
the results of Landau gauge from a different direction in the two dimensional
space of gauge parameters. 

\chapter{Landau gauge Yang-Mills theory on a four-torus \label{torus}}

There are a number of central aims connected to the investigation of the
Dyson--Schwinger equations on a four-torus. The {\it first} one is purely technical:
This allows to study finite volume effects also in the Dyson--Schwinger approach.
Monte-Carlo simulations on a lattice necessarily have to be done in a finite
volume. Therefore in the latter kind of approach infrared properties are only
accessible by extrapolations to an infinite volume. However, the avaliable data are
gained on volumes which differ at best by one order of magnitude 
due to limitations in computer time. We will see in the present
Dyson--Schwinger approach that available volumes cover several orders of
magnitude. And more importantly, one can compare to the results obtained 
in an infinite volume for several truncation schemes.

The {\it second} issue is the connection of analytical solutions obtained in the infrared limit
of the (truncated) Dyson--Schwinger equations to the numerical
solution obtained for finite momenta up to the ultraviolet. In chapter \ref{YM} we
have seen that not every analytical solution in the infrared connects to such a 
numerical solution. By imposing different infrared boundary conditions on our numerical
equations we could check, whether a given infrared solution is connected to a numerical solution 
for finite momenta and which is not. In this chapter we will verify this procedure by an even stronger
selection process: Due to the finite volume there is no input in the infrared and our
numerical solutions on a four-torus choose their infrared behaviour by themselves.
We will see, that the same infrared solutions are chosen than in the continuum.  

The {\it third} issue is that once we know how to treat
Dyson--Schwinger equations on a torus several interesting possibilities for
further investigations open up. Choosing an asymmetric four-torus might allow the
introduction of a non-vanishing temperature \cite{Landsman:1987uw} in a relatively simple way.  
Furthermore there is the possibility of topological obstructions on a
compact manifold. It is well known {\it e.g.} that a four-torus allows for a
non-vanishing Pontryagin index \cite{Leutwyler:1992yt}. Moreover one could think of
choosing twisted boundary conditions \cite{'tHooft:1979uj,Baal:1984}.

This chapter is organised as follows: In the first section we summarise
the truncation schemes of \cite{vonSmekal:1997is,vonSmekal:1998is} and
\cite{Atkinson:1998tu} in more detail. They serve as simple test cases to 
check the feasibility of Dyson-Schwinger calculations on a compact manifold.
In the second section we set up the Dyson-Schwinger equations
on a four-torus and discuss their numerical treatment. 
In the last section of this chapter we display our numerical
results for three different truncation schemes: the two summarised in the next section
and the new truncation scheme introduced already in chapter \ref{YM}. We discuss the finite volume 
effects of our solutions \cite{Fischer:2002eq} and perform a preliminary infinite 
volume extrapolation.    

\section{A summary of two truncation schemes employing angular approximations \label{old-trunc-sec}}

\begin{figure}
  \centerline{ \epsfig{file=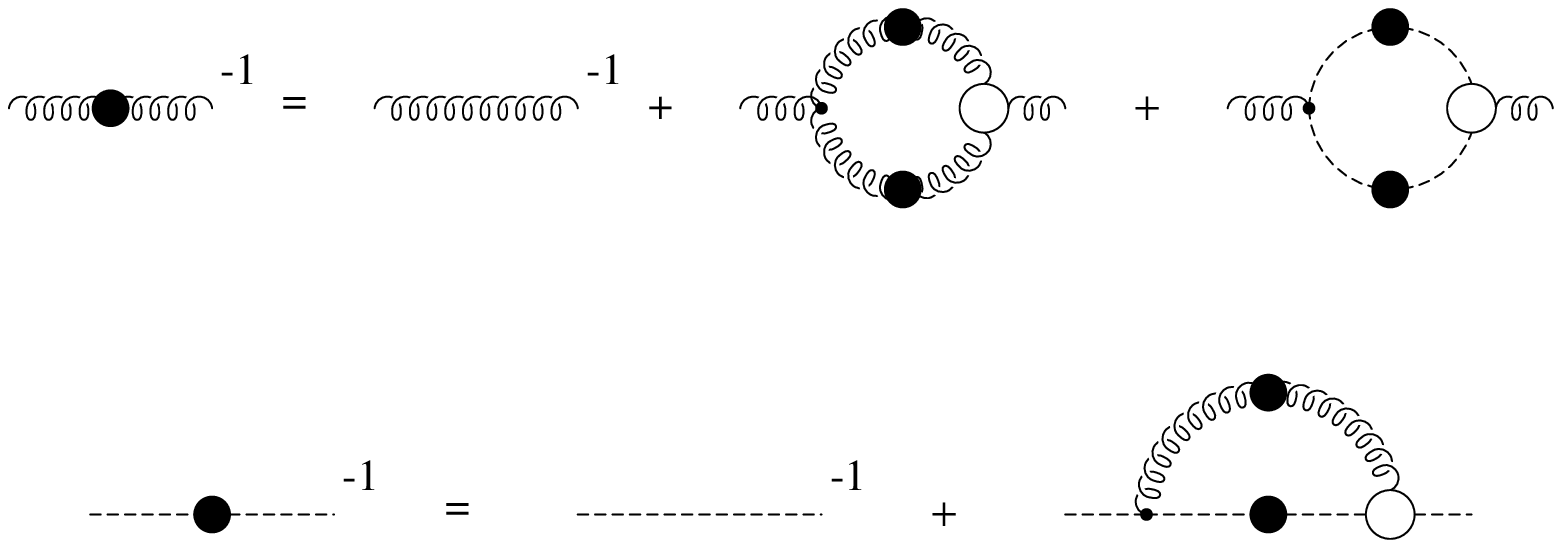,width=0.98\linewidth} }
  \vskip 3mm
  \caption{\sf Diagrammatic representation of the truncated gluon and ghost 
  Dyson--Schwinger equations according to the truncation schemes studied in this chapter.}
  \label{GluonGhost-tor}
\end{figure}

One central aim of the present investigation is to answer the question, how
the ghost and gluon propagators are affected by finite volume effects on a torus.
We will therefore compare solutions of the continuum DSEs with results obtained from 
the respective equations on the torus. To be able to distinguish genuine finite volume
effects from effects arising only in certain truncation schemes we will study not
only the scheme defined in chapter \ref{YM} but also the two previous
truncation schemes of refs.~\cite{vonSmekal:1997is,vonSmekal:1998is,Atkinson:1998tu} on a torus.
All three truncation schemes take into account the loops displayed
in Fig.~\ref{GluonGhost-tor}. However, contrary to our calculation in chapter \ref{YM}
the two previous schemes employ an angular approximation
in the loop integrals, {\it c.f.} the discussion in section \ref{gg-vertex-sec}.  
This angular approximation turned out to be good in the ultraviolet but less trustable
for small momenta. We therefore consider the system given in chapter \ref{YM} 
to capture more of the 
essential physics of the coupled ghost and gluon system. Nevertheless when it comes 
to a study of finite volume effects it seems helpful to employ all three truncation systems.

To proceed we will summarise the truncation schemes of 
refs.~\cite{vonSmekal:1997is,vonSmekal:1998is,Atkinson:1998tu} 
in the next two subsections. The main difference between these two 
truncation schemes is the treatment of the three-point functions.
Whereas in ref.~\cite{vonSmekal:1997is} approximate Slavnov--Taylor identities
have been employed to construct an {\it ansatz} for the vertices, in 
ref.~\cite{Atkinson:1998tu} bare three-point functions have been used. Amazingly,
though, both schemes provide results with identical qualitative infrared
behaviour: the gluon propagator vanishes in the infrared, the ghost propagator
is highly singular there, and the strong running coupling has an infrared fixed
point. 

\subsection{The dressed vertex truncation including the gluon loop \label{lorenz-trunc}}

In section \ref{gg-vertex-sec} we already discussed the full Slavnov-Taylor identity
for the ghost-gluon vertex, {\it c.f.} eq.~(\ref{ghost-gluon-STI}). Neglecting the
irreducible four-ghost term this identity and the corresponding one for the
three-gluon vertex are solved by the vertices \cite{vonSmekal:1997is}
\setlength{\jot}{3mm}
\beqa
\Gamma_\mu(p,q) &=& i q_\mu \left(\frac{G(k^2)}{G(q^2)} +\frac{G(k^2)}{G(p^2)} - 1\right) \,, 
\label{ghost-gluon-dressed}\\
\Gamma_{\mu \nu \rho}(p,q,k) &=& A_+(p^2,q^2,k^2)\, \delta_{\mu \nu}(p-q)_\rho 
+ A_-(p^2,q^2,k^2)\, \delta_{\mu \nu}(p+q)_\rho  \nonumber\\
&&+ 2 \frac{A_-(p^2,q^2,k^2)}{p^2-q^2}(\delta_{\mu \nu} \,
p.q - p_\nu q_\mu)(p-q)_\rho \,\,+\,\, \mbox{cyclic permut.} \,\,, \label{three-gluon-dressed}
\eeqa
with
\beq
A_\pm(p^2,q^2,k^2) = \frac{G(k^2)}{2} \left(\frac{G(q^2)}{G(p^2)Z(p^2)} \pm 
\frac{G(p^2)}{G(q^2)Z(q^2)} \right) \,.
\eeq
Here $k^2 = (p-q)^2$ denotes the gluon momentum in the ghost-gluon vertex $\Gamma_\mu$.
The three-gluon vertex $\Gamma_{\mu \nu \rho}$ 
is completely symmetric in the three momenta.
\setlength{\jot}{0mm}

The Lorenz indices in the gluon equation are contracted with 
the Brown-Pennington projector, eq.~(\ref{Rproj}).
In order to obtain the correct scaling behaviour of the gluon loop in the ultraviolet
the substitution
\begin{equation} 
Z_1 \:\: \longrightarrow \:\: {\cZ}_1(q^2) = 
\frac{G(q^2)^{(-2-6\delta)}}{Z(q^2)^{(1+3\delta)}} \, ,
\end{equation}
for the vertex renormalisation constant $Z_1$ has been introduced. Here the squared
momentum $q^2$ denotes the momentum integrated over in the gluon loop.
Recall the anomalous dimension of the ghost, $\delta=-9/44$, for an arbitrary number of
colours and zero flavours.

Furthermore in ref.~\cite{vonSmekal:1997is} the so called { \it modified angular approximations}
\beqa
Z(z),G(z) \simeq Z(x),G(x) &\longrightarrow& Z(y),G(y) \quad \mbox{if} \quad x \le y \,, \nonumber\\
Z(z),G(z) &\longrightarrow& Z(x),G(x) \quad \mbox{if} \quad x > y \,,
\label{ymax-mod} 
\eeqa
have been employed. The approximation in the first equation is designed
to eliminate a spurious term stemming from the vertex construction 
({\it c.f.} the discussion in subsection \ref{vertex-dressing}). Since the presence of this spurious
term causes a fatal inconsistency in the ghost equation \cite{Atkinson:1998tu}, this modified
angular approximation is a central ingredient in this truncation scheme.

Collecting all this together one arrives at the equations 
\beqa
\frac{1}{G(x)} &=& \tilde{Z}_3 - N_c \frac{g^2}{(2\pi)^4} \left\{  
\int_0^x d^4q \:\frac{{\sin}^2\Theta}{z^2}\: G(x)\: Z(x) 
+ \int_x^L d^4q \:\frac{{\sin}^2\Theta}{z^2}\: G(y)\: Z(y)\right\}, \hspace*{1cm}{ } \label{smekal-ghost}\\
\frac{1}{Z(x)} &=& Z_3 
+ \frac{N_c}{6} \frac{g^2}{(2\pi)^4}   
\int_0^L d^4q \:\frac{N(x,y,z)}{x y z} \: Z(y)^{-3\delta} \: G(y)^{(-1-6\delta)} 
\nonumber\\
&&+ \frac{N_c}{3} \frac{g^2}{(2\pi)^4} \left\{ 
\int_0^x d^4q \:\left(\frac{1-4 \cos^2\Theta}{x z} \: G^2(x) 
+ \frac{3 p\cdot q}{x y z} \: G(x) \: G(y) \right) \right. \nonumber\\
 && \hspace*{4.9cm}+ \left. \int_x^L d^4q \:\left(\frac{1-4 \cos^2\Theta}{x z}  
+ \frac{3 p\cdot q}{x y z} \right) G^2(y) \right\} \,, 
\label{smekal-gluon}
\eeqa
for the ghost and gluon dressing function \cite{vonSmekal:1997is}.
For ease of notation we have used the abbreviations $x:=p^2$, $y:=q^2$ and $z:=(p-q)^2$
for the squared momenta. Furthermore an $O(4)$--invariant momentum cutoff 
$L=\Lambda^2$ has been introduced. 
The integral kernel $N(x,y,z)$ is given by 
\beqa
N(x,y,z) &=& \frac{1}{4 x y z}
\left( 4z^4+32z^3 y+2z^3 x-26 z^2 x y-15 z^2 x^2-72 z^2 y^2+32 z y^3 \right. \nonumber\\
&& \left. -38 z y x^2+8 x^3 z - 26 z y^2 x+2 y^3 x+x^4+4 y^4-15 y^2 x^2+8 x^3 y \right) \,,
\hspace*{0.5cm}
\eeqa
and the angle $\Theta$ is defined by $z=(q-p)^2=x+y-2\sqrt{xy}\cos\Theta$. The equations
(\ref{smekal-ghost}) and (\ref{smekal-gluon})
will be implemented on a four-torus in section \ref{tor-sec}.

In the continuum one is now able to carry out the angular integrals analytically and solve
the equations along the lines described in ref.~\cite{vonSmekal:1997is,Hauck:1998sm,Hauck:1998fz}.
In the infrared the solutions $Z(x)$ and $G(x)$ behave power-like,
\begin{equation}
Z(x) = A x^{2\kappa} , \quad G(x) = B x^{-\kappa} ,
\label{power}
\end{equation}
with uniquely related coefficients $A$ and $B$. In this truncation scheme one 
obtains $\kappa \approx 0.92$. 
For three colours the running coupling approaches the fixed point
$\alpha(0) \approx 9.5$ in the infrared \cite{vonSmekal:1997is,vonSmekal:1998is}.

The dressed vertex truncation scheme of Hauck, Smekal and Alkofer has been the first one to 
include both the gluon and the ghost dressing function. Compared to the old
truncation of Mandelstam \cite{Mandelstam:1979xd}, which completely neglects the effects of
ghosts, this has been a major improvement. Whereas the Mandelstam equation is solved by a
gluon dressing function which diverges in the infrared, the inclusion of ghosts leads to the
qualitatively different picture of an infrared vanishing gluon dressing function and a diverging
ghost. This new picture, contradicting the old idea of infrared slavery, has been corroborated
since in other Dyson-Schwinger studies 
\cite{Atkinson:1998tu,Watson:2001yv,Zwanziger:2001kw,Lerche:2002ep}
as well as lattice calculations 
\cite{Langfeld:2001cz,Mandula:1999nj,Bonnet:2000kw,Bonnet:2001uh,Cucchieri:1999ky,Cucchieri:2000kw,Cucchieri:1998fy}.

One of the original strengths of the dressed vertex truncation scheme, namely the construction of the
vertices as solutions of approximate Slavnov-Taylor identities, is considered as a weakness in the
meantime. The so constructed vertex {\it ans\"atze} have been shown to be not in accordance with 
perturbation theory \cite{Watson:1999ha,watson} and, much worse, lead to inconsistent equations
when no angular approximations are employed. This is the central reason why we did not use the
vertices (\ref{ghost-gluon-dressed}) and (\ref{three-gluon-dressed}) 
in our truncation scheme of chapter \ref{YM}. 

\subsection{The bare vertex 'ghost-loop only' truncation \label{ghost-only-sec}}

In ref.~\cite{Atkinson:1998tu} also a bare ghost-gluon vertex has been used. In section
\ref{gg-vertex-sec} we have discussed at length why the bare vertex is capable of providing
reliable results even in the infrared region of momentum, where one would expect 
effects from non-perturbative vertex dressing to be most pronounced. In the gluon loop 
the authors of \cite{Atkinson:1998tu} use a bare three-gluon vertex {\it without}
modifying the vertex renormalisation constant $Z_1$. As this construction neither restores
a correct perturbative limit of the equations ({\it c.f.} our discussion in section \ref{truncation})
nor changes the infrared behaviour of the solutions, 
the authors themselves omit the gluon loop in the main part of their investigation. 

Substituting the tree-level ghost-gluon vertex for the dressed one and
neglecting the gluon loop the coupled system of equations 
(\ref{smekal-gluon})  reads 
\begin{eqnarray} 
\frac{1}{G(x)} &=& \tilde{Z}_3 - g^2N_c \int \frac{d^4q}{(2 \pi)^4}
\frac{K(x,y,z)}{xy}
G(y) Z(z) \; , \label{ghostbare-ab} \\ 
\frac{1}{Z(x)} &=& {Z}_3 + g^2\frac{N_c}{3} 
\int \frac{d^4q}{(2 \pi)^4} \frac{M(x,y,z)}{xy} G(y) G(z) \,. 
\label{gluonbare-ab} 
\end{eqnarray} 

Again we used the abbreviations $x:=p^2$, $y:=q^2$, $z:=(q-p)^2$, $s:=\mu^2$
and $L:=\Lambda^2$. 
The kernels $K$ and $M$ are already given in eqs. (\ref{new_kernels}).
The ghost equation (\ref{ghostbare-ab}) and the ghost loop (\ref{gluonbare-ab}) are 
identical to the ones of our truncation in section \ref{truncation}, provided
the projection parameter $\zeta$ is set to the special case
$\zeta=4$ ({\it c.f.} eqs. (\ref{Paproj}), (\ref{ghostbare}),
(\ref{gluonbare})). As the ghost loop is the leading part of the gluon equation,
both schemes share the same infrared behaviour for $\zeta=4$. 
Indeed, we have found the infrared solution $\kappa=1$ in subsection \ref{infrared} in accordance
with the infrared analysis in \cite{Atkinson:1998zc}, where no angular approximation
has been employed. However, we were not able to find a numerical solution for finite momenta
connecting to $\kappa=1$ in the infrared as has been detailed in section \ref{YM-results}.

Thus, similar to the truncation scheme summarised in the last subsection, one only gets
numerical solutions for finite momenta once an angular approximation has been employed.
The authors of \cite{Atkinson:1998tu} use the so called {\it ymax-approximation}
\beq
Z(z),G(z) \longrightarrow Z({\rm max}(x,y)),G({\rm max}(x,y)).
\label{ymax}   
\eeq
Upon angular integration the eqs.~(\ref{ghostbare-ab}), (\ref{gluonbare-ab}) are then simplified to
\begin{eqnarray}
\frac 1 {G(x)} &=& \widetilde Z_3(s,L) - \frac 9 4 \frac {g^2 N_c}{48\pi^2}
\left( Z(x) \int_0^x \frac{dy}{x} \frac{y}{x} G(y)
+ \int_x^L \frac{dy}{y} Z(y) G(y) \right) ,
\label{ghAB}\\
\frac 1 {Z(x)} &=& Z_3(s,L) + \frac {g^2 N_c}{48\pi^2}
\left( G(x) \int_0^x \frac{dy}{x} \left( -\frac{y^2}{x^2} + \frac{3y}{2x} 
\right) G(y) + \int_x^L \frac{dy}{2y} G^2(y)\right) .
\label{glAB}
\end{eqnarray} 
In the infrared the solutions of eqs.\ (\ref{ghAB},\ref{glAB})  behave
power-like, {\it c.f.\/} eq.\ (\ref{power}), with  $\kappa \approx 0.77$. 
The running coupling approaches the fix point
$\alpha(0)= 11.47$ in the infrared \cite{Atkinson:1998tu}.

The 'ghost-loop only' truncation scheme of Atkinson and Bloch successfully proved that bare vertices
result in the same qualitative behaviour of the ghost and gluon dressing functions in the infrared
as the dressed construction of the last subsection. This surprising result served as basis
for recent analytical investigations in the infrared, where either 
a bare ghost-gluon vertex \cite{Atkinson:1998zc,Zwanziger:2001kw} or a bare ghost-gluon vertex with
multiplicative corrections \cite{Lerche:2002ep,Lerche} has been employed. The weak point
of the 'ghost-loop only' truncation is its failure in the ultraviolet, where the absence of
the gluon loop leads to a contradiction with perturbation theory.

\section{Finite volume effects on a four-torus $T_4$ \label{tor-sec}}


From a technical point of view using a four-torus as the underlying manifold or
choosing (anti-)periodic boundary conditions on a hypercube is identical. The first question
to answer is therefore: are periodic or antiperiodic boundary conditions adequate for
the fields appearing in the coupled system of ghost and gluon Dyson-Schwinger equations.
This question is easily answered for the gluon field. Due to the bosonic nature of gluonic
excitations we have to use periodic boundary conditions for the gluons and subsequently
for the gluon dressing function $Z$. Although this is not obvious, the same is true
for the ghosts. The reason for this 'strange' behaviour of Grassmann fields is to be searched
in the definition of the Faddeev--Popov determinant, which due to its introduction
as gauge fixing device shares the symmetry properties of the gluon field
\cite{Hata:1980yr,Reinhardt:1996fs}.    
An easy way to see this is to check the BRS-transformation rule of the quark field $\Psi$, which
already has been given in eq.~(\ref{quark-brs}),
\beq
s \Psi = -ig t^a c^a \Psi \;. \\
\eeq
Here $t^a$ is a generator of $SU(N_c)$ and $s$ is the BRS-operator generating the transformation.
Since $s$ is continuous it does not change the symmetry properties of the fields.
We thus have antiperiodic behaviour on both sides of the equation due to the fermionic
nature of the quark field $\Psi$. This, however, is only consistent if the ghost field $c$
has periodic boundary conditions.

\begin{figure}[t]
\begin{center}
\epsfig{file=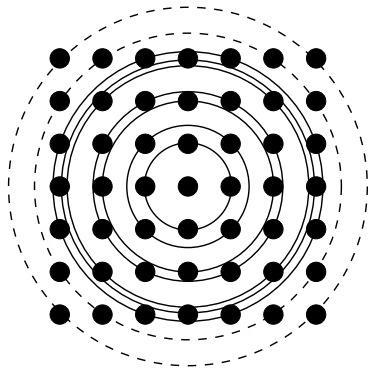,width=0.3\linewidth}
\end{center}
\caption{\sf \label{Latt}
Sketch of the momentum grid dual to the four-torus and the 
summation over complete hyperspheres indicated by fully drawn
circles. The hyperspheres depicted by dashed lines are not complete
due to the numerical ultraviolet cutoff in every direction of the grid.}
\end{figure}

With periodic boundary conditions on a hypercube with length $l$ in every direction 
the four-dimensional momentum integrals of our DSEs have to be substituted 
by a sum over four indices,
\begin{equation}
\int \frac {d^4q}{(2\pi )^4} \to \frac {1}{l^4} \sum _{j_1,j_2,j_3,j_4} \; .
\label{DefineSum}
\end{equation}
The quantities of interest, the gluon and ghost
renormalisation functions $Z(p^2)$ and $G(p^2)$ depend only on the
O(4) invariant squared momenta as all directions on the torus are treated on an
equal footing. This suggests to relabel the points on the
momentum grid not according to a Cartesian but a hyperspherical coordinate 
system,
\begin{equation} 
\frac {1}{l^4} \sum _{j_1,j_2,j_3,j_4} = \frac {1}{l^4} \sum _{j,m} \; ,
\label{SphericalSum} 
\end{equation}
where the index $j$ numbers the hyperspheres $q^2={const}$, i.e. the circles drawn in the
sketch in Fig.~\ref{Latt}. The index $m$, which numbers
the grid points on each hypersphere respectively, will be suppressed in the following. 

In the integrals to be discretised there appear three momenta, the external
momentum, labelled $p$, the loop momentum $q$ and for the
second propagator in the loop $k=q-p$. We will use the following notation:
\begin{eqnarray}
x:= p^2 \qquad &{\rm with}& \qquad x_i \in {\rm hypersphere} \; i \; ,
\nonumber\\
y:= q^2 \qquad &{\rm with}& \qquad y_j \in {\rm hypersphere} \; j \; ,
\nonumber\\
z:= k^2=(q-p)^2 \qquad &{\rm with}& \qquad z_n \in {\rm hypersphere} \; n \; .
\label{xyzDef}
\end{eqnarray} 
On the hypercubic momentum grid dual to the four-torus the momentum $k=q-p$
is located on the grid for every pair of grid momenta $p$ and $q$ as can
be seen from elementary vector operations. This is no longer true once we introduce a
momentum cutoff corresponding to a finite extent of the momentum grid.
We will detail below, how we treat the dressing functions at those momenta $z$ which are
larger than the cutoff.

For the moment let us think a little on the appropriate way to introduce such a cutoff on a grid.
Recall that an O(4) invariant cutoff $\Lambda$ has been introduced into all of the continuum
DSEs eqs.~(\ref{ghostbare}), (\ref{gluonbare}), 
(\ref{smekal-ghost}), (\ref{smekal-gluon}) and (\ref{ghostbare-ab}), (\ref{gluonbare-ab}). 
In order to compare the continuum results with the ones on the torus a
corresponding regularisation of the sums over grid momenta is required.
A first idea might be to simply cut the sums in each direction according to
$j_1,j_2,j_3,j_4=-N, \ldots ,0, \ldots N$. Such a
method, however, breaks O(4) invariance and introduces numerical 
errors, as will be shown in section \ref{tor-num-results}.
(Note that in lattice Monte-Carlo simulations the analysis of the
resulting data for O(4) invariance necessitates special kinds of cuts through
the lattice, see e.g. \cite{Bonnet:2000kw}.) As can be seen from Fig.~\ref{Latt} an 
O(4) invariant cutoff of the sums necessitates to neglect the 'edges':
The sum extends only over the fully drawn hyperspheres, and we omit the summation
over the dashed ones.


The main difference between the equations on the torus and the ones in the continuum
is the effective treatment in the infrared. The zero modes at $j=0$, which are not present
in the continuum, are neglected in all calculations presented here\footnote{ 
An estimate of their possible contribution is given in appendix \ref{zero-modes}.}.
The finite volume in coordinate space leads to a finite value of squared momentum for the first
hypersphere $j=1$. Thus one has not to worry about possible infrared
singularities. Furthermore the complicated matching procedure of the infrared integrals
to the remaining integration region in the continuum
is not necessary on a torus. On the other hand one might worry that the infrared part of
the loop integrals are crucial to obtain solutions at all \cite{Atkinson:1998tu}.
In appendix \ref{radial} we show that this is not the case. We reproduce the continuum
results in the 'ghost loop only' truncation scheme on a very coarse radial momentum grid 
without a single point in the infrared. This gives us first confidence that Dyson-Schwinger 
equations might be feasible on a torus.

Considering all this, one anticipates already at this level some
deviations in the infrared between the solutions obtained in these different
ways. After writing down the DSEs on a torus in the next two subsections we discuss
our numerical results. They demonstrate that using a torus as infrared cutoff works surprisingly well.

\subsection{Angular approximated DSEs on a torus}

Substituting the replacement rule $\int \frac{d^4q}{(2\pi)^4} \rightarrow
\frac{1}{l^4} \sum_j$ into the eqs.~(\ref{ghostbare-ab}), (\ref{gluonbare-ab}) 
and using the angular approximation
(\ref{ymax}) the DSEs in bare vertex ghost-loop only truncation read on a torus
\setlength{\jot}{3mm}
\begin{eqnarray}
\frac{1}{G(x_i)} &=&  \widetilde{Z}_3(s,L) - {g^2}{N_c} \frac{1}{l^4} \sum_j 
\frac{K(x_i,y_j,z_n)}{x_i y_j} G(y_j) Z({\rm max}(x_i,y_j))
 \; ,
\label{disGDSE}
\\
\frac{1}{Z(x_i)}&=& Z_3 (s,L) + {g^2} \frac{N_c}{3} 
\frac{1}{l^4} \sum _j \frac{M(x_i,y_j,z_n)}{x_i y_j} G(y_j) 
G({\rm max}(x_i,y_j))   \; . 
\label{disZDSE} 
\end{eqnarray} 
All arguments $x_i=(i 2\pi/l)^2$ and $y_j=(j 2\pi/l)^2$ of the dressing functions 
$G$ and $Z$ are on the momentum grid. However, note that $z_n=(q-p)^2$
might be larger than the ultraviolet cutoff even if $x_i$ and $y_j$ are not.
Nevertheless the kernels in eqs.~(\ref{disGDSE}), (\ref{disZDSE}) can be calculated 
straightforwardly according to the expressions in eqs.~(\ref{new_kernels}).

The corresponding equations for the dressed vertex truncation on a torus can be derived
analogously in a straightforward manner from eqs. (\ref{smekal-ghost}) and (\ref{smekal-gluon}).
However, as the expressions are quite lengthy and bring nothing new we
do not give their explicit form.
\setlength{\jot}{0mm}

\subsection{The novel truncation scheme on a torus}

With the replacement $\int \frac{d^4q}{(2\pi)^4} \rightarrow
\frac{1}{l^4} \sum_j$ discussed above, the eqs. (\ref{ghostbare}), (\ref{gluonbare}) for the novel
truncation scheme of chapter \ref{YM} read
on the torus:
\begin{eqnarray}
\frac{1}{G(x_i)} &=& \widetilde{Z}_3(s,L) - {g^2}{N_c}\frac{1}{l^4} \sum_j 
\frac{K(x_i,y_j,z_n)}{x_i y_j} G(y_j) Z(z_n) \; ,
\label{ghost_tor} \\
\frac{1}{Z(x_i)} &=& {Z}_3 (s,L) + {g^2} \frac{N_c}{3}
\frac{1}{l^4} \sum _j \frac{M(x_i,y_j,z_n)}{x_i y_j} G(y_j) G(z_n)
\label{gluon_tor}\\
&& \:\:\:\:\:\:\:\: + {g^2} \frac{N_c}{3}
\frac{1}{l^4} \sum_j \frac{\tilde{Q}(x_i,y_j,z_n)}{x_i y_j} 
G^{1-a/\delta-2a}(y_j) G^{1-b/\delta-2b}(z_n)
Z^{-a}(y_j) Z^{-b}(z_n) \; .\nonumber
\end{eqnarray}
The kernels $K$, $M$ and $\tilde{Q}$ are given in eqs.~(\ref{new_kernels}), (\ref{Q-tilde}).
As already stated $\sqrt{z}$ might be larger than the
ultraviolet cutoff $\sqrt{L}$ even if $\sqrt{x}$ and $\sqrt{y}$ are not.
Nevertheless the kernels can be evaluated straightforwardly. However,
if $z$ with $L<z<4L$ is the argument of a dressing function one has the choice of
two different methods. One way is to approximate $Z(z)$ and $G(z)$ by
$Z(L)$ and $G(L)$. Another more elaborate treatment consists of 
matching the corresponding perturbative ultraviolet tail to the function under 
consideration. We have applied both methods and found only very small quantitative 
differences.

\section{Renormalisation and results}

\subsection{The renormalisation scheme}

In order to obtain comparable results for the Dyson--Schwinger equations in
the continuum and on the torus we have to impose the same 
renormalisation conditions. This can be done in two ways: {\it First}, 
one can use the solutions of the continuum equations for fixed cutoff $\Lambda$ and
fixed renormalisation scale $\mu$ to calculate the corresponding renormalisation constants 
$Z_3(\mu^2,\Lambda^2)$ and $\tilde{Z}_3(\mu^2,\Lambda^2)$. These can subsequently be used
in the equations on the torus. {\it Second}, one can subtract the torus equations
at the squared momenta $s_G$ and $s_Z$ and trade the two
renormalisation constants for the values of the dressing functions at these
momenta, namely $Z(s_Z)$ and $G(s_G)$ ({\it c.f.} subsection \ref{subtraction}). 
These values are then taken from the continuum solution. 
If $s_Z$ and $s_G$ are sufficiently far in the ultraviolet region of
momentum, where finite volume effects play a minor role, the two procedures
lead to the same results within the limits of numerical accuracy.

In the bare vertex truncation scheme we have chosen 
the renormalisation condition $Z(\mu^2)=G(\mu^2)=1$ for the continuum equations. 
For the renormalisation point $\mu^2$ we took the same value as for the 
ultraviolet cutoff: $\Lambda^2=\mu^2=0.2$. 
Of course, this choice is by no means special and one is
completely free to choose the renormalisation point wherever one likes. We 
subtracted the continuum ghost equation at zero momentum and the 
gluon equation at the renormalisation point $\mu^2$. This is convenient as we are then
able to use the condition $Z(s_Z=\mu^2)=1$ directly as input in the
calculation. The second input is provided by the coefficient $A$ of the leading
order infrared expansion of the gluon dressing function, $Z_{IR}(x)=Ax^{2
\kappa}$. The condition $G(\mu^2)=1$ corresponds to $A=357.33$ in our
calculation. (Note that the coefficient $A$ is uniquely related to the 
corresponding coefficient $B$ of the ghost dressing function in the infrared, {\it c.f.}
the discussion in section \ref{infrared}.) The value of
the coupling at the renormalisation point is taken to be
$\alpha(\mu^2) = g^2/4\pi = 0.97$. Again this number is completely arbitrary provided one
stays in the interval $\alpha(0) > \alpha(\mu^2) \ge \alpha(\Lambda^2)$.
In the torus equations we furthermore used the values $Z_3(\mu^2=0.2,\Lambda^2=0.2)=0.9591$ and
$\tilde{Z}_3(\mu^2=0.2,\Lambda^2=0.2)=1.1034$, which have been 
determined from the continuum solution.

A physical momentum scale can be determined only in truncation schemes which 
provide the correct perturbative running of the coupling in the ultraviolet.
With the missing gluon-loop this is not the case in the 'ghost-loop
only' truncation scheme. We therefore have to stick to an internal momentum
scale without physical units in this case. The situation is different, however,
in the dressed vertex truncation scheme. Here we have fixed the momentum scale
by calculating the running coupling for the colour group SU(3) and using the
experimental value $\alpha(x)=0.118$ at $x=M_Z^2=(91.187 \, \mbox{GeV})^2$ 
to fix a physical scale.

In the numerical treatment of the continuum equations of the dressed vertex truncation scheme
the ghost equation is subtracted at zero
momentum and the gluon equation at the arbitrary finite momentum $s=1.048$ GeV$^2$. We solved
both equations similar to the method described in ref.~\cite{Hauck:1998sm},
especially we introduced also the auxiliary functions $F(x)$ and $R(x)$ as
defined in ref.~\cite{vonSmekal:1997is}. 
As input values serve the infrared expansion of $R(x)$, $R(x)=x^\kappa +\ldots$,
and the value $R(s)=0.8$ at the gluon subtraction point. For the calculations on the torus 
we determined the values $Z_3(\mu^2=M_Z^2,\Lambda^2=1.255 \,
\mbox{GeV}^2)=1.266$ and $\tilde{Z}_3(\mu^2=M_Z^2,\Lambda^2=1.255 \,
\mbox{GeV}^2)=0.966$ from the continuum solution.

For the truncation scheme of chapter \ref{YM} we have described in section \ref{subtraction}
how we obtain the continuum results. The corresponding torus
results have been calculated with torus equations both subtracted at the 
renormalisation point $\mu^2=1.9 \,\mbox{GeV}^2$. The input value for the 
gluon dressing function at this momentum is $Z(\mu^2)=0.83$. Requiring
$G(\mu^2)=1/\sqrt{Z(\mu^2)}$ then fixes the normalisation for $G(x)$. For the
value of the coupling at the renormalisation point we chose 
$\alpha(\mu^2)=0.97$ similar to the two other truncation schemes.  
\begin{figure}[th!]
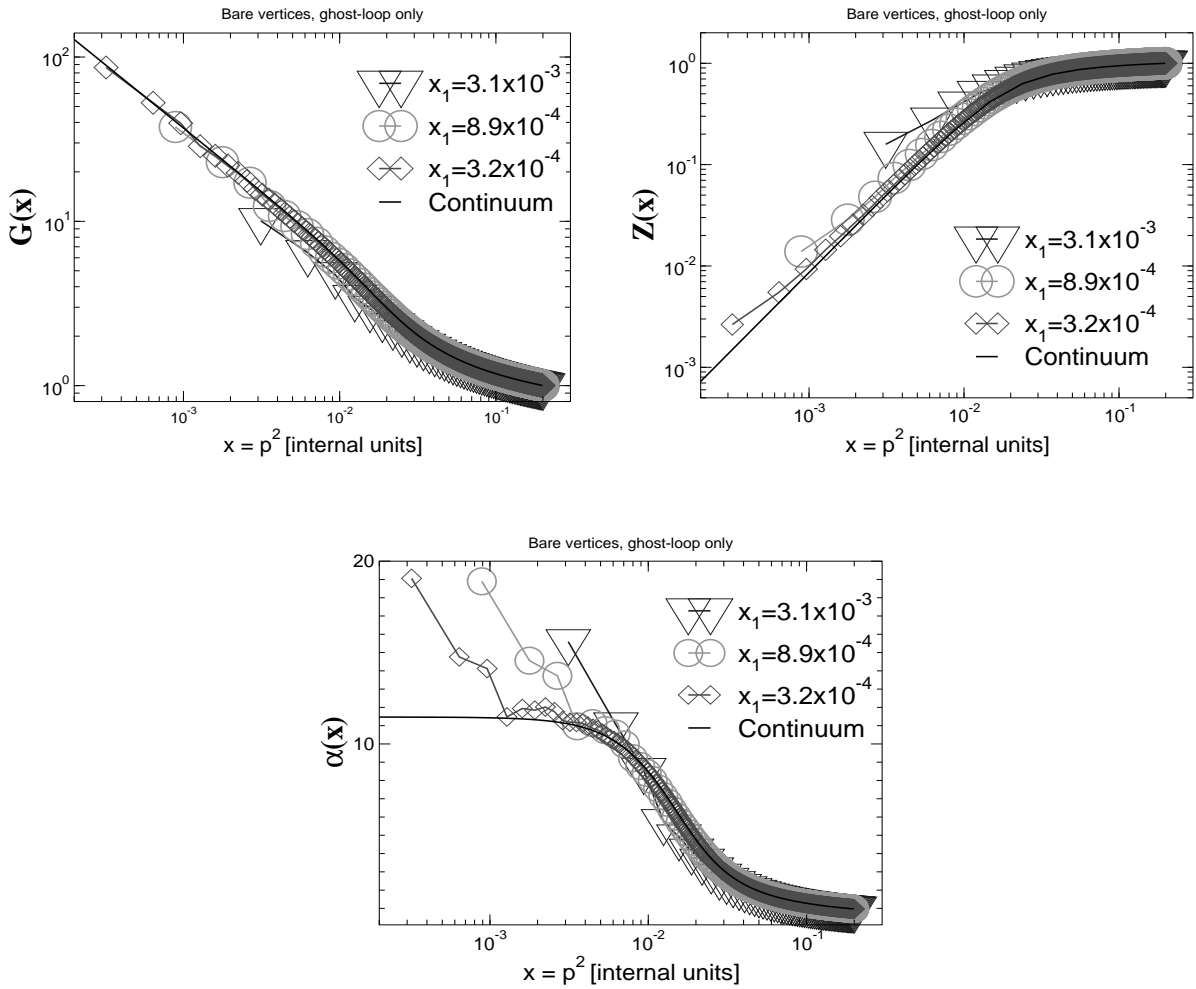

\vspace{0.5cm}
\centerline{
\epsfig{file=bare_ghostonly.g.eps,width=7.5cm,height=6cm}
\hspace{0.5cm}
\epsfig{file=bare_ghostonly.z.eps,width=7.5cm,height=6cm}
}
\vspace{1cm}
\centerline{
\epsfig{file=bare_ghostonly.a.eps,width=7.5cm,height=6cm}
}
\caption{\sf \label{bare.dat}
Shown are the ghost dressing function, $G(x)$, the gluon dressing function,
$Z(x)$, and
the running coupling, $\alpha(x)$, in the bare-vertex ghost-loop 
only truncation for
different momentum grid spacings corresponding to different 
finite volumes of a torus. The fully drawn lines labelled continuum
represent the respective results for continuous momenta.}
\end{figure}

\subsection{Numerical solutions \label{tor-num-results}}

\begin{figure}[th!]
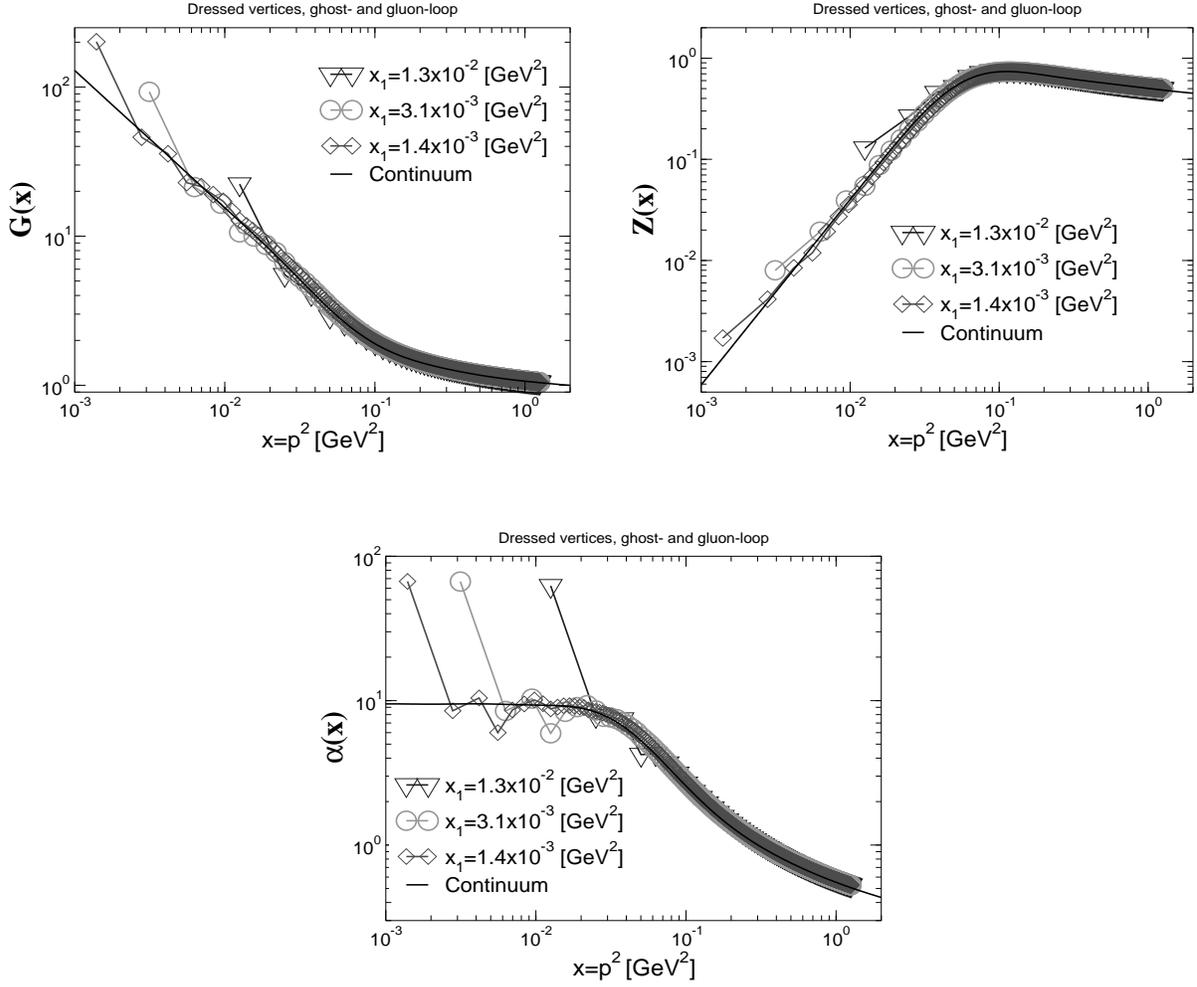

\vspace{0.5cm}
\centerline{
\epsfig{file=dressed_gluonincl.g.eps,width=7.5cm,height=6cm}
\hspace{0.5cm}
\epsfig{file=dressed_gluonincl.z.eps,width=7.5cm,height=6cm}
}
\vspace{1cm}
\centerline{
\epsfig{file=dressed_gluonincl.a.eps,width=7.5cm,height=6cm}
}
\caption{\sf \label{dressed.dat}
The same as Fig.~\protect{\ref{bare.dat}} for the dressed vertex truncation.}
\end{figure}

Our results for the ghost dressing function, the gluon dressing function and
the running coupling in the bare-vertex 'ghost-loop only' and the dressed vertex
truncation can be seen in Figs.~\ref{bare.dat} and \ref{dressed.dat}. 
In both truncation schemes we solved for three
different momentum spacings corresponding to different volumes in coordinate
space. To keep the cutoff identical for all the spacings within each truncation
scheme we have chosen three different grid sizes respectively. For the bare
vertex truncation they are $N=17^4, 31^4, 51^4$ and for the dressed vertex
truncation they are $N=21^4, 41^4, 61^4$.

The finite volume effects seen in both truncation schemes are very similar: 
Compared to the respective continuum solutions the ones obtained on a torus show deviations
for the first few lattice points in the infrared. For large momenta all
functions obtained on a torus approach the continuum ones. 
The biggest effect can be seen for the running coupling $\alpha$. 
As $\alpha(p^2)$ is proportional to the product $Z(p^2)G^2(p^2)$ the deviations of
the torus dressing functions from the continuum curves amplify in the infrared in
a somewhat erratic way, such that the data points at small momenta cannot be
connected by a smooth line. Comparing larger and smaller spacings of momentum
grids one clearly sees that the effect is always one of the first spheres on
the respective lattices and therefore moves to the infrared for smaller
spacings. 

The most important properties of the continuum solutions can still be 
found in the torus solutions despite some deviations in the infrared. Going
from larger to smaller spacings a power-like behaviour of the dressing functions
in the infrared with the correct exponents can still be inferred. For the truncation
scheme with dressed vertices the gluon dressing function on the torus has the
same shape and the same height of the bump in the bending region of the curve.
This is a first central result of our investigation: {\it Employing a torus as
infrared regularisation is possible.}

\begin{figure}[t]
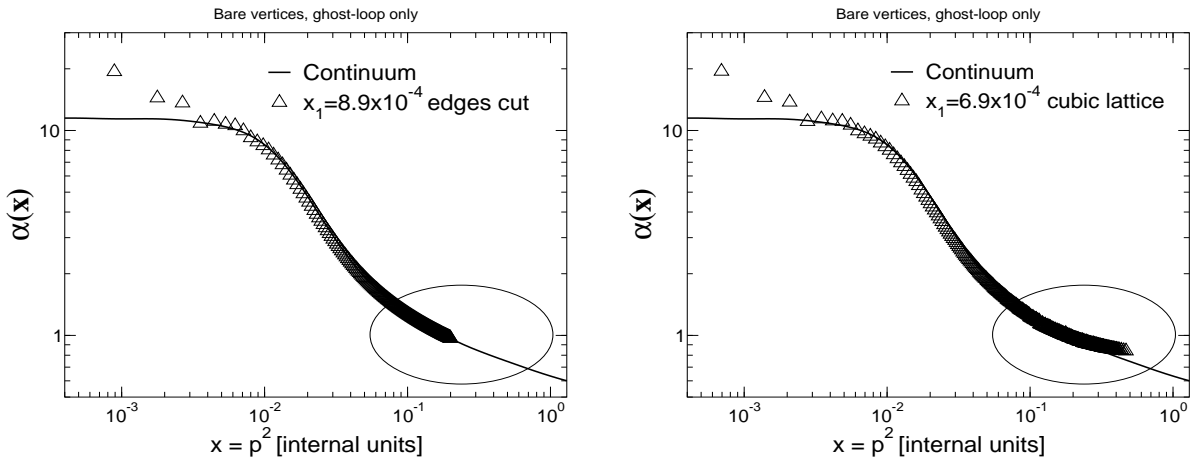

\vspace{0.5cm}
\centerline{
\epsfig{file=a.tor.latt.pap.eps,width=7.5cm,height=6cm}
\hspace{0.5cm}
\epsfig{file=a.tor.latt2.pap.eps,width=7.5cm,height=6cm}
}
\caption{\sf \label{cut.dat}
Here we compare the running coupling of the bare vertex ghost-loop only truncation
for different treatments of the lattice cut-off, {\it c.f.} section \ref{tor-sec}.
The torus curve on the left panel is calculated on a grid where the edges have
been cut, whereas the coupling on the right panel is obtained on a full hypercubic
grid. The ellipses mark the region where the latter treatment is inferior.}
\end{figure}

Before we move on to discuss the results in our novel truncation scheme we show what happens, 
if the four dimensional momentum grid is not cut at the edges according to the prescription
discussed in section \ref{tor-sec}. In Fig.~\ref{cut.dat} we compare  
the running coupling of the bare vertex 'ghost-loop only' truncation
for different treatments of the lattice cut off. The torus curve in the left panel is 
calculated on a grid with edges cut, 
whereas the coupling on the right panel is obtained on a full hypercubic
grid. Clearly one observes sizeable deviations in the ultraviolet behaviour of the
latter curve compared to the continuum result. As explained in section \ref{tor-sec} this
is a result of breaking rotational symmetry on the lattice. On the inner, full hyperspheres
we have the discrete rotational symmetry group Z(4) for very small discrete steps,
whereas these steps become larger and larger for those hyperspheres which contain
points on the edges of the lattice. Cutting these edges obviously 
improves the ultraviolet behaviour of our solutions. 

\begin{figure}[t]
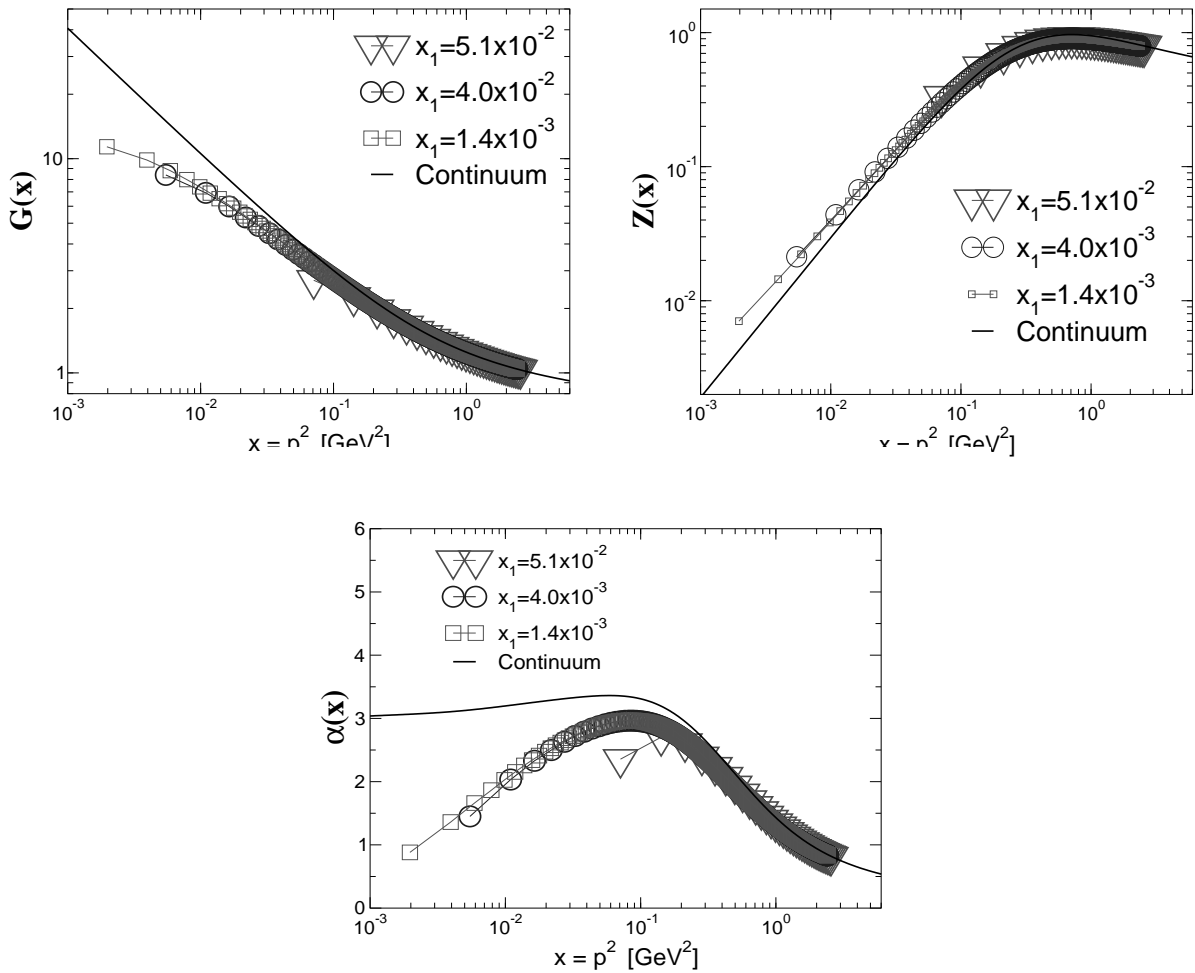

\vspace{0.5cm}
\centerline{
\epsfig{file=newtrunc.main.g.eps,width=7.5cm,height=6cm}
\hspace{0.5cm}
\epsfig{file=newtrunc.main.z.eps,width=7.5cm,height=6cm}
}

\vspace{0.8cm}
\centerline{
\epsfig{file=newtrunc.main.a.eps,width=7.5cm,height=6cm}
}

\caption{\sf \label{new-tor.dat}
Here we present the ghost dressing function, $G(x)$, the gluon dressing function, $Z(x)$, and
the running coupling, $\alpha(x)$, in the truncation scheme of chapter \ref{YM} 
for different volumes using a transverse projector, {\it i.e.} $\zeta=1$.}
\end{figure}

The numerical results for our novel truncation scheme of chapter \ref{YM} 
with a transverse projector, $\zeta=1$,
are shown in Fig.~\ref{new-tor.dat}. At the time those results were found  
they have been the first numerical solutions at all in a truncation scheme without any
angular approximations \cite{Fischer:2002eq}. In the meantime, as we have seen in 
chapter \ref{YM}, solutions for continuous momenta are also available \cite{Fischer:2002hn}. 
Both calculations are compared in Fig.~\ref{new-tor.dat}. 
For the presented torus solutions on three different volumes lattice sizes of $N=13^4,
43^4,71^4$ have been used\footnote{Note that the scale for the torus
results shown in Fig.~\ref{new-tor.dat} is slightly different than in reference 
\cite{Fischer:2002eq}. There an extrapolation to the mass of the Z-boson has been used, whereas
here we simply take the scale which has been determined for the continuum result.}. 
As mentioned in the last subsection we did check cutoff effects by extrapolating
the propagator functions at $z > \Lambda^2$ with a logarithmic tail with the 
correct anomalous dimensions. The results as compared to the ones obtained by 
simply setting $Z(z)=Z(\Lambda^2)$ and $G(z)=G(\Lambda^2)$ for  
all $z>\Lambda^2$ change by less than one per mille.


\begin{figure}[t]
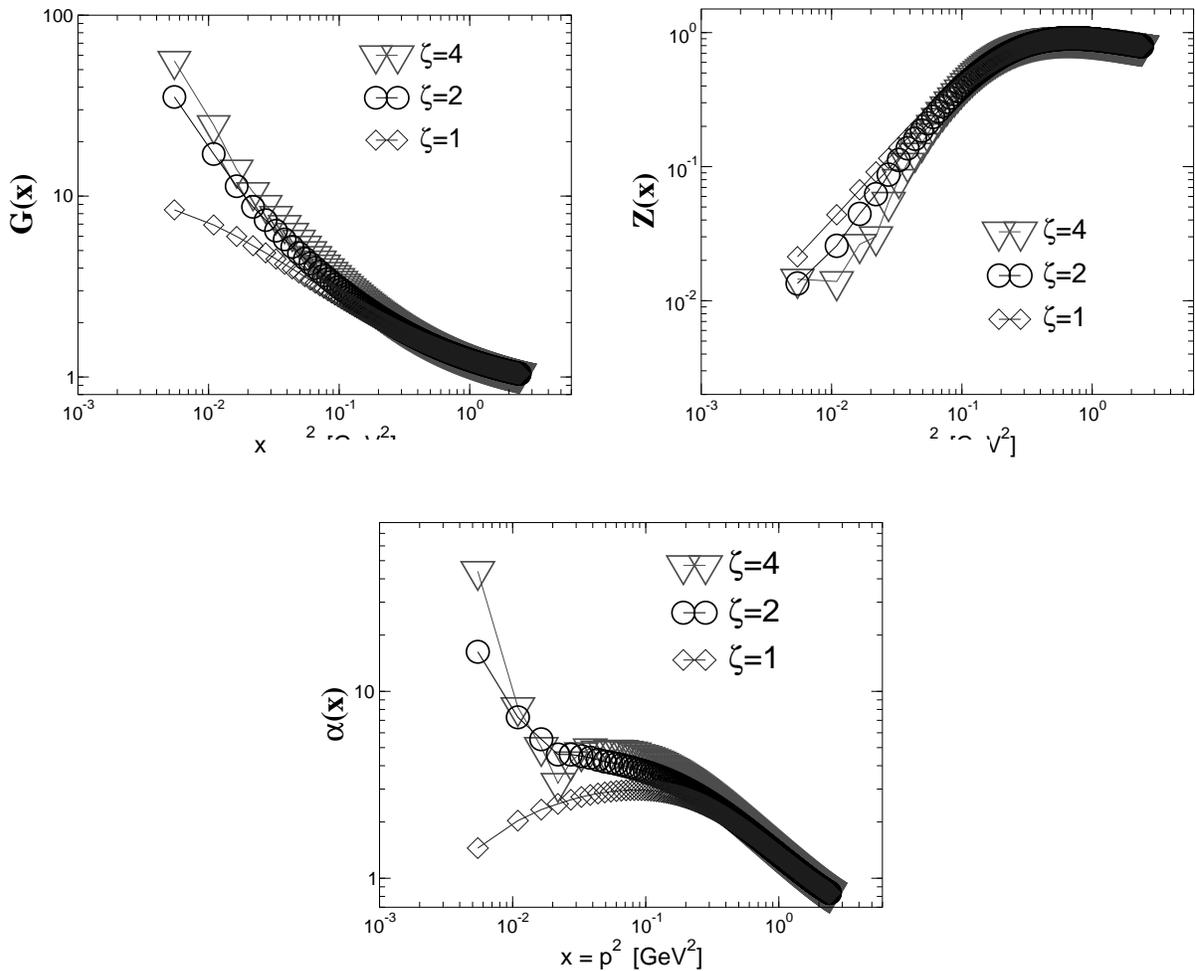

\vspace{0.5cm}
\centerline{
\epsfig{file=newtrunc.proj.g.eps,width=7.5cm,height=6cm}
\hspace{0.5cm}
\epsfig{file=newtrunc.proj.z.eps,width=7.5cm,height=6cm}
}

\vspace{0.8cm}
\centerline{
\epsfig{file=newtrunc.proj.a.eps,width=7.5cm,height=6cm}
}
\caption{\sf \label{proj-tor.dat}
The same as Fig.~\protect{\ref{new-tor.dat}} for different projectors. The corresponding
solutions in the continuum are given in Fig.~\ref{proj.dat}.}
\end{figure}


In section \ref{infrared} we raised the question, which one out of two analytical solutions
of the continuum DSEs in the infrared is connected to a full numerical solution for finite
momenta. We have answered this question partially in section \ref{YM-ren-results} where we 
showed that a numerical solution exists which connects to the infrared solution $\kappa=0.595$.
However, in a numerical calculation in the continuum the analytical infrared solutions are
used as input in the infrared and reproduced self-consistently. Thus one might argue that we showed
the existence but not the uniqueness of the numerical solution with $\kappa=0.595$.
On a torus we have no input in the infrared. Therefore a torus solution favours automatically
the physical solution in the infrared, as we believe the infinite volume limit to be smooth.
The results shown in Fig.~\ref{new-tor.dat} are clearly in 
agreement with the power behaviour $\kappa=0.595$ 
already found in the continuum and disfavours the second solution, $\kappa \approx 1.3$. 
The infrared critical exponent as calculated in refs.\ 
\cite{Zwanziger:2001kw,Lerche:2002ep}, $\kappa=0.595$, thus has been verified.

The gluon dressing function in Fig.~\ref{new-tor.dat}
is remarkably stable against changes of the volume  and
approaches more and more the expected power solution for small momenta, although
this process seems to be very slow.  For
the ghost dressing function one observes deviations of the first points
in the infrared: An extraction of the correct infrared critical exponent from
the numerical solution for the ghost function is hardly possible. Only some
points come close to the analytical value of the continuum before the curve
starts bending down again for small momenta. For the extracted value of the
running coupling in the infrared this leads to a distinct mismatch to what one
is to expect on the basis of analytical results.

At first sight the fact that the power solution for the ghost dressing function
could be not reproduced numerically to a reasonable precision may seem
disappointing. Nevertheless these numerical results themselves show that the
ghost dressing function is highly singular in the infrared. This reflects the
long-range correlation of ghosts in Landau gauge. Therefore one should expect
the ghost dressing  function  to be the one  affected most by a finite volume.
On the contrary the gluon dressing function vanishes in the infrared and
consequently it is much less affected by a finite volume. We expect the ghost
dressing function together with the running coupling to approach more and more
the correct power solution in the infrared as lattice spacings are decreased
and lattice sizes are increased.

\begin{figure}[t]
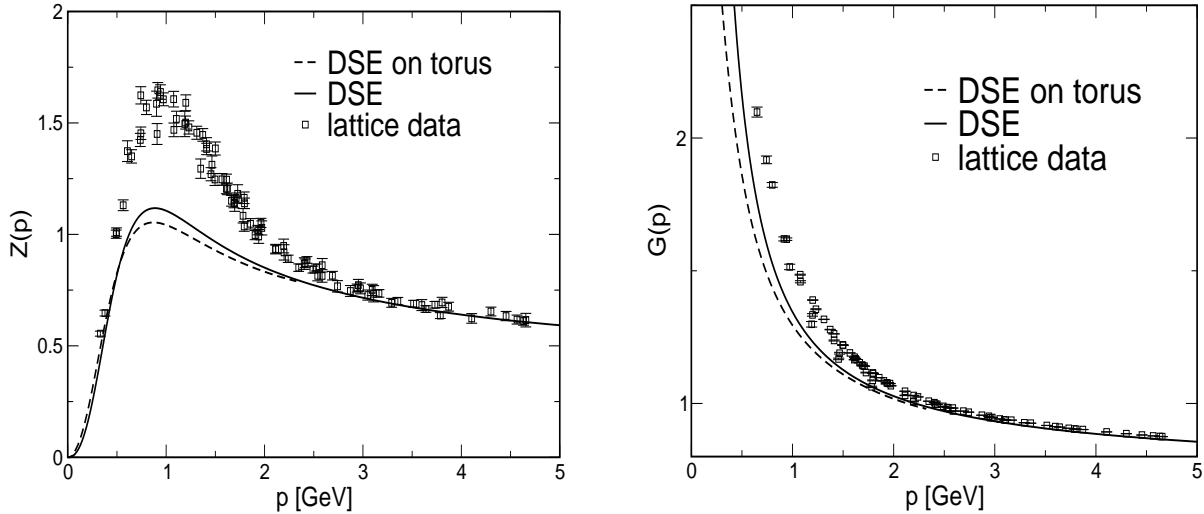

\vspace{0.5cm}
\centerline{
\epsfig{file=latcomp.glue.eps,width=7.4cm,height=6.8cm}
\hspace{0.8cm}
\epsfig{file=latcomp.ghost.eps,width=7.4cm,height=6.8cm}
}
\caption{\sf \label{lattice-tor.dat}
Results on the torus compared to recent lattice results \cite{Langfeld:2002dd,Bloch:2002we}. 
As the torus points are very
close to each other on a linear momentum scale we did not resolve the torus curves into
single points.}
\end{figure}

Furthermore, we add a remark on the transversality of the gluon propagator,
{\it c.f.} the respective discussion in section \ref{gluonghost}.
Our numerical results on a torus for different values of the parameter $\zeta$ can be seen in
Fig.~\ref{proj-tor.dat}. Although the solutions show the expected dependence
on the form of the projector this dependence is not too drastic and in general
the behaviour of these different solutions is very similar.
For the gluon dressing function one observes that the more $\zeta$ grows the
greater is the deviation from the pure power behaviour and correspondingly
from the continuum solutions, which have been shown in subsection \ref{YM-results}. 
The points at small momenta cannot be connected by a smooth line any more. 
Based on the infrared analysis in subsection \ref{infrared} one might anticipate
that $\kappa$ should approach the value $\kappa=0.5$ more and more as $\zeta$
grows until there is a jump to the solution from $\kappa=0.5^+$ to $\kappa=1$
as the Brown--Pennington limit $\zeta=4$ is reached. We do not observe such
a qualitative jump in our solutions on a torus. The solution shown for
$\zeta=4$  is approached smoothly when $\zeta$ approaches this limit. 
This clearly indicates that the solution $\kappa=1$ might not exist at all
if one removes the torus as a regulator.

Finally we compare our results of the new truncation scheme to recent SU(2)
lattice calculations \cite{Bloch:2002we,Langfeld:2002dd}, {\it c.f.} the corresponding
comparison between lattice and DSE results in section \ref{YM-results}. 
The two graphs in 
Fig.~(\ref{lattice-tor.dat}) suggest that 
the differences between our solutions in the continuum and on a torus are much smaller 
than the difference to the lattice result, at least for intermediate momenta. 
We explained in chapter \ref{YM} that this difference can be attributed to full two-loop contributions
which are missed out in the DSE calculations but are certainly present
in lattice Monte-Carlo simulations.
For small momenta we see that our calculations on a torus employ much larger volumes than is possible 
in lattice simulations up to now.
A thorough investigation of the infinite volume limit of the DSE solutions on a torus
including a careful analysis of possible cut-off effects is certainly desirable and under way.

\section{Summary}

In this chapter we have presented numerical solutions of truncated systems of
Dyson--Schwinger equations for the gluon and ghost propagators in Landau gauge
SU(N) Yang--Mills theories. We have employed a four-torus, {\it i.e.}~a
compact space-time manifold, as an infrared regulator. Apart from the
infrared finite volume effects encountered on such a manifold we found
solutions on a torus which are very close to the ones obtained in the continuum 
formulation for various truncation schemes. Thus a central result of this
chapter is: Dyson-Schwinger equations on a torus are feasible.

For small momenta we found the expected finite volume effects in the solutions of
the DSEs on a torus. These deviations from the continuum results are somewhat erratic
for the two truncation schemes employing an angular approximation in the sense that
the values of the dressing functions on the momentum grid cannot be connected by a 
straight line. For our new truncation scheme without angular 
approximations this is different: systematic deviations from the continuum solutions
occur. We find small deviations for the gluon
dressing function and larger differences for the ghost dressing function on a torus 
compared to the continuum one. This is what one expects, since the infrared dominant ghost
correlation is long ranged and should therefore be more affected by the finite volume than the 
gluon correlation. 

Employing a torus as infrared regulator has certain advantages compared to the continuum
formulation. In the numerical calculation of the continuum DSEs one has to match
the analytically determined solution in the infrared carefully to the region where
the dressing functions are determined numerically. On a torus the finite volume in coordinate
space leads to a finite value of the smallest squared momentum encountered in the calculations.
Thus one has not to worry about infrared singularities. Furthermore it is much easier to
obtain convergence in the numerical iteration process. Historically we solved the
truncation scheme introduced in chapter \ref{YM} on the torus first and only subsequently
in the continuum using the torus solutions as start values in the iteration process.
Thus the torus formulation is an important technical tool. 

Furthermore we used the Dyson-Schwinger equations on a torus to corroborate a result of 
chapter \ref{YM}: only one out of two analytical solutions from the infrared analysis of 
the continuum equation is connected to numerical results. For the transverse projection of the gluon 
equation this is the solution corresponding to $\kappa=0.595$. Such a statement is stronger if
it is inferred from solutions on a torus than from corresponding continuum results. The reason
is that we have to use the analytical infrared solution as self consistent input in the continuum
calculations to obtain numerical stability. This is not the case on a torus.

\chapter{The coupled system of quark, gluon and ghost DSEs \label{quark}}

In this chapter we will enlarge our focus from pure Yang-Mills theory to Landau gauge QCD. 
We will investigate the coupled system of Dyson-Schwinger 
equations for the ghost, gluon {\it and} the quark propagators.
In the quark DSE we will study the mechanism by which physical quark masses
are generated even though the bare quark masses in the Lagrangian are zero. This is
a genuine effect of Strong QCD. It is well known that for vanishing
bare masses the renormalised masses remain zero at each order of perturbation theory.
  
A thorough study of the infrared phenomena in the quark sector of QCD requires 
a continuum formulation. Lattice simulations of dynamical chiral symmetry 
breaking \cite{Skullerud:2000un,Skullerud:2001aw,Bonnet:2002ih,Bonnet:2002dx,Bowman:2002bm} 
have to deal with finite volume effects and in addition have to extrapolate 
from finite to zero quark mass. It is not possible to put
massless quarks on a finite lattice.  
A recently performed study \cite{Bonnet:2002ih} in the overlap formalism {\it e.g.}
employs masses in the range of $m_0=(126-734) \, \mbox{MeV}$. These values 
suggest that even the most elaborate extrapolation method to zero quark mass 
needs guidance and check from continuum results. 

Apart from the phenomenon of mass generation we are interested in
quark confinement. Single quark states have non-vanishing colour charge and  
are therefore not contained in the physical part of the state space of QCD.
In subsection \ref{positivity} we argued for a positive (semi-)definite metric
in this physical subspace, whereas the remaining state 
space of QCD contains negative norm states as well. Consequently, 
negative norm contributions to the quark propagator are theoretical evidence
for quark confinement.  

The quark propagator is an important ingredient for many phenomenological models
(see \cite{Roberts:2000aa,Alkofer:2000wg,Roberts:1994dr} and references therein). 
Thus, the quark DSEs have been studied extensively. Various {\it ans\"atze} for 
the gluon interaction in the quark equation have been explored. The resulting
quark propagators have been used in bound state calculations
based on the Bethe-Salpeter equations for mesons 
(see {\it e.g.} \cite{Tandy:1997qf,Bender:1996bb,Maris:1997tm,Maris:1998hd,Tandy:2001qk})
or Faddeev equations for baryons 
\cite{Hellstern:1997pg,Oettel:1998bk,Bloch:1999ke,Bloch:1999rm,Alkofer:1999jf,Oettel:1999gc,Oettel:1999bu,Oettel:2000jj,Ahlig:2000qu,Alkofer:2001qj,Fischer:2001qx,Oettel:2002wf}.
One of the central aims of this thesis is to provide a solution for the 
quark propagator which incorporates the effects from the ghost and gluon DSEs directly 
and not via model assumptions.

This chapter is organised as follows: In the next section we will construct suitable {\it ans\"atze} 
for the quark-gluon vertex such that two important properties of the full quark DSE are reproduced: 
the independence of the generated quark mass from the renormalisation point and the asymptotic 
matching of the DSE-solutions to the results of perturbation theory. Fortunately, the corresponding 
DSEs for the fermions of QED are well studied 
(a short overview is given {\it e.g.} in \cite{Pennington:1998cj}). 
We will dwell on these results and construct non-Abelian generalisations of Abelian vertices, 
which have the desired properties. 

In the second section we present solutions for the quenched system of quark, ghost and gluon
DSEs, {\it i.e.} we neglect the quark-loop in the gluon equation. A corresponding approximation is 
frequently used in lattice simulations. Comparing our solutions with recent lattice results
\cite{Bonnet:2002ih} we find very good agreement for the quark propagator.

We then proceed to the unquenched case and incorporate the quark-loop into our truncation scheme 
for the ghost and gluon DSE from chapter \ref{YM}. We present solutions for the full coupled 
system of DSEs for the quark, ghost and gluon propagators in the last section of this chapter.
Compared to the quenched case we will find only moderate differences for the number of light flavours
$N_f \le 3$ \cite{Fischer}.
 
\section{The quark Dyson-Schwinger equation \label{quark-dse-sec}}

\begin{figure}[t]
\vspace{0.5cm}
\centerline{
\epsfig{file=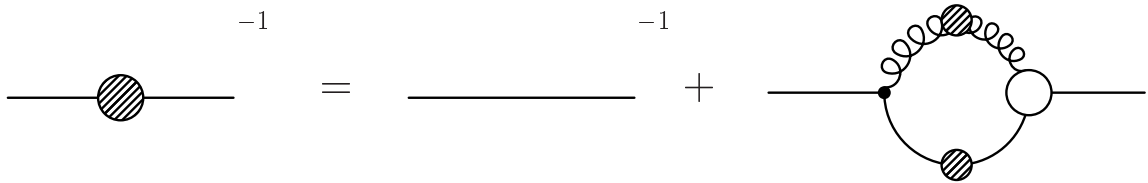}
}
\caption{\sf \label{quark_dse} A diagrammatical representation of the quark Dyson-Schwinger equation.
}
\end{figure}
We outlined the derivation of the Dyson-Schwinger equation for the quark propagator $S(p)$ 
at the end of section \ref{DSE-sec}. The renormalised equation in flat Euclidean space-time
is given by
\beq
S^{-1}(p) = Z_2 \, S^{-1}_0(p) + \frac{g^2}{16\pi^4}\, Z_{1F}\, C_F\, \int d^4q\,
\gamma_{\mu}\, S(q) \,\Gamma_\nu(q,k) \,D_{\mu \nu}(k) \,,
\label{quark1}
\eeq
with the momentum routing $k=q-p$. The factor $C_F=(N_c^2-1)/2N_c$ in front of the integral 
stems from the colour trace of the loop. The symbol $\Gamma_\nu(q,k)$ denotes the full
quark-gluon vertex. A diagrammatical representation of the equation is given in Fig.~\ref{quark_dse}.

Suppressing colour indices the quark and gluon propagators in Landau gauge are given by
\setlength{\jot}{3mm}
\beqa
S(p) &=& \frac{1}{-i \pslash A(p^2) + B(p^2)} \:\:\: 
= \:\:\: A^{-1}(p^2)\frac{i \pslash + M(p^2)}{p^2 + M^2(p^2)} \,, \\
S_0(p) &=& \frac{1}{-i \pslash + m_0} 
\,, \\
D_{\mu \nu}(p) &=& \left(\delta_{\mu \nu} - \frac{p_\mu p_\nu}{p^2} \right) \frac{Z(p^2)}{p^2} \,, 
\eeqa 
where the quark mass function $M$ is defined as $M(p^2)=B(p^2)/A(p^2)$.
For obvious reasons the dressing function $A(q^2)$ is frequently called 'vector self energy'
and the dressing function $B(q^2)$ 'scalar self energy'. Furthermore the inverse of the
vector self energy, $1/A(q^2)$, is denoted as 'quark wave function renormalisation'.
The bare quark propagator $S_0(p^2)$ contains the unrenormalised quark mass $m_0(\Lambda^2)$ which 
depends on the cutoff $\Lambda$ of the theory. The bare mass is related to the
renormalised mass $m_R(\mu^2)$ via the renormalisation constant $Z_m$: 
\beq
m_0(\Lambda^2)=Z_m(\mu^2,\Lambda^2) \: m_R(\mu^2) \,. \label{mass-ren} 
\eeq
Here $\mu^2$ is the squared renormalisation point.
\setlength{\jot}{0mm}

The renormalised and unrenormalised vector self energy, $A$ and $A_0$, are related by 
\beq
A_0^{-1}(p^2,\Lambda^2) = Z_2(\mu^2,\Lambda^2) \: A^{-1}(p^2,\mu^2).
\label{Aren}
\eeq
In Landau gauge the quantum corrections to the vector self energy are
finite. Correspondingly $Z_2(\mu^2,\Lambda^2)$ stays finite when the cutoff is sent to infinity and we
have $0 < Z_2(\mu^2,\Lambda^2) < 1$. Furthermore the ghost-gluon
vertex is not ultraviolet divergent in Landau gauge and we can choose $\tilde{Z}_1=1$, 
{\it c.f.} section \ref{gg-vertex-sec}. The Slavnov-Taylor identity
for the quark-gluon vertex renormalisation factor $Z_{1F}$ thus simplifies, 
\beq
Z_{1F}=\frac{\tilde{Z}_1 Z_2}{\tilde{Z}_3} = \frac{Z_2}{\tilde{Z}_3}.
\eeq

Previous studies of the quark equation in the so called Abelian approximation (an overview is given in
ref.~\cite{Roberts:2000aa}) as well as the recent investigation 
in ref.~\cite{Bloch:2002eq} assume implicit cancellations between
the full quark-gluon vertex, the dressed gluon propagator and the integral over the kernel of the DSE.
Furthermore in the tensor structure of the quark-gluon vertex only a term proportional to $\gamma_\mu$
is employed. 

In this thesis we do not have to rely on implicit cancellations since  
we calculated explicit solutions for the dressed gluon and ghost propagators, 
{\it c.f.} section \ref{YM-results}.
We will also construct explicit non-perturbative {\it ans\"atze} for the quark-gluon vertex
including different tensor structures than $\gamma_\mu$.
The advantages of such a treatment are obvious: every building block of the 
equation is explicitly given and well under control. We are able to assess the
expedience of different vertex constructions\footnote{Results for the quark-gluon vertex on
the lattice have been obtained in ref.~\cite{Skullerud:2002ge}. However, at present the error bars
from such simulations are too large to use the lattice results as guideline in the construction of our
{\it ans\"atze}.}. Furthermore the inclusion of several tensor structures in the vertex is   
supposed to be important in bound state calculations of scalar meson masses \cite{Alkofer:2002bp,Tandy}. 
Finally we hope providing an explicit construction of all parts of the quark equation leads in turn to 
more thorough statements on quark confinement.

In the following we assume an effective non-Abelian quark-gluon vertex of the form  
\beq
\Gamma_\nu(q,k) = V_\nu^{abel}(p,q,k) \, W^{\neg abel}(p,q,k),
\label{vertex-ansatz}
\eeq
with $p$ and $q$ denoting the quark momenta and $k$ the gluon momentum.
The non-Abelian factor $W^{\neg abel}$ multiplies an Abelian part $V_\nu^{abel}$, which carries 
the tensor structure of the vertex. This {\it ansatz} is motivated by the aim to respect
gauge invariance as much as possible on the present level of truncation. 
The Slavnov-Taylor identity (STI) for the quark-gluon vertex is given by \cite{Marciano:1978su} 
\beq
(-i) \: G^{-1}(k^2) \: k_\mu \: \Gamma_\mu(q,p) = S^{-1}(p) \: H(q,p) - H(q,p) \: S^{-1}(q),
\label{quark-gluon-STI}
\eeq
with the ghost dressing function $G(k^2)$ and the ghost-quark scattering kernel $H(q,p)$. 
At present the non-perturbative behaviour of the ghost-quark scattering kernel 
is unknown. Therefore we cannot solve the STI explicitly.
However, comparing the structure of eq.~(\ref{quark-gluon-STI}) with the corresponding
Ward identity of QED,
\beq
(-i) \: k_\mu \: \Gamma_\mu^{QED}(q,p) = S^{-1}(p) -  S^{-1}(q),
\label{QED-quark-gluon-STI}
\eeq
we are able to infer some information: Whereas the ghost fields of QED decouple from the theory and
consequently do not show up in the Ward identity, there is an explicit factor of $G^{-1}(k^2)$ 
on the left hand side of eq.~(\ref{quark-gluon-STI}). We therefore suspect the quark-gluon vertex of 
QCD to contain an additional factor of $G(k^2)$ compared to the fermion-photon vertex of QED. 
Some additional ghost dependent structure seems necessary to account for the ghost-quark scattering
kernel on the right hand side of eq.~(\ref{quark-gluon-STI}). For simplicity we assume the 
whole ghost dependence of the vertex to be contained in a non-Abelian factor 
multiplying an Abelian tensor
structure\footnote{For quenched calculations and in the context of angular approximated DSEs a
similar strategy has already been adopted in refs.~\cite{Smekal,Ahlig}.}. 

The Abelian part of the vertex, $V_\nu^{abel}$, can be adopted from QED. The Ward identity 
(\ref{QED-quark-gluon-STI}) has long been solved \cite{Ball:1980ay,Ball:1980ax}. 
Furthermore transverse parts of the fermion-photon vertex
have been fixed by Curtis and Pennington\footnote{This so called Curtis-Pennington 
vertex has been used extensively in the study of the fermion DSE in QED 
\cite{Burden:1993gy,Atkinson:1994mz,Bashir:1994az,Hawes:1996mw,Hawes:1997ig,Kizilersu:2001pd}.} 
to satisfy multiplicative renormalisability in the Abelian fermion DSE 
for all linear covariant gauges \cite{Curtis:1990zs,Curtis:1993py}. 

The non-Abelian factor $W^{\neg abel}$ is chosen such that the complete quark equation fulfils two
conditions: 
\begin{itemize}
\item[(i)]
The quark mass function $M(p^2)$ should be independent of the renormalisation point $\mu^2$.
\item[(ii)] The anomalous dimension $\gamma_{m}$ of the mass function known from resummed perturbation theory should be
recovered in the ultraviolet. 
\end{itemize}
In the course of this section we will prove the vertex {\it ansatz}
\setlength{\jot}{2mm}
\beqa
W^{\neg abel}(p,q,k) &=&  G^2(z,s) \:\tilde{Z}_3(s,L)\: 
\frac{\left(G(z,s) \tilde{Z}_3(s,L) \right)^{-2d-d/\delta}}
{\left(Z(z,s) {Z}_3(s,L)\right)^d} \,,\label{vertex_nonabel}\\
V^{abel}_\nu(p,q,k) &=& \Gamma_\nu^{CP}(p,q,k) \nonumber\\ &=&
\frac{A(x,s)+A(y,s)}{2} \gamma_\nu \nonumber\\
&& + \, \frac{A(x,s)-A(y,s)}{2(x-y)} (\pslash+\qslash)(p+q)_\nu + i \frac{B(x,s)-B(y,s)}{x-y} (p+q)_\nu \nonumber\\
&& + \, \frac{A(x,s)-A(y,s)}{2} \left[(x-y)\gamma_\nu - (\pslash-\qslash)(p+q)_\nu \right] \times\nonumber\\
&& \hspace*{4.5cm}  \frac{x+y}{(x-y)^2+(M^2(x)+M^2(y))^2}\,,
\label{vertex_CP}
\eeqa
with the new parameter $d$ to satisfy the conditions (i) and (ii).
To ease notation we used the abbreviations $x=p^2$, $y=q^2$ and $z=(p-q)^2$ for the
squared momenta, also $s=\mu^2$ for the squared renormalisation point and $L=\Lambda^2$ for the squared cutoff
of the theory. The anomalous dimension $\delta$ of the ghost propagator is 
$\delta = -9 \:N_c/(44 \:N_c-8 \: N_f)$ at one loop order for $N_c$ colours and $N_f$ flavours. 
The Abelian part of the vertex is given by the Curtis-Pennington (CP) 
vertex $\Gamma_\nu^{CP}(p,q,k)$. 
\setlength{\jot}{0mm}

From a systematic point of view the 
newly introduced parameter $d$ in the non-Abelian part of the vertex is completely arbitrary.
Our numerical results, however, will indicate that values around the somewhat natural 
choice $d=0$ match best with lattice simulations, {\it c.f.} subsection \ref{quenched-num}. 

If we would not care about (Abelian) gauge invariance we could also employ
the much simpler vertex 
\beqa
W^{\neg abel}(p,q,k) &=&  G^2(z,s) \:\tilde{Z}_3(s,\Lambda)\: 
\frac{\left(G(z,s) \: \tilde{Z}_3(s,L) \right)^{-2d-d/\delta}}
{\left(Z(z,s) \: {Z}_3(s,L)\right)^d} \nonumber\\
V^{abel}_\nu(p,q,k) &=& Z_2(s,L) \, \gamma_\nu
\label{vertex_bare}
\eeqa
where we have taken the bare Abelian vertex, $\gamma_\nu$, multiplied with an extra factor of $Z_2$. 
In Landau gauge this construction also satisfies the conditions (i) and (ii), as will be shown
in the next two subsections\footnote{In the numerical treatment we will 
additionally employ a vertex where the transverse part of the CP-vertex is left out, {\it i.e.}
a generalised Ball-Chiu (BC) vertex. In Landau gauge such a vertex also satisfies the conditions (i) and (ii).}.

\subsection{Multiplicative renormalisability of the quark equation \label{subsec-MR}}

To proceed we substitute the vertex {\it ansatz} (\ref{vertex_bare}) into the quark equation (\ref{quark1}).
By taking the Dirac trace once with and once without multiplying the equation with $\pslash$
we project out the mass function $M(x)$ and the vector self energy $A(x)$. We arrive at
\setlength{\jot}{1mm}
\beqa
M(x)A(x,s) &=& Z_2(s,L) \: m_0(L) + \frac{Z_2(s,L)}{3\pi^3} \int d^4q \Bigg\{
\frac{\alpha(z)}
{z \:(y+M^2(y))} \: Z_2(s,L) \: A^{-1}(y,s) \times  \nonumber\\
&&\hspace*{4.5cm}  \frac{\left(G(z,s) \: \tilde{Z}_3(s,L) \right)^{-2d-d/\delta}}
{\left(Z(z,s) \: {Z}_3(s,L)\right)^d} \: 3M(y) \Bigg\} \,,  \label{bare-dse-m}\\
A(x,s) &=& Z_2(s,L) + \frac{Z_2(s,L)}{3\pi^3} \int d^4q \Bigg\{
\frac{\alpha(z)}
{x \:z \:(y+M^2(y))} \: Z_2(s,L) \: A^{-1}(y,s) \times  \nonumber\\
&&\hspace*{1cm}  \frac{\left(G(z,s) \: \tilde{Z}_3(s,L) \right)^{-2d-d/\delta}}
{\left(Z(z,s) \: {Z}_3(s,L)\right)^d} 
\left(-z + \frac{x+y}{2} + \frac{(x-y)^2}{2 \: z} \right) \Bigg\} \,, \nonumber\\
\label{bare-dse-a}
\eeqa
where we have used the definition of the running coupling $\alpha$ in Landau gauge
\beq
\alpha(x) = \frac{g^2}{4 \pi} \: Z(x,s)G^2(x,s) = \alpha(s) \: Z(x,s) \:G^2(x,s) \,,
\eeq
\setlength{\jot}{0mm}
{\it c.f.} eq.~(\ref{coupling}). 

The behaviour of eqs.~(\ref{bare-dse-m}), (\ref{bare-dse-a}) under renormalisation can be
explored by changing the renormalisation point $s=\mu^2$ to a new point $t=\nu^2$.
We first note that the factor stemming from the non-Abelian 
part of the quark-gluon vertex is not affected by such a change:
\beq
\frac{\left(G(z,s) \: \tilde{Z}_3(s,L) \right)^{-2d-d/\delta}}
{\left(Z(z,s) \: {Z}_3(s,L)\right)^d} =
\frac{\left(G(z,t) \: \tilde{Z}_3(t,L) \right)^{-2d-d/\delta}}
{\left(Z(z,t) \: {Z}_3(t,L)\right)^d} \,. \nonumber
\eeq
This can be seen easily with the help of the relations
\beqa
G_0(x,L) &=& G(x,s) \: \tilde{Z}_3(s,L) \,, \\
Z_0(x,L) &=& Z(x,s) \: {Z}_3(s,L) \,,
\label{ZGren}
\eeqa
between the unrenormalised and renormalised ghost and gluon dressing function, {\it c.f.} eq.~(\ref{zg-ren}). 
Furthermore the running coupling $\alpha(z)$ is independent of the 
renormalisation point, {\it c.f.} the discussion in subsection \ref{coupling-sec}. 
From eq.~(\ref{Aren}) we infer 
\beq
Z_2(t,L) \: A^{-1}(x,t) = Z_2(s,L) \: A^{-1}(x,s) \,.
\eeq
With the renormalisation condition $A(t,t)=1$ we have  
\beq
Z_2(t,L) = Z_2(s,L)\: A^{-1}(t,s) \,,
\label{Z_2}
\eeq
and subsequently
\beq
A(x,t) = A(x,s) \:A^{-1}(t,s) \,.
\label{A_R}
\eeq
Substituting eqs.~(\ref{Z_2}) and (\ref{A_R}) into the Dyson-Schwinger equations (\ref{bare-dse-m})
we find the mass function $M(x)$ to be independent of the renormalisation point,
{\it i.e.} condition (i) is satisfied. 
Note that without the extra factor of $Z_2$ in the Abelian part of the vertex (\ref{vertex_bare}) 
we would violate this condition.

\begin{figure}[t]
\vspace{0.5cm}
\centerline{
\epsfig{file=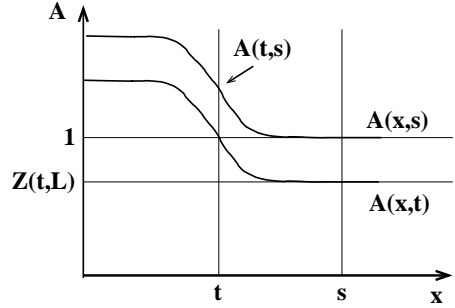,height=4cm}
}
\caption{\sf \label{Afig} Sketch of a finite renormalisation from a perturbative point $s$ 
to a non-perturbative point $t$ for the vector self energy $A$. 
}
\end{figure}
Before we examine the case of the more sophisticated Curtis-Pennington type vertex (\ref{vertex_CP}),
we mention two important points. {\it First}, according to perturbation theory 
we have $A(x \rightarrow \infty,s) \rightarrow 1$ and $Z_2(s,L) \rightarrow 1$ for large 
renormalisation points $s$. However, this is just a special case of the general relation
\beq
A(x \rightarrow \infty,s) \rightarrow Z_2(s,L) \label{A-limit}
\eeq
which can be inferred from eqs.~(\ref{Z_2}) and (\ref{A_R}).
In Fig.~\ref{Afig} we sketch the vector self energy renormalised at two different points $s$ and $t$,
with $s$ in the perturbative and $t$ in the non-perturbative region of momentum. 
We will see a similar picture when we present our numerical solutions in section \ref{quenched-num}.

{\it Second}, the alert reader might worry about the appearance of the ghost and gluon
renormalisation factors $Z_3(L)$ and $\tilde{Z}_3(L)$ in the interaction kernel
of the quark equation. Certainly the renormalised functions $M(x)$ and $A(x)$ should not depend on the  
cutoff of the integral. The balance of cutoff dependent quantities in the equation 
is controlled by various factors of $Z_2(L)$ and $Z_m(L)$. We have to take care not to disturb this
balance by the vertex {\it ansatz}. Thus the non-Abelian part of our quark-gluon vertex
contains such powers of $Z_3(L)$ and $\tilde{Z}_3(L)$ that the cutoff dependence of these quantities
cancel. This can be easily checked using the scaling behaviour
\beqa
Z_3(s,L) &=& \left(\frac{\alpha(L)}{\alpha(s)}\right)^{\gamma} \,, \nonumber\\
\tilde{Z}_3(s,L) &=& \left(\frac{\alpha(L)}{\alpha(s)}\right)^{\delta} \,, 
\label{Z-scaling}
\eeqa
of the renormalisation factors for $L \rightarrow \infty$ and the 
relation $\gamma+2\delta+1=0$, {\it c.f.} eq.~(\ref{gd1}).

Along the same lines as for the bare vertex construction we prove condition (i) for the Curtis-Pennington type 
vertex. Plugging eqs.~(\ref{vertex_nonabel}), (\ref{vertex_CP}) into the quark equation (\ref{quark1}) 
and projecting onto $M(x)$ and $A(x)$ we arrive at 
\setlength{\jot}{3mm}
\beqa
M(x)A(x,s) &=& Z_2(s,L) m_0(L) + \frac{Z_2(s,L)}{3\pi^3} \int d^4q 
\frac{\alpha(z)}
{z(y+M^2(y))}\frac{\left(G(z,s) \tilde{Z}_3(s,L) \right)^{-2d-d/\delta}}
{\left(Z(z,s) {Z}_3(s,L)\right)^d} \nonumber\\
&&\times A^{-1}(y,s) \: \left[\frac{3}{2}(A(x,s)+A(y,s))M(y) \right.\nonumber\\
&&\hspace*{2cm} + \frac{1}{2}(\Delta A\:M(y)-\Delta
B)\left(-z+2(x+y)-\frac{(x-y)^2}{z}\right)
\nonumber\\
&&\hspace*{2cm}\left.+\frac{3}{2}(A(x,s)-A(y,s))M(y)\Omega(x,y)(x-y)\right] \label{CP-M}
\eeqa
\beqa
A(x,s) &=& Z_2(s,L) + \frac{Z_2(s,L)}{3\pi^3} \int d^4q 
\frac{\alpha(z)}
{xz(y+M^2(y))} \frac{\left(G(z,s) \tilde{Z}_3(s,L) \right)^{-2d-d/\delta}}
{\left(Z(z,s) {Z}_3(s,L)\right)^d} \nonumber\\
&&\times A^{-1}(y,s)\:\left[\left(-z+\frac{x+y}{2}+\frac{(x-y)^2}{2z}\right)\frac{A(x,s)+A(y,s)}{2}\right.\nonumber\\
&&\hspace*{2cm}-\left(\frac{\Delta A}{2}
(x+y)+\Delta
B\:M(y)\right)\left(-\frac{z}{2}+(x+y)-\frac{(x-y)^2}{2z}\right)\nonumber\\
&&\hspace*{2cm}+\left.
\frac{3}{2}(A(x,s)-A(y,s))M(y)\Omega(x,y)\left(\frac{x^2-y^2}{2}-z\frac{x-y}{2}\right)\right]. 
\nonumber\\
\label{CP-A}
\eeqa
Here we have used the abbreviations
\beqa
\Delta A &=& \frac{A(x,s)-A(y,s)}{x-y},\nonumber\\ 
\Delta B &=& \frac{B(x,s)-B(y,s)}{x-y},\nonumber\\
\Omega(x,y) &=& \frac{x+y}{(x-y)^2+(M^2(x)+M^2(y))^2}.\nonumber
\eeqa
With the help of the relations (\ref{Z_2}) and (\ref{A_R}) we find the quark equations (\ref{CP-M}), (\ref{CP-A}) 
to be consistently renormalised if and only if the scalar self energy 
behaves like 
\beq
B(x,t) = B(x,s) \, A^{-1}(t,s)
\eeq 
and the mass function 
\beq
M(x)=B(x,s)/A(x,s)
\eeq
is independent of the renormalisation point. 
\setlength{\jot}{0mm}

Summing up the results of this subsection we have shown that both vertex constructions, 
eqs.~(\ref{vertex_CP}) and (\ref{vertex_bare}), lead to a renormalisation point independent
mass function as is required by condition (i) of section \ref{quark-dse-sec}. The same is true for
a Ball-Chiu type vertex, which is the Curtis-Pennington construction (\ref{vertex_CP}) without
the transverse term proportional to $\Omega(x,y)$. 
Note that in different gauges than Landau gauge only the Curtis-Pennington construction would
satisfy condition (i), similar to QED \cite{Curtis:1990zs,Curtis:1993py}.

\subsection{Ultraviolet analysis of the quark equation \label{quark-UV-analysis}}

In this subsection we will show that the {\it ans\"atze} (\ref{vertex_CP}) and (\ref{vertex_bare}) for
the quark-gluon vertex both lead to the correct perturbative limit of the 
quark mass function $M(x)$. We first examine the case of the bare vertex construction, 
eq.~(\ref{bare-dse-m}).

The ghost and gluon dressing functions $G$ and $Z$ are slowly varying for large momenta 
according to their perturbative limit given in eqs.~(\ref{gluon_uv}), (\ref{ghost_uv}).
For loop momenta $y$ larger than the external momentum $x$ we are therefore justified to employ 
the angular approximation $G(z),Z(z) \rightarrow G(y),Z(y)$, {\it c.f.} subsection \ref{ultraviolet}.
Furthermore there is a region $x_0<y<x$ where the approximation $G(z),Z(z) \rightarrow G(x),Z(x)$ 
is adequate. We are then able to carry out the angular integrals in eq.~(\ref{bare-dse-m})  
with the help of the formulae given in appendix \ref{angular}. If we additionally take the
external momentum $x$ to be large enough, then all masses in the denominators become negligible since 
the integral is dominated by loop momenta $y \approx x$. We then obtain
\beqa
M(x)A(x,s) &=& Z_2(s,L) m_0(L) \nonumber\\
&&\hspace*{0cm}+ \, \frac{Z_2(s,L)}{\pi} \frac{\alpha(x)}
{x}\: \int_{x_0}^x dy \: Z_2(s,L) \: A^{-1}(y,s) \:
\frac{\left(G(z,s) \tilde{Z}_3(s,L) \right)^{-2d-d/\delta}}
{\left(Z(z,s) {Z}_3(s,L)\right)^d} M(y) \nonumber\\ 
&&\hspace*{0cm}+ \, \frac{Z_2(s,L)}{\pi} \int_x^L dy
\frac{\alpha(y)}
{y} \: Z_2(s,L) \: A^{-1}(y,s)\: 
\frac{\left(G(z,s) \tilde{Z}_3(s,L) \right)^{-2d-d/\delta}}
{\left(Z(z,s) {Z}_3(s,L)\right)^d} M(y) \,,\nonumber\\ 
\label{bare-dse-m-UV1} 
\eeqa
where the integral from $y=0$ to $y=x_0$ has already been neglected.

For large momenta $y>x_0$ the wave function renormalisation $A^{-1}$ and the renormalisation factor $Z_2$ 
cancel each other according to eq.~(\ref{A-limit}). Furthermore we use the perturbative limit of
the ghost and gluon dressing functions ({\it c.f.} eqs.~(\ref{gluon_uv}), (\ref{ghost_uv}))
\setlength{\jot}{2mm}
\beqa
G(z) &=& G(s)\left[\omega\log\left(\frac{z}{s}\right)+1\right]^\delta \,, \nonumber\\
Z(z) &=& Z(s)\left[\omega\log\left(\frac{z}{s}\right)+1\right]^\gamma \,, 
\label{Z_UV}
\eeqa
with $\omega=\beta_0\alpha(s)/(4 \pi)=(11N_c-2N_f)\alpha(s)/(12 \pi)$.
If we additionally substitute the scaling behaviour of the renormalisation 
constants $Z_3$ and $\tilde{Z}_3$, 
eqs.~(\ref{Z-scaling}),
and exploit the relation $\gamma+2\delta+1=0$, eq.~(\ref{gd1}),  
we arrive at
\setlength{\jot}{0mm}
\beqa
M(x) &=&  m_0(L) 
+ \frac{1}{\pi} \frac{\alpha(x)}{x} \int_{x_0}^x dy   M(y) 
+ \frac{1}{\pi} \int_x^L dy \frac{\alpha(y)}{y}  M(y) \,. 
\label{bare-dse-m-UV2}
\eeqa

This well known equation describes the ultraviolet behaviour of the quark mass function. 
The classification of its solutions has been clarified by Miransky \cite{Miransky:1985ib,Gusynin:1987fu},
employing the perturbative form of the running coupling,
\beq
\alpha(y) = \alpha(s)\left[\omega\log\left(\frac{y}{s}\right)+1\right]^{-1} \,.
\label{alpha-UV}
\eeq
In the chiral limit, $m_0(L) = 0$, we obtain the so called {\it regular asymptotic} form,  
\beq
M(x) = \frac{2 \pi^2 \gamma_m}{3}
\frac{-\langle \bar{\Psi}\Psi\rangle}{x \left(\frac{1}{2} \ln(x/\Lambda^2_{QCD})\right)^{1-\gamma_m}} \,.
\label{chiral-M_UV}
\eeq
Here $\langle \bar{\Psi}\Psi\rangle$ denotes the chiral condensate which is discussed in more detail
in subsection \ref{f-pi-sec}. In the case of non-vanishing bare quark mass, $m_0(L) \ne 0$, the equation
(\ref{bare-dse-m-UV2}) is solved by the {\it irregular asymptotic} form,  
\beq
M(x) = M(s)\left[\omega\log\left(\frac{x}{s}\right)+1\right]^{-\gamma_m} \,.
\label{M_UV}
\eeq
In this case we furthermore find
\beqa
\gamma_m &=& \frac{12}{11N_c-2N_f} \,, \\
m_0(L) &=& M(s) \left[\omega\log\left(\frac{L}{s}\right)+1\right]^{-\gamma_m} \,,
\label{m_0_UV} 
\eeqa
in accordance with resummed perturbation theory.

We thus have shown that
the bare vertex construction (\ref{vertex_bare}) admits a solution for the mass function $M(x)$,
which has the correct perturbative behaviour for large momenta. A similar analysis is possible for 
the DSE with the Curtis-Pennington type vertex, eq.~(\ref{CP-M}).
As the vector self energy goes to a constant in the limit of large momenta, 
eq.~(\ref{A-limit}), all terms proportional to $A(x)-A(y)$ are suppressed in this limit.
Furthermore, according to the perturbative expression (\ref{M_UV}) the $\Delta B$-term contributes 
at most subleading logarithmic corrections in eq.~(\ref{CP-M}). The first term in the brackets
reduces to the bare vertex form because $A(x,s) \approx A(y,s)$ for large momenta $x,y$.
Thus we obtain the same ultraviolet limit from eq.~(\ref{CP-M}) than for the bare vertex construction.   
This is certainly also the case if a Ball-Chiu type vertex is employed. 

\section{The quark propagator in quenched QCD \label{quenched}}

In the course of this section we will compare our results for three different vertex types, which
share the non-Abelian part proposed in eq.~(\ref{vertex_nonabel}) but differ in their Abelian 
parts. We will employ the bare vertex, eqs.~(\ref{vertex_bare}), and the Curtis-Pennington (CP)
type vertex, eqs.~(\ref{vertex_CP}). Furthermore we use a Ball-Chiu (BC) type construction, which employs only
the first three terms of the CP-vertex. In Landau gauge all these vertex {\it ans\"atze} satisfy the 
conditions (i) and (ii) formulated in section \ref{quark-dse-sec}. 
In order to compare the different vertex types on a quantitative level  
we will calculate the pion decay constant $f_\pi$ and the chiral condensate
from the respective solutions for the quark mass function.   

\subsection{Pion decay constant, chiral condensate and quark masses \label{f-pi-sec}}

The pion decay constant is calculated from the Bethe-Salpeter equation (BSE), which describes the
pion as bound state of quark and antiquark \cite{Maris:1998hd}. Apart from the dressed quark propagator
the BSE involves couplings between quarks and gluons. On the level of the quark DSE we have substituted
the full quark-gluon vertex by an vertex {\it ansatz}. However, at present it is only known for certain 
cases how such a vertex {\it ansatz} in the quark DSE translates to the corresponding quark-gluon coupling 
in the Bethe-Salpeter equation \cite{Bender:2002as,detmold}. No method is known up to know to derive the 
corresponding BSE for dressed quark-gluon vertices as the BC- or CP-vertex constructions.

We thus have to rely on the approximation 
\beqa
f_\pi^2 &=& -\frac{N_c}{4 \pi^2} \int dy \: y \: \frac{M(y) A^{-1}(y)}{(y + M^2(y))^2} 
\left(M(y) - \frac{y}{2} \frac{dM(y)}{dy}\right) \,,
\label{f-pi-eq}
\eeqa
which incorporates only the effects of the leading pion Bethe-Salpeter amplitude in the chiral limit \cite{Roberts:1994dr}. 
From a comparison of the relative size of the amplitudes in model calculations \cite{Maris:1997tm,Alkofer:2002bp}
one concludes that the approximation (\ref{f-pi-eq}) should lead to an underestimation of $f_\pi$ by roughly 
ten percent.

The {\it renormalisation point independent} chiral condensate, $\langle \bar{\Psi}\Psi\rangle$, 
can be extracted from the ultraviolet behaviour of the quark mass function in the chiral limit
({\it c.f.} eq.~(\ref{chiral-M_UV})):
\beq
M(x) \, \stackrel{x \to L}{\longrightarrow} \, \frac{2 \pi^2 \gamma_m}{3}
\frac{-\langle \bar{\Psi}\Psi\rangle}{x \left(\frac{1}{2} \ln(x/\Lambda^2_{QCD})\right)^{1-\gamma_m}} \,.
\label{ch-fit}
\eeq
Recall $x=p^2$ and $L=\Lambda^2$, where $\Lambda$ is the cutoff of our theory not to be confused with 
the scale $\Lambda_{QCD}$, which is to be taken from a fit to the running coupling, {\it c.f.}
eqs.~(\ref{fitA}), (\ref{fitB}). 

The {\it renormalisation point 
dependent} chiral condensate $\langle \bar{\Psi}\Psi\rangle_\mu$ can be calculated via \cite{Maris:1998hd}
\beq
-\langle \bar{\Psi}\Psi\rangle_\mu := Z_2(s,L) \, Z_m(s,L) \, N_c \,\mbox{tr}_D \int 
\frac{d^4q}{(2\pi)^4} S_{ch}(q^2,s) \,,
\label{ch-cond}
\eeq
where the trace is over Dirac indices, $S_{ch}$ is the quark propagator in the chiral limit
and the squared renormalisation point is denoted by $s=\mu^2$. 
To one-loop order both expressions for the condensate are connected by
\beq
\langle \bar{\Psi}\Psi\rangle_\mu = \left(\frac{1}{2}\ln(\mu^2/\Lambda^2_{QCD})\right)^{\gamma_m}
\langle \bar{\Psi}\Psi\rangle \,, 
\label{ch-loop}
\eeq
with the anomalous dimension $\gamma_m$ of the quark mass function.

For the calculation of the chiral condensate we first have to determine the mass 
renormalisation constant $Z_m(s,L)$. Recall the formal structure of the mass equation (\ref{bare-dse-m}),
which is given as
\beq
M(x)A(x,s) = Z_2(s,L) \,Z_m(s,L) \,m_R(s)  + Z_2(s,L) \, \Pi_M(x,s) \,, 
\label{m_mu}
\eeq
where $\Pi_M(x,s)$ represents the dressing loop. In order to extract $Z_m(s,L)$ from this equation
we have to clarify the meaning of $m_R(s)$ which is related to the unrenormalised mass by 
\beq
m_0(L)=Z_m(s,L) \,m_R(s),
\eeq
{\it c.f.} eq.~(\ref{mass-ren}). 
Evaluating eq.~(\ref{m_mu}) at the perturbative momentum $x=s$ the matter seems
clear. We achieve consistency with eqs.~(\ref{M_UV}) and (\ref{m_0_UV}), if
\beqa
m_R(s) &=& M(s) \label{mR} \,,\\
Z_m(s,L) &=& \left[\omega\log\left(\frac{L}{s}\right)+1\right]^{-\gamma_m} \,, 
\eeqa
which is indeed the correct perturbative scaling of the renormalisation constant $Z_m$ \cite{Roberts:2000aa}.  

Certainly one could implicitly {\it define the finite parts} of $Z_m$ such that 
the relation (\ref{mR}) holds in general for all renormalisation points $s$. 
Then the parameter $m_R$ in the renormalised QCD-Lagrangian would already know about dynamical 
symmetry breaking. However, as the mass parameters of QCD are supposed to be generated 
in the electroweak sector of the standard model one could equally well argue that it is more systematic to
exclude the effect of mass generation by strong interaction from $m_R$. 
 
In our numerical calculations we will choose $s$ to be sufficiently large, therefore 
eq.~(\ref{mR}) is valid anyway. Then $Z_m$ is determined by
\setlength{\jot}{2mm}
\beqa
Z_m(s,L) &=&  \frac{M(x)\, A(x,s) - Z_2(s,L) \, \Pi_M(x,s)}{Z_2(s,L)\, M(s)} \nonumber\\
&=& \frac{1}{Z_2(s,L)} - \frac{\Pi_M(s,s)}{M(s)} \,. 
\eeqa  
For the last equation we have set $x=s$ and have used the renormalisation condition $A(s,s)=1$.
\setlength{\jot}{0mm}

In the numerical calculations we have to specify the masses $m_R(s)$ as input. Choosing a 
perturbative renormalisation point $s$ allows one to evolve the masses $m_R(s)$ to
a different scale $t$ by
\beq
m_R(t) = m_R(s) \left(\frac{\ln(s/\Lambda_{QCD}^2)}{\ln(t/\Lambda_{QCD}^2)}\right)^{\gamma_m}. 
\eeq 
For $t=(2 \,\mbox{GeV})^2$ typical values for the masses of the light quarks are given
by the Particle Data Group \cite{Hagiwara:2002pw}:
\beq
\frac{1}{2}(m_{u}+m_{d})(2 \,\mbox{GeV}) \approx 4.5 \mbox{MeV}, \hspace*{1cm} 
m_{s}(2 \,\mbox{GeV}) \approx 100 \mbox{MeV}.
\label{masses-pdg} 
\eeq
We will use similar masses in our calculations.
  
\subsection{Renormalisation scheme and numerical method \label{ren-quenched}}

In the quark equation we employ a MOM regularisation scheme\footnote{In quenched $\mbox{QED}_4$ this
technique has already been used in \cite{Kizilersu:2001pd,Kizilersu:2000qd}.}
similar to the one used in the ghost and gluon equations in chapter \ref{YM}.
The formal structure of the quark equation is given by
\beqa
A(x,s) &=& Z_2(s,L) + Z_2(s,L) \, \Pi_A(x,s) \,, \\
M(x)A(x,s) &=& Z_2(s,L) \, Z_m(s,L) \, m_R(s)  + Z_2(s,L) \, \Pi_M(x,s) \,. 
\label{m-formal}
\eeqa
We eliminate $Z_2$ from the first equation by isolating it on the left hand side and subtracting 
the same equation for $x=s$. 
With
\beq
\frac{1}{Z_2(s,L)} = \frac{1}{A(x,s)} + \frac{1}{A(x,s)}\Pi_A(x,s)\,, \label{Z2-formal}
\eeq
we then have
\beq
\frac{1}{A(x,s)} = 1 - \frac{1}{A(x,s)}\Pi_A(x,s) + \Pi_A(s,s)\,, \label{A-formal}
\eeq
using the renormalisation condition $A(s,s)=1$. 
In each iteration step we determine the vector self energy $A(x)$ from eq.~(\ref{A-formal})
and subsequently $Z_2$ from eq.~(\ref{Z2-formal})\footnote{A check for
the numerics is to determine $Z_2$ at different momenta $x=p^2$ and search for
an artificial momentum dependence of $Z_2$. In our calculations we find $Z_2$ to be independent
of $p^2$ to an excellent degree.}. For the mass function $M(x)$ we use 
\beq
M(x)A(x,s) = Z_2(s,L) \, \Pi_M(x,s) 
\eeq
in the chiral limit and the subtracted equation
\beq
M(x)A(x,s) = M(s) + Z_2(s,L)\, \Pi_M(x,s) - Z_2(s,L)\, \Pi_M(s,s)
\eeq
if chiral symmetry is broken explicitly, {\it i.e.} $m_0 \not= 0$.

For the numerical iteration we employ a Newton method
and represent the dressing functions $A(x)$ and $M(x)$ with the help of Chebychev polynomials.
Furthermore, we use a numerical infrared cutoff $\epsilon$, which is taken small enough for 
the numerical results to be independent of $\epsilon$.
Numerical difficulties arise in the case of the Curtis-Pennington
type vertex and even more for the Ball-Chiu construction. If the external momentum $x$ and the loop
momentum $y$ are both small and close to each other then the derivative-like terms
\beq
\Delta A = \frac{A(x)-A(y)}{x-y}, \hspace*{2cm} \Delta B = \frac{B(x)-B(y)}{x-y},
\eeq
are hard to evaluate accurately. Although the functions $A(x)$ and $B(x)$ are constant
in the infrared and consequently should have derivatives close to zero
one encounters large values for $\Delta A$ and $\Delta B$ due to numerical 
inaccuracies in $A$ and $B$. 
In order to evaluate $\Delta A$ and $\Delta B$ much more precisely at small momenta we
fit the expressions
\beq
A(x) = \frac{A(0)}{1+a_1 \,(x/\Lambda_{QCD}^2)^{a_2}}\,, \hspace*{1cm}
B(x) = \frac{B(0)}{1+b_1 \,(x/\Lambda_{QCD}^2)^{b_2}}\,, \label{ab-fit}
\eeq
with the parameters $a_1$, $a_2$, $b_1$ and $b_2$ to the numerically evaluated functions. 
The scale $\Lambda_{QCD}=0.714 \,\mbox{GeV}$ has been determined from our fits to the running
coupling in section \ref{YM-results}.
For $x-y$ smaller than a suitable matching point we calculate the terms $\Delta A$ and $\Delta B$ 
from the fits. This procedure eliminates the numerical errors in the derivative terms and 
smoothes the numerical results considerably. 
In the case of the Ball-Chiu type vertex the iteration process does not converge unless we use
these fits.

The renormalisation condition employed in the ghost-gluon system of equations is $G^2(s)Z(s)=1$ with
$\alpha(s)=0.2$ at the squared renormalisation point $s=\mu^2$. Furthermore we choose the
transversal tensor, $\zeta=1$, to contract the gluon equation, {\it c.f.} chapter \ref{YM}. 
The physical scale in the quenched calculations is taken directly from the Yang-Mills results of
section \ref{YM-results}, {\it i.e.} we use the experimental value $\alpha(M_Z^2)=0.118$ of 
the running coupling at the mass of the Z-boson to fix the scale. 

\subsection{Numerical results \label{quenched-num}}
\begin{figure}[th!]
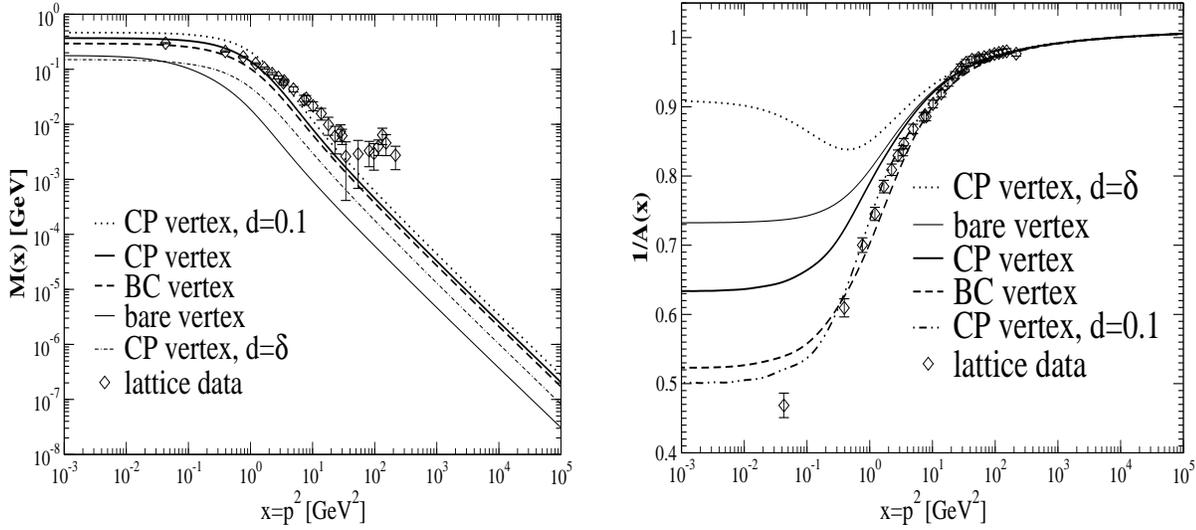

\vspace{0.7cm}
\centerline{
\epsfig{file=qM.vert.eps,width=7.5cm,height=7cm}
\hspace{0.5cm}
\epsfig{file=qA.vert.eps,width=7.5cm,height=7cm}
}
\caption{\sf \label{qpic1.dat} The mass function $M(x)$ and the inverse vector self energy 
$1/A(x)$ of a chiral quark are shown. We compare the results
for five different vertices with lattice data taken from ref.~\cite{Bonnet:2002ih}. 
}
\end{figure}

\begin{table}
\begin{tabular}{c|c|c|c|c|c|c|c|c}
& M(0)& $f_\pi$ & $(-\langle \bar{\Psi}\Psi\rangle)^{1/3}$ 
&$(-\langle \bar{\Psi}\Psi\rangle)^{1/3}$
&&&\\ 
&[MeV]&[MeV]&[MeV] (calc.)&[MeV] (fit)
&$a_1$&$a_2$&$b_1$&$b_2$\\
\hline
\mbox{bare vertex} &177&38.5&162&160&3.05&0.99&0.06&1.00\\
\mbox{CP d=$\delta$}&150&50.5&223&225&-&-&-&-\\
\mbox{BC-vertex} &293&62.6&276&284&1.10&0.99&0.29&0.92\\
\mbox{CP-vertex}&369&78.7&303&300&0.83&0.99&0.20&1.00\\
\mbox{CP d=0.1}&464&87.5&334&330&0.79&0.99&0.34&0.95
\end{tabular}
\caption{\sf \label{qtable} The mass $M(0)$, the pion decay constant $f_\pi$
calculated with eq.~(\ref{f-pi-eq}), the renormalisation point independent chiral condensate 
calculated with eqs.~(\ref{ch-cond}) and (\ref{ch-loop}), and the condensate obtained by 
fitting the expression (\ref{ch-fit}) to
the chiral mass function in the ultraviolet for all four vertex types. Recall $\delta=-9/44$
in quenched approximation.
If not stated otherwise the parameter $d$ in the vertex construction is taken to be $d=0$.
For the case of the CP-vertex with $d=\delta$ we did not get good fits in the infrared.}
\end{table}
In Fig.~\ref{qpic1.dat} we give our numerical solutions for the quark mass function
and the inverse vector self energy in the chiral limit. We compare results obtained with 
five different {\it ans\"atze} for the quark-gluon vertex.
For the generalised CP-vertex we investigate the 'natural' case $d=0$, the value $d=\delta=-9/44$,
already adopted in refs.~\cite{Smekal,Ahlig}, and the value
$d=0.1$. Furthermore we employed the bare vertex construction and a Ball-Chiu type vertex.
The corresponding masses at the momentum $p^2=0$, the pion decay constant $f_\pi$, 
the renormalisation point independent chiral condensate and the fit parameters for the
functions (\ref{ab-fit}) are displayed in table \ref{qtable}.

The numerical results for the mass function all have a characteristic plateau
in the infrared and show the regular asymptotic behaviour for large momenta, {\it c.f.} eq.~(\ref{ch-fit}).
The bare vertex construction and the CP type vertex with $d=\delta$ both generate masses much 
smaller than typical phenomenological values of $300-400$ MeV. The BC- and the CP-type construction with
$d=0$ provide good results, whereas the choice $d=0.1$ leads to a somewhat large mass.
The lattice calculations taken from ref.~\cite{Bonnet:2002ih} favour masses around $300$ MeV
with the caveat that they are obtained by an extrapolation from sizeable bare quark masses
to the chiral limit\footnote{This extrapolation is intricate, as can be inferred from
the data points in the ultraviolet region of momentum, where the expected
regular asymptotic behaviour in the chiral limit is not reproduced 
by the lattice data.}.
The numerical solutions for the wave function renormalisation $1/A$ can be seen in the right diagram of
Fig.~\ref{qpic1.dat}. Whereas the ultraviolet asymptotic behaviour of all vertex constructions is similar 
we observe sizeable differences for small momenta. Again the bare vertex construction and 
the CP-vertex with $d=\delta$ are clearly disfavoured by the lattice data.

\begin{figure}[th!]
\vspace{0.5cm}
\centerline{
\epsfig{file=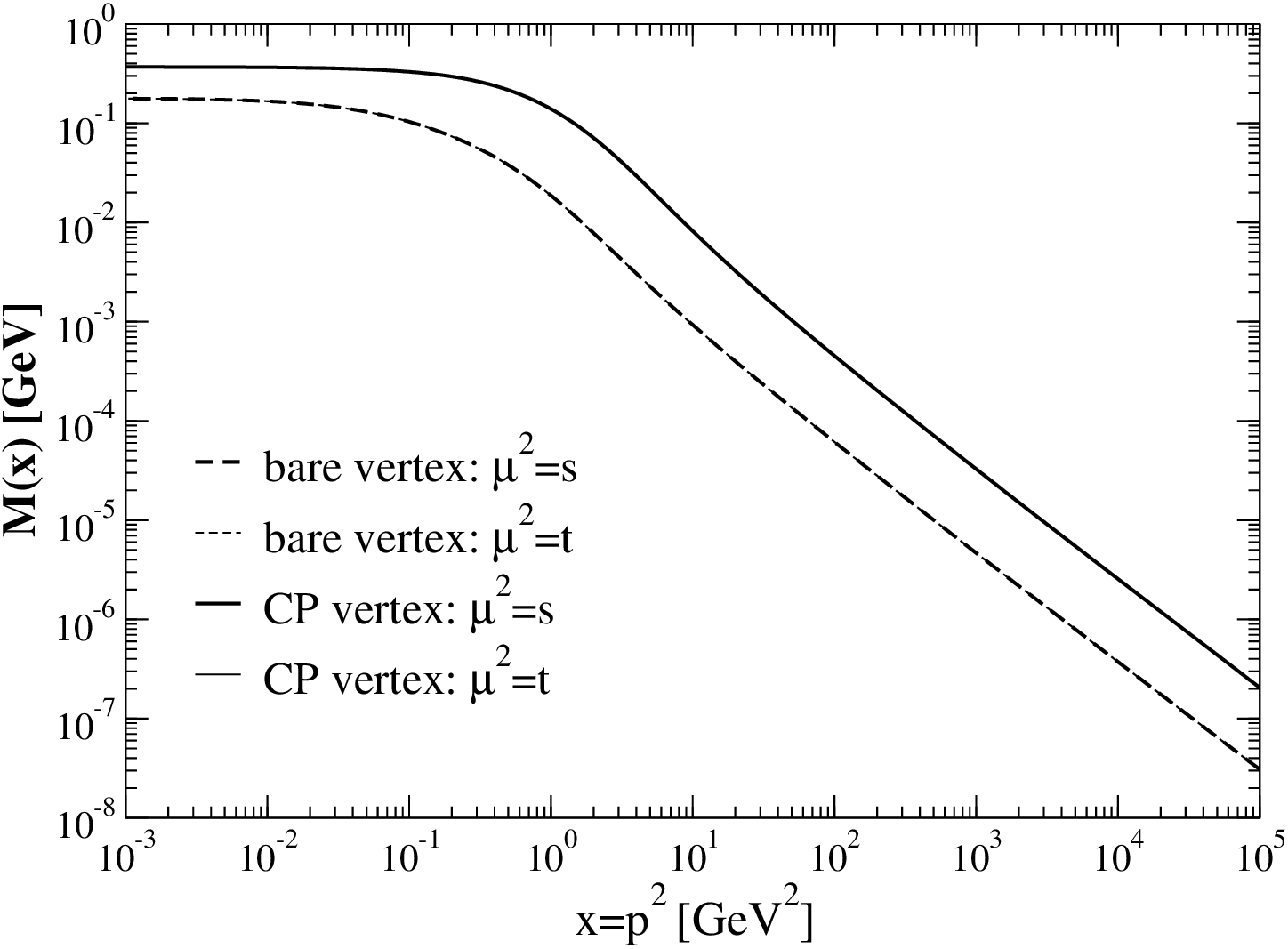,width=7.5cm,height=6.0cm}
\hspace{0.5cm}
\epsfig{file=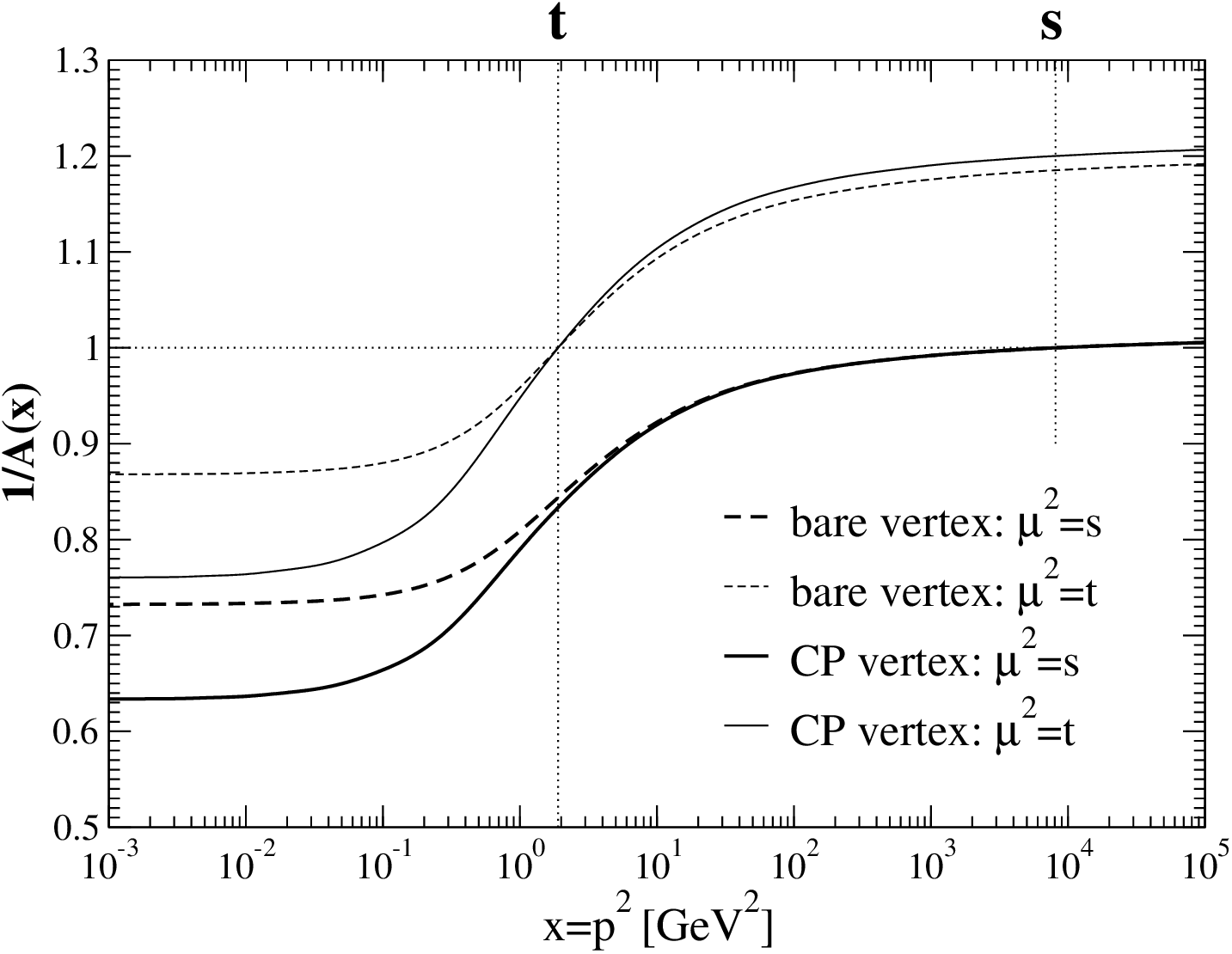,width=7.5cm,height=6.0cm}
}
\caption{\sf \label{qpic2.dat}  The behaviour of $M(x)$ and $1/A(x,s)$ 
under a change of the renormalisation point from $\mu^2=s$ to $\mu^2=t$.}
\vspace{1cm}
\centerline{
\epsfig{file=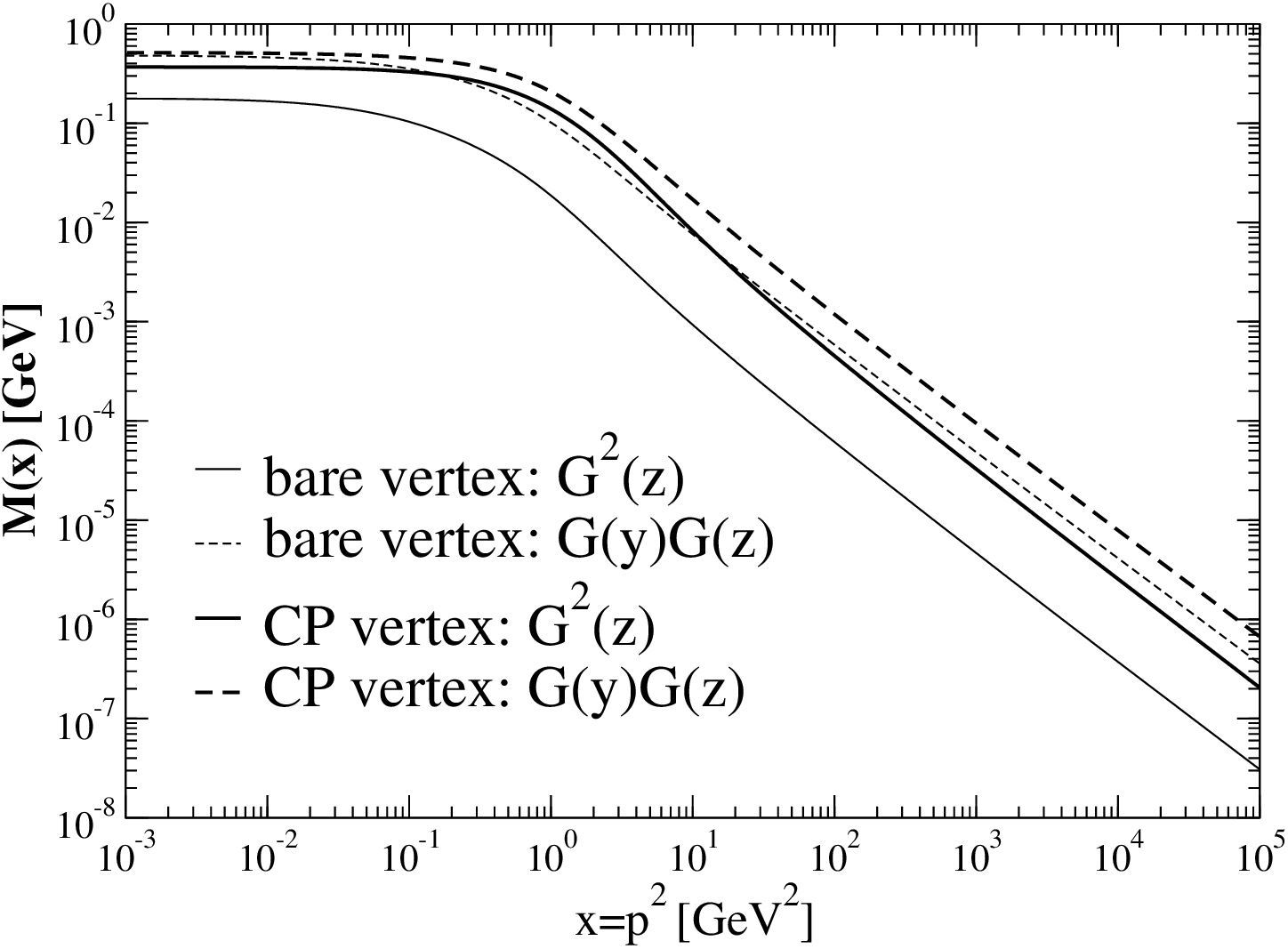,width=7.5cm,height=6.0cm}
\hspace{0.5cm}
\epsfig{file=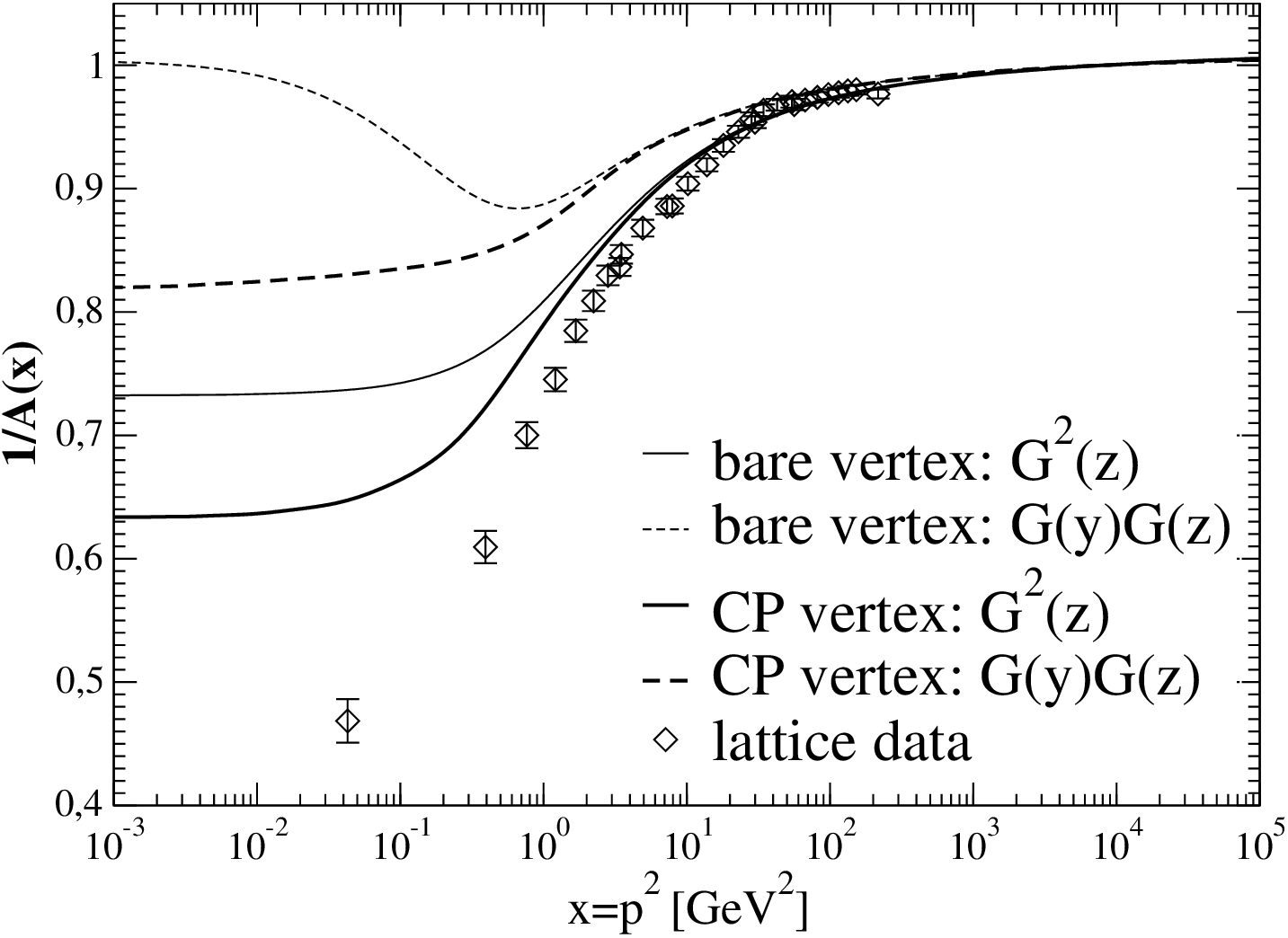,width=7.5cm,height=6.0cm}
}
\caption{\sf \label{qpic3.dat} In these diagrams we show that a change of arguments, 
$G^2(z) \rightarrow G(z)G(y)$, in the non-Abelian part of the quark-gluon 
vertex is disfavoured by lattice data.} 
\end{figure}

Our approximate calculation of the pion decay constant should 
underestimate the experimental value $f_\pi= 93 \,\mbox{MeV}$ by about ten percent, 
{\it c.f.} the discussion below eq.~(\ref{f-pi-eq}). We thus have best results for
the CP-vertex construction with $d=0$ and $d=0.1$.
Furthermore we obtain very good agreement between the two different methods to 
extract the chiral condensate, {\it c.f.} subsection \ref{f-pi-sec}. Compared to the value
$(-\langle \bar{\Psi}\Psi\rangle)^{1/3} = 227 \, \mbox{MeV}$
from the phenomenological study summarised in ref.~\cite{Roberts:2000hi} most of our results are larger. 

Our favourite vertex for further investigations will be the CP-type construction. 
Although the bare vertex construction is by no means
capable to reproduce phenomenological values of $M(0)$ and $f_\pi$ we will keep
this vertex for the sake of comparison. However, we discard the BC-type vertex due to 
numerical problems in the infrared.

Apart from the case $d=\delta$ we obtain very good fits for the scalar and vector self energy, $A(x)$ and $B(x)$, 
for small momenta. The results for the fit parameters can be found in table \ref{qtable}. It is 
interesting to note that the exponents $a_2$ and $b_2$ in the fit functions of eq.~(\ref{ab-fit}) are
all fitted  very close to one. Such a behaviour might turn out to be crucial for the continuation of the
quark propagator to negative $p^2$, {\it i.e.} timelike momenta, in future work\footnote{There are several
possibilities how the quark propagator might look like for negative $p^2$ \cite{Oehme:1980ai,Oehme:1995pv}. 
Confinement seems to require
that there are no poles on the negative $p^2$-axis. This would prevent a K\"all\'en-Lehmann spectral 
representation \cite{Bjorken:1979dk} which is mandatory for a propagator representing a physical particle.
In phenomenological models two different {\it ans\"atze} for the quark propagator have been explored:
an exponential form with a pole at $p^2 \rightarrow -\infty$ (see {\it e.g.} 
\cite{Maris:1997tm,Maris:1998hd,Alkofer:1999rn})
and a form proposed by Stingl \cite{Stingl:1996nk} with complex conjugate poles in the negative half of the
$p^2$-plane \cite{Ahlig:2000xm,Ahlig:2000qu}. It has been shown, however, that the exponential form is ruled
out by scattering processes 
\cite{Ahlig:2000qu}.}. 

A technical point is illustrated in Fig.~\ref{qpic2.dat}. Here we display 
the dressing functions calculated with the bare vertex and the CP-vertex construction ($d=0$) for
two different renormalisation points, $s=8100 \, \mbox{GeV}^2$ and $t=1.9 \, \mbox{GeV}^2$.
Although these two points differ by three orders of magnitude the resulting mass functions 
for each vertex are indistinguishable in the plot. The wave function renormalisation $1/A(x,s)$ 
is multiplied by a constant factor when we change from $s$ to $t$, {\it c.f.} eq.~(\ref{A_R}). 
We thus find the behaviour expected from the discussion in subsection \ref{subsec-MR}.

In Fig.~\ref{qpic3.dat} we compare the results for two different momentum assignments in the non-Abelian
part of the vertex. We employ 
\beq
W^{\neg abel}(x,y,z) = G^2(z,s) \:\tilde{Z}_3(s,L) \,, \label{gz-quad}
\eeq
in accordance with the $d=0$ case of eq.~(\ref{vertex_nonabel}) and the modified assignment
\beq
W^{\neg abel}(x,y,z) = G(z,s)\:G(y,s) \:\tilde{Z}_3(s,L). \label{gy-gz}
\eeq
Both expressions behave similar under finite renormalisation and become equal in the limit of large 
momenta. Therefore both expressions lead to a mass function satisfying the conditions
(i) and (ii) formulated in section \ref{quark-dse-sec}. However, as can be seen from the plot,
the lattice data for the wave function renormalisation $1/A$ clearly disfavour the second momentum assignment. 
This will be important in the next section, where we investigate the quark-loop in the gluon equation.

\begin{figure}[t!]
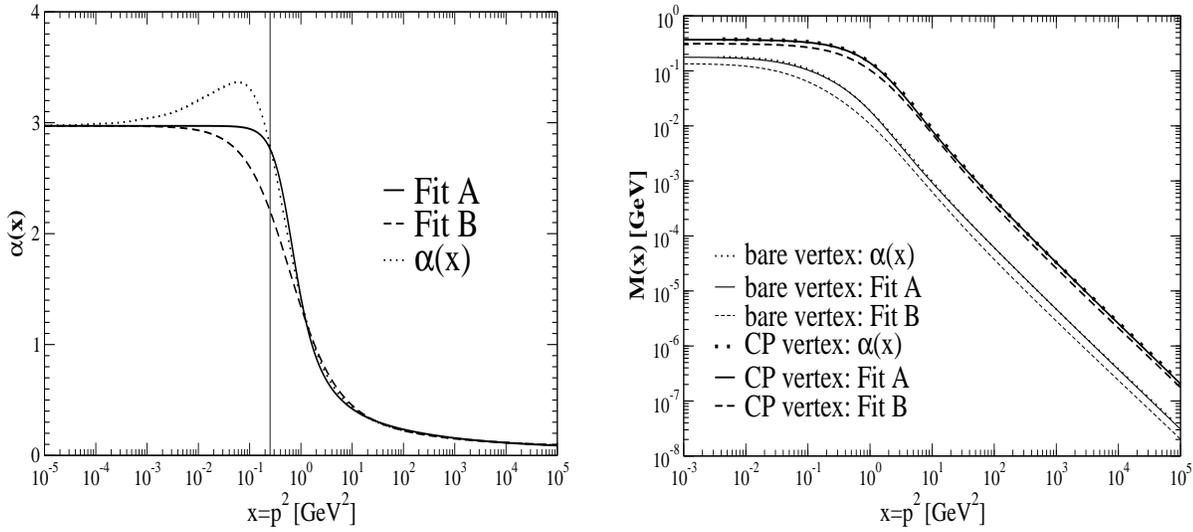

\vspace{1cm}
\centerline{
\epsfig{file=alpha.fit.eps,width=7.5cm,height=7cm}
\hspace{0.5cm}
\epsfig{file=qM.fit.eps,width=7.5cm,height=7cm}
}
\caption{\sf \label{qpic4.dat} Results for three different forms of the running coupling in the 
quark equation: The running coupling calculated in chapter \ref{YM} and the two fits given in 
eqs.~(\ref{fitA}), (\ref{fitB}). 
}
\end{figure}
Fig.~\ref{qpic4.dat} compares results for the bare vertex and the CP-type construction
for three different forms of the running coupling in the interaction kernel of the quark equation.
The two fit-functions, 'Fit A' and 'Fit B', have been given in eqs.~(\ref{fitA}), (\ref{fitB})
Furthermore we used the running coupling calculated from the quenched ghost and gluon DSEs in section
\ref{YM-results}. Although there is the (presumably) artificial bump at $p^2 \approx 0.1 \, \mbox{GeV}^2$
in the running coupling, the mass functions obtained from the DSE-result and from 'Fit A' are
virtually indistinguishable. 'Fit B', however leads to somewhat smaller masses.
This observation suggests
that nearly all the dynamically generated mass is produced from the integration strength above $p=500$ MeV, 
as is
indicated by the vertical line in the plot of the running coupling. This is a favourable result as it 
would have been very unsatisfying if the artificial bump contributed a considerable amount to the quark 
mass function. 

\begin{figure}[t!]
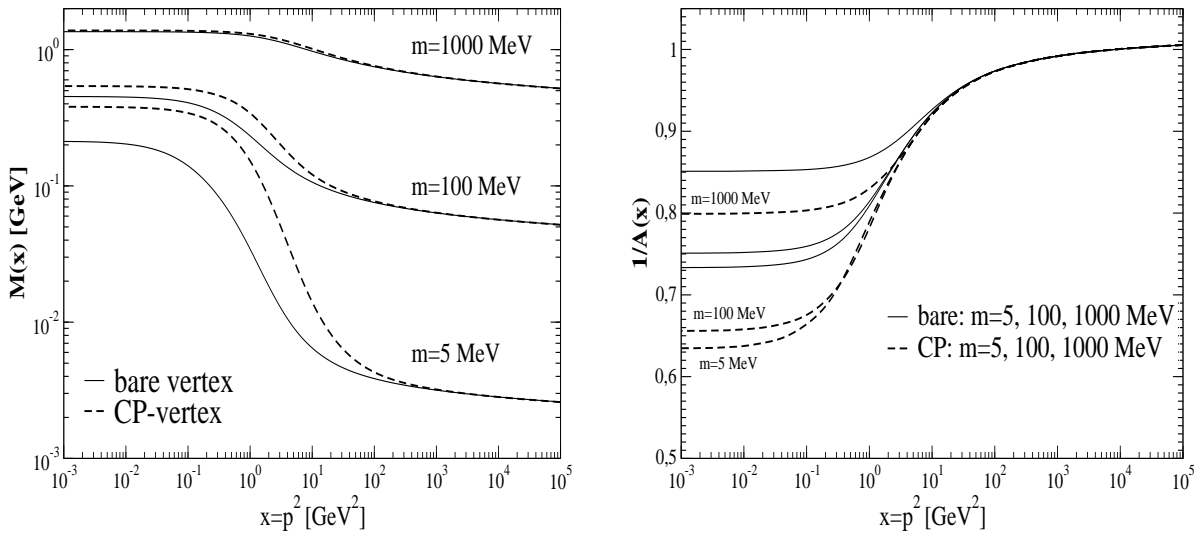

\vspace{1cm}
\centerline{
\epsfig{file=qM.mass.eps,width=7.5cm,height=7cm}
\hspace{0.5cm}
\epsfig{file=qA.mass.eps,width=7.5cm,height=7cm}
}
\caption{\sf \label{qpic5.dat} These diagrams show our results when three different bare
quark masses are employed. In the diagram on the right small quark masses correspond to small
values for $1/A$ in the infrared. 
}
\end{figure}
Finally we observe the effects of explicit chiral symmetry breaking in the plots of 
Fig.~\ref{qpic5.dat}. We give results for three different quark masses, $m(2 \,\mbox{GeV})=5$ MeV, 
$m(2 \,\mbox{GeV})=100$ MeV and $m(1 \,\mbox{GeV})=1000$ MeV. 
These values correspond roughly to the ones given by the Particle Data Group 
for the up/down-quark, the strange-quark
and the charm-quark \cite{Hagiwara:2002pw}.
For small momenta we note again that the dressed vertex generates more mass in the quark equation
than the bare vertex construction. This effect becomes much less dominant for the heavy quarks, where 
more and more of the infrared mass stems from explicit chiral symmetry breaking and not from 
dynamical mass generation. Furthermore in accordance with our analysis in subsection \ref{quark-UV-analysis}
we observe the same ultraviolet behaviour of the mass function for both vertex constructions. 

\section{Incorporating the quark-loop in the gluon equation}

\begin{figure}[t]
\vspace{0.5cm}
\centerline{
\epsfig{file=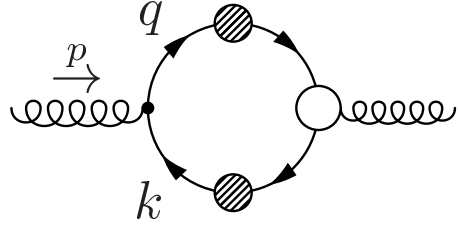,width=6.0cm,height=4cm}
}
\caption{\sf \label{quarkloop.dat} Diagrammatical representation of the quark-loop in the gluon equation.
}
\end{figure}

In section \ref{YM-results} and in the last section we have discussed results for the ghost, gluon and 
quark propagators in quenched approximation. Now we go one step further and investigate the 
unquenched equations, {\it i.e.} we include the back-reaction of the quarks on the 
ghost-gluon system. To this end we incorporate the quark-loop in the truncation
scheme developed in chapter \ref{YM}. Due to our experience with the ghost- and gluon-loop
this might seem to be a straightforward task. However, 
we will encounter some additional problems which are related to the quark-gluon vertex.

The contribution of the quark-loop to the gluon equation is given by 
\beq
 \Pi^{quark}_{\mu \nu} = -\frac{g^2 N_f}{2 (2 \pi)^4} \,Z_{1F} \,\int d^4q \:\: 
 \mbox{Tr}\left\{ \gamma_\mu \,S(q)\, \Gamma_\nu(q,k) \,S(k)
 \right\},
\eeq
where $x=p^2$ is the external gluon momentum and $y=q^2$ and $z=k^2=(q-p)^2$ are the
squared momenta of the two quarks running in the loop. The trace is over Dirac indices. 

In section \ref{quark-dse-sec} we have proposed an effective quark-gluon vertex $\Gamma_\nu(q,k)$
with Abelian and non-Abelian parts such that 
the quark equation is multiplicatively renormalisable and one-loop perturbation theory is 
recovered for large momenta. However, this construction is not capable to account as well for
the one-loop behaviour of the unquenched gluon equation unless we switch 
the arguments of the non-Abelian part $W^{\neg abel}$, eq.~(\ref{vertex_nonabel}),
to different momenta. In the quark equation such a change of momentum leads to
results clearly disfavoured by the lattice data, {\it c.f.} Fig.~\ref{qpic3.dat}. We therefore 
have to use different momentum assignments for the quark-loop and the quark equation.
Certainly, this is a deficiency which has to be resolved by a more elaborate vertex construction
in future work. The aim of the present study, however, is to present an effective construction 
which captures essential properties of the theory.

Taking care of symmetries we propose the following {\it ansatz} for the 
non-Abelian part of the quark-gluon vertex in the quark-loop: 
\beqa
W^{\neg abel}_{quark-loop}(x,y,z) &=& G(y) G(z) \tilde{Z}_3(L) 
\frac{\left(G(y) \tilde{Z}_3(L) \right)^{-d-d/(2\delta)}}
{\left(Z(y) {Z}_3(L)\right)^{d/2}}
\frac{\left(G(z) \tilde{Z}_3(L) \right)^{-d-d/(2\delta)}}
{\left(Z(z) {Z}_3(L)\right)^{d/2}}. \nonumber\\
\eeqa
Here $x=p^2$ is the squared gluon momentum, $y=q^2$ and $z=k^2=(q-p)^2$ are the squared quark momenta,
and $L=\Lambda^2$ is the squared cutoff.
The Abelian part of the vertex is already symmetric
with respect to the quark momenta.

Plugging the Curtis-Pennington type vertex into the quark-loop and contracting the free
Lorenz-indices with the tensor ({\it c.f.} eq.~(\ref{Paproj}))
\beq
{\mathcal P}^{(\zeta )}_{\mu\nu} (p) = \delta_{\mu\nu} - 
\zeta \frac{p_\mu p_\nu}{p^2} 
\, , \label{tensor1} 
\eeq
we obtain
\setlength{\jot}{2mm}
\beqa
 \Pi_{quark} &=&-\frac{g^2 N_f}{(2 \pi)^4} Z_2 \int d^4q 
\frac{G(y)}{y + M^2(y)} 
\frac{G(z)}{z + M^2(z)} 
 \frac{ 
(G(y)\:G(z)\: \tilde{Z}_3^2(L))^{-d-d/(2\delta)}}{ (Z(y)\:Z(z) \:Z_3^2(L))^{d/2}}
 \times \nonumber\\
&&\times A^{-2}(y)A^{-2}(z)\:\left\{\frac{A(y)+A(z)}{2} \left(W_1(x,y,z) A(y)A(z) + W_2(x,y,z) B(y)B(z)\right)
\right. \nonumber\\
&&\hspace*{3cm}+ \frac{A(y)-A(z)}{2(y-z)} \left(W_3(x,y,z) A(y)A(z) + W_4(x,y,z) B(y)B(z) \right) \nonumber\\
&&\hspace*{3cm}+ \frac{B(y)-B(z)}{y-z} \left(W_5(x,y,z) A(y)B(z) + W_6(x,y,z) B(y)A(z) \right) \nonumber\\
&&\hspace*{-0.8cm}\left. + \frac{(A(y)-A(z)) (y+z)}{2((y-z)^2 + (M^2(y)+M^2(z))^2)} 
\left(W_7(x,y,z)A(y)A(z) + W_8(x,y,z)B(y)B(z) \right) \right\}, \nonumber\\  
\label{quarkloop}
\eeqa
with the kernels
\beqa
W_1(x,y,z) &=&\frac{\zeta z^2}{3 x^2} + z\left(\frac{2-\zeta}{3x} - \frac{2\zeta y}{3x^2}\right)
              -\frac{2}{3}+\frac{(2-\zeta)y}{3x}  + \frac{\zeta y^2}{3x^2} \,, \\
W_2(x,y,z) &=& \frac{2 (4-\zeta)}{3x} \,, \\	       
W_3(x,y,z) &=& \frac{\zeta z^3}{3 x^2} - z^2\left(\frac{1+\zeta}{3x} + \frac{\zeta y}{3x^2}\right)
              +z \left(\frac{1}{3} + \frac{(2\zeta-6)y}{3x}-\frac{\zeta y^2}{3x^2}  \right) \nonumber\\
	    &&+\frac{y}{3} 
	      - \frac{(\zeta+1)y^2}{3x} + \frac{\zeta y^3}{3x^2} \,, \\ 
W_4(x,y,z) &=& \frac{-2 \zeta z^2}{3x^2} + z \left(\frac{4}{3x}+\frac{4 \zeta y}{3x^2}  \right) 
              -\frac{2}{3} + \frac{4y}{3x} - \frac{2 \zeta y^2}{3 x^2}  \,,\\ 
W_5(x,y,z) &=&  \frac{\zeta z^2}{3x^2} -z\left( \frac{1+\zeta}{3x}+\frac{2\zeta y}{3x^2}\right)
              + \frac{1}{3} + \frac{(\zeta-3) y}{3x} + \frac{\zeta y^2}{3x^2}  \,,\\ 
W_6(x,y,z) &=& \frac{\zeta z^2}{3x^2} - z \left(\frac{3-\zeta}{3x}+\frac{2\zeta y}{3x^2} \right)
              + \frac{1}{3}  + \frac{(-\zeta-1) y}{3x}+ \frac{\zeta y^2}{3x^2}  \,,\\
W_7(x,y,z) &=&-\frac{z^2}{x} + z + \frac{y^2}{x} - y  \,,\\ 
W_8(x,y,z) &=& 2 \left( - \frac{z}{x} + \frac{y}{x} \right) \,.
\eeqa
Note that the symmetry factor $1/2$ and a factor $1/(3x)$ from the left hand side of the gluon equation
have already been absorbed in the kernels. 
From this expression the corresponding one for the bare vertex construction can be read off easily by setting
$W_{3-8}=0$ and replacing the remaining factor $(A(y)+A(z))/2$ in eq.~(\ref{quarkloop}) by unity.

\subsection{Ultraviolet analysis of the quark-loop}

In section \ref{ultraviolet} we encountered quadratic divergences in the gluon equation
for $\zeta \not=4$. These quadratic divergences occur in the ghost- and 
gluon-loop of the gluon equation and show up in the quark-loop as well.
To identify such terms we expand the dressing functions
in the integrand of the quark-loop around large loop momenta $y$ with the difference $(z-y)$ still 
larger than any quark mass. 
To leading order this expansion amounts in the replacements 
\beqa
G(z) &\rightarrow& G(y), \nonumber\\
A(z) &\rightarrow& A(y), \nonumber\\
\frac{A(y) - A(z)}{y-z}  &\rightarrow& A^\prime(y), \nonumber\\
\frac{B(y) - B(z)}{y-z}  &\rightarrow& B^\prime(y), \nonumber\\
\frac{(A(y)-A(z)) (y+z)}{2((y-z)^2 + (M^2(y)+M^2(z))^2)} &\rightarrow& 
\frac{A^\prime(y) (y+z)}{2(y-z)},  
\eeqa
with the derivatives $A^\prime$ and $B^\prime$.
Note that the first two equations are identical to the angular approximation employed previously
in subsection \ref{ultraviolet}. 
For large momenta $x$ and $z$ the denominators in eq.~(\ref{quarkloop})
simplify and the angular integrals are trivially performed using the integrals given in appendix 
\ref{angular}. We arrive at 
\setlength{\jot}{3mm}
\beqa
 \Pi_{quark}^{UV} &=&-\frac{g^2 N_f}{16 \pi^2} Z_2 \int dy \:G^2(y)\: 
 \frac{G(y)^{-2d-d/\delta}}{ Z(y)^{d}} \: A^{-2}(y) \times \nonumber\\
 && \hspace*{1.9cm}
\left\{A(y) \left(\frac{-2}{3y}+\frac{4-\zeta}{3x}
 + \frac{2 (4-\zeta)}{3xy} M^2(y)\right)
\right. \nonumber\\
&& \hspace*{1.9cm}+ \frac{A^\prime(y)}{2} \left(\frac{1}{3}+\frac{-2(4-\zeta)y}{3x}
  + \left(\frac{-2}{3y} + \frac{2(4-\zeta)}{3x} \right) M^2(y) \right) \nonumber\\
&& \hspace*{1.9cm}+ B^\prime(y) M(y) \left( \frac{2}{3y} - \frac{4-\zeta}{3x} \right)
\left. + \frac{A^\prime(y)}{2} 
\left(\frac{4y}{x}-1 + \frac{4}{x}M^2(y) \right) \right\}. \hspace*{1.1cm}  
\label{UV_quarkloop}
\eeqa
Keeping in mind a factor $(1/y)$ hiding in the derivatives we are now able to identify three
quadratically divergent terms: $(4-\zeta)/3x$ in the second line, 
$-2(4-\zeta)y/3x$ in the third line and $4 y/x$ in the last line. The first two
of them are proportional to $(4-\zeta)$ and reminds us of similar terms occurring in eq.~(\ref{gluon-UV}).
These terms are artefacts of the regularisation and will be subtracted from the kernels.
However, we encounter the additional $\zeta$-independent quadratic divergent term $4 y/x$
originating from the transverse part of the Curtis-Pennington vertex.
Such a term is already known from corresponding studies in QED \cite{Bloch}. 
Although first suggestions have been made how the 
Curtis-Pennington vertex should be modified to avoid this problem \cite{Pennington:1998cj}, 
a convincing solution has not been found yet. In this thesis we therefore 
choose the pragmatic strategy of
subtracting this term by hand together with the other quadratically divergent parts.

Moreover we subtract all further terms proportional to $(4-\zeta)$. Although these terms are not 
quadratically divergent they are artefacts of the regularisation. We then obtain
a $\zeta$-independent expression for the quark loop at large momenta. Together
with the corresponding expressions for the ghost-loop and the gluon-loop, eq.~(\ref{gluon-UV}),
we obtain a transversal gluon propagator in the ultraviolet as it should be the case
in Landau gauge.

Collecting all modifications together we arrive at the new kernels
\setlength{\jot}{2mm}
\beqa
\widetilde{W}_1(x,y,z) &=& W_1(x,y,z) - \frac{(y+z)(4-\zeta)}{6x}, \\
\widetilde{W}_2(x,y,z) &=& 0, \\ 
\widetilde{W}_3(x,y,z) &=& W_3(x,y,z) + \frac{2zy(4-\zeta)}{3x}, \\ 
\widetilde{W}_4(x,y,z) &=& W_4(x,y,z) - \frac{(y+z)(4-\zeta)}{3x}, \\
\widetilde{W}_5(x,y,z) &=& W_5(x,y,z) - \frac{(y+z)(4-\zeta)}{6x}, \\
\widetilde{W}_6(x,y,z) &=& W_6(x,y,z) - \frac{(y+z)(4-\zeta)}{6x}, \\
\widetilde{W}_7(x,y,z) &=& W_7(x,y,z) - \frac{(y-z)(y+z)}{x},\\
\widetilde{W}_8(x,y,z) &=& W_8(x,y,z).
\eeqa   
Note that the subtracted terms are chosen to preserve 
the symmetry of the kernels with respect to the squared quark momenta $y$ and $z$.

Without quadratic divergences we are in a position to extract the leading 
logarithmic divergence of the quark-loop. With modified kernels the ultraviolet limit of the 
quark-loop is given by
\beqa
 \Pi_{quark}^{UV} &=&-\frac{g^2 N_f}{16 \pi^2} Z_2 \int dy \:G^2(y)\: 
 \frac{G(y)^{-2d-d/\delta}}{ Z(y)^{d}} \: A^{-2}(y) \times \nonumber\\
 && \hspace{2.5cm}
\left\{A(y) \frac{-2}{3y} 
   + \frac{A^\prime(y)}{2} \left(\frac{1}{3}
  + \frac{-2}{3y} M^2(y)\right)  \right.\nonumber\\
&& \hspace{2.5cm}+ B^\prime(y) M(y)  \frac{2}{3y}
\left. + \frac{A^\prime(y)}{2} 
\left(-1+\frac{4}{x}M^2(y)\right)  \right\}.  
\label{UV2_quarkloop}
\eeqa
Similar to the situation in the DSE for the quark mass function, {\it c.f.} subsection \ref{quark-UV-analysis},
the leading ultraviolet term is the first term in the curly brackets. Substituting the ultraviolet limit
of the vector self energy, eq.~(\ref{A-limit}), the ghost and gluon dressing functions, eqs.~(\ref{Z_UV}),
and choosing the perturbative renormalisation condition $G(s)=Z(s)=1$ we arrive at
\beq
\Pi_{quark}^{UV} = \frac{2 N_f}{3 (2\delta+1) \omega} \frac{g^2}{16 \pi^2} \left\{
\left[ \omega \log\left(\frac{L}{s}\right)+1 \right]^{2\delta+1} -
\left[ \omega \log\left(\frac{x}{s}\right)+1 \right]^{2\delta+1} \right\}. \label{quark-anom}
\eeq
Here we have $\omega=\beta_0\alpha(s)/(4 \pi)=(11N_c-2N_f)\alpha(s)/(12 \pi)$.
Moreover $\delta$ is the anomalous dimension of the ghost propagator which is related to the 
corresponding
anomalous dimension of the gluon by $\gamma + 2 \delta +1=0$, {\it c.f.} eq.~(\ref{gd1}).
Combining the expression (\ref{quark-anom}) with the results for the ghost and gluon loop, 
eq.~(\ref{gluoneq_uv}) we obtain the anomalous dimensions 
\beqa
\gamma &=& \frac{-13 N_c + 4 N_f}{22 N_c - 4 N_f} \\
\delta &=& \frac{-9 N_c}{44 N_c - 8 N_f}
\label{anom_dim}
\eeqa
which are in accordance to one-loop perturbation theory for arbitrary numbers of colours $N_c$ and flavours $N_f$.
\setlength{\jot}{0mm}

\subsection{Infrared analysis of the quark-loop \label{ir-quark}}

In section \ref{infrared} we employed the power law {\it ansatz}
\beq
Z(x) = A x^{2\kappa} , \hspace*{1cm} G(x) = B x^{-\kappa},
\eeq
for the ghost and gluon dressing functions at small momenta $x$. Substituting this {\it ansatz} into
the gluon equation we have found the ghost loop to be proportional to $x^{-2\kappa}$, dominating the
gluon-loop in the infrared. When we compare our expression for the quark-loop, eq.~(\ref{quarkloop}), 
with the ghost loop in eq.~(\ref{gluonbare}) we find two ghost dressing functions, $G(y)$ and $G(z)$, in
each loop respectively.
However, as the momenta $y$ and $z$ in the denominators of the quark loop are negligible compared to the
quark masses in the infrared we anticipate that the quark-loop is less
divergent than the ghost-loop for small momenta, provided the parameter $d$ is small. 

An explicit calculation along the lines of our analysis in subsection \ref{infrared}
and appendix \ref{app-diag} shows that the quark loop is proportional to 
$x^{-2\kappa+2+\kappa d/\delta}$
in the infrared. Therefore {\it the quark loop is suppressed for small momenta} 
provided the parameter $d$ fulfils the condition
\beq
d < \frac{-2\delta}{\kappa}.
\eeq
As we have $\kappa \approx 0.5953$ and $\delta=-1/4$ for $N_c=3$ and $N_f=3$ we find the condition 
$d<0.84$,
which is satisfied for all quark-gluon vertices employed in our calculation.
From a numerical point of view we encounter serious instabilities in the quark and the gluon equation
once $d$ is taken to be larger than $d \approx 0.2$. 

We conclude that the quark-loop does not change the infrared 
behaviour of the ghost and gluon dressing functions found in section \ref{infrared}. In pure 
Yang-Mills theory as well as in QCD we thus have an infrared finite or vanishing gluon propagator and a
ghost propagator which is more divergent than a simple pole. The Kugo-Ojima
confinement criterion and Zwanziger's horizon condition are both fulfilled not only in pure Yang-Mills 
theory but also in QCD. This is a central result of this chapter.

\section{Numerical results \label{res-unquenched}}

The numerical treatment of the integrals in the quark, ghost and gluon equations has been described 
in subsections \ref{subtraction} and \ref{ren-quenched}.
The iteration process is done for the ghost-gluon system and the quark equations 
separately: we first iterate the $N_f$ mutually uncoupled quark systems until convergence is achieved,
feed the output into the ghost and gluon system, iterate until the ghost-gluon system converges, 
feed the output back into the quark equations and so on, until complete convergence of all 
equations is achieved. Similar to the quenched calculations we used $\alpha(s)=0.2$ at the
renormalisation point $s=\mu^2$, and a transverse tensor to contract the gluon equation, $\zeta=1$.

In contrast to section \ref{quenched} we fix the physical scale of the system not by the condition
$\alpha(M_Z^2)=0.118$ but by adjusting the pion decay constant to the experimental value.
This choice has an important advantage: Whereas the asymptotic behaviour of the running coupling
depends strongly on $N_f$ via the coefficient $\beta_0$ ({\it c.f.} appendix \ref{alpha}), 
the pion decay constant turns
out to be almost independent of $N_f$. In order to be able to compare different constructions for
the quark-gluon vertex we fix the scale by requiring $f_\pi=85$ MeV  for the
CP-construction with $d=0.1$ and keep this scale when employing other vertices\footnote{ 
Recall from subsection \ref{f-pi-sec} that the approximation (\ref{f-pi-eq}) for the pion decay constant does not
include the effects of subdominant pion Bethe-Salpeter amplitudes. Therefore the value $f_\pi=85$ MeV 
is chosen to allow for corrections compared to the experimental value $f_\pi=93$ MeV.}. 

\begin{table}
\begin{tabular}{c||c|c||c|c||c|c||c|c||c|c}
&\multicolumn{2}{c||}{M(0)} & \multicolumn{2}{c||}{$f_\pi$} &  
\multicolumn{2}{c||}{$(-\langle \bar{\Psi}\Psi\rangle)^{1/3}$}&
\multicolumn{2}{c||}{$\alpha(M_Z)$}&\multicolumn{2}{c}{$\Lambda_{QCD}^{MOM}$}\\ 
& \multicolumn{2}{c||}{[MeV]} & \multicolumn{2}{c||}{[MeV]} &  
\multicolumn{2}{c||}{[MeV]}&
\multicolumn{2}{c||}{-}&\multicolumn{2}{c}{[MeV]}\\ \hline
vertex & qu.&unqu.&qu.&unqu.&qu.&unqu.&qu.&unqu.&qu.&unqu.\\ \hline
\mbox{bare d=0} &172&169&37.4&37.2&155&162&0.116&0.144&694&718\\
\mbox{CP d=$\delta$}&146&153&49.1&49.8&219&215&0.116&0.138&694&578\\
\mbox{CP d=0}&358&346&76.5&76.2&291&295&0.116&0.141&694&644\\
\mbox{CP d=0.1}&451&425&85.0&85.0&321&325&0.116&0.143&694&669
\end{tabular}
\caption{\sf \label{unqtable} A comparison between the quenched (qu.) and unquenched (unqu.)
results for the quark mass M(0), the pion decay constant $f_\pi$,
the renormalisation point independent chiral condensate, the running coupling at the mass 
of the Z-boson and $\Lambda_{QCD}^{MOM}$ for different vertices and values of the  
parameter $d$. The unquenched calculations are done for $N_f=3$ chiral quarks. 
Furthermore we have $\delta=-9N_c/(44N_c-8N_f)=-0.25$ in the present case.}
\end{table}
\begin{figure}[th!]
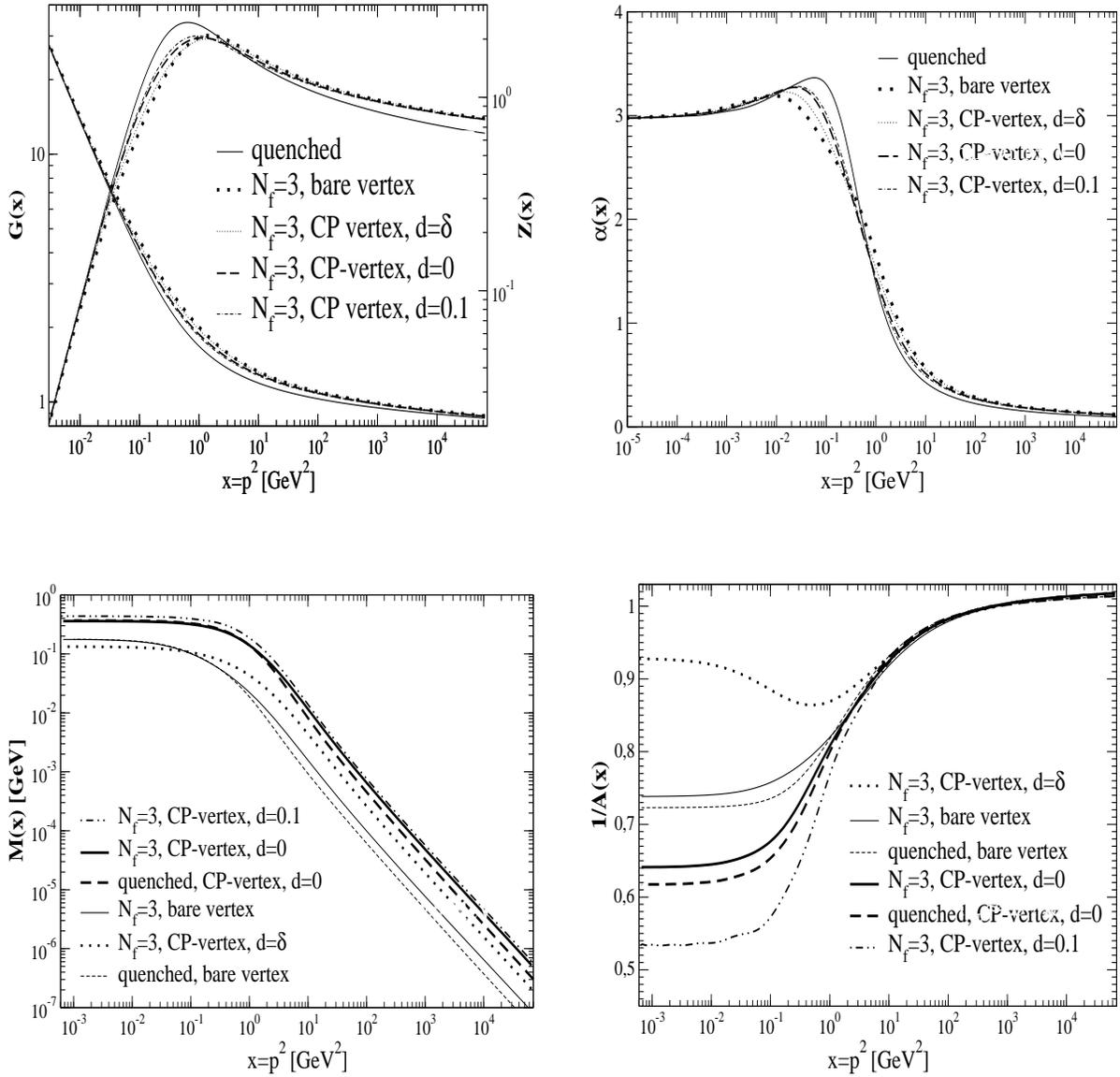

\vspace{0.6cm}
\centerline{
\epsfig{file=zg.1.eps,width=7.5cm,height=7cm}
\hspace{0.5cm}
\epsfig{file=alpha.1.eps,width=7.5cm,height=7cm}
}
\vspace{1.2cm}
\centerline{
\epsfig{file=M.1.eps,width=7.5cm,height=7cm}
\hspace{0.5cm}
\epsfig{file=1A.1.eps,width=7.5cm,height=7cm}
}
\caption{\sf \label{pic1.dat} Displayed are the ghost and gluon dressing function, $Z$ and $G$, 
the running coupling $\alpha$, the quark mass function $M$ and the inverse vector self energy
$1/A$. The calculations are done quenched and unquenched with $N_f=3$ quarks in the chiral limit. 
}
\end{figure}
\begin{figure}[th!]
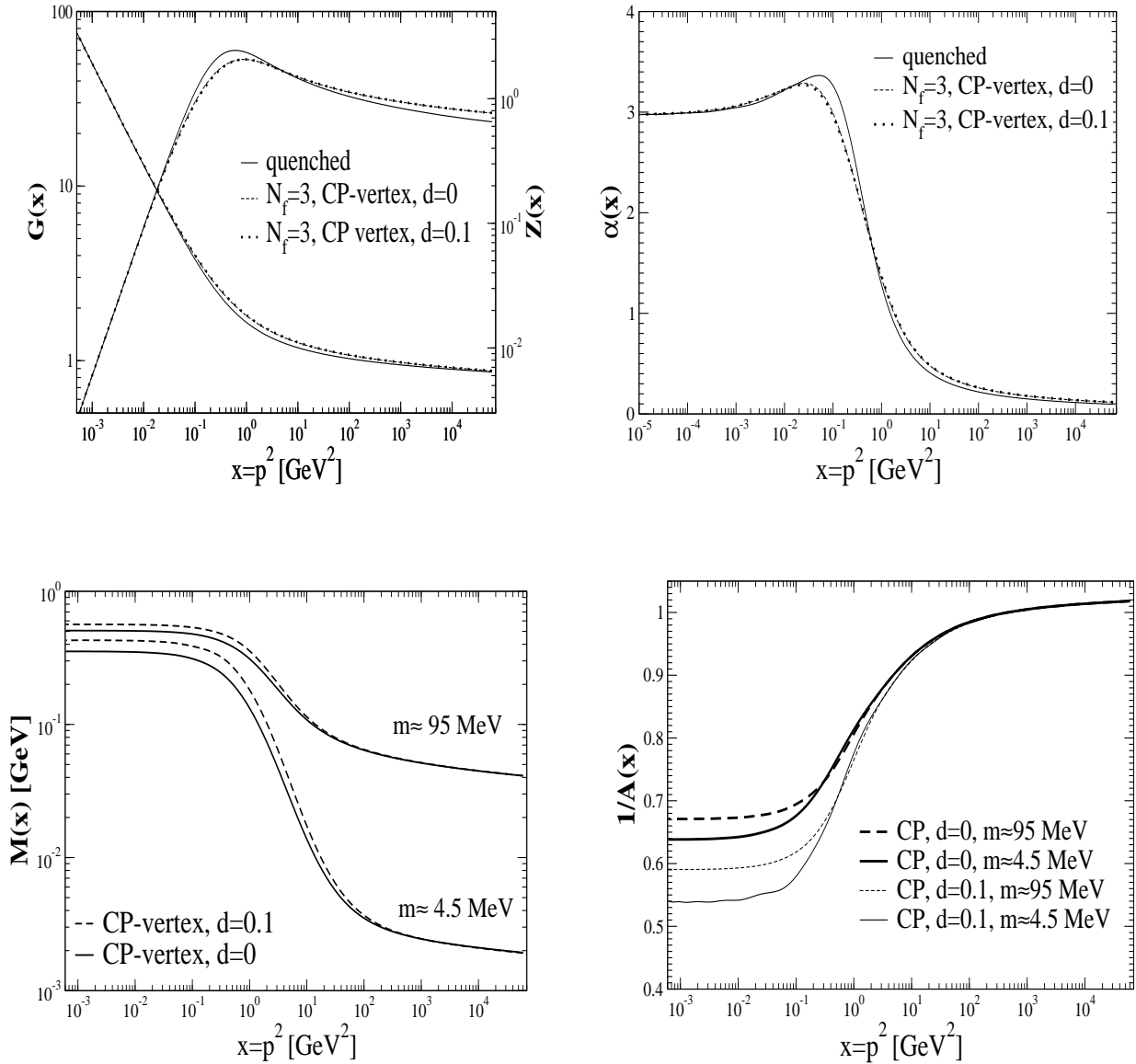

\vspace{0.4cm}
\centerline{
\epsfig{file=zg.2.eps,width=7.5cm,height=7cm}
\hspace{0.5cm}
\epsfig{file=alpha.2.eps,width=7.5cm,height=7cm}
}
\vspace{1.3cm}
\centerline{
\epsfig{file=M.2.eps,width=7.5cm,height=7cm}
\hspace{1cm}
\epsfig{file=1A.2.eps,width=7.5cm,height=7cm}
}
\caption{\sf \label{pic2.dat} Results from the unquenched calculation with $N_f=3$ massive quarks.
We used the renormalised masses $m_{u/d}(2\,\mbox{GeV}) \approx 4.5$ MeV and 
$m_{s}(2\,\mbox{GeV}) \approx 95$ MeV. 
}
\end{figure}

In table \ref{unqtable} we compare results for the quenched and unquenched system of 
equations\footnote{The scale of the quenched
results has been transformed to our modified scale. Both scales are very close to each other
as can be seen by comparison with table \ref{qtable}.}, {\it c.f.} section \ref{quenched}.  
The quark mass, the pion decay constant and the chiral condensate differ only slightly for
each vertex construction respectively. The only sizeable difference occurs in the running coupling.
As expected from perturbation theory the unquenched running for $N_f = 3$ results in
larger values of the running coupling at $p^2=(M_Z)^2$ compared to the quenched case $N_f=0$.
We obtain $\alpha(M_z) \approx 0.140$, which is somewhat larger than usually quoted values from
experiment\footnote{However, such large values are not yet excluded by experiment. A recent analysis
of experimental data from $\tau$-decay suggests $\alpha(M_z) \approx 0.129$ \cite{Geshkenbein:2002ri}.}.
If we increase the number of flavours in our calculation we encounter large numerical uncertainties 
and do not obtain convergence for $N_f \ge 5$. 
  
All employed vertex constructions allow for nontrivial solutions of the quark equation corresponding to
dynamical chiral symmetry breaking. However, similar to the quenched case the bare vertex construction 
and the CP-type vertex with $d=\delta$ generate much too small quark masses compared with typical
phenomenological values\footnote{This is in accordance with previous quenched \cite{Smekal} and unquenched
\cite{Ahlig} investigations of these cases in a truncation scheme with angular 
approximations in the Yang-Mills sector.}. For $d=0$ we obtain good results for the quark mass, 
the pion decay constant and the chiral condensate, whereas the choice $d=0.1$ leads to somewhat 
large values. It is interesting to note that $d=0$ of all values is preferred as in this case the
quark equation resembles most the fermion equation of QED.

In Fig.~\ref{pic1.dat} we display the ghost, gluon and quark dressing functions corresponding to 
the unquenched and two representative quenched cases in table \ref{unqtable}. 
We find different anomalous dimensions in the ultraviolet corresponding due to the 
change from $N_f=0$ to $N_f=3$, {\it c.f.} eqs.~(\ref{Z_UV}), (\ref{chiral-M_UV}), (\ref{anom_dim}).
As expected from the infrared analysis in subsection
\ref{ir-quark} the back-reaction of the quark-loop in 
the gluon equation does not affect the infrared behaviour of the ghost and gluon dressing functions.
Consequently the infrared fixed point of the running coupling is the same as in pure Yang-Mills 
theory. Thus the Kugo-Ojima confinement criterion and 
Zwanziger's horizon condition, {\it c.f.} section \ref{conf}, are satisfied in quenched 
and unquenched Landau gauge QCD.

Our results for the case of explicitly broken chiral symmetry are shown in Fig.~\ref{pic2.dat}.
We choose $N_f=3$ with renormalised quark masses corresponding to 
$m_{u/d}(2 \,\mbox{GeV})=(4.4 - 4.6)$ MeV and $m_{s}(2 \,\mbox{GeV})=(90-95)$ MeV.
These masses are well in the range suggested by the Particle Data Group \cite{Hagiwara:2002pw}.
Compared to the chiral case the behaviour of the ghost and gluon dressing
functions hardly changes. For the quark mass function we obtain the irregular asymptotic solution in 
the ultraviolet as expected from the analysis in subsection \ref{quark-UV-analysis}.  

Furthermore we provide fits to our results for the quark propagator employing the
fit functions\footnote{For simplicity we did not include the regular asymptotic term in the fit 
function for the quark mass and we did not care about logarithmic corrections to the ultraviolet 
behaviour of the wave function renormalisation. Nevertheless both fit functions work very well.}
\setlength{\jot}{3mm}
\beqa
M(x) &=& \frac{1}{g_1+(x/\Lambda^2_{QCD})^{g_2}} \Bigg(g_1 \, M(0) + \nonumber\\
&& \left.\hspace*{2cm}\hat{m} \left[\frac{2}{\ln(x/\Lambda^2_{QCD})}-
\frac{2}{(x/\Lambda^2_{QCD})-1}\right]^{\gamma_m} 
(x/\Lambda^2_{QCD})^{g_2} \right)\,, \hspace*{1cm}
\\
\left[A(x)\right]^{-1} &=& \frac{\left[A(0)\right]^{-1} + h_1\,(x/\Lambda^2_{QCD}) 
+ h_2 \,(x/\Lambda^2_{QCD})^2}{1 + h_3\,(x/\Lambda^2_{QCD}) + h_4 \,(x/\Lambda^2_{QCD})^2}\,, 
\eeqa
with $x=p^2$ and the six parameters $g_1,g_2,h_1,h_2,h_3,h_4$. We used the renormalisation point independent current-quark
mass $\hat{m}$, which is related to the renormalised mass $M(s)$ by
\beq
\hat{m}=M(s)\, \left(\frac{1}{2}\ln[s/\Lambda^2_{QCD}]\right)^{\gamma_m}\,,
\eeq
to one loop order. For the running coupling, the ghost and the gluon dressing function we use the 
form 'Fit B', given in eq.~(\ref{fitB}) and the fit functions from eqs.~(\ref{zg-fit}). 
In table \ref{fit-table} we give our values for all parameters as well as the numerical results for
$M(0)$ and $[A(0)]^{-1}$. Note that the scale $\Lambda_{QCD}^{MOM}$
is different to the corresponding scale in the chiral limit due to the different ultraviolet behaviour
of the quark-loop when quarks with non-vanishing bare masses are employed. When plotted the fits are
virtually indistinguishable from our results in Fig.~\ref{pic2.dat}.
\setlength{\jot}{0mm}

\begin{table}[t]
\begin{tabular}{c||c|c|c|c|c||c|c|c|c|c}
&$\Lambda^{MOM}_{QCD}$&a&b&c&d&$m_R$&$\hat{m}$&M(0) &$g_1$&$g_2$ \\
&[MeV]                & & & & &[MeV]&[MeV]    &[MeV]& &  \\\hline
\mbox{CP d=0}  &600&1.35&1.04&1.36&1.81&4.4&4.6&354&2.30&1.29\\
               &   &    &    &    &    &90 &96 &507&3.42&1.09\\
\mbox{CP d=0.1}&620&1.32&1.06&1.42&1.78&4.5&4.6&429&2.22&1.28\\
               &   &    &    &    &    &95 &96 &564&3.80&1.14\\
\end{tabular}

\vspace*{1cm}
\begin{tabular}{c||c|c|c|c|c|c}
&$m_R$&$\left[A(0)\right]^{-1}$&$h_1$&$h_2$&$h_3$&$h_4$\\
&[MeV]&    &     &&     &     \\\hline
\mbox{CP d=0}  &4.4&0.638&0.503&5.263$\times 10^{-3}$&0.544&5.202$\times 10^{-3}$\\
               &90 &0.671&0.305&1.420$\times 10^{-3}$&0.321&1.402$\times 10^{-3}$\\
\mbox{CP d=0.1}&4.5&0.538&0.448&2.740$\times 10^{-3}$&0.477&2.706$\times 10^{-3}$\\
               &95 &0.590&0.303&1.051$\times 10^{-3}$&0.317&1.037$\times 10^{-3}$\\
\end{tabular}
\caption{\sf \label{fit-table} Parameters for the fits to the unquenched results with $N_f=3$, 
$\alpha(s)=0.2$, $\delta=-0.25$, $\gamma_m=12/27$ and $\beta_0=27/3$.}
\end{table}
\begin{figure}[th]
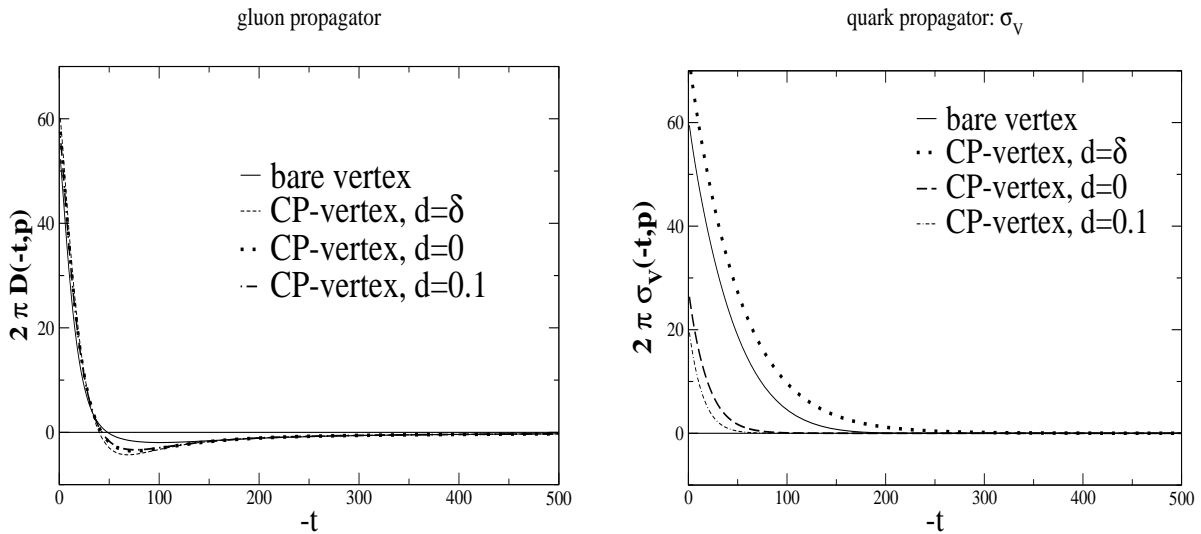

\vspace{0.5cm}
\centerline{
\epsfig{file=pos.glue.eps,width=7.5cm,height=7cm}
\hspace{0.5cm}
\epsfig{file=pos.quarkA.eps,width=7.5cm,height=7cm}
}
\caption{\sf \label{pic3.dat} Here we display the one dimensional Fourier transforms of the
gluon propagator, $D(-t,\vec{p}^2)$, and the vector part of the quark propagator, 
$\sigma_V(-t,\vec{p}^2)$.
We observe violation of reflection positivity for the gluon propagator but not for the quark propagator. 
}
\end{figure}

Unquenched lattice calculations employing dynamical quarks are complex and time consuming \cite{Aoki:2000kp}.
To our knowledge such simulations for the propagators of QCD have not yet been performed.
From our results in the Dyson-Schwinger approach we do not expect drastic differences between
quenched and unquenched propagators on the lattice. 

Finally, we investigate positivity violations in the gluon and quark propagators. Recall from our discussion
in subsection \ref{positivity} that the condition 
\beq
\int_0^\infty dt \; dt^\prime \; \bar{f}(t^\prime,\vec{p}) \; S(-(t+t^\prime),\vec{p}) \; \label{pos-res}
 f(t,\vec{p}) \; < 0 
\eeq
is sufficient for a given propagator $S$ to violate the reflection positivity axiom
of Euclidean quantum field theory. Here $f$ are complex valued test functions, {\it c.f.} subsection 
\ref{positivity}. The one-dimensional Fourier transform $S(t,\vec{p})$ of the propagator $S(p_0,\vec{p})$ 
is given by
\beq
S(t,\vec{p}) := \int \frac{dp_0}{2 \pi} \,S(p_0,\vec{p}) \,e^{i p_0 t}.
\eeq
Provided there is a region around $t_0$ where $S(-t_0,\vec{p}) <0$ one can choose a
real test function $f(t)$ which peaks strongly at $t_0$ to show positivity violation.
In Fig.~\ref{pic3.dat} we display the Fourier transform of the nontrivial part $D(p^2)=Z(p^2)/p^2$
of the gluon propagator and the vector part $\sigma_V(p^2)=A(p^2)/(p^2 A(p^2) + B(p^2)$ of the
quark propagator (we chose $\vec{p}^2=0$). Clearly the Fourier transform of the gluon propagator is negative
on a large interval. The resulting positivity violation for the transverse gluon propagator in Landau gauge
can be interpreted as a signal for gluon confinement, {\it c.f.} subsection \ref{positivity}.  
Note that the ghost propagator violates positivity trivially for $G(p^2)>0$ as can be seen 
from the definition in eq.~(\ref{Gh-prop})\footnote{Already the bare ghost propagator violates
positivity. This is not surprising due to the unphysical spin-statistic relation of the ghost field.}. 
From the Fourier transforms of $\sigma_V(p^2)$ displayed in the second diagram of Fig.~\ref{pic3.dat} 
we cannot conclude positivity violation for the quark propagator. As we have an indefinite metric in the
unphysical part of the state space of QCD this result is not in contradiction with quark confinement.
In addition we have to keep in mind that eq.~(\ref{pos-res}) is a sufficient 
but not a necessary condition for the violation of positivity. Further investigations of the quark propagator
will be done in future work.

\section{Summary}

In this chapter we have presented solutions of the (truncated) Dyson-Schwinger equations for the propagators of 
Landau gauge QCD. We first concentrated on the Dyson-Schwinger equation for the quark
propagator. We proposed several {\it ans\"atze} for the quark-gluon vertex which consist of an Abelian
part carrying the tensor structure of the vertex and a non-Abelian multiplicative correction. Our guiding
principles for the construction of these vertices have been two important conditions
on the truncated quark equation: it should be multiplicatively renormalisable and recover perturbation
theory for large external momenta. We showed that the resulting quark mass function is independent of the
renormalisation point and has the correct asymptotic behaviour for large momenta. 

In the quark equation both the ghost and gluon dressing function show up at least implicitly. 
In the quenched approximation we employ our solutions
of the ghost and gluon Dyson-Schwinger equations in the truncation scheme of chapter \ref{YM}. In the unquenched 
calculations we include the back-reaction of the quarks on the ghost and gluon system and solve
the quark, gluon and ghost Dyson-Schwinger equations self-consistently.

All our solutions show dynamical chiral symmetry breaking. However, only carefully constructed 
vertex {\it ans\"atze} have been able to generate masses in the typical phenomenological range 
of $300-400$ MeV. Constructions with an Abelian part satisfying the Abelian Ward-identity are superior
to other vertex {\it ans\"atze}. We obtained very good results for the quark mass, the pion 
decay constant and the chiral condensate by employing a generalised Curtis-Pennington vertex. 
In the chiral limit both, the quark mass function and the vector self energy are close to recently 
obtained lattice results. This agreement confirms the quality of our truncation and in turn
it shows that chiral extrapolation on the lattice works well. 

In the unquenched case including the quark-loop in the gluon equation with $N_f=3$ light quarks 
we obtain only small corrections compared to the quenched calculations. 
In particular the quark-loop turns out to be suppressed in the gluon equation for small momenta. 
We thus showed on the level of our truncation that the Kugo-Ojima confinement criterion and 
Zwanziger's horizon condition are satisfied in Landau gauge QCD. 

Furthermore we searched for positivity
violations in the gluon and quark propagators. We confirmed previous findings that the gluon
propagator shows violation of reflection positivity. Thus the gluon is not contained
in the physical state space of QCD. We did not find similar violations for the quark propagator.
It is an open question whether these violations exist and if not, how else confinement shows up
in the quark propagator.

\chapter{Conclusions and outlook}

The central objects of interest in this thesis have been the
two-point Green's functions of QCD, {\it i.e.} the ghost, gluon and
quark propagators. We have presented solutions for the corresponding set
of coupled Dyson-Schwinger equations, employing {\it ans\"atze} for the
ghost-gluon vertex, the three-gluon vertex and the quark-gluon vertex.
These {\it ans\"atze} have been constructed such that important constraints 
from general principles are satisfied: both, the running coupling and 
the quark mass function are independent of the renormalisation point. 
Furthermore, we obtained the correct one-loop anomalous dimensions for 
all propagators.

We have been able to achieve considerable progress by overcoming the angular 
approximations used in previous studies. These approximations have proven
to be quantitatively unreliable for small momenta. Our solutions corroborate 
previous findings of a vanishing gluon propagator in the infrared,
whereas the ghost propagator is highly singular at $p^2=0$. In the momentum 
range covered by lattice simulations the lattice data and our results agree 
very well even on a quantitative level. Differences 
occur for intermediate momenta where the Dyson-Schwinger equations suffer 
from neglecting genuine two-loop diagrams. An important aim in future studies 
is the inclusion of these diagrams.

For the running coupling we obtain a fixed point in the infrared. Together, an 
infrared finite coupling and an infrared vanishing gluon propagator disprove the
old idea of {\it infrared slavery}. Instead the ghost propagator turns out to be
the dominant degree of freedom in the infrared, at least in Landau gauge.
This is in accordance with Zwanziger's horizon condition and the Kugo-Ojima 
confinement scenario. A strongly infrared diverging ghost propagator signals 
an unbroken global colour charge. In the Kugo-Ojima scenario this charge is used
to show that the physical states of QCD are colourless. However, as two other 
central assumptions of the scenario, the existence of a well defined BRS charge 
and the violation of the cluster decomposition property, are currently unproven
a complete verification of the Kugo-Ojima scenario is still lacking.

First steps towards a solution of the Dyson-Schwinger equations for the 
ghost and gluon propagators in 
general gauges have been performed in this thesis. We investigated the infrared behaviour
of the ghost and gluon dressing functions in these gauges employing a bare vertex
truncation. In all linear covariant gauges we find identical results to Landau gauge.
For general ghost-antighost symmetric gauges, however, we do not find power solutions
in the infrared when bare vertices are employed. 

Another piece of progress achieved in this thesis
is the solution of Dyson-Schwinger equations on a torus, {\it i.e.} for 
periodic boundary conditions for the fields. 
For various truncation schemes these solutions suffer only mildly from
finite volume effects in the infrared and are very close to the continuum 
results for intermediate and large momenta. From a numerical point of view
we found it to be easier to obtain solutions on a torus than for continuous 
momenta. The reason is the property of the finite volume torus to act as
regulator in the infrared. Thus one avoids all problems with infrared singularities 
encountered in the continuum formulation of the Dyson-Schwinger equations. 
Our treatment of the Dyson-Schwinger equations 
on a torus is an ideal starting point for further investigations, provided the role of the 
zero modes can be clarified further. {\it First}, we are in a position to study
finite volume effects in more detail by comparing solutions on different volumes to
the continuum results. {\it Second}, by varying the extension
of one direction in space-time the inclusion of finite temperature effects
might be accomplished in a relatively easy way. Here the main qualitative question arises 
about the fate of the Kugo--Ojima confinement criterion at the 
deconfinement transition. {\it Finally}, changing from periodic to twisted
boundary conditions we hope to be able to include topological effects in the
Dyson-Schwinger equation approach. 

The quark propagator is the basic input in many phenomenological models
which describe mesons and baryons as bound states of quarks and gluons.
In this context the Dyson-Schwinger equation for the quark propagator is
{\it the} link between underlying QCD and the model frameworks. By solving 
the unquenched set of quark, gluon and ghost Dyson-Schwinger equations
in different truncation schemes 
we have provided an important step in connecting these models with
the fundamental theory. We obtained dynamical chiral symmetry breaking
in the quark equation and found masses in the order of phenomenological values. 
Similar to previous studies in quenched QED we had to build the 
quark-gluon vertex carefully along general principles to obtain satisfying results.    
A central aim in future work will be to extend these calculations to the
corresponding Bethe-Salpeter equation for mesons and thus make further 
contact with experiment.

\noindent
{\bf \Large Acknowledgements}
\thispagestyle{empty}

\vspace*{1cm}
With pleasure I take the opportunity to thank the people contributing
in essential ways to this thesis.

First of all I am very grateful to Reinhard Alkofer for very useful
advice, stimulating discussions, support and encouragement throughout this work.
Furthermore I would like to thank Hugo Reinhardt for support
and helpful discussions. With pleasure I recall the stimulating 
atmosphere in his group. I am grateful to Kurt Langfeld
for communicating and elucidating his lattice results, Jacques Bloch for
useful hints concerning the numerics, Lorenz von Smekal and Peter Watson
for valuable discussions, and Sebastian Schmidt for useful advice,
encouragement and support. I am indebted to Oliver Schr\"oder for 
critical reading of the manuscript.

Furthermore I am indepted to Peter Tandy and Daniel Zwanziger for their warm hospitality and
helpful discussions.

Finally I would like to thank my friends and family and especially my wife 
Elisabeth.
 
This work has been supported by the DAAD and the DFG under contracts Al 279/3-3, 
Al 279/3-4, Re 856-4/1 and GRK683 (European graduate school Basel--T\"ubingen).


\appendix
\chapter{Notations, conventions and decompositions}
\section{Euclidean space conventions \label{gamma}}

In Euclidean space-time we use the metric
$g_{\mu \nu} = \delta_{\mu \nu}$
and Hermitian Dirac matrices $\gamma_\mu$ related to the standard Minkowski ones \cite{Itzykson:1980rh}
by
\beqa
\gamma^j &=& -i \gamma^j_M\,, \hspace{1cm} j=1..3\,, \\
\gamma^4 &=& \gamma^0_M \,.
\eeqa
The Dirac matrices satisfy the Clifford algebra
\beq
\{\gamma_{\mu},\gamma_{\nu} \} = 2 \delta_{\mu \nu}.
\eeq
 
\section{Perturbative running coupling \label{alpha}}

The non-perturbative definition of the running coupling in Landau gauge 
is given by ({\it c.f.} eq.~(\ref{coupling}) in section \ref{coupling-sec}):
\beq
\alpha(p^2) = \alpha(\mu^2) \,Z(p^2,\mu^2)\, G^2(p^2,\mu^2). 
\eeq
Here $Z$ is the dressing function of the full gluon propagator, $G$ the
corresponding dressing function for the ghost propagator and $\mu^2$ is the squared
renormalisation point. Since we have the renormalisation condition
\beq
Z(\mu^2,\mu^2)\, G^2(\mu^2,\mu^2)=1 \,,
\eeq
the momentum dependence of the running coupling
is the same as the dependence of the renormalised coupling $\alpha(\mu^2)=g^2(\mu^2)/4\pi$
on the renormalisation point $\mu^2$. 

For large renormalisation points $\mu^2$
the behaviour of $\alpha(\mu^2)$ can be calculated from perturbation theory.
To three-loop order the Particle Data Group gives the expression \cite{Hagiwara:2002pw}
\setlength{\jot}{2mm}
\beqa
\alpha(\mu^2) &=& \frac{4 \pi}{\beta_0 \ln(\mu^2/\Lambda_{QCD}^2)}\left[
1- \frac{2\beta_1}{\beta_0^2} \frac{\ln\left[\ln(\mu^2/\Lambda_{QCD}^2)\right]}{\ln(\mu^2/\Lambda_{QCD}^2)}
+\frac{4\beta_1^2}{\beta_0^4\ln^2(\mu^2/\Lambda_{QCD}^2)} \right.\nonumber\\
&&\hspace*{3cm} \left. \times \left(\left(\ln\left[\ln(\mu^2/\Lambda_{QCD}^2)\right] -\frac{1}{2} \right)^2
+\frac{\beta_0\beta_2}{8\beta_1^2} - \frac{5}{4}\right)\right],
\eeqa
in the $\bar{\mbox{MS}}$ renormalisation scheme. The coefficients of the $\beta$-function are defined by
\beqa
\mu \frac{\partial \alpha}{\partial \mu} &=& 2 \beta(\alpha) = -\frac{\beta_0}{2\pi}
\alpha^2 - \frac{\beta_1}{4 \pi^2} \alpha^3 
- \frac{\beta_2}{64 \pi^3} \alpha^4 - \ldots, \label{beta-def}\\
\beta_0 &=& 11 - \frac{2}{3}N_f \,, \\
\beta_1 &=& 51 - \frac{19}{3}N_f \,, \\
\beta_2 &=& 2857 - \frac{5033}{9}N_f + \frac{325}{27}N_f^2 \,,
\eeqa
where $N_c$ is the number of colours and $N_f$ is the number of flavours.

\section{Definitions and decompositions of correlation functions
 \label{Def-app}}

In this section we give definitions and conventions for some correlation functions
needed in this thesis. Further definitions are given at the appropriate places in the main 
body of this thesis.

\subsection{Ghost, gluon and quark propagators}

The full ghost, gluon and quark propagators in coordinate space are defined as
\beqa
\langle \bar{c}^a(x) {c}^b(y) \rangle &=& 
\frac{\delta^2 W}{\delta \sigma^a(x) \delta \bar{\sigma}^b(y)}
=D_G^{ab}(x-y), \\
\langle A_\mu^a(x) A_\nu^b(y) \rangle &=& 
\frac{\delta^2 W}{\delta J_\mu^a(x) \delta J_\nu^b(y)}
=D_{\mu \nu}^{ab}(x-y), \\
\langle \bar{\Psi}^a(x) {\Psi}^b(y) \rangle &=&
\frac{\delta^2 W}{\delta \eta^a(x) \delta \bar{\eta}^b(y)}
=S^{ab}(x-y). 
\eeqa
The inverse bare propagators in coordinate space are easily derived from the quadratic part of the 
action $\cS=\int dz \cL$ with the Lagrangian $\cL$ given in eq.~(\ref{Lagrangian}). One obtains 
\beqa
\left[D_G^{(0) ab}(x-y)\right]^{-1} &=& \frac{\delta^2 \cS}{\delta \bar{c}^a(x) c^b(y)} 
 \hspace*{3mm}=\delta^{ab} \partial^2  \delta(x-y) \,, \\
\left[D_{\mu \nu}^{(0) ab}(x-y)\right]^{-1} &=&\frac{\delta^2 \cS}{\delta A_\mu^a(x) A_\nu^b(y)} 
= \delta^{ab}\left(- \partial^2 \delta_{\mu \nu} + \left(1-\frac{1}{\lambda}\right)\partial_\mu 
\partial_\nu \right) \, \delta(x-y)\,, \hspace*{1cm} \\
\left[S^{(0) ab}(x-y) \right]^{-1} &=& \frac{\delta^2 \cS}{\delta \bar{\Psi}^a(x) \Psi^b(y)} = 
\delta^{ab} (\dslash^x+m_0) \, \delta(x-y) \,,  
\eeqa
with the gauge parameter $\lambda$ and the bare quark mass $m_0$.
After Fourier transformation the corresponding expressions in momentum space are given by
\beqa
\left[D_G^{(0) ab}(p)\right]^{-1} &=& -\delta^{ab} p^2 \,, \\
\left[D_{\mu \nu}^{(0) ab}(p)\right]^{-1} &=& \left(\delta_{\mu \nu} 
     - \left(1-\frac{1}{\lambda}\right) \frac{p_\mu p_\nu}{p^2}\right)p^2 \,, \\
\left[S^{(0) ab}(p) \right]^{-1} &=& \delta^{ab} \left(-i \pslash + m_0 \right) \,. 
\eeqa

\subsection{The ghost-gluon vertex}

\begin{figure}[t]
\centerline{
\epsfig{file=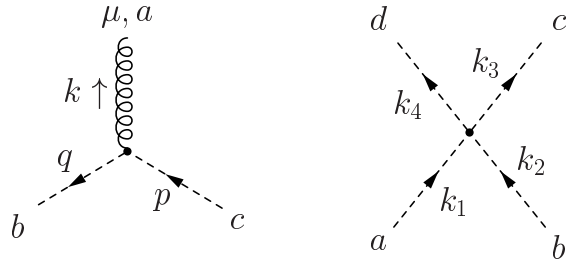,width=7.5cm}
}
\caption{ \sf \label{TL-vertices}
Momentum routing for the tree level ghost-gluon and four-ghost vertices.}
\end{figure}

The tree level ghost-gluon vertex, $\Gamma_\mu^{(0) abc}$, is derived from the ghost-gluon 
part of the action,
\beq
\cS_{ghost-gluon} =  \int d^4x^\prime \left\{ -i\left(1-\frac{\alpha}{2} \right) g f^{abc} 
\left(\partial^\mu \bar{c}^a \right) A^{c}_\mu c^b
+ i\frac{\alpha}{2} g f^{abc} \bar{c}^a A_\mu^c \partial^\mu c^b \right\} \,.
\eeq
For general values of the gauge parameters $\alpha$ and $\lambda$ the vertex is given by
\beqa
\Gamma_\mu^{(0) abc}(x,y,z) &=& \frac{\delta^3 \cS_{ghost-gluon}}{\delta A_\mu^a(x) \delta \bar{c}^b(y) \delta c^c(z)} \nonumber\\
&&\hspace*{-1cm}= -g f^{abc} \left[ i\left(1-\frac{\alpha}{2}\right) \left(\partial_\mu^z \delta^4(z-y) \right) \delta^4(z-x)
+ i\frac{\alpha}{2} \partial_\mu^z\left(\delta^4(z-y) \delta^4(z-x) \right)  \right]. \nonumber\\
\eeqa
Using the momentum conventions of Fig.~\ref{TL-vertices} the Fourier transformed 
bare ghost-gluon vertex reads 
\beqa
\Gamma_\mu^{(0) abc}(k,p,q) &=& \int d^4 [x y z] \, \Gamma_\mu^{abc}(x,y,z) \, e^{i(k\cdot x+q\cdot y-p\cdot z)} \nonumber\\
&=& g f^{abc}\, (2 \pi)^4 \,\delta^4(k+q-p) \,\left[ \left(1-\frac{\alpha}{2}\right)q_\mu 
+ \frac{\alpha}{2} p_\mu \right] \,,
\eeqa
where the abbreviation $d^4x\, d^4y\, d^4z =: d^4[xyz]$ has been introduced.
Note the symmetry of the vertex in the ghost momenta $p_\mu$ and $q_\mu$ if $\alpha=1$.
For convenience we define a reduced ghost-gluon vertex function $\Gamma_\mu^{(0)}(p,q)$ by
\beqa
\Gamma_\mu^{(0) abc}(k,p,q) &=& g f^{abc} \, (2 \pi)^4 \, \delta^4(k+q-p)\,  \Gamma_\mu^{(0)}(p,q)\,, \nonumber\\
\Gamma_\mu^{(0)}(p,q) &=& \left[ \left(1-\frac{\alpha}{2}\right)q_\mu + \frac{\alpha}{2} p_\mu \right].
\eeqa
Furthermore we will need the full one-particle irreducible ghost-gluon vertex in coordinate space,
given by
\beq
\Gamma_\mu^{abc}(x,y,z) = 
\frac{\delta^3 \Gamma}{
\delta A^a_\mu(x) \delta \bar{c}^b(y) \delta {c}^c(z)}.
\eeq

\subsection{The four-ghost vertex}

The bare four-ghost vertex $\Gamma_{4gh}^{(0) abcd}$ is derived from the four-ghost part of the action,
\beq
\cS_{4gh} =  \int d^4x^\prime \left\{\frac{\alpha}{2} \left(1-\frac{\alpha}{2}\right) \frac{\lambda}{2}
g^2 f^{ace} f^{bde} \bar{c}^a  \bar{c}^b c^c c^d \right\} \,.
\eeq
One obtains
\beqa
\Gamma_{4gh}^{(0) abcd}(x,y,z,w) &=& \frac{\delta^4 \cS_{4gh}}{\delta \bar{c}^a(x) \delta \bar{c}^b(y) 
\delta c^c(z)\delta c^d(w)} \nonumber\\
&=&  \frac{\alpha}{2} \left(1-\frac{\alpha}{2}\right) \lambda 
g^2 f^{abe} f^{cde} \delta^4(x-y) \delta^4(y-z) \delta^4(z-w).
\eeqa
With the momentum conventions given in Fig.~\ref{TL-vertices} the Fourier transformed 
bare four-ghost vertex is given by 
\beqa
\Gamma_{4gh}^{(0) abcd}(k_1,k_2,k_3,k_4) &=& \frac{\alpha}{2} \left(1-\frac{\alpha}{2}\right) \lambda 
g^2 f^{abe} f^{cde} (2\pi)^4 \delta^4(k_1+k_2-k_3-k_4).
\eeqa
Again we define a reduced vertex function $\Gamma_{4gh}^{(0)}$ for convenience:
\beqa
\Gamma_{4gh}^{(0) abcd}(k_1,k_2,k_3,k_4) &=& 
g^2 f^{abe} f^{cde} (2\pi)^4 \delta^4(k_1+k_2-k_3-k_4) \Gamma_{4g}^{(0)}\nonumber\\
\Gamma_{4gh}^{(0)} &=& \frac{\alpha}{2} \left(1-\frac{\alpha}{2}\right) \lambda. \label{barefourghostvertex} 
\eeqa 
The full four-ghost vertex in coordinate space is formally given by
\beq
\Gamma^{abcd}(x,y,z) = 
\frac{\delta^4 \Gamma}{\delta \bar{c}^a(x) \delta \bar{c}^b(y) \delta {c}^c(z) \delta {c}^d(y)}.
\eeq

\subsection{The other vertices \label{other-vert}}

The quark-gluon vertex, the three-gluon vertex and the four-gluon vertex do not show up
in the derivation of the ghost Dyson-Schwinger equation in appendix \ref{qcd-appendix}.
We therefore refrain from giving the expressions for these vertices in coordinate space
but merely state the corresponding expressions for the bare vertices in momentum space.
All vertices given in this subsection are reduced vertices {\it i.e.} we have suppressed a factor $(2\pi)^4$ and a
delta function for momentum conservation.

The bare quark-gluon vertex in momentum space is given by
\beq
\Gamma^{(0) a}_\mu(k,q) = i g t^a \gamma_\mu \,,
\eeq
where $t^a$ is a generator of $SU(N_c)$ gauge transformations.

For the bare three-gluon vertex in momentum space one finds 
\beq
\Gamma^{(0) abc}_{\mu \nu \rho}(p,q,k) = g f^{abc} \left( \delta_{\mu \nu} \, (p-q)_\rho + \delta_{\nu \rho} \, (q-k)_\mu
+ \delta_{\rho \mu} \, (k-p)_\nu \right)  \,, \label{three-gluon-app}
\eeq
where $f^{abc}$ is the structure constant of the gauge group.
The three-gluon vertex is completely symmetric with respect to the three gluon momenta. 

The four-gluon vertex is given by
\beqa
\Gamma^{(0) abcd}_{\mu \nu \alpha \beta}(k_1,k_2,k_3,k_4) &=& g^2 \left\{ 
 f^{eab}f^{ecd} [\delta_{\mu \alpha} \, \delta_{\nu \beta} - \delta_{\mu \beta} \, \delta_{\nu \alpha}]
+f^{ebc}f^{ead} [\delta_{\mu \nu} \, \delta_{\alpha \beta} - \delta_{\mu \alpha} \, \delta_{\nu \beta}] \right.\nonumber\\
&&\left.
\hspace*{3mm}+f^{ebd}f^{eac} [\delta_{\mu \nu} \, \delta_{\alpha \beta} - \delta_{\nu \alpha} \, \delta_{\mu \beta}]
\right\} \,,
\eeqa
and is symmetric in the Lorentz indices of the four attached gluons. 

\subsection{Decomposition of the connected ghost-gluon correlation function}

For the derivation of the ghost Dyson-Schwinger equation from the generalised Lagrangian
(\ref{Lagrangian-gen}) performed in appendix \ref{qcd-appendix} we will need some decompositions
of connected correlation functions. These decompositions are derived in this and the next subsection.
We will use the generating functional of connected Green's functions, $W$, and the effective action, $\Gamma$,
see appendix \ref{qcd-appendix} for details. Furthermore we will exploit the fact that the fields and sources
can be written as functional derivatives of $W$ and $\Gamma$:
\beqa
\frac{\delta W}{\delta \sigma} &=& \bar{c}, \hspace*{1cm}\frac{\delta W}{\delta \bar{\sigma}} = c, 
\hspace*{1cm}\frac{\delta W}{\delta J_\mu} = A_\mu, \nonumber\\
\frac{\delta \Gamma}{\delta c} &=& \bar{\sigma}, \hspace*{1cm}\frac{\delta \Gamma}{\delta \bar{c}} = \bar{\sigma}, 
\hspace*{1cm}\frac{\delta \Gamma}{\delta A_\mu} = J_\mu. 
\eeqa
The sign conventions have been chosen such that
derivatives with respect to $\bar{c}$ and $\bar{\sigma}$ are left derivatives whereas
the ones with respect to ${c}$ and ${\sigma}$ are right derivatives.
 
With the help of the matrix relation
\beq
\frac{\delta \chi^{-1}}{\delta \phi}  = - \chi^{-1} \frac{\delta \chi}{\delta \phi} \chi^{-1},
\eeq
and the identity
\beqa
\delta(y-x)\delta^{ab}  
=\int d^4z \frac{\delta \bar{\sigma}^b(y)}{\delta \bar{c}^d(z)}\frac{\delta \bar{c}^d(z)}{\delta \bar{\sigma}^a(x)}
&=&\int d^4z \frac{\delta^2 \Gamma}{\delta \bar{c}^d(z) \delta c^b(y)} \frac{\delta^2 W}
{\delta \bar{\sigma}^a(x) \delta \sigma^d(z)} \,, 
\eeqa
we decompose the connected ghost-gluon correlation function, 
$\langle A^a_\mu(x) \bar{c}^b(y) c^c(z) \rangle$, in the following way:
\beqa
\langle A^a_\mu(x) \bar{c}^b(y) c^c(z) \rangle &=& 
\frac{\delta^3 W}{\delta J^a_\mu(x) \delta \bar{\sigma}^b(y) \delta {\sigma}^c(z)} \nonumber\\
&&\hspace*{-3cm}= \frac{\delta}{\delta J^a_\mu(x)} \left[ 
\frac{\delta^2 \Gamma}{\delta \bar{c}^b(y) \delta {c}^c(z)}\right]^{-1} \nonumber\\ 
&&\hspace*{-3cm}= 
\int d^4u_1 \frac{\delta A^d_\nu(u_1)}{\delta J^a_\mu(x)}\frac{\delta}{\delta A^d_\nu(u_1)}
 \left[ \frac{\delta^2 \Gamma}{\delta \bar{c}^b(y) \delta {c}^c(z)}\right]^{-1} \nonumber\\
&&\hspace*{-3cm}= \int d^4[u_1u_2u_3] \frac{\delta^2 W}{\delta J^a_\mu(x) \delta J^d_\nu(u_1)} \:
\frac{\delta^2 W}{\delta \bar{\sigma}^b(y) \delta {\sigma}^e(u_2)} \: \frac{\delta^3 \Gamma}{
\delta A^d_\nu(u_1) \delta \bar{c}^e(u_2) \delta {c}^f(u_3)} \:
\frac{\delta^2 W}{\delta \bar{\sigma}^f(u_3) \delta {\sigma}^c(z)}\nonumber\\ 
&&\hspace*{-3cm}=\int d^4 [u_1u_2u_3] D_{\mu \nu}^{ad}(x-u_1) D_G^{eb}(u_2-y) 
\Gamma^{def}_\nu(u_1,u_2,u_3)D_G^{cf}(u_3-z).
 \label{ghgldecomp}
\eeqa
Here we used the abbreviation $d^4[u_1u_2u_3]:=d^4u_1\, d^4 u_2 \, d^4u_3$ and the definitions of the
gluon propagator $D_{\mu \nu}$, the ghost propagator $D_G$ and the ghost-gluon vertex $\Gamma_\nu$
given in previous subsections.

\subsection{Decomposition of connected four-ghost correlation function}

Furthermore we need the decomposition of the four-ghost correlation function into
one-particle irreducible parts. We start at a stage where the sources are still present and set them
to zero at the end of the derivation.
We first give the decomposition of the connected ghost-antighost-ghost
three-point function
\beqa
\langle \bar{c}^b(y) {c}^c(z) \bar{c}^d(w) \rangle &=& 
\frac{\delta^3 W}{\delta \bar{\sigma}^b(y) \delta \sigma^c(z) \delta \bar{\sigma}^d(w)} \nonumber\\
&&\hspace*{-3cm}= \frac{\delta}{\delta \bar{\sigma}^b(y)} \left[ \frac{\delta^2 \Gamma}
{\delta \sigma^c(z) \delta \bar{\sigma}^d(w)}\right]^{-1} \nonumber\\
&&\hspace*{-3cm}=\int d^4u_1 \frac{\delta A^e_\nu(u_1)}{\delta \bar{\sigma}^b(y)}\frac{\delta}{\delta A^e_\nu(u_1)}
 \left[ \frac{\delta^2 \Gamma}{\delta \sigma^c(z) \delta \bar{\sigma}^d(w)}\right]^{-1} \nonumber\\
&&\hspace*{-3cm}= \int d^4[u_1u_2u_3] \frac{\delta^2 W}{\delta \bar{\sigma}^b(y) \delta J^e_\nu(u_1)} \:
\frac{\delta^2 W}{\delta \sigma^c(z) \delta \bar{\sigma}^f(u_2)} \: \frac{\delta^3 \Gamma}{
\delta A^e_\nu(u_1) \delta {c}^f(u_2) \delta \bar{c}^g(u_3)} \:
\frac{\delta^2 W}{\delta \sigma^g(u_3) \delta \bar{\sigma}^d(w)} \,. \nonumber\\
\eeqa
Then we decompose the connected four-ghost Green's function:
\beqa
\langle {c}^a(x) \bar{c}^b(y) {c}^c(z) \bar{c}^d(w) \rangle &=& \frac{\delta^4 W}
{\delta \sigma^a(x) \delta \bar{\sigma}^b(y) \delta \sigma^c(z) \delta \bar{\sigma}^d(w)} \nonumber\\
&=& \frac{\delta}{\delta \sigma^a(x)}
\int d^4[u_1u_2u_3] \frac{\delta^2 W}{\delta \bar{\sigma}^b(y) \delta J^e_\mu(u_1)} \:
\frac{\delta^2 W}{\delta \sigma^c(z) \delta \bar{\sigma}^f(u_2)} \nonumber\\
&& \hspace*{1.5cm}\times\frac{\delta^3 \Gamma}{
\delta A^e_\mu(u_1) \delta {c}^f(u_2) \delta \bar{c}^g(u_3)} \:
\frac{\delta^2 W}{\delta \sigma^g(u_3) \delta \bar{\sigma}^d(w)}.\hspace*{1cm}
\eeqa
Carrying out the remaining derivative gives four terms. The two terms where the derivative acts on the second and
on the last propagator vanish, because the term $\frac{\delta W}{\delta \bar{\sigma}^b(y) \delta J^e_\nu(u_1)}$
vanishes when the sources are set to zero. The contribution where the derivative acts on the first propagator
can be treated using eq.~(\ref{ghgldecomp}). In the expression  with the derivative acting on the vertex we
use
\beqa
-\frac{\delta^2 W}{\delta \bar{\sigma}^b(y) \delta J^e_\mu(u_1)} 
\frac{\delta^4 \Gamma}{\delta \sigma^a(x) \delta A^e_\mu(u_1) \delta {c}^f(u_2) \delta \bar{c}^g(u_3)}
&=& \frac{\delta^4 \Gamma}{\delta \sigma^a(x) \delta \bar{\sigma}^b(y) \delta {c}^f(u_2) \delta \bar{c}^g(u_3)} 
\nonumber\\
&&\hspace*{-7cm}= \int d^4u_4\frac{\delta^2 W}{\delta \sigma^a(x) \bar{\sigma}^e(u_4)}\:\:
\frac{\delta^4 \Gamma}{\delta c^e(u_4) \delta \bar{\sigma}^b(y) \delta {c}^f(u_2) \delta \bar{c}^g(u_3)}
\nonumber\\
&&\hspace*{-7cm}= \int d^4[u_4u_5]\frac{\delta^2 W}{\delta \sigma^a(x) \bar{\sigma}^e(u_4)}\:\: 
\frac{\delta^2 W}{\delta \bar{\sigma}^b(y) \delta{\sigma}^h(u_5)}\:\:
\frac{\delta^4 \Gamma}{\delta c^e(u_4) \delta \bar{c}^h(u_5) \delta {c}^f(u_2) \delta \bar{c}^g(u_3)}
\nonumber\\
&&\hspace*{-7cm}= -\int d^4[u_4u_5]\frac{\delta^2 W}{\delta \sigma^a(x) \bar{\sigma}^e(u_4)} \:\:
\frac{\delta^2 W}{\delta{\sigma}^h(u_5) \delta \bar{\sigma}^b(y)}\:\:
\frac{\delta^4 \Gamma}{ \delta \bar{c}^g(u_3)\delta \bar{c}^h(u_5)\delta c^e(u_4) \delta {c}^f(u_2)}.
\nonumber\\
\eeqa
Collecting all this together we arrive at
\beqa
\langle {c}^a(x) \bar{c}^b(y) {c}^c(z) \bar{c}^d(w) \rangle &=&
\int d^4[u_1u_2u_3u_4u_5u_6] \left\{ 
\frac{\delta^2 W}{\delta J^e_\mu(u_1)\delta J^f_\nu(u_4)} \:\:
\frac{\delta^2 W}{\delta \sigma^a(x) \delta \bar{\sigma}^g(u_5)} \right. \nonumber\\
&&\hspace*{1cm} \times
\frac{\delta^3 \Gamma}{\delta A^f_\nu(u_4) \delta {c}^g(u_5) \delta \bar{c}^h(u_6)}\:\:
\frac{\delta^2 W}{\delta \sigma^h(u_6) \delta \bar{\sigma}^b(y)}\:\:
\frac{\delta^2 W}{\delta \sigma^c(z) \delta \bar{\sigma}^i(u_2)} \nonumber\\
&&\hspace*{1cm} \left. \times
\frac{\delta^3 \Gamma}{\delta A^e_\mu(u_1) \delta {c}^i(u_2) \delta \bar{c}^j(u_3)}\:\:
\frac{\delta^2 W}{\delta \sigma^j(u_3) \delta \bar{\sigma}^d(w)}\right\}\nonumber\\
&-& \int d^4[u_1u_2u_3u_4u_5] \left\{
\frac{\delta^2 W}{\delta \sigma^a(x) \bar{\sigma}^e(u_4)} \:\:
\frac{\delta^2 W}{\delta{\sigma}^h(u_5) \delta \bar{\sigma}^b(y)} \right.\nonumber\\
&&\hspace*{1cm}\times
\frac{\delta^2 W}{\delta \sigma^c(z) \delta \bar{\sigma}^f(u_2)}\:\:
\frac{\delta^4 \Gamma}{ \delta \bar{c}^g(u_3)\delta \bar{c}^h(u_5)\delta c^e(u_4) \delta {c}^f(u_2)} 
\nonumber\\
&&\hspace*{1cm}\times\left. \frac{\delta^2 W}{\delta \sigma^g(u_3) \delta \bar{\sigma}^d(w)}\right\}.
\eeqa
Interchanging some Grassmann fields in the correlations and using the definitions for the propagators and
vertices given in the previous subsections we arrive at
\beqa
\langle \bar{c}^b(y) \bar{c}^d(w) {c}^a(x) {c}^c(z)  \rangle &=& 
\int d^4[u_1u_2u_3u_4u_5u_6] \left\{ 
D_{\mu \nu}^{ef}(u_1-u_4)  \, D_G^{ag}(x-u_5) \right. \nonumber\\
&&\hspace*{1cm}\times\Gamma^{fhg}_\nu(u_4,u_6,u_5) D_G^{hb}(u_6-y) \, D_G^{ci}(z-u_2) \nonumber\\
&&\hspace*{1cm}\left.\times \Gamma^{eji}_\mu(u_1,u_3,u_2) \, D_G^{jd}(u_3-w) \right\}\nonumber\\
&+& \int d^4[u_1u_2u_3u_4u_5] \left\{
D_G^{ae}(x-u_4) \, D_G^{hb}(u_5-y) \right.\nonumber\\
&&\hspace*{1cm}\left. \times
D_G^{cf}(z-u_2) \, \Gamma_{4gh}^{hgef}(u_5,u_3,u_4,u_2) \, D_G^{gd}(u_3-w) \right\}, \hspace*{1cm}
\eeqa
which is the decomposition of the four-ghost correlation used in appendix \ref{qcd-appendix}.

\chapter{The derivation of the ghost Dyson-Schwinger equation in general
ghost-antighost symmetric gauges \label{qcd-appendix}}

In this appendix we will derive the ghost Dyson-Schwinger equation from the generalised
Lagrangian (\ref{Lagrangian-gen}). As there are no direct couplings between quarks and ghosts contained
in the Lagrangian it is sufficient to employ only the ghost-gluon part. We start by 
transforming the ghost-gluon part of the Lagrangian into a more suitable form using
partial integrations under the assumption of vanishing fields at infinity. We obtain
\beqa
\cal{L} &=& \frac{1}{2} A_\mu^a \left( -\partial^2 \delta_{\mu \nu} + \left(1-\frac{1}{\lambda}\right)
\partial_\mu \partial_\nu \right) A_\nu^a 
- g f^{abc} \left(\partial_\mu A_\nu^a\right) A_\mu^b A_\nu^c \nonumber\\
&&+\frac{g^2}{4} f^{abe} f^{cde} A_\mu^a A_\nu^b A_\mu^c A_\nu^d + \bar{c}^a \partial^2 c^a 
+ \frac{\alpha}{2} \left(1-\frac{\alpha}{2}\right) \frac{\lambda}{2}
g^2 f^{ace} f^{bde} \bar{c}^a  \bar{c}^b c^c c^d \nonumber\\
&&+ i\left(1-\frac{\alpha}{2} \right) g f^{abc}  \bar{c}^a \partial_\mu \left(A_\mu^c c^b\right)
+ i\frac{\alpha}{2} g f^{abc} \bar{c}^a A_\mu^c \partial_\mu c^b.
\eeqa
The partition function of the theory is given by
\beq
Z[J,\sigma,\bar{\sigma}] = \int {\cal D} [A \bar{c} c] \exp\left\{-\int d^4z \, {\cal{L}} + \int d^4z \, 
\left(A^a J^a+\bar{\sigma}c+\bar{c}\sigma \right)\right\}
\eeq
with the sources $J$, $\sigma$ and $\bar{\sigma}$ of the gluon, antighost and ghost fields, respectively.
The action is given by $\cS[J,c,\bar{c}]=\int d^4z\, {\cal{L}}$.
The generating functional of connected Green's functions, $W[J,\sigma,\bar{\sigma}]$,
is defined as the logarithm of the partition function. The functional Legendre transform of $W$ is 
the effective action
\beq
\Gamma[A,\bar{c},c] = -W[J,\sigma,\bar{\sigma}] + \int d^4z \left(A^a J^a+\bar{\sigma}c+\bar{c}\sigma \right) ,
\eeq
which is the generating functional of one-particle irreducible vertex functions.
The fields and sources can be written as functional derivatives of the respective generating functionals
in the following way
\beqa
\frac{\delta W}{\delta \sigma} &=& \bar{c}, \hspace*{1cm}\frac{\delta W}{\delta \bar{\sigma}} = c, 
\hspace*{1cm}\frac{\delta W}{\delta J_\mu} = A_\mu, \nonumber\\
\frac{\delta \Gamma}{\delta c} &=& \bar{\sigma}, \hspace*{1cm}
\frac{\delta \Gamma}{\delta \bar{c}} = \bar{\sigma}, 
\hspace*{1cm}\frac{\delta \Gamma}{\delta A_\mu} = J_\mu .
\label{derivatives}
\eeqa
The sign conventions have been chosen such that
derivatives with respect to $\bar{c}$ and $\bar{\sigma}$ are left derivatives whereas
the ones with respect to ${c}$ and ${\sigma}$ are right derivatives,
\beq
\frac{\delta}{\delta\left( \bar{\sigma},\bar{c}\right)} := \mbox{left derivative} \hspace*{1cm}
\frac{\delta}{\delta\left( {\sigma},{c}\right)} := \mbox{right derivative}.
\eeq

Given that the functional integral is well-defined, the Dyson-Schwinger equation for the ghost propagator
is derived from the observation that the integral of a total derivative vanishes. We take the derivative 
with respect to the antighost field and obtain
\beqa
0 &=& \int {\cal D} [A \bar{c} c] \frac{\delta}{\delta \bar{c}} \exp\left\{-\int d^4z \, {\cal{L}} + \int d^4z 
\, \left(A^a J^a+\bar{\sigma}c+\bar{c}\sigma \right)\right\} \nonumber\\
&=& \int {\cal D} [A \bar{c} c] \left(-\frac{\delta S\left[A,c,\bar{c}\right]}{\delta \bar{c}} + \sigma \right)
 \exp\left\{-\int d^4z\, {\cal{L}} + \int d^4z \, 
\left(A^a J^a+\bar{\sigma}c+\bar{c}\sigma \right)\right\}\nonumber\\
&=& \left(-\frac{\delta S\left[\frac{\delta}{\delta J},\frac{\delta}{\delta \bar{\sigma}}
,\frac{\delta}{\delta \sigma}\right]}{\delta \bar{c}} + \sigma \right)Z[J,\sigma,\bar{\sigma}].
\eeqa
Now we use the relations (\ref{derivatives}) and apply a further functional derivative with
respect to the source $\sigma^b(y)$. We arrive at
\beq
0 = \left(-\frac{\delta S}{\delta \bar{c}^c(z)} \bar{c}^b(y) + \sigma^c(z) \bar{c}^b(y) + \delta(z-y)\delta_{cb}
\right)Z[J,\sigma,\bar{\sigma}]
\eeq
with explicit colour indices and space-time arguments. Setting the sources equal to zero we obtain the
ghost Dyson-Schwinger equation
\beq
\left\langle \frac{\delta S}{\delta \bar{c}^c(z)} \bar{c}^b(y) 
\right\rangle = \delta(z-y) \delta_{cb}.
\label{ghostdse}
\eeq
The derivative is easily calculated
\beqa
\frac{\delta S}{\delta \bar{c}^c(z)} &=&  \partial^2 c^c(z)
+ \frac{\alpha}{2} \left(1-\frac{\alpha}{2}\right) \frac{\lambda}{2}
g^2 f^{cde} f^{fge} \bar{c}^d(z) c^f(z) c^g(z) \nonumber\\
&&+ i\left(1-\frac{\alpha}{2} \right) g f^{cde} \partial_\mu \left(
A^e_\mu(z) c^d(z)\right)
+ i\frac{\alpha}{2} g f^{cde} A^e_\mu(z) \partial_\mu 
c^d(z).
\label{ghostder}
\eeqa
Whereas in the covariant formalism full and connected {\it three-point functions} are the same, 
the {\it four-point correlations} 
have to be decomposed into disconnected and connected parts. For the four-ghost correlation
function this results in
\beqa 
\langle \bar{c}^b(y) \bar{c}^d(z) c^f(z) c^g(z)\rangle &=& \langle \bar{c}^b(y)c^g(z)\rangle \, \langle \bar{c}^d(z) c^f(z)\rangle
 - \langle \bar{c}^b(y)c^f(z)\rangle \, \langle \bar{c}^d(z)c^g(z)\rangle \nonumber\\
&+& \langle \bar{c}^b(y) \bar{c}^d(z) c^f(z) c^g(z)\rangle_{conn.} \,.
\eeqa
Keeping in mind the Grassmann nature of the ghost and antighost fields we then obtain
\beqa
-\delta(z-y) \delta_{cb}&=&\partial^2 \langle  \bar{c}^b(y)  c^c(z) \rangle 
+ \frac{\alpha}{2} \left(1-\frac{\alpha}{2}\right) \frac{\lambda}{2}
g^2 f^{cde} f^{fge} \left\{ \langle \bar{c}^b(y) \bar{c}^d(z) c^f(z) c^g(z)\rangle \right. \nonumber\\ 
&&\left.+\left( \langle \bar{c}^b(y)c^g(z)\rangle \, \langle \bar{c}^d(z) c^f(z)\rangle
 - \langle \bar{c}^b(y)c^f(z)\rangle \, \langle \bar{c}^d(z)c^g(z)\rangle \right) \right\}\nonumber\\ 
&+& \left(1-\frac{\alpha}{2} \right) g f^{cde} \langle \bar{c}^b(y)\partial_\mu \left(
A^e_\mu(z) c^d(z)\right)\rangle
+ \frac{\alpha}{2} g f^{cde} \langle \bar{c}^b(y) A^e_\mu(z) \partial_\mu c^d(z) \rangle \,, 
\nonumber\\ \label{DSEa}
\eeqa
where all correlations are connected Green's functions.
We now use the relation
\beqa
\delta(y-x)\delta^{ab} = \frac{\delta \bar{\sigma}^b(y)}{\delta \bar{\sigma}^a(x)}
=\int d^4z \frac{\delta \bar{\sigma}^b(y)}{\delta \bar{c}^d(z)}\frac{\delta \bar{c}^d(z)}{\delta \bar{\sigma}^a(x)}
&=&\int d^4z \frac{\delta^2 \Gamma}{\delta \bar{c}^d(z) \delta c^b(y)} \frac{\delta^2 W}
{\delta \bar{\sigma}^a(x) \delta \sigma^d(z)} \nonumber\\
&=:& \int d^4z \left[D_G^{db}(z-y) \right]^{-1} D_G^{ad}(x-z) \nonumber\\
\eeqa
and multiply eq.~(\ref{DSEa}) 
with $-[D_G^{ac}(x-z)]^{-1}=\left[ \langle \bar{c}^c(z) c^a(x) \rangle \right]^{-1}$. We arrive at 
\beqa
[D_G^{ab}(x-y)]^{-1}
&=&\partial^2 \delta(x-y) \delta^{ab} \nonumber\\ 
&-& \frac{\alpha}{2} \left(1-\frac{\alpha}{2}\right) \frac{\lambda}{2}
g^2 f^{cde} f^{fge} \int d^4z \,[D_G^{ac}(x-z)]^{-1}  
\left\{ \langle \bar{c}^b(y) \bar{c}^d(z) c^f(z) c^g(z)\rangle \right.\nonumber\\
&&\hspace*{2.4cm}+\, \left.\langle \bar{c}^b(y)c^g(z)\rangle\,\langle \bar{c}^d(z) c^f(z)\rangle
 - \langle \bar{c}^b(y)c^f(z)\rangle\, \langle \bar{c}^d(z)c^g(z)\rangle \right\} \nonumber\\ 
&-& i\left(1-\frac{\alpha}{2} \right) g f^{cde} \int d^4z \, [D_G^{ac}(x-z)]^{-1}
\langle \bar{c}^b(y)\partial_\mu \left(
A^e_\mu(z) c^d(z)\right)\rangle \nonumber\\
&-& i\frac{\alpha}{2} g f^{cde} \int d^4z \, [D_G^{ac}(x-z)]^{-1}\langle \bar{c}^b(y) A^e_\mu(z) 
\partial_\mu c^d(z) \rangle. 
\nonumber\\ \label{DSEb}
\eeqa
Before we decompose the connected Green's functions into one particle irreducible ones we
have to take care of the space-time derivatives. Noting that
\beqa
\partial_\mu^z \frac{\delta^2 W}{\delta J^c_\mu(z) \sigma^d(z)} &=& -\int d^4u \partial_\mu^u\left(\delta(u-z)   
\right) \frac{\delta^2 W}{\delta J^c_\mu(u) \sigma^d(u)} \nonumber\\
&=& -\int d^4[uv] \partial_\mu^u\left(\delta(u-z)   
\right) \delta(u-v)\frac{\delta^2 W}{\delta J^c_\mu(v) \sigma^d(u)} 
\eeqa
with the abbreviation $d^4u \, d^4v =: d^4[uv]$, and 
\beqa
\frac{\delta }{\delta J^c_\mu(z)} \partial_\mu^z \frac{\delta W}{\sigma^d(z)}
&=& \int d^4[uv]\, \delta(u-z) \delta(u-v) 
\frac{\delta }{\delta J^c_\mu(v)} \partial_\mu^u \frac{\delta W}{\sigma^d(u)} \nonumber\\
&=& -\int d^4[uv]\, \partial_\mu^u\left(\delta(u-z)   
 \delta(u-v)\right)\frac{\delta^2 W}{\delta J^c_\mu(v) \sigma^d(u)} 
\eeqa
we can replace the derivative terms by the bare ghost-gluon vertex defined in appendix \ref{Def-app}.
The tadpole term can be treated in the following way:
\beqa
\int d^4z [D_G^{ac}(x-z)]^{-1} 
 f^{cde} f^{fge} \left\{\langle \bar{c}^b(y)c^g(z)\rangle\,\langle \bar{c}^d(z) c^f(z)\rangle
 - \langle \bar{c}^b(y)c^f(z)\rangle\,\langle \bar{c}^d(z)c^g(z)\rangle \right\}
 \nonumber\\
&&\hspace*{-14cm} =2 \int d^4z\, [D_G^{ac}(x-z)]^{-1} 
 f^{cde} f^{fge} \left\{\langle \bar{c}^b(y)c^g(z)\rangle\,\langle \bar{c}^d(z) c^f(z)\rangle
 \right\}
 \nonumber\\
&&\hspace*{-14cm} =2 \int d^4[zuv]\, [D_G^{ac}(x-z)]^{-1} \delta(z-u)\,\delta(u-v)\,
 f^{cde} f^{fge} \,D_G^{gb}(z-y) \,D_G^{fd}(v-u)
 \nonumber\\
&&\hspace*{-14cm} =2 \int d^4[uv]\, \delta(x-y)\,\delta(z-u)\,\delta(u-v)\, f^{bde} f^{fae}\, D_G^{fd}(v-u).
\eeqa
Plugging the expressions for the ghost-gluon loop and the one for the tadpole into eq.~(\ref{DSEb}) 
and using the expression for
the bare four-ghost vertex given in appendix \ref{Def-app} we obtain
\beqa
[D_G^{ab}(x-y)]^{-1}
&=&\partial^2 \delta(x-y) \delta^{ab} - \int d^4[uv]\, \Gamma_{4gh}^{(0) bdfa}(x,u,v,y) \, D_G^{fd}(v-u) \nonumber\\
&&+ \frac{\alpha}{2} \left(1-\frac{\alpha}{2}\right) \frac{\lambda}{2}
g^2 f^{cde} f^{fge} \times \nonumber\\
&& \hspace*{0.5cm}\int d^4[zuv] \, \delta(z-u)\, \delta(u-v) \,[D_G^{ac}(x-z)]^{-1}\:  
\langle \bar{c}^b(y) \bar{c}^d(z) c^f(u) c^g(v)\rangle \nonumber\\ 
&&- \int d^4[zuv] \, \Gamma^{(0) cde}_\mu (z,u,v) \, [D_G^{ac}(x-z)]^{-1} \, \langle \bar{c}^b(y)
A^e_\mu(v) c^d(u)\rangle \,.
\nonumber\\ \label{DSEd}
\eeqa
 
To decompose the connected Green's functions into one-particle irreducible ones we use the
relations
\beqa
\langle A^e_\mu(v) \bar{c}^b(y) c^d(u)\rangle &=&
\int d^4 [z_1z_2z_3]\, D_{\mu \nu}^{ef}(v-z_1) \,D_G^{bg}(y-z_2) \, 
\Gamma_\nu^{fhg}(z_1,z_3,z_2) \, D_G^{hd}(u-z_3)\nonumber\\ 
\\
\langle \bar{c}^b(y) \bar{c}^d(z) c^f(u) c^g(v)\rangle  &=& 
\int d^4[u_1u_2u_3u_4u_5u_6]\, \left\{ 
D_{\mu \nu}^{ek}(u_1-u_4) \, D_G^{fl}(u-u_5) \right. \nonumber\\
&&\hspace*{1cm}\times \,
\Gamma^{khl}_\nu(u_4,u_6,u_5) \, D_G^{hb}(u_6-y) \,D_G^{gi}(v-u_2) \nonumber\\
&&\hspace*{1cm}\left. \times \,
\Gamma^{eji}_\mu(u_1,u_3,u_2) \, D_G^{jd}(u_3-z) \right\}\nonumber\\
&-& \int d^4[u_1u_2u_3u_4u_5]\, \left\{
D_G^{fe}(u-u_4) \,D_G^{hb}(u_5-y) \right.\nonumber\\
&&\hspace*{1.0cm}\left. \times \,
D_G^{gi}(v-u_2) \, \Gamma_{4gh}^{jhei}(u_3,u_5,u_4,u_2) \, D_G^{jd}(u_3-z) \right\} \,, \hspace*{1.5cm}
\eeqa
which have been derived in appendix \ref{Def-app}.

Substituting these expressions into eq.~(\ref{DSEd}) we arrive at the final expression for the 
ghost Dyson-Schwinger equation in coordinate space:
\beqa
[D_G^{ab}(x-y)]^{-1}
&=&[D_G^{(0) ab}(x-y)]^{-1} \nonumber\\ 
&-&  \int d^4[uv]\, \Gamma_{4gh}^{(0) bdfa}(x,u,v,y) \, D_G^{fd}(v-u) \nonumber\\
&-& \frac{1}{2} \int d^4[zuvu_1u_2u_3u_4u_5]\, \Gamma^{(0) bdgf}_{4gh}(y,z,v,u)\, 
D_{\mu \nu}^{ek}(u_1-u_4)\, D_G^{fl}(u-u_5) \nonumber\\
&&  \hspace*{1cm} \times\, \Gamma^{kal}_\nu(u_4,x,u_5)\, D_G^{gi}(v-u_2) \,\Gamma^{eij}_\mu(u_1,u_3,u_2)
\, D_G^{jd}(u_3-z) \nonumber\\ 
&-& \frac{1}{2} \int d^4[zuvu_1u_2u_3u_4] \Gamma^{(0) bdgf}_{4gh}(y,z,v,u)\, 
D_G^{fe}(u-u_4)\nonumber\\
&& \hspace*{1cm} \times\,
D_G^{gi}(v-u_2) \,\Gamma_{4gh}^{jaei}(u_3,x,u_4,u_2) \,D_G^{jd}(u_3-z) \nonumber\\ 
&-& \int d^4[zuvz_1z_2z_3]\, \Gamma^{(0) bde}_\mu (y,u,v) \, 
D_{\mu \nu}^{ef}(v-z_1) \,\Gamma_\nu^{fha}(z_1,z_3,x)D_G^{hd}(u-z_3)
\nonumber\\ \label{DSEe}
\eeqa
where an additional minus signs arises from the interchange of the colour indices $f$ and $g$ in the
bare four-ghost vertices and from the interchange of $j$ and $i$ in the ghost-gluon vertex.

After performing a Fourier transformation we obtain the respective expression in momentum space
\beqa
[D_G(p)]^{-1}
&=&[D_G^{(0)}(p)]^{-1} \nonumber\\ 
&+& \left( -N_c \right) \frac{g^2}{(2\pi)^4} \int d^4q \: \Gamma_{4gh}^{(0)} \:  D_G(q) \nonumber\\
&+& \left( \frac{-N_c^2}{2} \right) \frac{1}{2} \frac{g^4}{(2\pi)^8} \int d^4[q_1q_2] \: \Gamma^{(0)}_{4gh}\:  
D_{\mu \nu}(p-q_1) \: D_G(q1) \nonumber\\
&&  \hspace*{1cm} \times \, \Gamma_\nu(p,q_1)\:  D_G(q_2)\: \Gamma_\mu(-p+q_1+q_2,q_2)\: D_G(p-q_1-q_2) \nonumber\\ 
&-& \left( -N_c^2 \right)  \frac{1}{2} \frac{g^4}{(2\pi)^8}\int d^4[q_1q_2] \Gamma^{(0)}_{4gh} \:
D_G(q_1) \:D_G(p-q_1-q_2) \:\Gamma_{4gh}(p,q_1,q_2)\: D_G(q_2) \nonumber\\ 
&+& \left( -N_c \right) \frac{g^2}{(2\pi)^4}\int d^4q \:\Gamma^{(0)}_\mu (p,q) \: 
D_{\mu \nu}(p-q) \:\Gamma_\nu(q,p)D_G(q)
\nonumber\\ \label{DSE-ghost}
\eeqa
where the colour traces have been carried out and the reduced vertices defined in 
appendix \ref{Def-app} have been used.

\chapter{Methods to solve DSEs in flat Euclidean space-time}

\section{Angular integrals \label{angular}}
Working in Euclidean space-time the four dimensional integrals in the loops of Dyson-Schwinger equations 
can be transformed to hyperspherical coordinates as 
\beq
\int d^4q \:\ldots = \frac{1}{2} \int_0^\infty dq^2 q^2 \int_0^{2 \pi} d\Phi 
\int_0^\pi d\Psi \sin(\Psi) \int_0^\pi d\theta
\sin^2(\theta) \:\ldots .
\eeq
We choose the external momentum $p_\mu$ to point in the four-direction,
enclosing the angle $\theta$ with the loop momentum $q_\mu$. All integrands are then functions
of the squared external momentum, the squared loop momentum and the angle $\theta$ only.
The other two angular integrals can then be
performed trivially yielding a factor of $4\pi$. We thus have the relation
\beq
\int d^4q \, f(p^2,q^2,\theta) = 2\pi \int_0^\infty dq^2 q^2 \int_0^\pi d\theta
\sin^2(\theta) \, f(p^2,q^2,\theta),
\eeq
which is used frequently throughout this thesis.

Sometimes the remaining angular integral can be performed as well using the integration formulae
\beqa
\int_0^\pi d\theta \frac{\sin^2(\theta)}{z^2} &=& \frac{\pi}{2} 
       \left[\frac{\Theta(x-y)}{x(x-y)} + \frac{\Theta(y-x)}{y(y-x)} \right]  \label{first}\\
\int_0^\pi d\theta \frac{\sin^2(\theta)}{z} &=& \frac{\pi}{2} 
       \left[\frac{\Theta(x-y)}{x} + \frac{\Theta(y-x)}{y} \right]  \label{second}\\
\int_0^\pi d\theta \sin^2(\theta) &=&  \frac{\pi}{2} \\
\int_0^\pi d\theta \sin^2(\theta) \: z &=&  \frac{\pi}{2} (x+y) \\
\int_0^\pi d\theta \sin^2(\theta) \: z^2 &=&  \frac{\pi}{2} \left((x+y)^2 + xy \right)
\eeqa
where we have used the abbreviations $x:=p^2$, $y:=q^2$ and the squared momentum $z$ is defined as 
$z=(p-q)^2 = x+y-2\sqrt{xy} \cos(\theta)$.
A derivation of eqs.~(\ref{first}), (\ref{second}) can be found {\it e.g.} in \cite{Bloch:2002eq}.      

\section{Tensor integrals \label{tensor}}

The explicit expression for the scalar bubble integral $I$, defined in eq.~(\ref{sc-bubble}), 
can be easily evaluated in Euclidean space-time
using the Feynman-parameterisation (see {\it e.g.} ref.~\cite{Peskin:1995ev}). 
With the squared momenta $x=p^2$, $y=q^2$ and $z=(p-q)^2$ the result is given by
\beqa
I(a,b,p) &:=& \int d^4q \frac{1}{y^a z^b} \label{sc-bubble}\\ 
&=& \pi^2 \,x^{2-a-b}\, \frac{\Gamma(2-a)\,\Gamma(2-b)\,\Gamma(a+b-2)}
{\Gamma(a)\,\Gamma(b)\,\Gamma(4-a-b)}\label{sc} \,.
\eeqa
The corresponding tensor integrals can be reduced to scalar integrals by extracting combinations of 
momenta $p_\mu$ and the symmetric tensor $\delta_{\mu \nu}$ according to the symmetry 
properties of the integrand:
\beqa
J_\mu(a,b,p)&:=&\int d^4q \frac{q_\mu}{y^a z^b}\hspace*{0.7cm} =J_1(a,b,p) \,\, p_\mu  \,,\\
K_{\mu \nu}(a,b,p)&:=&\int d^4q \frac{q_\mu q_\nu}{y^a z^b}\hspace*{0.6cm}\,=K_1(a,b,p)\,\,p_\mu p_\nu   
+ K_2(a,b,p) \,\,x \, \delta_{\mu \nu} \,, \\
L_{\mu\nu\rho}(a,b,p)&:=&\int d^4q \frac{q_\mu q_\nu q_\rho}{y^a z^b}\hspace*{0.3cm}\,
=L_1(a,b,p)\,\,p_\mu p_\nu p_\rho  \nonumber\\
&&\hspace*{2.7cm}\, +\,L_2(a,b,p)\,\, x \,\left(p_\mu \,\delta_{\nu \rho}+p_\nu \,\delta_{\rho \mu} 
+ p_\rho \,\delta_{\mu \nu} \right) \,,\\
M_{\mu\nu\rho\sigma}(a,b,p)&:=&\int d^4q \frac{q_\mu q_\nu q_\rho q_\sigma}{y^a z^b}=
M_1(a,b,p)\,\,p_\mu p_\nu p_\rho p_\sigma   \nonumber\\
&&\hspace*{2.8cm}+ \,M_2(a,b,p)\,\,x\,\left(\delta_{\mu \nu} \,p_\rho p_\sigma + \delta_{\mu \rho} \,p_\nu p_\sigma +
\delta_{\mu \sigma} \,p_\rho p_\mu  +\right. \nonumber\\
&&\hspace*{5.6cm}\left. \,\delta_{\nu \rho} \,p_\mu p_\sigma 
+\delta_{\nu \sigma}\, p_\rho p_\mu + \delta_{\rho \sigma} \,p_\mu p_\nu \right) \nonumber\\
&&\hspace*{2.8cm}+\,M_3(a,b,p)\,\,
x^2\, \left(\delta_{\mu \nu}\,\delta_{\rho \sigma}+\delta_{\mu \rho}\,\delta_{\nu \sigma}+
\delta_{\mu \sigma}\,\delta_{\rho \nu} \right)\,. \hspace*{1cm}
\eeqa
The scalar integrals in these expressions are calculated by contracting them with appropriate tensors,
writing all scalar products in terms of squared momenta $x,y$ and $z$
and applying eq.~(\ref{sc}). One arrives at 
\beqa
J_1&=& \pi^2 \frac{\Gamma(3-a)\,\Gamma(2-b)\,\Gamma(a+b-2)}
{\Gamma(a)\,\Gamma(b)\,\Gamma(5-a-b)}\, x^{2-a-b}       \,,  \\
K_1&=& \pi^2 \frac{\Gamma(4-a)\,\Gamma(2-b)\,\Gamma(a+b-2)}
{\Gamma(a)\,\Gamma(b)\,\Gamma(6-a-b)}\, x^{2-a-b}  \,, \\
K_2&=& \pi^2 \frac{\Gamma(3-a)\,\Gamma(3-b)\,\Gamma(a+b-2)}
{\Gamma(a)\,\Gamma(b)\,\Gamma(6-a-b)} \,\frac{1}{2(-3+a+b)}\, x^{2-a-b} \,,\\
L_1&=& \pi^2 \frac{\Gamma(5-a)\,\Gamma(2-b)\,\Gamma(a+b-2)}
{\Gamma(a)\,\Gamma(b)\,\Gamma(7-a-b)} \,x^{2-a-b}\,,\\
L_2&=& \pi^2 \frac{\Gamma(4-a)\,\Gamma(3-b)\,\Gamma(a+b-2)}
{\Gamma(a)\,\Gamma(b)\,\Gamma(7-a-b)}\,\frac{1}{2(-3+a+b)} \,x^{2-a-b}\,,\\
M_1&=& \pi^2 \frac{\Gamma(6-a)\,\Gamma(2-b)\,\Gamma(a+b-2)}
{\Gamma(a)\,\Gamma(b)\,\Gamma(8-a-b)} \,x^{2-a-b}\,,\\
M_2&=& \pi^2 \frac{\Gamma(5-a)\,\Gamma(3-b)\,\Gamma(a+b-2)}
{\Gamma(a)\,\Gamma(b)\,\Gamma(8-a-b)}\,\frac{1}{2(-3+a+b)} \, x^{2-a-b}\,,\\
M_3&=& \pi^2 \frac{\Gamma(4-a)\,\Gamma(4-b)\,\Gamma(a+b-2)}
{\Gamma(a)\,\Gamma(b)\,\Gamma(8-a-b)}\,\frac{1}{4(-3+a+b)(-4+a+b)}\, x^{2-a-b} \,.\hspace*{1cm}
\eeqa

\section{Analytic expressions for some diagrams in bare vertex approximation \label{app-diag}}

In this appendix we give explicitly the expressions for some diagrams needed for our investigation in
section \ref{infrared-a}. All algebraic manipulations have been done using the program FORM 
\cite{Vermaseren:2000nd}. Our {\it ans\"atze} for the small momentum behaviour of the
ghost dressing function $G$, the transversal gluon dressing function $Z$ and the longitudinal 
gluon dressing function $L$ are the power laws 
\beq
G(x) = Bx^\beta, \:\:\:\:
Z(x) = Ax^\sigma, \:\:\:\:
L(x) = Cx^\delta,
\label{ansatz}
\eeq 
where we have used the abbreviation $x=p^2$.

\begin{figure}
\centerline{
\epsfig{file=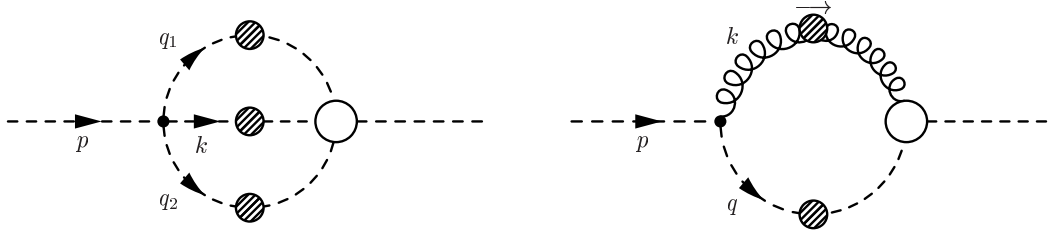,width=14cm}
}
\caption{ \sf \sf \label{ghostsunset}
Momentum routing for the sunset and for the dressing diagram in the ghost equation.}
\end{figure}

We first evaluate the sunset diagram in the ghost equation given diagrammatically in Fig.~\ref{ghostsunset}.
With the bare four-ghost vertex given in eq.~(\ref{barefourghostvertex}) and the abbreviations for the
squared momenta
$x=p^2$, $y_1=(q_1)^2$, $y_2=(q_2)^2$, $z_1=(p-q_1)^2$ and $z_2=(p-q_1-q_2)^2$ the sunset
diagram reads
\beq
U^{sun} = \frac{N_c^2 \,g^4 \,\tilde{Z}_4}{2 \,(2\pi)^8} \left(\frac{\alpha}{2}\left(1-\frac{\alpha}{2}\right)\lambda\right)^2
\int d^4q_1 \, \frac{B \,(y_1)^\beta}{x \,y_1} \,\int d^4q_2 \,\frac{B^2 \,(y_2)^\beta \,(z_2)^\beta}{y_2 \,z_2} .
\eeq
The factor $1/x$ in the first integral stems from the left hand side of the ghost equation. 
We now integrate the
inner loop with the help of formula (\ref{sc}) and obtain
\beq
U^{sun} = \frac{N_c^2\, g^4 \,\tilde{Z}_4 \,B^3}{512 \,\pi^6} \left(\frac{\alpha}{2}\left(1-\frac{\alpha}{2}\right)\lambda\right)^2
\frac{\Gamma^2(1+\beta)\,\Gamma(-2\beta)}{\Gamma^2(1-\beta)\,\Gamma(2+2\beta)}\,
\int d^4q_1  \,\frac{(y_1)^\beta}{x\, y_1}\, (z_1)^{2\beta} \,, 
\eeq
where $z_1$ is the total squared momentum flowing through the integrated loop. The second integration is done
in the same way. We arrive at
\beqa
U^{sun} &=& x^{3\beta} \,\frac{N_c^2 \,g^4 \,\tilde{Z}_4\, B^3}{512 \,\pi^4} \left(\frac{\alpha}{2}\left(1-\frac{\alpha}{2}\right)\lambda\right)^2
\frac{\Gamma^3(1+\beta)\,\Gamma(-3\beta-1)}{\Gamma^3(1-\beta)\,\Gamma(3+3\beta)}\nonumber\\
&:=& x^{3\beta} (U^\prime)^{sun} \,.
\eeqa
As each integration step eats up the two squared momenta in the denominators of the integral kernels only 
powers of $x$ to the anomalous dimensions of the dressing functions in the loop (here $3\beta$ from
three ghost propagators) survive. This mechanism works in the same way for all diagrams and explains 
the pattern in the eqs.~(\ref{gheq1}), (\ref{gl_trans1}) and (\ref{gl_long1}) of section \ref{infrared-a}.

Next we evaluate the two contributions in the gluon equation needed for the argument below eq.~(\ref{gl_long2}). 
The explicit expressions for the kernels of two-loop gluon diagrams are rather lengthy but the calculation
is done along the same lines as in the ghost sunset diagram above. Therefore we just give
the final results:
\beqa
V^{squint}_{TTTT} &=& x^{4\sigma} \, \frac{-27 \, g^4 \, N_c^2 \, Z_4 \, A^4}{4096 \, \pi^4}\,  \frac{\Gamma(-1-4\sigma)\:
\Gamma(1/2-\sigma)\:
\Gamma(3\sigma)\: \Gamma^2(1+\sigma)}{\Gamma(4-3\sigma)\:\Gamma^2(2-\sigma)\:\Gamma(3/2-\sigma)\:\Gamma(4+4\sigma)}
\times \nonumber\\
&& 2^{-4 \sigma}\:(-1+3\sigma)\:(10+\sigma-66\sigma^2+63\sigma^3)\:(5+43\sigma+47\sigma^2) \nonumber\\
&:=& x^{4\sigma}\, (V^\prime)^{squint}_{TTTT} \,, \\
W^{sun}_{LLL} &=&x^{3\delta}\, \frac{g^4 \, N_c^2 \, Z_4\,  C^3}{1536 \, \pi^4} \, \frac{1}{(1+3\delta)} 
\frac{\Gamma^3(1+\delta)\:\Gamma(1-3\delta)}{\Gamma^3(2-\delta)\: 
\Gamma(3+3\delta)} \:\: \lambda^3   \nonumber\\
&:=& x^{3\delta} \, (W^\prime)^{sun}_{LLL} \,.
\eeqa 

Finally we calculate that part in the dressing diagram of the ghost equation which contains the
longitudinal part of the gluon propagator
for the special case $\alpha=0,2$, where $L(x)=1$. These are the linear covariant gauges. With the
momentum assignments $x=p^2$, $y=q^2$ and $z=k^2=(p-q)^2$ the longitudinal part of the diagram is given by
\beqa
U^{dress}_L &=& -\frac{N_c \, g^2 \,\tilde{Z}_1 }{(2\pi)^4} \int d^4q\, q_\mu \,\lambda \frac{k_\mu k_\nu}{z^2}p_\nu 
\frac{B y^\beta}{x\,y}
\eeqa
where again the factor $1/x$ stems from the left hand side of the equation. Writing the kernel in terms of
squared momenta we obtain
\beqa
U^{dress}_L &=& -\frac{N_c g^2 \,\tilde{Z}_1}{(2\pi)^4} \int d^4q  \: \lambda \: B \: y^\beta \:
\left(\frac{1}{2z^2} - \frac{x}{4y z^2} - \frac{y}{4xz^2} + \frac{1}{4xy} \right). 
\eeqa
After integration we obtain
\beqa
U^{dress}_L &=& -\frac{N_c \,g^2 \,\tilde{Z}_1}{16 \,\pi^2} \: \lambda \: B \, x^\beta \:
\left(\frac{1}{2} - \frac{1}{4} - \frac{1}{4} + 0 \right) \Gamma(0) \\
&=&0. 
\eeqa
Although $\Gamma(0)$ is formally divergent this contribution vanishes due to vanishing coefficients. 

\section{Numerical methods for flat Euclidean space-time \label{numerics}}

Here we detail the numerical method we employed to solve the coupled
system of ghost and gluon Dyson-Schwinger equations, (\ref{ghostbare}) and (\ref{gluonbare}).
For the convenience of the reader we display the equations again:
\begin{eqnarray} 
\frac{1}{G(x)} &=& \tilde{Z}_3 - g^2N_c \int \frac{d^4q}{(2 \pi)^4}
\frac{K(x,y,z)}{xy}
G(y) Z(z) \; , \label{ghostbare1} \\ 
\frac{1}{Z(x)} &=& {Z}_3 + g^2\frac{N_c}{3} 
\int \frac{d^4q}{(2 \pi)^4} \frac{M(x,y,z)}{xy} G(y) G(z) \nonumber\\
&&+ 
 g^2 \frac{N_c}{3} \int \frac{d^4q}{(2 \pi)^4} 
\frac{Q(x,y,z)}{xy} \frac{G(y)^{(1-a/\delta-2a)}}{Z(y)^{a}}
\frac{G(z)^{(1-b/\delta-2b)}}{Z(z)^{b}} 
 \; .
\label{gluonbare1} 
\end{eqnarray} 
Recall $x=p^2$, $y=q^2$ and $z=(q-p)^2=x+y-2\sqrt(xy) \cos\Theta$. The integral over
the loop momentum $q$ is transformed to four-dimensional hyperspherical coordinates.
Two of the four integrals are then trivial
and yield a factor of $4\pi$. The remaining angular
integral and the radial integral have to be performed with the help of
numerical routines. We use the Gauss-Legendre quadrature rule
described in \cite{press}. 
To achieve high accuracy we split the radial loop integral
into three parts, $y \in [0,\epsilon^2]$, $y \in (\epsilon^2,x]$ and  $y \in
(x,x_{UV}]$. The second split 
is necessary as the integrands are not smooth at the boundary $x$ and too much
accuracy would be lost, if one uses a quadrature rule that spans the whole
region $(\epsilon^2,x_{UV}]$. 

According to
the value of their argument the dressing functions $Z$ as well as $G$ have to be
handled differently. In the infrared region, $y,z \in [0,\epsilon^2]$, $Z$ and
$G$ behave like powers and are replaced according to eq.~(\ref{zg-power-kappa}).
Recall that the approximation by leading powers in the infrared is justified by
the analysis of subleading contributions at the end of subsection \ref{infrared}.
The infrared matching point $\epsilon^2$ is chosen sufficiently low. In physical
units we have $\epsilon^2=(0.55 \,
\mbox{MeV})^2$ in our calculations. In the high momentum regime, $y \in
(x,x_{UV}]$, arguments $z$ occur which are larger than the numerical cutoff
$x_{UV}$. There we approximate the respective dressing functions by the
expressions
\begin{eqnarray}
Z(z) &=& Z(l) \left[ \omega \log\left(\frac{z}{l}\right)+1 \right]^\gamma  \; ,
\label{gluon_uv1}\\
G(z) &=& G(l) \left[ \omega \log\left(\frac{z}{l}\right)+1 \right]^\delta  \; .
\label{ghost_uv1}
\end{eqnarray}
according to the one loop behaviour of the solutions as has been detailed in subsection
\ref{ultraviolet}. Here
$\omega = 11N_c\alpha(l)/12\pi$ and the squared momentum $l$ is a perturbative scale. We chose 
$l=(174 \,\mbox{GeV})^2$ to be slightly lower than the numerical cutoff $x_{UV}=(177 \,\mbox{GeV})^2$.   
To be able to perform the angular integrations for momenta $[\epsilon^2,x_{UV}]$
we expand the dressing functions in Chebychev 
polynomials and solve the coupled system of equations for the expansion
coefficients using a Newton iteration
method. Details of this technique can be found in appendix B of ref.\ \cite{Atkinson:1998tu}.

\section{One-loop scaling \label{one-loop}}

In the framework of the truncation scheme presented in section \ref{truncation}
we have shown that the substitution
\begin{equation} 
Z_1 \rightarrow {\cZ}_1(x,y,z;s,L) = 
\frac{G(y)^{(1-a/\delta-2a)}}{Z(y)^{(1+a)}}
\frac{G(z)^{(1-b/\delta-2b)}}{Z(z)^{(1+b)}}
\end{equation}
for the gluon vertex renormalisation constant $Z_1$ together with a bare three-gluon vertex
yields the correct one-loop
scaling of the gluon loop in the gluon Dyson-Schwinger equation. This is
true for any values $a$ and $b$. Of course, in a full treatment of the
coupled ghost-gluon system ${\cZ}_1(s,L)$ would be independent
of momentum. Therefore a choice of $a$ and $b$ which keeps ${\cZ}_1$ as
weakly varying as possible seems the most reasonable one.
\begin{figure}
\begin{center}
\epsfig{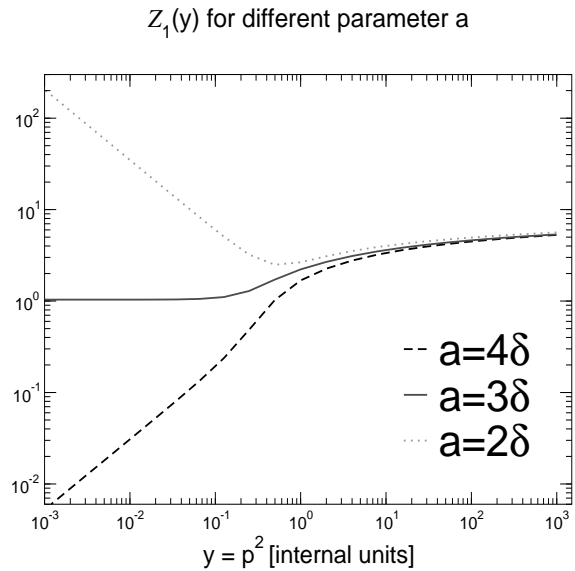}
\end{center}
\caption{ \sf \label{z1.dat}
The $y$-dependence of the function ${\cZ}_1(y,z)$ for different
values of the parameter $a$. Only the choice $a=3\delta$ leads to 
momentum independence in the infrared. Due to the symmetry of the {\it ansatz} for
${\cal Z}_1(y,z)$ the $z$-dependence is the same for $b=a$.
}
\end{figure}

This choice can be inferred using the scaling of the dressing functions
extracted from the renormalisation group equation, see ref.~\cite{vonSmekal:1997is}
for details. The dressing functions can be expressed as
\begin{eqnarray}
Z(x) &=& \left(\frac{\alpha(x)}{\alpha(s)}\right)^{1+2\delta}R^2(x) \; ,
 \nonumber\\
G(x) &=& \left(\frac{\alpha(x)}{\alpha(s)}\right)^{-\delta}R^{-1}(x) \; ,
\end{eqnarray}
where the running coupling provides the correct one loop scaling in the ultraviolet.
Consequently the function $R(x)$ approaches unity for high momenta.
Furthermore, from the known infrared behaviour of $Z(x)$, $G(x)$ and
$\alpha(x)$ (c.f. subsection \ref{infrared}) one infers that $R(x)$ 
is proportional to $x^\kappa$ in the infrared. Writing ${\cZ}_1$ in terms of $\alpha(x)$
and $R(x)$ yields
\begin{equation} 
{\cZ}_1(x,y,z;s,L) = 
\left(\frac{\alpha(\mu^2)}{\alpha(y)}\right)^{1+3\delta} R^{-3+a/\delta}(y)
\left(\frac{\alpha(\mu^2)}{\alpha(z)}\right)^{1+3\delta} R^{-3+b/\delta}(z).
\end{equation}
In the perturbative region $R(y),R(z) \rightarrow 1$ and the function ${\cZ}_1$ is
therefore slowly varying for any $a$ and $b$ 
according to the logarithmic behaviour of the running coupling 
$\alpha$. In the infrared, however, $\alpha$ approaches its fixed point while
the functions $R$ behaves like a power. Consequently the choice $a=b=3\delta$ 
guarantees the weakest momentum dependence of ${\cZ}_1$ which is illustrated
in Fig.~\ref{z1.dat}.
Shown is the $y$-dependence of the function
${\cZ}_1(x,y,z=y;s,L)$. (Note also that due to the symmetry 
${\cZ}_1(x,y,z;s,L)={\cZ}_1(x,z,y;s,L)$ and the absence of an explicit
$x$-dependence this is sufficient to demonstrate
its momentum dependence.)
In the perturbative momentum regime the function ${\cZ}_1$ does not vary
with the parameter $a$. So all three choices give the same logarithmic running
in momentum as required to give the correct one loop scaling behaviour of the
integral. In the infrared, however, a change in  $a$ gives rise to substantial
changes in the behaviour of ${\cZ}_1$, with only the choice $a=3\delta$
leading to a constant. 

\chapter{Discretisation and finite volume effects}

\section{Radial discretised DSEs \label{radial}}

In this appendix we study possible discretisation errors in the Dyson-Schwinger 
equations\footnote{I owe Jacques Bloch the idea to the study performed in this section.}.
We investigate a discretised version of the angular approximated DSEs in the ghost-loop only 
truncation scheme discussed in subsection \ref{ghost-only-sec}. 
From the infrared analysis of these DSEs one finds power laws for the ghost and gluon dressing functions. 
These power laws are used in the numerical treatment of eqs.~(\ref{ghAB}), (\ref{glAB}) to solve the 
integrals from zero momentum to an infrared matching point $\epsilon^2$. It has been
claimed that these integrals are
crucial to find numerical solutions for the ghost and gluon propagators \cite{Atkinson:1998tu}.
In this section we will show that they are not. We will introduce a prescription to discretise
the radial integrals in the DSEs such that the smallest momentum showing up in the equations is nonzero. 
Despite the infrared divergence of the ghost dressing function we reproduce the
solutions from the formulation with continuous momenta without having to include the region in the very infrared.
This is very important as otherwise the attempt to solve Dyson-Schwinger
equations on a four-torus, performed in chapter \ref{torus}, might be hopeless from the very beginning. 

For the discretisation of eqs.~(\ref{ghAB}), (\ref{glAB}) we use the same prescription for the radial
integral as one gets for all four Cartesian momentum directions if employing a finite volume. Note that
contrary to the finite volume case there is no easy geometrical interpretation for a discretised radial
integral. 

With $l$ denoting the inverse spacing of the radial momentum $q$, we thus substitute
\beq
\int dq^2 = \int dq \: 2q \: \longrightarrow \: \left(\frac{2\pi}{l}\right)
 \sum_j 2 \left( \frac{2\pi}{l}j \right) 
=  \left(\frac{2\pi}{l}\right)^2 \sum_j 2j \,,
\eeq
into eqs.~(\ref{ghAB}), (\ref{glAB}). We obtain
\beqa
\frac{1}{G(x_i)} &=& \tilde{Z}_3 - \frac{9}{4} \frac{g^2 N_c}{48 \pi^2} \left(\frac{2\pi}{l}\right)^2 \left[ 
Z(x_i) \sum_{j=1}^{i-1} \frac{2j}{x_i} \frac{y_j}{x_i} G(y_j) 
+ \sum_{j=i}^{N} \frac{2j}{y_j} Z(y_j) G(y_j) 
\right]  \,, \label{ghost-rad} \\
\frac{1}{Z(x_i)} &=& Z_3 +  \frac{g^2 N_c}{48 \pi^2} 
\left( \frac{2\pi}{l}\right)^2 \left[ 
 G(x_i) \sum_{j=1}^{i-1} \frac{2j}{x_i} \left( -\frac{y_j^2}{x_i^2} + \frac{3y_j}{2x_i} \right) G(y_j) 
+ \sum_{j=i}^{N} \frac{2j}{2y_j} G^2(y_j) 
\right] \,. \nonumber\\ \label{gluon-rad}
\eeqa

The squared momenta $x_i$ are given as $x_i=\left(i\,2\pi/l\right)^2$. The largest momentum $x_N$
corresponds to the numerical cutoff $L=\Lambda^2$ in the continuous version, eqs.~(\ref{ghAB}), (\ref{glAB}).
The momentum $x_0 = 0$ is discarded in the calculation.
We solve the discretised eqs.~(\ref{ghost-rad}), (\ref{gluon-rad}) by iteration employing the Newton method. 
To directly compare the continuous solution with the discretised one we
use the same cut-off and renormalise at the same momentum. This is most easily done by taking
$Z_3(\mu^2,\Lambda^2)$ and $\tilde{Z}_3(\mu^2,\Lambda^2)$ in eqs.~(\ref{ghost-rad}), (\ref{gluon-rad}) 
from the continuum solution.

It is quite amusing that there is a second way to solve eqs.~(\ref{ghost-rad}), (\ref{gluon-rad}),
which corresponds to the conversion of the continuous integral equations (\ref{ghAB}) and (\ref{glAB})
into differential equations\footnote{Note that such a conversion is only possible for 
angular approximated DSEs.}, see ref.~\cite{Atkinson:1998tu}:
Subtract eqs.~(\ref{ghost-rad}), (\ref{gluon-rad}) from themselves for two different arbitrary momenta 
$x_i$ and $x_{s}$. One obtains  
\beqa
\frac{1}{G(x_i)} &=& \frac{1}{G(x_s)} - \frac{9}{4} \frac{g^2 N_c}{48 \pi^2} 
\left(\frac{2\pi}{l}\right)^2 \left[ 
Z(x_i) \sum_{j=1}^{i-1} \frac{2j}{x_i} \frac{y_j}{x_i} G(y_j)
-Z(x_s) \sum_{j=1}^{s-1} \frac{2j}{x_s} \frac{y_j}{x_s} G(y_j) \right.\nonumber\\ 
&&\hspace*{6cm} + \left. \sum_{j=min(i,s)}^{min(i-1,s-1)} \frac{2j}{y_j} Z(y_j) G(y_j) 
\right]  \,, \\
\frac{1}{Z(x_i)} &=& \frac{1}{Z(x_s)} +  \frac{g^2 N_c}{48 \pi^2} 
\left( \frac{2\pi}{l}\right)^2 \left[ 
 G(x_i) \sum_{j=1}^{i-1} \frac{2j}{x_i} \left( -\frac{y_j^2}{x_i^2} + \frac{3y_j}{2x_i} \right) G(y_j) 
\right. \nonumber\\ 
&&\hspace*{1cm} \left. - G(x_s) \sum_{j=1}^{s-1} \frac{2j}{x_s} \left( -\frac{y_j^2}{x_s^2} + \frac{3y_j}{2x_s} \right) G(y_j)
+ \sum_{j=min(i,s)}^{min(i-1,s-1)} \frac{2j}{2y_j} G^2(y_j) 
\right] \,. 
\eeqa

\begin{figure}[t]
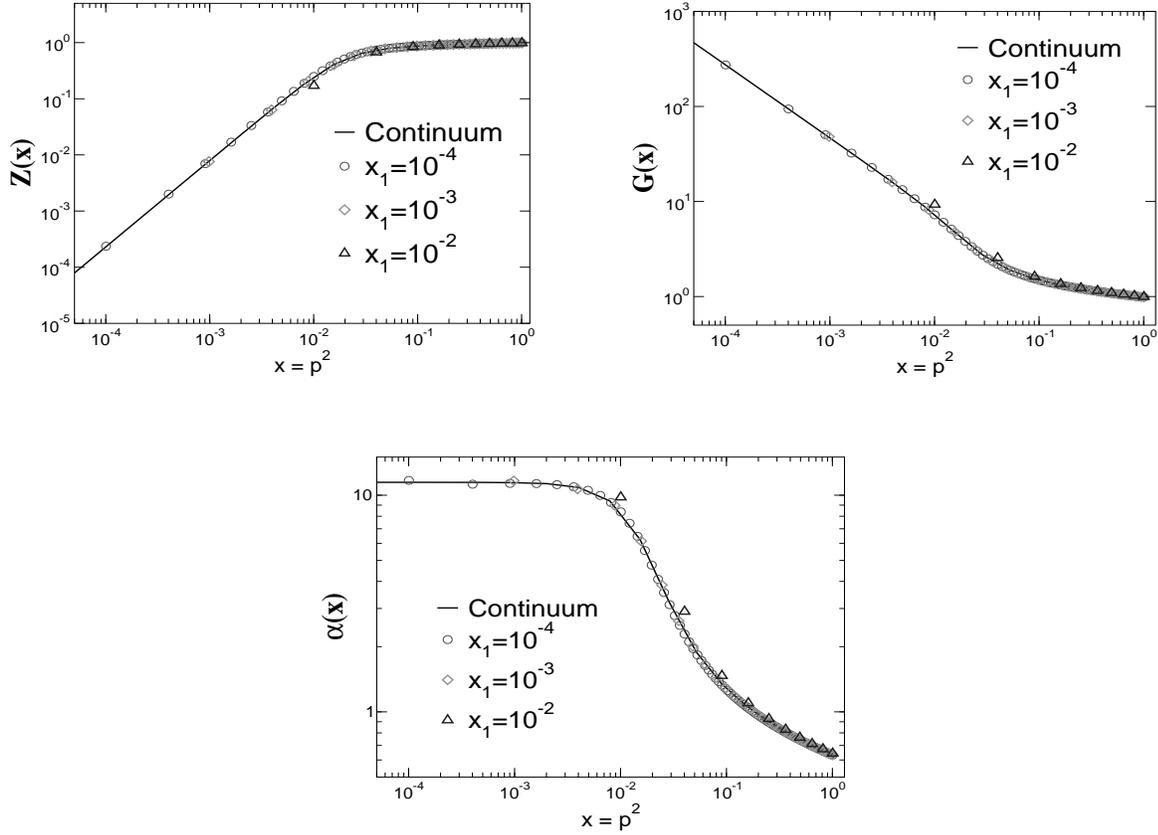

\vspace{0.5cm}
\centerline{
\epsfig{file=z.rad.pap.eps,width=7.0cm,height=5.0cm}
\hspace{1cm}
\epsfig{file=g.rad.pap.eps,width=7.0cm,height=5.0cm}
}
\vspace{1cm}
\centerline{\epsfig{file=a.rad.pap.eps,width=7cm,height=5.0cm}}
\caption{ \sf \sf \label{fg.rad.eps}
Comparison of the gluon dressing function $Z(x)$, the ghost dressing function $G(x)$ and the running
coupling $\alpha(x)$ obtained from the DSEs with continuous momenta and from the discretised version.
For the radial grid of the discretised version we used three different momentum spacings.}
\end{figure}
If we now set $s=1$ the sums in the equations run
from $j=1..i$. Given the values for $Z(x_1)$ and $G(x_1)$ 
the equations for $Z(x_2)$ and $G(x_2)$ are then simple quadratic equations which can
be solved easily. Given these solutions we can solve the equations for $i=3$
and so on. The complete solution is thereby built point for point from the infrared to
the ultraviolet region of momentum. The advantage of this method is that one does not have to use
a numerical iteration procedure to solve the equations. Of course, the specification of 
the input values $Z(x_1)$ and $G(x_1)$ corresponds to a certain choice of the
renormalisation constants $Z_3(\mu^2,\Lambda^2)$ and
$\tilde{Z}_3(\mu^2,\Lambda^2)$, {\it i.e.} to the boundary restriction from a certain choice of renormalisation
point. The disadvantage of this method is thus that
one has to improve iteratively on the input values $Z(x_1)$ and $G(x_1)$ until 
the same $Z_3(\mu^2,\Lambda^2)$ and
$\tilde{Z}_3(\mu^2,\Lambda^2)$ are obtained as for the continuous integral equations. 
Both ways of solving eqs.~(\ref{ghost-rad}), (\ref{gluon-rad}) certainly yield the same results. 

Our solutions for the dressing functions on the radial momentum grid are compared with
the solutions from the continuum in Fig.~\ref{fg.rad.eps}. For the
dressing functions and the running coupling we find nearly the same results with both methods. 
There are only small discretisation errors for our largest momentum spacing. For smaller
spacings these errors become more and more irrelevant and we thus conclude that there is a 
smooth transition to the continuum as the spacing more and more decreases.
 
\section{The influence of zero modes on the solutions of DSEs on a torus \label{zero-modes}}

An important point when formulating the Dyson--Schwinger equations  on the
torus could be the treatment of the zero modes.  In addition, on the torus an
infrared analysis like the one in flat Euclidean space-time is not possible,
and one is left with the problem how the dressing functions behave at 
vanishing momenta. Guided by the intuition that especially the
long ranged modes should be affected by the  finite volume we assume in the
following $Z(x \rightarrow 0) = 0$ just like in the continuum and $G(x
\rightarrow 0) =const$ if zero modes are neglected. Phrased otherwise we
assume that the zero modes are the missing ingredient to ensure the correct
infinite volume limit for the torus results. Therefore, if on tori of different
volumes $G(x= 0)$ shows no sign of becoming divergent, the infrared enhancement
seen in $G(x \rightarrow 0)$ or in the flat space-time results has to be due to
the torus zero modes of gluons and ghosts.

Therefore, in this appendix, we will show that the  assumption $G(x= 0) <
\infty$ does not lead to a contradiction in the equations on the torus if zero
modes are neglected. To this end we focus on the truncation scheme without
angular approximations.  First we rewrite eqs.~(\ref{ghostbare}), (\ref{gluonbare}) as
\begin{eqnarray} 
\frac{1}{G(x)} &=& Z_3 - g^2N_c \int \frac{d^4q}{(2 \pi)^4}
\frac{K(x,y,z)}{xy}
G(y) Z(z) \; , \label{ghostbare_app} \\ 
\frac{1}{Z(x)} &=& \tilde{Z}_3 + g^2\frac{N_c}{3} 
\int \frac{d^4q}{(2 \pi)^4} \frac{M(x,y,z)}{xy} G(y) G(z) \nonumber\\
&&\hspace*{0.6cm}+\, g^2 \frac{N_c}{3} \int \frac{d^4q}{(2 \pi)^4} 
\frac{Q(x,y,z)}{xy} \frac{G(y)^{-2-6\delta}}{Z(y)^{3\delta}} 
\frac{G(z)^{-2-6\delta}}{Z(z)^{3\delta}} \; .
\label{gluonbare_app} 
\end{eqnarray} 
According to Appendix \ref{one-loop} 
we have chosen $a=b=3\delta$, where $\delta=-9/44$, the
anomalous dimension of the ghost. The kernels have the form:
\begin{eqnarray}
K(x,y,z) &=& \frac{1}{z^2}\left(-\frac{(x-y)^2}{4}\right) + 
\frac{1}{z}\left(\frac{x+y}{2}\right)-\frac{1}{4} = xy\frac{\sin^2\Theta}{z^2}
\; , \\
M(x,y,z) &=& \frac{1}{z} \left( \frac{\zeta-2}{4}x + 
\frac{y}{2} - \frac{\zeta}{4}\frac{y^2}{x}\right)
+\frac{1}{2} + \frac{\zeta}{2}\frac{y}{x} - \frac{\zeta}{4}\frac{z}{x}\; ,  \\
Q^\prime(x,y,z) &=& \frac{1}{z^2} 
\left( \frac{1}{8}\frac{x^3}{y} + x^2 -\frac{19-\zeta}{8}xy + 
\frac{5-\zeta}{4}y^2
+\frac{\zeta}{8}\frac{y^3}{x} \right)\nonumber\\
&& +\frac{1}{z} \left( \frac{x^2}{y} - \frac{15+\zeta}{4}x-
\frac{17-\zeta}{4}y+\zeta\frac{y^2}{x}\right)\nonumber\\
&& - \left( \frac{19-\zeta}{8}\frac{x}{y}-\frac{3-4\zeta}{2}+
\frac{9\zeta}{4}\frac{y}{x} \right) \nonumber\\
&& + z\left(\frac{\zeta}{x}+\frac{5-\zeta}{4y}\right) + z^2\frac{\zeta}{8xy}
\; .
\label{new_kernels_app}
\end{eqnarray}

We first analyse the behaviour
of the integrands in the limit $y \rightarrow 0$ for finite momenta $x$. 
Then $Z(z) \rightarrow Z(x)$ and $G(z) \rightarrow G(x)$ and 
the kernels times the respective dressing functions are 
to appropriate order in momentum $y$:
\begin{eqnarray}
\frac{G(y) Z(z)}{xy}K(x,y,z) &\rightarrow& G(0)Z(x)\frac{\sin^2\Theta}{x^2}\; , 
\label{kernel_y1}\\
\frac{G(y) G(z)}{xy}M(x,y,z) &\rightarrow& G(0)G(x) \frac{1}{x^2} 
\left(1+(\zeta-2)\cos^2\Theta \right) \; , 
\label{kernel_y2}\\
\frac{G(y)^{-2-6\delta}G(y)^{-2-6\delta}}{Z(y)^{3\delta}Z(z)^{3\delta}xy} 
Q^\prime(x,y,z) &\rightarrow& \frac{G(0)^{-2-6\delta}G(x)^{-2-6\delta}}
{Z(0)^{3\delta}Z(x)^{3\delta}xy} 
\left(\frac{\zeta\cos^2\Theta}{xy}+...\right) \; .
\label{kernel_y3}  
\end{eqnarray}
Furthermore, $z=x+y-2\sqrt{xy}\cos\Theta$ 
has been used and terms proportional to $\cos\Theta$ have been dropped, 
as they either integrate to zero in the continuum or cancel each other in the 
sums on the torus. Each of the expressions 
(\ref{kernel_y1}), (\ref{kernel_y2}), (\ref{kernel_y3}) is then the appropriate
term for $j=0$ on the right hand side of the Dyson-Schwinger equations on the 
torus. Clearly one observes that only a finite ghost mode $G(0)$ avoids 
trouble with divergences. This is especially true for the kernel $Q^\prime$ of the 
gluon loop, as $Z^{-3\delta}(y\rightarrow 0)$ is more singular than
the simple pole, so this kernel vanishes for small momenta $y$. The other two 
expressions (\ref{kernel_y1}) and (\ref{kernel_y2}) are finite. One is then 
left with the ambiguous quantities $\sin^2\Theta$ and $\cos^2\Theta$ which will 
be replaced by their integrals from zero to $2\pi$ in the calculation at the end 
of this section. The arbitrariness of this procedure is considerably moderated by the 
observation that any number plugged in for the trigonometric functions yields the 
same qualitative result at the end of this section. 
\begin{figure}[t]
\vspace*{0.3cm}
\centerline{
\epsfig{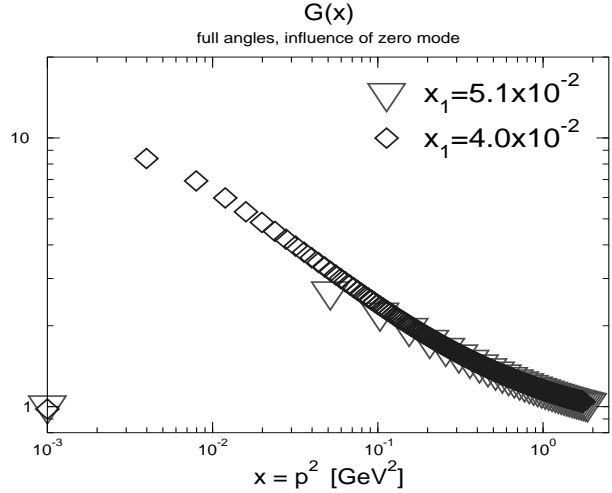}
}
\caption{ \sf \label{null.dat}
$G(0)$ compared with the results for finite momentum $x$ on the torus.
For convenience we have kept the logarithmic momentum scale and plotted the zero modes
on the left border of the figure.}
\end{figure}

Second, we take the limit $z\rightarrow 0$, which on the
torus is identical to $\Theta \rightarrow 0$. The ghost kernel
$\sin^2\Theta/z^2$ alone would certainly diverge as $\Theta \rightarrow 0$, 
but taking into account the power law behaviour $Z(z) \sim z^{2\kappa}$
for the gluon dressing function the integrand is zero in this limit. This is valid for
$\kappa > 0.5$, which is in agreement with the infrared analysis in the continuum.
We therefore may omit the points $z=0$ in the ghost equation.  
The situation is different in the gluon equation where
the kernel of the ghost loop has a finite limit $z\rightarrow 0$: $M(x,x,0)/xy=(\zeta+1)/(2x^2)$. 
Therefore with a finite ghost dressing function $G(0)$ the points $z=0$ in the ghost loop
contribute but no divergences occur.
In the gluon loop the kernel $Q^\prime$ multiplied by the dressing functions
approaches zero for vanishing momentum $z$ due to the power law behaviour of 
$Z^{-3\delta}(z\rightarrow 0)$. 

To obtain a definite value for $G(0)$ we now investigate the behaviour of
the eqs.~(\ref{ghostbare_app}), (\ref{gluonbare_app})
in the limit $x\rightarrow0$. The integrands are then given by 
\begin{eqnarray}
\frac{G(y) Z(z)}{xy}K(x,y,z) &\rightarrow& G(y)Z(y)\frac{\sin^2\Theta}{y^2}\; ,\\
\frac{G(y) G(z)}{xy}M(x,y,z) &\rightarrow& G(y)G(y)\left(\frac{1-\zeta\cos^2\Theta}{xy}
+... \right)\; ,
\label{kernel_x}
\end{eqnarray}
where the kernel $Q^\prime$ is of no interest, as we know the gluon loop to be subleading
in the infrared. Clearly, the kernel $M$ of the ghost loop in the gluon 
equation is now singular for $x\rightarrow 0$, 
corresponding to a vanishing gluon dressing function in 
the infrared. This result confirms our working hypothesis that the gluon mode 
$Z(0)$ is not affected by the finite volume of the torus. 
The integrand of the ghost equation is finite up to the point $y=0$. There
the pole in the kernel is cancelled by the behaviour of the
gluon dressing function $Z(y) \sim y^{2\kappa}$ resulting in a zero for
vanishing momentum $x$ and $y$.

We therefore arrive at a consistent
set of equations for the ghost propagator at $p^2=0$ and a vanishing gluon $Z(0)$.
In Fig.~\ref{null.dat} we show the results for the ghost dressing function
gained on two different volumes on the torus. Within numerical accuracy the values 
of $G(0)$ are the same
for the two volumes. Obviously terms with high loop momentum $y$ 
contribute most to the right hand side of the ghost equation for vanishing momentum $x$.
Furthermore, one observes that the actual value of $G(0)$ is not in accordance with
an extrapolation of the ghost curves to the infrared. There is also no sizeable 
change of the gluon and the ghost 
dressing function when $G(0)$ is set to zero by hand. This has been done in all 
calculations performed in chapter \ref{torus}.

Having shown that $G(0)< \infty$ on the torus even in the infinite-volume limit
and assuming that the torus should provide a reasonable infrared
regularisation  of physics in flat space-time we conclude that the divergence
of $G(0)$  is very probably due to the torus zero modes of gluons and ghosts.
Noting furthermore that a diverging  $G(0)$ is related to Zwanziger's
horizon condition and the Kugo--Ojima confinement criterion this indicates a
direct relation between zero modes, the Gribov horizon and confinement.

\bibliographystyle{utcaps.bst}
\bibliography{dissref}
\end{document}